\documentclass[12pt,a4paper]{article}
\usepackage[usenames,dvipsnames]{xcolor}
\usepackage{jheppub}  
\usepackage{amssymb} 
\usepackage{amsmath}
\usepackage{amsthm}
\usepackage{mathtools}
\usepackage{amsfonts}    
\usepackage{dsfont}
\usepackage{pdfpages}
\usepackage{verbatim}
\usepackage{tensor}
\usepackage{mathrsfs}
\usepackage{siunitx}

\usepackage{hyperref}
\usepackage{tikz}
\usetikzlibrary{patterns,decorations.pathreplacing, decorations.markings,calc,shapes.misc,decorations.pathmorphing}

\hypersetup{
    pdfencoding=unicode,
	colorlinks=true,
	urlcolor=Maroon,
	linkcolor=RoyalBlue,
	citecolor=Maroon,
	pdftitle={Notes on crossing transformations of Virasoro conformal blocks},
	pdfauthor={Lorenz Eberhardt},
	pdfdisplaydoctitle=true,
	pdfstartview=FitH,
	linktocpage=true
}

\newcommand{\ba}{\begin{align}}

\newcommand{\be}{\begin{equation}}
\newcommand{\ee}{\end{equation}}
\def\bd{\begin{tikzpicture}}
\def\ed{\end{tikzpicture}}

\newtheorem{conj}{Conjecture}

\newcommand{\ket}[1]{\left| #1\right\rangle}

\DeclareMathOperator\tr{tr}

\renewcommand\d{\text{d}}

\newcommand\CC{\mathbb{C}}
\newcommand\ZZ{\mathbb{Z}}
\newcommand\RR{\mathbb{R}}
\newcommand\QQ{\mathbb{Q}}

\newcommand\PP{\mathbb{P}}

\newcommand\KK{\mathbb{K}}
\newcommand\NN{\mathbb{N}}

\newcommand\GL{\text{GL}}

\newcommand\PSL{\text{PSL}}

\newcommand\SL{\text{SL}}

\newcommand\Diff{\text{Diff}}

\newcommand\Ad{\text{Ad}}
\newcommand\Li{\text{Li}}

\let\S\relax
\DeclareMathOperator\S{S}

\let\H\relax
\DeclareMathOperator\H{H}

\newcommand\id{\mathds{1}}

\DeclareMathOperator\Map{Map}

\DeclareMathOperator\Gal{Gal}

\let\Re\relax
\DeclareMathOperator\Re{Re}
\let\Im\relax
\DeclareMathOperator\Im{Im}
\DeclareMathOperator*{\Res}{Res}

\newcommand{\sixjnorm}[6]{\begin{vmatrix}
#1 & #2 & #3 \\
#4 & #5 & #6
\end{vmatrix}
}

\newcommand{\sixj}[6]{\begin{Bmatrix}
#1 & #2 & #3 \\
#4 & #5 & #6
\end{Bmatrix}
}

\allowdisplaybreaks[1]

\title{Notes on crossing transformations of Virasoro conformal blocks}

\author{Lorenz Eberhardt} 
\affiliation{School of Natural Sciences, Institute for Advanced Study, \\
\hspace*{0.3cm}Einstein Drive 1, Princeton,  NJ 08540, USA}
\emailAdd{elorenz@ias.edu}

\abstract{We carefully bootstrap the crossing kernels of Virasoro conformal blocks from first principles. 
Our approach emphasizes the Hilbert space structure of the space of Virasoro conformal blocks which makes the consistency of crossing transformations obvious.
We give a pedagogical explanation of the necessary background about Virasoro conformal blocks and special functions.
We also explain several applications and prove new results about the crossing kernels.
}

\begin{document}

\maketitle

% Make math in all titles bold
\makeatletter
\g@addto@macro\bfseries{\boldmath}
\makeatother
%end code

%%%%%%%%%%%%%%%%%%%%%%%%%%%%%%%%%%%%%%%%%%%%%%%%%%%%%%%%%%%%%%%

\section{Introduction}
Virasoro symmetry plays a ubiquitous role in modern theoretical physics. Every correlation function in a conformal field theory can be expanded into conformal blocks -- the basic building blocks of the theory. Knowing the structure constants of a theory together with the conformal blocks lets one write down all correlation functions of the theory.
There is no particularly explicit formula for Virasoro conformal blocks with generic parameters; all known expressions are either series expansions or recursive representations, see e.g.\ \cite{Belavin:1984vu, Zamolodchikov:1984eqp,Zamolodchikov_recursion2, Alba:2010qc, Perlmutter:2015iya}. This makes it challenging to study conformal blocks directly.

Virasoro conformal blocks have surprisingly rich connections to many different subjects. Let us name a few. There exist probabilistic constructions of the conformal blocks \cite{Guillarmou:2020wbo, Ghosal:2020sis, Ghosal} that make the Dotsenko-Fateev construction of conformal blocks in the Coulomb gas formalism precise \cite{Dotsenko:1984nm}. 
As such, conformal blocks at genus 0 are closely related to hypergeometric integrals.
Conformal blocks are also sensitive to the algebraic geometry of the moduli space of Riemann surfaces $\mathcal{M}_{g,n}$ \cite{Friedan:1986ua}.
They transform non-trivially under crossing transformations which gives rise to a projective representation of the mapping class group of the surface.
This data allows one also to define certain 3d TQFTs \cite{Teschner:2005bz} which in turn are related to three-dimensional quantum gravity \cite{Collier:2023fwi}.
Conformal blocks can be mapped to certain quantities in 4d $\mathcal{N}=2$ gauge theories via the AGT correspondence \cite{Alday:2009aq}, which also gives rise to new unexpected combinatorial expansions of conformal blocks and connections to integrable hierarchies \cite{Alba:2010qc, Gamayun:2012ma}.
The representation theory of the Virasoro algebra is related to a quantum group known as the modular double \cite{Faddeev:1999fe}. This was used by Ponsot and Teschner to infer the crossing properties of Virasoro conformal blocks \cite{Ponsot:1999uf, Ponsot:2000mt}.

2d CFTs are heavily constrained by crossing symmetry. One can use different channels for the conformal block expansion and the statement of crossing symmetry means that the final correlation function is independent of which channel one uses. There is only a handful of CFTs for which crossing symmetry is under analytic control. Insisting on Virasoro symmetry, these are only minimal models and (spacelike) Liouville theory. There are other models for which there are convincing numerical demonstrations of crossing symmetry such as timelike Liouville theory (aka imaginary Liouville theory or Liouville theory with $c \le 1$) and non-analytic Liouville theory \cite{Ribault:2015sxa}.

Thus in order to construct CFTs axiomatically, it is crucial to understand the crossing properties of conformal blocks in detail. It is perhaps surprising that even though there are no simple expressions for the conformal blocks themselves, there is a much more explicit expression for the crossing kernel which describes the behaviour of conformal blocks under crossings. In the most studied case of the conformal block on the four-punctured sphere, the crossing kernel takes the schematic form
\be 
\begin{tikzpicture}[baseline={([yshift=-.5ex]current bounding box.center)}, xscale=.8, yscale=.7]
        \draw[very thick, Maroon, bend left=30] (-1.8,1.2) to (-.6,0);
        \draw[very thick, Maroon, bend right=30] (-1.8,-1.2) to (-.6,0);
        \draw[very thick, Maroon, bend right=30] (2.2,1.2) to (.6,0);
        \draw[very thick, Maroon, bend left=30] (2.2,-1.2) to (.6,0);
        \draw[very thick, Maroon] (-.6,0) to (.6,0);
        \fill[Maroon] (-.6,0) circle (.07);
        \fill[Maroon] (.6,0) circle (.07);
        \draw[very thick] (-2,1.2) circle (.2 and .5);
        \draw[very thick] (-2,-1.2) circle (.2 and .5);
        \begin{scope}[shift={(2,1.2)}, xscale=.4]
            \draw[very thick] (0,-.5) arc (-90:90:.5);
            \draw[very thick, dashed] (0,.5) arc (90:270:.5);
        \end{scope}
        \begin{scope}[shift={(2,-1.2)}, xscale=.4]
            \draw[very thick] (0,-.5) arc (-90:90:.5);
            \draw[very thick, dashed] (0,.5) arc (90:270:.5);
        \end{scope}
        \begin{scope}[xscale=.4]
            \draw[very thick, RoyalBlue] (0,-.9) arc (-90:90:.9);
            \draw[very thick, dashed, RoyalBlue] (0,.9) arc (90:270:.9);
        \end{scope}
        \draw[very thick, out=0, in=180] (-2,1.7) to (0,.9) to (2,1.7);
        \draw[very thick, out=0, in=180] (-2,-1.7) to (0,-.9) to (2,-1.7);
        \draw[very thick, out=0, in=0, looseness=2.5] (-2,-.7) to (-2,.7);
        \draw[very thick, out=180, in=180, looseness=2.5] (2,-.7) to (2,.7);
        \node at (-2,1.2) {3};
        \node at (2.4,1.2) {2};            
        \node at (2.4,-1.2) {1};
        \node at (-2,-1.2) {4}; 
        \node at (0,.3) {$P$};
        \end{tikzpicture}        
    \hspace{-.3cm} = \int_0^\infty\!\!  \d P'\    \mathbb{F}_{P,P'} \!\begin{bmatrix}
                P_3 & P_2 \\
                P_4 & P_1
    \end{bmatrix}
    \ 
    \begin{tikzpicture}[baseline={([yshift=-.5ex]current bounding box.center)}, xscale=.8, yscale=.7]
        \draw[very thick, Maroon, bend left=10] (-1.8,1.2) to (0,.6);
        \draw[very thick, Maroon, bend right=10] (-1.8,-1.2) to (0,-.6);
        \draw[very thick, Maroon, bend right=10] (2.2,1.2) to (0,.6);
        \draw[very thick, Maroon, bend left=10] (2.2,-1.2) to (0,-.6);
        \draw[very thick, Maroon] (0,-.6) to (0,.6);
        \fill[Maroon] (0,.6) circle (.07);
        \fill[Maroon] (0,-.6) circle (.07);
        \draw[very thick] (-2,1.2) circle (.2 and .5);
        \draw[very thick] (-2,-1.2) circle (.2 and .5);
        \begin{scope}[shift={(2,1.2)}, xscale=.4]
            \draw[very thick] (0,-.5) arc (-90:90:.5);
            \draw[very thick, dashed] (0,.5) arc (90:270:.5);
        \end{scope}
        \begin{scope}[shift={(2,-1.2)}, xscale=.4]
            \draw[very thick] (0,-.5) arc (-90:90:.5);
            \draw[very thick, dashed] (0,.5) arc (90:270:.5);
        \end{scope}
        \begin{scope}[yscale=.4]
            \draw[very thick, RoyalBlue] (-.97,0) arc (-180:0:.97);
            \draw[very thick, dashed, RoyalBlue] (.97,0) arc (0:180:.97);
        \end{scope}
        \draw[very thick, out=0, in=180] (-2,1.7) to (0,.9) to (2,1.7);
        \draw[very thick, out=0, in=180] (-2,-1.7) to (0,-.9) to (2,-1.7);
        \draw[very thick, out=0, in=0, looseness=2.5] (-2,-.7) to (-2,.7);
        \draw[very thick, out=180, in=180, looseness=2.5] (2,-.7) to (2,.7);
        \node at (-2,1.2) {3};
        \node at (2.4,1.2) {2};            
        \node at (2.4,-1.2) {1};
        \node at (-2,-1.2) {4};   
        \node at (.3,0) {$P'$};
    \end{tikzpicture} \ , \label{eq:crossing of four-punctured sphere}
\ee
where the red graph on the surface specifies the conformal block channel under consideration. Here $P$, $P'$ and $P_j$ are the Liouville momenta associated to the internal and external lines, related to the conformal weight as $\Delta=\frac{c-1}{24}+P^2$. Crossing properties of conformal blocks on arbitrary surfaces can then be formalized in terms of a non-rational generalization of the Moore-Seiberg construction \cite{Moore:1988qv}.

In a seminal work, Ponsot and Teschner made an explicit proposal for the crossing kernel $\mathbb{F}$ appearing in \eqref{eq:crossing of four-punctured sphere} \cite{Ponsot:1999uf, Ponsot:2000mt}. Their derivation involves a particular connection of the Virasoro representation theory to quantum groups and the crossing kernel can be identified as a 6j symbol of that quantum group. Their function for $\mathbb{F}$ involves a Mellin-Barnes type integral of products of double Gamma functions familiar from the DOZZ formula \cite{Dorn:1994xn, Zamolodchikov:1995aa}. In this approach, it is also difficult to prove consistency of the construction on arbitrary surfaces. Consistency has been proven by now by translating the statements to quantum Teichm\"uller theory where consistency is known to hold \cite{Chekhov:1999tn, Kashaev:1998fc, Teschner:2013tqy}.

The crossing kernels were studied in many different papers. Let us mention a few. After the initial works by Ponsot and Teschner \cite{Ponsot:1999uf, Ponsot:2000mt}, 
the crossing kernel was applied to the Liouville bootstrap in \cite{Teschner:2001rv, Ponsot:2001ng}.
A different `chiral bootstrap' for the crossing kernel was explained in \cite{Teschner:2001rv, Teschner:2003en}.
The special case for $c=1$ and its relation to Painlev\'e VI was discussed in \cite{Iorgov:2013uoa}. 
The kernel was reconsidered in \cite{Teschner:2012em, Teschner:2013tqy} with an eye towards supersymmetric gauge theories from the AGT correspondence.
The relation to quantum Teichm\"uller theory was explained in \cite{Nidaiev:2013bda, Collier:2023fwi}.
The implications for the Virasoro bootstrap and generalized Cardy limits were explored in \cite{Collier:2018exn, Kusuki:2018wpa, Collier:2019weq}. The crossing kernels can be used to solve the Schwarzian theory in one dimension via dimensional reduction \cite{Mertens:2017mtv}.
The kernel was also considered by Nemkov \cite{Nemkov:2014rsa, Nemkov:2015zha} where an infinite sum representation for the modular kernel was given and by Roussillon \cite{Roussillon:2020lyc}, where the relation to the Ruijsenaars hypergeometric function was emphasized.

\medskip

The purpose of these notes is to give a careful rederivation using only basic CFT techniques. Our methods are not new, we essentially use Teschner's analytic bootstrap argument to fix the answer \cite{Teschner:1995yf, Ponsot:1999uf}. However we demonstrate that this bootstrap problem has a unique solution, which was so far missing in the literature. This argument is rather subtle and we use difference Galois theory to achieve this.
We also endow the space of Virasoro conformal blocks with an inner product that has appeared before in three-dimensional quantum gravity \cite{Verlinde:1989ua, Collier:2023fwi}. Virasoro conformal blocks with $\Delta \ge \frac{c-1}{24}$ form a complete delta-function normalizable basis of the space of conformal blocks. In this approach it becomes obvious that this set of Virasoro conformal blocks needs to close under crossing transformations. This shows that the crossing kernels on the sphere \eqref{eq:crossing of four-punctured sphere} and other surfaces have to exist and they necessarily satisfy the Moore-Seiberg consistency conditions. Hence our approach makes it manifest that the crossing kernels satisfy those consistency conditions. Mathematically speaking, our derivation is not completely rigorous and we have stated the gaps as two conjectures, see Conjectures~\ref{conj:timelike Liouville exists} and \ref{conj:complete basis}. However these conjectures are essentially obvious from a physicists' point of view.

Since the literature on the subject is rather convoluted, we try to give a pedagogical presentation of Virasoro conformal blocks. In particular, we include a lot of background material on different perspectives on conformal blocks. In Section~\ref{sec:Virasoro conformal blocks}, we start by reviewing the necessary background and explain how the bootstrap problem for the crossing kernel is set up. We then solve the bootstrap problem in Section~\ref{sec:bootstrapping the crossing kernels} and give the unique expressions for the kernel. This requires a number of properties of the relevant special functions which we explain in more detail in Appendix~\ref{app:special functions} and \ref{app:hypergeometric functions}. The appendices are more complete than they need to be in the hope that they will be useful elsewhere. We also prove many of the properties of the appearing special functions.
We then explain several applications and further properties of the crossing kernels in Section~\ref{sec:further properties and applications}. We explain in detail that they give rise a unitary projective representation of the mapping class group of the surface. We also prove that the representation is faithful. We explain the relation of conformal blocks with the quantization of Teichm\"uller space and hyperbolic geometry.

\paragraph{Omissions.} We follow a particular path towards the Virasoro crossing kernels. Essentially three different other approaches have been considered in the literature, which we mention for completeness but do not discuss in the rest of the manuscript:
\begin{enumerate}
\item Construction from the 6j symbols of the modular double $\mathcal{U}_q(\mathfrak{sl}(2,\mathbb{R}))$. This construction uses the equivalence of Virasoro representation theory with the representation theory of $\mathcal{U}_q(\mathfrak{sl}(2,\mathbb{R}))$ and originally let Ponsot and Teschner in \cite{Ponsot:2000mt} to the explicit form for the crossing kernel.
\item Construction from free field realization. A second route makes use of a direct construction of chiral Liouville vertex operators in terms of free fields \cite{Teschner:2001rv, Teschner:2003en}. The crossing kernel then follows by an explicit free field calculation.
\item Construction from probability theory. In this framework, conformal blocks are defined through the Gaussian free field \cite{Ghosal}, which is then used to rigorously set up the bootstrap problem that we also discuss in Section~\ref{sec:bootstrapping the crossing kernels}. 
\end{enumerate}

\section{Virasoro conformal blocks and the Moore-Seiberg construction} \label{sec:Virasoro conformal blocks}
\subsection{Definition of conformal blocks} \label{subsec:definition of conformal blocks}
We consider throughout the paper conformal blocks on a Riemann surface $\Sigma_{g,n}$ of genus $g$ and $n$ punctures.

\paragraph{Abstract characterization.} Virasoro conformal blocks on $\Sigma_{g,n}$ provide a complete basis for the Virasoro Ward identities. Conformal blocks are to first approximation functions depending on the complex structure on the surface.
This should be made more precise in two ways. First, a conformal block is not a single-valued function on the moduli space of Riemann surfaces, but transforms non-trivially e.g.\ when moving one puncture around another puncture. It should thus to second approximation be thought of as a function on the universal cover of the moduli space of Riemann surfaces, which is known as Teichm\"uller space $\mathcal{T}_{g,n}$. This is however still not quite correct. Teichm\"uller space can be defined as the space of all metrics on $\Sigma_{g,n}$, up to \emph{small} diffeomorphisms and Weyl rescalings,
\be  
    \mathcal{T}_{g,n}=\frac{\text{Metrics on }\Sigma_{g,n}}{\Diff_0(\Sigma_{g,n}) \times \text{Weyl}(\Sigma_{g,n})}\ . \label{eq:definition Teichmuller space}
\ee
Here, small diffeomorphisms are all diffeomorphisms that can be continuously deformed to the identity. In other word, $\Diff_0(\Sigma_{g,n})$ is the connected component of $\Diff(\Sigma_{g,n})$ containing the identity diffeomorphism. However, conformal blocks have a conformal anomaly and are thus not quite invariant under Weyl rescalings, but transform with the usual anomaly factor.
This means that they don't descend to functions on Teichm\"uller space, but rather to holomorphic sections of a particular holomorphic line bundle $\mathscr{L}$. Moreover, vertex operators transform with conformal weight $\Delta$ and thus the line bundle $\mathscr{L}$ depends besides the central charge $c$ also on the conformal weights $\Delta_1,\dots,\Delta_n$. We should mention that we could of course trivialize the line bundle $\mathscr{L}$ over Teichm\"uller space since Teichm\"uller space is in fact contractible and hence any line bundle is trivial. However, any choice of such a trivialization is \emph{not} canonical.

One can make this somewhat more concrete as follows. It makes sense to ask about the curvature $F(\mathscr{L})$ of the line bundle $\mathscr{L}$. We then have the formula
\be 
\frac{F(\mathscr{L})}{2\pi}=\frac{c\, \omega_{\text{WP}}(\ell_1,\dots,\ell_n)}{48\pi^2}\ , \label{eq:line bundle conformal blocks curvature}
\ee
where $\omega_{\text{WP}}(\ell_1,\dots,\ell_n)$ is the Weil-Petersson symplectic form on $\mathcal{T}_{g,n}$. In the presence of punctures, the Weil-Petersson form appearing is the Weil-Petersson form on a surface with conical defects or geodesic boundaries, depending on whether the respective conformal weight satisfies $\Delta<\frac{c}{24}$ or $\Delta>\frac{c}{24}$. The geodesic boundary length $\ell$ is related to the conformal weight as follows,
\be 
\Delta=\frac{c}{24}\left(1+\frac{\ell^2}{4\pi^2} \right)\ . \label{eq:conformal weight geodesic length relation}
\ee
For states with $\Delta<\frac{c}{24}$, the geodesic length is replaced by $\ell=i (2\pi-\alpha)$, where $\alpha$ is the defect angle.
This fact has appeared in the literature under various guises, see \cite{Quillen, Friedan:1986ua, Zograf1, Zograf2, Verlinde:1989ua}. We explain it in some more detail in Section~\ref{subsec:line bundle conformal blocks}. Given the close connection to hyperbolic geometry that results from this fact, we will draw all surfaces in this work with geodesic boundaries, even though we sometimes refer to them as punctures.

\paragraph{Behaviour under degeneration.} Abstractly, a conformal block is a particular section of the holomorphic line bundle $\mathscr{L}$ over Teichm\"uller space. However, there is an infinite number of such sections since Teichm\"uller space is non-compact. 
To fully characterize the block, we specify boundary conditions. A convenient way to specify such a boundary condition is to set the value at the boundary of Teichm\"uller space.  To specify a point in Teichm\"uller space (as opposed to moduli space), it is useful to employ the graphical notation advocated by Moore and Seiberg \cite{Moore:1988qv}.

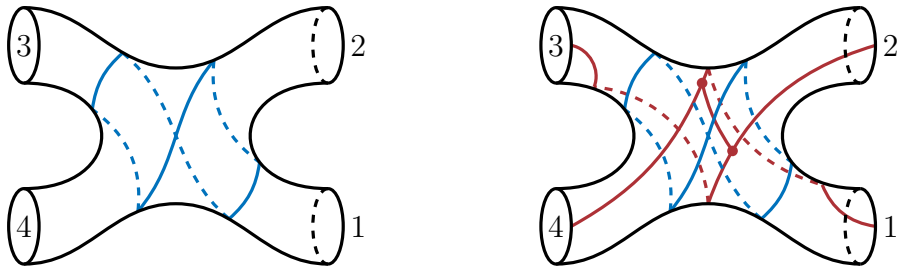
\begin{figure}[ht]
    \centering
    \begin{tikzpicture}
    \begin{scope}
        \draw[very thick, RoyalBlue, in=50, out=-130] (.5,1) to (-.5,-1);
        \draw[very thick, RoyalBlue, bend left=20] (-1.1,.35) to (-.7,1.1);
        \draw[very thick, RoyalBlue, bend left=20] (1.1,-.35) to (.7,-1.1);
        \draw[very thick, RoyalBlue, dashed, bend right=30] (-.5,-1) to (-1.1,.35);
        \draw[very thick, RoyalBlue, dashed, bend right=30] (.5,1) to (1.1,-.35);
        \draw[very thick, RoyalBlue, dashed, out=-40, in=140] (-.7,1.1) to (.7,-1.1);
        \draw[very thick] (-2,1.2) circle (.2 and .5);
        \draw[very thick] (-2,-1.2) circle (.2 and .5);
        \begin{scope}[shift={(2,1.2)}, xscale=.4]
            \draw[very thick] (0,-.5) arc (-90:90:.5);
            \draw[very thick, dashed] (0,.5) arc (90:270:.5);
        \end{scope}
        \begin{scope}[shift={(2,-1.2)}, xscale=.4]
            \draw[very thick] (0,-.5) arc (-90:90:.5);
            \draw[very thick, dashed] (0,.5) arc (90:270:.5);
        \end{scope}
        \draw[very thick, out=0, in=180] (-2,1.7) to (0,.9) to (2,1.7);
        \draw[very thick, out=0, in=180] (-2,-1.7) to (0,-.9) to (2,-1.7);
        \draw[very thick, out=0, in=0, looseness=2.5] (-2,-.7) to (-2,.7);
        \draw[very thick, out=180, in=180, looseness=2.5] (2,-.7) to (2,.7);
        \node at (-2,1.2) {3};
        \node at (2.4,1.2) {2};            
        \node at (2.4,-1.2) {1};
        \node at (-2,-1.2) {4}; 
    \end{scope}
    \begin{scope}[shift={(7,0)}]
        \draw[very thick, Maroon, bend right=20] (-1.8,-1.2) to (0,.9);
        \draw[very thick, Maroon, bend right=30, dashed] (0,.9) to (1.5,-.65);
        \draw[very thick, Maroon, bend right=30] (1.5,-.65) to (2.2,-1.2); 
        \draw[very thick, Maroon, bend right=27] (2.2,1.2) to (0,-.9);
        \draw[very thick, Maroon, bend right=40, dashed] (0,-.9) to (-1.5,.65);
        \draw[very thick, Maroon, bend right=50] (-1.5,.65) to (-1.8,1.2); 
        \draw[very thick, Maroon, bend right=10] (-.08,.7) to (.32,-.2);
        \fill[Maroon] (-.08,.7) circle (.07);
        \fill[Maroon] (.32,-.2) circle (.07);
        \draw[very thick, RoyalBlue, in=50, out=-130] (.5,1) to (-.5,-1);
        \draw[very thick, RoyalBlue, bend left=20] (-1.1,.35) to (-.7,1.1);
        \draw[very thick, RoyalBlue, bend left=20] (1.1,-.35) to (.7,-1.1);
        \draw[very thick, RoyalBlue, dashed, bend right=30] (-.5,-1) to (-1.1,.35);
        \draw[very thick, RoyalBlue, dashed, bend right=30] (.5,1) to (1.1,-.35);
        \draw[very thick, RoyalBlue, dashed, out=-40, in=140] (-.7,1.1) to (.7,-1.1);
        \draw[very thick] (-2,1.2) circle (.2 and .5);
        \draw[very thick] (-2,-1.2) circle (.2 and .5);
        \begin{scope}[shift={(2,1.2)}, xscale=.4]
            \draw[very thick] (0,-.5) arc (-90:90:.5);
            \draw[very thick, dashed] (0,.5) arc (90:270:.5);
        \end{scope}
        \begin{scope}[shift={(2,-1.2)}, xscale=.4]
            \draw[very thick] (0,-.5) arc (-90:90:.5);
            \draw[very thick, dashed] (0,.5) arc (90:270:.5);
        \end{scope}
        \draw[very thick, out=0, in=180] (-2,1.7) to (0,.9) to (2,1.7);
        \draw[very thick, out=0, in=180] (-2,-1.7) to (0,-.9) to (2,-1.7);
        \draw[very thick, out=0, in=0, looseness=2.5] (-2,-.7) to (-2,.7);
        \draw[very thick, out=180, in=180, looseness=2.5] (2,-.7) to (2,.7);
        \node at (-2,1.2) {3};
        \node at (2.4,1.2) {2};            
        \node at (2.4,-1.2) {1};
        \node at (-2,-1.2) {4}; 
    \end{scope}
    \end{tikzpicture} 
    \caption{Left: A pair of pants decomposition of the four-punctured sphere. Right: The same pair of pants decomposition with additional decoration keeping track of the Dehn twists.}
    \label{fig:four-punctured sphere pair of pants decomposition}
\end{figure}

Fix a surface $\Sigma_{g,n}$. A pair of pants decomposition of $\Sigma_{g,n}$ cuts the surface  into $2g-2+n$ three-punctured spheres (or pair of pants) that are glued along $3g-3+n$ non-intersecting curves on the surface. For example, for a four-punctured sphere, a valid pair of pants decomposition is given on the left in Figure~\ref{fig:four-punctured sphere pair of pants decomposition}. We drew a slightly complicated pair of pants decomposition to emphasize the fact that there are many pair of pants decompositions besides the obvious ones. Fixing a pair of pants decomposition is not yet sufficient to specify a point in Teichm\"uller space. Indeed, let us recall that by definition moduli space is the quotient of Teichm\"uller space by the mapping class group
\be 
\Map(\Sigma_{g,n})=\Diff(\Sigma_{g,n})/\Diff_0(\Sigma_{g,n})\ .
\ee
We can apply a Dehn twist along the curve specifying the pair of pants decomposition. This means that we rotate the left side of the curve against the right side of the curve until we fully rotated by $2\pi$. This is a large diffeomorphism and hence gives a non-trivial element in the mapping class group. By construction, this does not change the pair of pants decomposition. To keep track of this additional data, one decorates the surface further with a dual trivalent graph, where every vertex lies in one pair of pants. Such a graph is drawn on the right side of Figure~\ref{fig:four-punctured sphere pair of pants decomposition}. Under a Dehn twist, this graph obviously changes and thus such a decoration specifies a channel for the conformal block.\footnote{Specifying a dual graph is naively actually slightly too much data, since it specifies actually an element of the central extension of the mapping class group. We come back to this point in Section~\ref{subsec:representation mapping class group}.}

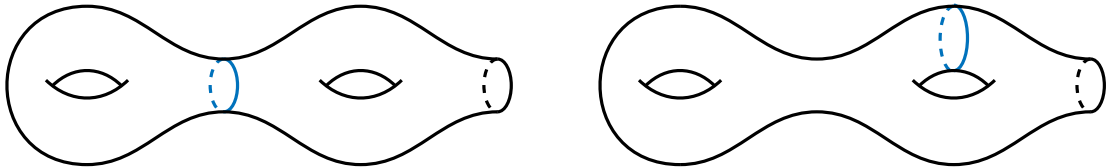
\begin{figure}[ht]
    \centering
    \begin{tikzpicture}[xscale=.6, yscale=.7]
        \begin{scope}
                \draw[very thick, RoyalBlue] (-3,-.5) arc (-90:90:.3 and .5);
                \draw[very thick, RoyalBlue, dashed] (-3,.5) arc (90:270:.3 and .5);
                \draw[very thick] (3,-.5) arc (-90:90:.3 and .5);
                \draw[very thick, dashed] (3,.5) arc (90:270:.3 and .5);
        		\draw[very thick, in=180, out=0] (-6,-1.5) to (-3,-.5) to (0,-1.5) to (3,-.5);
        		\draw[very thick, in=180, out=0] (-6,1.5) to (-3,.5) to (0,1.5) to (3,.5);
                \draw[very thick, out=180, in=180, looseness=2] (-6,1.5) to (-6,-1.5);
        		\draw[very thick, bend right=40] (-.9,0.1) to (.9,.1);
        		\draw[very thick, bend left=40] (-.75,0) to (.75,0);
        		\draw[very thick, bend right=40] (-6.9,0.1) to (-5.1,.1);
        		\draw[very thick, bend left=40] (-6.75,0) to (-5.25,0);
        \end{scope}
        \begin{scope}[shift={(13,0)}]
                \draw[very thick, RoyalBlue] (0,.28) arc (-90:90:.3 and .62);
                \draw[very thick, RoyalBlue, dashed] (0,1.5) arc (90:270:.3 and .62);
                \draw[very thick] (3,-.5) arc (-90:90:.3 and .5);
                \draw[very thick, dashed] (3,.5) arc (90:270:.3 and .5);
        		\draw[very thick, in=180, out=0] (-6,-1.5) to (-3,-.5) to (0,-1.5) to (3,-.5);
        		\draw[very thick, in=180, out=0] (-6,1.5) to (-3,.5) to (0,1.5) to (3,.5);
                \draw[very thick, out=180, in=180, looseness=2] (-6,1.5) to (-6,-1.5);
        		\draw[very thick, bend right=40] (-.9,0.1) to (.9,.1);
        		\draw[very thick, bend left=40] (-.75,0) to (.75,0);
        		\draw[very thick, bend right=40] (-6.9,0.1) to (-5.1,.1);
        		\draw[very thick, bend left=40] (-6.75,0) to (-5.25,0);
        \end{scope}
    \end{tikzpicture}
    \caption{The two types of degenerations of surfaces. Pinching the blue curve on the left results in a once-punctured torus and a twice-punctured torus which are connected at a single point (node). Pinching the blue curve on the right results in a three-punctured torus.}
    \label{fig:surface degenerations}
\end{figure}

We can now talk about the behaviour of conformal blocks when we pinch one of the curves specifying the pair of pants decomposition. In this limit, the surface either degenerates into two surfaces $\Sigma_{g,n} \to \Sigma_{g_1,n_1+1} \cup \Sigma_{g_2,n_2+1}$ with $g=g_1+g_2$ and $n=n_1+n_2$ or into one surface $\Sigma_{g,n} \to \Sigma_{g-1,n+2}$. The two additional punctures correspond to the two sides of the pinched curve, see Figure~\ref{fig:surface degenerations}.
In either case we can find a local complex coordinate $q$ in moduli space such that $q=0$ corresponds to the degenerate surface (the boundary divisor of moduli space). We then require that a conformal block $\mathcal{F}_{g,n}$\footnote{We will fix a precise labelling convention for conformal blocks later and be somewhat schematic for now.} behaves in the limit $q \to 0$ to leading order as
\be 
\mathcal{F}_{g,n} \sim  \mathcal{F}_{g_1,n_1+1}\mathcal{F}_{g_2,n_2+1}\quad \text{or}\quad  \mathcal{F}_{g-1,n+2}\ . \label{eq:conformal block degeneration}
\ee
Here the conformal weight associated to the two additional punctures (the nodes of the surface) is $\Delta$. It can be thought of as the conformal weight of the primary field running through the curve that we pinched. 
Eq.~\eqref{eq:conformal block degeneration} recursively fixes a basis of conformal blocks by specifying the behaviour under degeneration. Eventually, we reduce everything back to three-punctured spheres, whose conformal blocks are uniquely fixed by conformal symmetry,
\be 
\begin{tikzpicture}[baseline={([yshift=-.5ex]current bounding box.center)}, scale=.8]
        \draw[very thick, Maroon] (0,-1.2) to (0,-.2);
        \fill[Maroon] (0,-.2) circle (.07);
        \draw[very thick, Maroon, bend left=20] (0,-.2) to (-1,.8);
        \draw[very thick, Maroon, bend right=20] (0,-.2) to (1,.8);
        \node at (-.4,.4) {1};
        \node at (.4,.4) {2};
        \node at (.17,-.55) {3};
        \draw[very thick, out=270, in=90] (1.5,1) to (.5,-1);
        \draw[very thick, out=270, in=90] (-1.5,1) to (-.5,-1);
        \draw[very thick, out=270, in=270, looseness=1.5] (-.5,1) to (.5,1);
        \draw[very thick] (-1,1) circle (.5 and .2); 
        \draw[very thick] (1,1) circle (.5 and .2);
        \begin{scope}[shift={(0,-1)}, yscale=.4]
            \draw[very thick] (-.5,0) arc (-180:0:.5);
            \draw[dashed, very thick] (-.5,0) arc (180:0:.5);
        \end{scope}
    \end{tikzpicture}=z_{21}^{-\Delta_1-\Delta_2+\Delta_3}z_{31}^{-\Delta_1-\Delta_3+\Delta_2}z_{32}^{-\Delta_3-\Delta_2+\Delta_1}\ . \label{eq:conformal block three-punctured sphere}
\ee
Of course this formula has still many branches, but they are precisely specified by the trivalent graph that we draw on the three-punctured sphere.
This logic allows us to identify pictures such as the one on the right-hand side of Figure~\ref{fig:four-punctured sphere pair of pants decomposition} with a conformal block, once we label all internal and external lines with conformal weights. The dependence on the moduli is left implicit.

Descendants appear if we expand to subleading orders in $q$, but importantly all subleading terms in $q$ are uniquely specified by conformal symmetry and thus the conformal block is fully determined. Concretely, a conformal block is nothing but a resolution of the identity as follows. Let $\Psi$ be a primary state of conformal weight $\Delta$. Let then $\Psi^{(n)}_i$ for $i=1,\dots,|\mathbb{P}_n|$ be an enumeration of all the level $n$ descendants. Here $|\mathbb{P}_n|$ denotes the number of partitions of $n$, corresponding to the number of level $n$ descendants in a generic Virasoro representation. Let us choose the $\Psi^{(n)}_i$ for simplicity such that they form an orthonormal basis of the descendants, i.e.\ $\langle \Psi^{(n)}_i\, | \, \Psi^{(n)}_j \rangle= \delta_{ij}$. Then e.g.\ the $g=1$, $n=2$ conformal block admits a series expansion as follows
\be 
    \begin{tikzpicture}[baseline={([yshift=-.5ex]current bounding box.center)}]
        \begin{scope}[xscale=.7, yscale=.7]
        			\draw[very thick, Maroon, bend right=30] (2.2,1.2) to (.6,0);
        			\draw[very thick, Maroon, bend left=30] (2.2,-1.2) to (.6,0);
        			\draw[very thick, Maroon] (-.6,0) to (.6,0);
        			\fill[Maroon] (-.6,0) circle (.07);
        			\fill[Maroon] (.6,0) circle (.07);
                    \draw[very thick, Maroon, out=135, in=90, looseness=1.2] (-.6,0) to (-3.3,0);
                    \draw[very thick, Maroon, out=-135, in=-90, looseness=1.2] (-.6,0) to (-3.3,0);
                    \draw[very thick, out=180, in=180, looseness=2] (-2,1.7) to (-2,-1.7);
                    \draw[very thick, bend right=40] (-2.9,0.1) to (-1.1,.1);
        		    \draw[very thick, bend left=40] (-2.75,0) to (-1.25,0);
        			\begin{scope}[shift={(2,1.2)}, xscale=.4]
            			\draw[very thick] (0,-.5) arc (-90:90:.5);
            			\draw[very thick, dashed] (0,.5) arc (90:270:.5);
        			\end{scope}
        			\begin{scope}[shift={(2,-1.2)}, xscale=.4]
            			\draw[very thick] (0,-.5) arc (-90:90:.5);
            			\draw[very thick, dashed] (0,.5) arc (90:270:.5);
        			\end{scope}
        			\begin{scope}[xscale=.4]
            			\draw[very thick, RoyalBlue] (0,-.9) arc (-90:90:.9);
            			\draw[very thick, dashed, RoyalBlue] (0,.9) arc (90:270:.9);
        			\end{scope}
                    \draw[very thick, RoyalBlue] (-2.75,0) arc (0:180:.61 and .3);
                    \draw[very thick, dashed, RoyalBlue] (-2.75,0) arc (0:-180:.61 and .3);
        			\draw[very thick, out=0, in=180] (-2,1.7) to (0,.9) to (2,1.7);
        			\draw[very thick, out=0, in=180] (-2,-1.7) to (0,-.9) to (2,-1.7);
        			\draw[very thick, out=180, in=180, looseness=2.5] (2,-.7) to (2,.7);
           \node at (2.45,-1.2) {$1$};
           \node at (2.45,1.2) {$2$};
           \node at (0,.3) {$\Delta$};
           \node at (-3.5,.6) {$\Delta\!{}'$};
		\end{scope}
   \end{tikzpicture}\hspace{-.3cm}=\sum_{n \ge 0} \sum_i 
          \begin{tikzpicture}[baseline={([yshift=-.5ex]current bounding box.center)}]
          \begin{scope}[xscale=.7, yscale=.7]
        			\draw[very thick, Maroon] (-.6,0) to (.2,0);
        			\fill[Maroon] (-.6,0) circle (.07);
\draw[very thick, Maroon, out=135, in=90, looseness=1.2] (-.6,0) to (-3.3,0);
                    \draw[very thick, Maroon, out=-135, in=-90, looseness=1.2] (-.6,0) to (-3.3,0);
                    \draw[very thick, out=180, in=180, looseness=2] (-2,1.7) to (-2,-1.7);
                    \draw[very thick, bend right=40] (-2.9,0.1) to (-1.1,.1);
        		    \draw[very thick, bend left=40] (-2.75,0) to (-1.25,0);
        			\begin{scope}[xscale=.4]
            			\draw[very thick] (0,-.5) arc (-90:90:.5);
            			\draw[very thick, dashed] (0,.5) arc (90:270:.5);
        			\end{scope}
                    \draw[very thick, RoyalBlue] (-2.75,0) arc (0:180:.61 and .3);
                    \draw[very thick, dashed, RoyalBlue] (-2.75,0) arc (0:-180:.61 and .3);
        			\draw[very thick, out=0, in=180] (-2,1.7) to (0,.5);
        			\draw[very thick, out=0, in=180] (-2,-1.7) to (0,-.5);
           \node at (0,1) {$\Psi^{(n)}_i$};
           \node at (-3.5,.6) {$\Delta\!{}'$};
		\end{scope}
   \end{tikzpicture}
   \times 
       \begin{tikzpicture}[baseline={([yshift=-.5ex]current bounding box.center)}]
        \begin{scope}[xscale=.7, yscale=.7]
        			\draw[very thick, Maroon, bend right=30] (2.2,1.2) to (.6,0);
        			\draw[very thick, Maroon, bend left=30] (2.2,-1.2) to (.6,0);
        			\draw[very thick, Maroon] (.2,0) to (.6,0);
        			\fill[Maroon] (.6,0) circle (.07);
        			\begin{scope}[shift={(2,1.2)}, xscale=.4]
            			\draw[very thick] (0,-.5) arc (-90:90:.5);
            			\draw[very thick, dashed] (0,.5) arc (90:270:.5);
        			\end{scope}
        			\begin{scope}[shift={(2,-1.2)}, xscale=.4]
            			\draw[very thick] (0,-.5) arc (-90:90:.5);
            			\draw[very thick, dashed] (0,.5) arc (90:270:.5);
        			\end{scope}
        			\draw[very thick, out=0, in=180] (0,.5) to (2,1.7);
        			\draw[very thick, out=0, in=180] (0,-.5) to (2,-1.7);
        			\draw[very thick, out=180, in=180, looseness=2.5] (2,-.7) to (2,.7);
           \draw[very thick] (0,0) circle (.2 and .5);
           \node at (2.45,-1.2) {$1$};
           \node at (2.45,1.2) {$2$};
           \node at (0,1) {$\Psi^{(n)}_i$};
		\end{scope}
   \end{tikzpicture}\ . \label{eq:general conformal block power series expansion}
\ee
The conformal blocks appearing on the right are conformal blocks where the external states are not primary vertex operators. Using Virasoro symmetry, once can always reduce them to conformal blocks of primary vertex operators only. This expression is a power series expansion in $q$. It should converge in a neighborhood of the degeneration. This has recently been proven rigorously in the case of the four-punctured sphere \cite{Guillarmou:2020wbo}. We make this equation very concrete below for the case of the four-punctured sphere and the once-punctured torus. See \cite{Cho:2017oxl} for more complicated topologies.

\paragraph{Explicit characterization for the four-punctured sphere.}
The previous discussion is somewhat abstract and is often useful to have more explicit formulas. We will discuss the well-studied case of a conformal block on the four-punctured sphere, see e.g.\ \cite{Belavin:1984vu}. We can choose the flat metric on the plane as usual, which removes issues of the conformal anomaly. We put $z_1=1$, $z_2=\infty$, $z_3=0$ and $z_4=z$, the unique remaining cross-ratio.

We choose $0<z<1$ and choose the branch cuts of the conformal block to be $z \in (-\infty,0] \cup [1,\infty)$. We consider the conformal block in the limit $z \to 0$. This is well-adapted to the standard pair of pants decomposition
\be 
    \begin{tikzpicture}[baseline={([yshift=-.5ex]current bounding box.center)}]
        \begin{scope}[xscale=.7, yscale=.7]
        \draw[very thick, Maroon, bend left=30] (-1.8,1.2) to (-.6,0);
        			\draw[very thick, Maroon, bend right=30] (-1.8,-1.2) to (-.6,0);
        			\draw[very thick, Maroon, bend right=30] (2.2,1.2) to (.6,0);
        			\draw[very thick, Maroon, bend left=30] (2.2,-1.2) to (.6,0);
        			\draw[very thick, Maroon] (-.6,0) to (.6,0);
        			\fill[Maroon] (-.6,0) circle (.07);
        			\fill[Maroon] (.6,0) circle (.07);
        			\draw[very thick] (-2,1.2) circle (.2 and .5);
        			\draw[very thick] (-2,-1.2) circle (.2 and .5);
        			\begin{scope}[shift={(2,1.2)}, xscale=.4]
            			\draw[very thick] (0,-.5) arc (-90:90:.5);
            			\draw[very thick, dashed] (0,.5) arc (90:270:.5);
        			\end{scope}
        			\begin{scope}[shift={(2,-1.2)}, xscale=.4]
            			\draw[very thick] (0,-.5) arc (-90:90:.5);
            			\draw[very thick, dashed] (0,.5) arc (90:270:.5);
        			\end{scope}
        			\begin{scope}[xscale=.4]
            			\draw[very thick, RoyalBlue] (0,-.9) arc (-90:90:.9);
            			\draw[very thick, dashed, RoyalBlue] (0,.9) arc (90:270:.9);
        			\end{scope}
        			\draw[very thick, out=0, in=180] (-2,1.7) to (0,.9) to (2,1.7);
        			\draw[very thick, out=0, in=180] (-2,-1.7) to (0,-.9) to (2,-1.7);
        			\draw[very thick, out=0, in=0, looseness=2.5] (-2,-.7) to (-2,.7);
        			\draw[very thick, out=180, in=180, looseness=2.5] (2,-.7) to (2,.7);
           \node at (2.4,-1.2) {$1$};
           \node at (2.4,1.2) {$2$};
           \node at (-2,-1.2) {$4$};
           \node at (-2,1.2) {$3$};
           \node at (0,.3) {$\Delta$};
		\end{scope}
   \end{tikzpicture}\ .
\ee
We often us numbers to label the internal lines with the understanding that the corresponding external conformal weights are given by $\Delta_j$ and the internal one by $\Delta$. Here, $z$ plays the role of the general local variable $q$ that we considered above.
To leading order in $z \to 0$, the conformal block behaves as $z^{-\Delta_3-\Delta_4+\Delta}$, which follows from the structure of the operator product expansion.
If we write down the power series expansion in $z$, it has the following structure
\begin{align}
    \begin{tikzpicture}[baseline={([yshift=-.5ex]current bounding box.center)}]
        \begin{scope}[xscale=.7, yscale=.7]
        \draw[very thick, Maroon, bend left=30] (-1.8,1.2) to (-.6,0);
        			\draw[very thick, Maroon, bend right=30] (-1.8,-1.2) to (-.6,0);
        			\draw[very thick, Maroon, bend right=30] (2.2,1.2) to (.6,0);
        			\draw[very thick, Maroon, bend left=30] (2.2,-1.2) to (.6,0);
        			\draw[very thick, Maroon] (-.6,0) to (.6,0);
        			\fill[Maroon] (-.6,0) circle (.07);
        			\fill[Maroon] (.6,0) circle (.07);
        			\draw[very thick] (-2,1.2) circle (.2 and .5);
        			\draw[very thick] (-2,-1.2) circle (.2 and .5);
        			\begin{scope}[shift={(2,1.2)}, xscale=.4]
            			\draw[very thick] (0,-.5) arc (-90:90:.5);
            			\draw[very thick, dashed] (0,.5) arc (90:270:.5);
        			\end{scope}
        			\begin{scope}[shift={(2,-1.2)}, xscale=.4]
            			\draw[very thick] (0,-.5) arc (-90:90:.5);
            			\draw[very thick, dashed] (0,.5) arc (90:270:.5);
        			\end{scope}
        			\begin{scope}[xscale=.4]
            			\draw[very thick, RoyalBlue] (0,-.9) arc (-90:90:.9);
            			\draw[very thick, dashed, RoyalBlue] (0,.9) arc (90:270:.9);
        			\end{scope}
        			\draw[very thick, out=0, in=180] (-2,1.7) to (0,.9) to (2,1.7);
        			\draw[very thick, out=0, in=180] (-2,-1.7) to (0,-.9) to (2,-1.7);
        			\draw[very thick, out=0, in=0, looseness=2.5] (-2,-.7) to (-2,.7);
        			\draw[very thick, out=180, in=180, looseness=2.5] (2,-.7) to (2,.7);
           \node at (2.4,-1.2) {$1$};
           \node at (2.4,1.2) {$2$};
           \node at (-2,-1.2) {$4$};
           \node at (-2,1.2) {$3$};
           \node at (0,.3) {$\Delta$};
		\end{scope}
   \end{tikzpicture}\hspace{-.3cm}=\sum_{n \ge 0} \sum_{\nu_\text{L},\, \nu_\text{R} \in \mathbb{P}_n} \  
          \begin{tikzpicture}[baseline={([yshift=-.5ex]current bounding box.center)}]
          \begin{scope}[xscale=.7, yscale=.7]
        \draw[very thick, Maroon, bend left=30] (-1.8,1.2) to (-.6,0);
        			\draw[very thick, Maroon, bend right=30] (-1.8,-1.2) to (-.6,0);
        			\draw[very thick, Maroon] (-.6,0) to (.2,0);
        			\fill[Maroon] (-.6,0) circle (.07);
        			\draw[very thick] (-2,1.2) circle (.2 and .5);
        			\draw[very thick] (-2,-1.2) circle (.2 and .5);
        			\begin{scope}[xscale=.4]
            			\draw[very thick] (0,-.5) arc (-90:90:.5);
            			\draw[very thick, dashed] (0,.5) arc (90:270:.5);
        			\end{scope}
        			\draw[very thick, out=0, in=180] (-2,1.7) to (0,.5);
        			\draw[very thick, out=0, in=180] (-2,-1.7) to (0,-.5);
        			\draw[very thick, out=0, in=0, looseness=2.5] (-2,-.7) to (-2,.7);
           \node at (-2,-1.2) {$4$};
           \node at (-2,1.2) {$3$};
           \node at (.4,1) {$\mathbf{L}_{-\nu_\text{L}}\Psi$};
		\end{scope}
   \end{tikzpicture} \hspace{-.8cm}
   \times \mathbf{M}_{\nu_\text{L},\nu_\text{R}} \times 
   \hspace{-.8cm}
       \begin{tikzpicture}[baseline={([yshift=-.5ex]current bounding box.center)}]
        \begin{scope}[xscale=.7, yscale=.7]
        			\draw[very thick, Maroon, bend right=30] (2.2,1.2) to (.6,0);
        			\draw[very thick, Maroon, bend left=30] (2.2,-1.2) to (.6,0);
        			\draw[very thick, Maroon] (.2,0) to (.6,0);
        			\fill[Maroon] (.6,0) circle (.07);
        			\begin{scope}[shift={(2,1.2)}, xscale=.4]
            			\draw[very thick] (0,-.5) arc (-90:90:.5);
            			\draw[very thick, dashed] (0,.5) arc (90:270:.5);
        			\end{scope}
        			\begin{scope}[shift={(2,-1.2)}, xscale=.4]
            			\draw[very thick] (0,-.5) arc (-90:90:.5);
            			\draw[very thick, dashed] (0,.5) arc (90:270:.5);
        			\end{scope}
        			\draw[very thick, out=0, in=180] (0,.5) to (2,1.7);
        			\draw[very thick, out=0, in=180] (0,-.5) to (2,-1.7);
        			\draw[very thick, out=180, in=180, looseness=2.5] (2,-.7) to (2,.7);
           \draw[very thick] (0,0) circle (.2 and .5);
           \node at (2.4,-1.2) {$1$};
           \node at (2.4,1.2) {$2$};
           \node at (-.5,1) {$\mathbf{L}_{-\nu_\text{R}} \Psi$};
		\end{scope}
   \end{tikzpicture}\ . \label{eq:conformal block four-punctured sphere abstract expansion}
\end{align}
This is the same formula as \eqref{eq:general conformal block power series expansion}, but for the four-punctured sphere and without assuming that the basis of descendants is orthonormal. The notation is as follows.
$\nu=(\nu_1,\nu_2,\dots,\nu_k)$ with $\nu_1 \ge \nu_2 \ge \nu_3 \ge \dots \ge \nu_k$ with $n=\sum_i \nu_i$ is a partition of $n$. We denoted by $\mathbb{P}_n$ the set of all partitions. It specifies a descendant at level $n$ via the following action of the Virasoro modes
\be 
    \mathbf{L}_{-\nu} \equiv L_{-\nu_k} L_{-\nu_{k-1}} \cdots L_{-\nu_1} \Psi
\ee
on the primary state. $\mathbf{M}_{\nu_\text{L},\nu_\text{R}}$ denotes the inverse of the Kac-matrix,
\be 
(\mathbf{M}^{-1})_{\nu_\text{L},\nu_\text{R}}=\langle \mathbf{L}_{-\nu_\text{L}} \Psi\, |\, \mathbf{L}_{-\nu_\text{R}} \Psi \rangle\ . \label{eq:Kac matrix definition}
\ee
Since the fields in the two-point function are inserted at $0$ and $\infty$, the insertion of $\mathbf{L}_{-\nu_\text{L}}\Psi$ should be at $\infty$ on the left three-punctured sphere, while  $\mathbf{L}_{-\nu_\text{R}}\Psi$ should be inserted at $0$.

We should note that \eqref{eq:conformal block four-punctured sphere abstract expansion} is manifestly basis independent. Even though we exhibited an explicit basis in the descendants, we could have taken any other basis.

To make this formula concrete, it remains to evaluate the three-point functions. Let us compute the three-point function on the right. Recall that the state $\mathbf{L}_{-\nu_\text{R}} \Psi$ is inserted at $0$. We have 
\begin{align} 
    \langle L_{-\nu} V_{\Delta}(0) V_{\Delta_1}(1) V_{\Delta_2}(\infty) \rangle&= \langle \Psi_2 \, | \, V_{\Delta_1}(1)\, L_{-\nu} \, |\, \Psi \rangle  \\
    &=\Res_{x=0} (x^{-\nu+1}-x) \langle \Psi_2 \, | \, V_{\Delta_1}(1)\, T(x) \, |\, \Psi \rangle \nonumber\\
    &\qquad+\Delta_{21} \langle \Psi_2 \, | \, V_{\Delta_1}(1)\, |\, \Psi_{21} \rangle \\
    &=-\Big[\Res_{x=1}+\Res_{x=\infty}\Big]\ (x^{-\nu+1}-x) \langle \Psi_2 \, | \, V_{\Delta_1}(1) \, T(x)\,  |\, \Psi \rangle \nonumber\\
    &\qquad+\Delta \langle \Psi_2 \, | \, V_{\Delta_1}(1)\, |\, \Psi \rangle \\
    &=(\nu \Delta_1-\Delta_2+\Delta)\langle V_{\Delta_{21}}(0) V_{\Delta_1}(1) V_{\Delta_2}(\infty) \rangle\ . \label{eq:three-point function removing L generator}
\end{align}
Here we frequently used the operator product expansion of primary fields.
Iterating this procedure gives
\be 
       \begin{tikzpicture}[baseline={([yshift=-.5ex]current bounding box.center)}]
        \begin{scope}[xscale=.7, yscale=.7]
        			\draw[very thick, Maroon, bend right=30] (2.2,1.2) to (.6,0);
        			\draw[very thick, Maroon, bend left=30] (2.2,-1.2) to (.6,0);
        			\draw[very thick, Maroon] (.2,0) to (.6,0);
        			\fill[Maroon] (.6,0) circle (.07);
        			\begin{scope}[shift={(2,1.2)}, xscale=.4]
            			\draw[very thick] (0,-.5) arc (-90:90:.5);
            			\draw[very thick, dashed] (0,.5) arc (90:270:.5);
        			\end{scope}
        			\begin{scope}[shift={(2,-1.2)}, xscale=.4]
            			\draw[very thick] (0,-.5) arc (-90:90:.5);
            			\draw[very thick, dashed] (0,.5) arc (90:270:.5);
        			\end{scope}
        			\draw[very thick, out=0, in=180] (0,.5) to (2,1.7);
        			\draw[very thick, out=0, in=180] (0,-.5) to (2,-1.7);
        			\draw[very thick, out=180, in=180, looseness=2.5] (2,-.7) to (2,.7);
           \draw[very thick] (0,0) circle (.2 and .5);
           \node at (2.4,-1.2) {$1$};
           \node at (2.4,1.2) {$2$};
           \node at (-.5,1) {$\mathbf{L}_{-\nu_\text{R}} \Psi$};
		\end{scope}
   \end{tikzpicture}=\prod_{j=1}^k \left(\nu_{\text{R},j} \Delta_1-\Delta_2+\Delta+\sum_{m <j} \nu_{\text{R},m} \right)\ . \label{eq:right three punctured sphere conformal block}
\ee
Similarly, one computes
\be 
          \begin{tikzpicture}[baseline={([yshift=-.5ex]current bounding box.center)}]
          \begin{scope}[xscale=.7, yscale=.7]
        \draw[very thick, Maroon, bend left=30] (-1.8,1.2) to (-.6,0);
        			\draw[very thick, Maroon, bend right=30] (-1.8,-1.2) to (-.6,0);
        			\draw[very thick, Maroon] (-.6,0) to (.2,0);
        			\fill[Maroon] (-.6,0) circle (.07);
        			\draw[very thick] (-2,1.2) circle (.2 and .5);
        			\draw[very thick] (-2,-1.2) circle (.2 and .5);
        			\begin{scope}[xscale=.4]
            			\draw[very thick] (0,-.5) arc (-90:90:.5);
            			\draw[very thick, dashed] (0,.5) arc (90:270:.5);
        			\end{scope}
        			\draw[very thick, out=0, in=180] (-2,1.7) to (0,.5);
        			\draw[very thick, out=0, in=180] (-2,-1.7) to (0,-.5);
        			\draw[very thick, out=0, in=0, looseness=2.5] (-2,-.7) to (-2,.7);
           \node at (-2,-1.2) {$4$};
           \node at (-2,1.2) {$3$};
           \node at (.4,1) {$\mathbf{L}_{-\nu_\text{L}}\Psi$};
		\end{scope}
   \end{tikzpicture}\hspace{-.7cm} =z^{-\Delta_3-\Delta_4+\Delta+n} \prod_{j=1}^k \left(\nu_{\text{L},j} \Delta_4-\Delta_3+\Delta+\sum_{m <j} \nu_{\text{L},m} \right)\ . \label{eq:left three punctured sphere conformal block}
\ee
Plugging \eqref{eq:right three punctured sphere conformal block}, \eqref{eq:left three punctured sphere conformal block} and the definition of the Kac-Matrix \eqref{eq:Kac matrix definition} into the abstract expansion \eqref{eq:conformal block four-punctured sphere abstract expansion} gives a very explicit formula for the power series expansion of conformal blocks.

We should mention that the power series definition of conformal blocks is not a very efficient method to compute them numerically. There are rapidly converging recursion relations known as the Zamolodchikov recursion relations that are far better adapted to that purpose \cite{Zamolodchikov:1984eqp, Zamolodchikov_recursion2}.

\paragraph{Explicit characterization for the once-punctured torus.} The other case in which we can write a simple power series definition of the conformal blocks is the once-punctured torus. It has some new features that will become important below. We have abstractly
\be 
 \begin{tikzpicture}[baseline={([yshift=-.5ex]current bounding box.center)}]
        \draw[very thick, Maroon] (0,-.7) to (0,0);
        \fill[Maroon] (0,0) circle (.07);
        \draw[very thick, Maroon, out=140, in=-60] (0,0) to (-1,.8);
        \draw[very thick, Maroon, out=120, in=180] (-1,.8) to (0,1.8);
        \draw[very thick, Maroon, out=40, in=240] (0,0) to (1,.8);
        \draw[very thick, Maroon, out=60, in=0] (1,.8) to (0,1.8);
        \node at (-.4,.6) {$\Delta$};
        \node at (0,-1) {$\Delta_0$};
        \draw[very thick, out=270, in=90] (1.5,1) to (.5,-.5);
        \draw[very thick, out=270, in=90] (-1.5,1) to (-.5,-.5);
        \draw[very thick, out=90, in=180] (-1.5,1) to (0,2.2);
        \draw[very thick, out=0,in=90] (0,2.2) to (1.5,1);
        \draw[very thick, bend left=70] (-.7,1) to (.7,1);
        \draw[very thick, bend right=70] (-.8,1.2) to (.8,1.2);
        \begin{scope}[shift={(0,-.5)}, yscale=.4]
            \draw[very thick] (-.5,0) arc (-180:0:.5);
            \draw[dashed, very thick] (-.5,0) arc (180:0:.5);
        \end{scope}
        \begin{scope}[shift={(0,1.8)}, xscale=.4]
            \draw[very thick, RoyalBlue] (0,-.4) arc (-90:90:.4);
            \draw[dashed, very thick, RoyalBlue] (0,.4) arc (90:270:.4);
        \end{scope}
    \end{tikzpicture} 
    =\sum_{n \ge 0} \sum_{\nu_\text{L},\, \nu_\text{R} \in \mathbb{P}_n} \mathbf{M}_{\nu_\text{L},\nu_\text{R}} \times \hspace{-.4cm}
    \begin{tikzpicture}[baseline={([yshift=-.5ex]current bounding box.center)}]
            \draw[very thick, Maroon] (0,-1.2) to (0,-.2);
            \fill[Maroon] (0,-.2) circle (.07);
            \draw[very thick, Maroon, bend left=20] (0,-.2) to (-1,.8);
            \draw[very thick, Maroon, bend right=20] (0,-.2) to (1,.8);
            \node at (-1,1.5) {$\mathbf{L}_{-\nu_\text{L}} \Psi$};
            \node at (1,1.5) {$\mathbf{L}_{-\nu_\text{R}} \Psi$};
            \node at (0,-1.5) {$\Delta_0$};
            \draw[very thick, out=270, in=90] (1.5,1) to (.5,-1);
            \draw[very thick, out=270, in=90] (-1.5,1) to (-.5,-1);
            \draw[very thick, out=270, in=270, looseness=1.5] (-.5,1) to (.5,1);
            \draw[very thick] (-1,1) circle (.5 and .2); 
            \draw[very thick] (1,1) circle (.5 and .2);
            \begin{scope}[shift={(0,-1)}, yscale=.4]
                \draw[very thick] (-.5,0) arc (-180:0:.5);
                \draw[dashed, very thick] (-.5,0) arc (180:0:.5);
        \end{scope}
    \end{tikzpicture} . \label{eq:conformal block once-punctured torus abstract expansion}
\ee
Here, $\mathbf{M}_{\nu_\text{L},\nu_\text{R}}$ is again the inverse of the Kac matrix.
It remains to compute the three-point block on the right. To do so, it is convenient to parametrize the torus via its modulus $\tau$. We also set $q=\mathrm{e}^{2\pi i \tau}$. The degeneration corresponds to $q \to 0$. 

To make the formula concrete, we again work out the three-punctured sphere block. To map this to the geometry of the plane, we apply the standard exponential map.
This gives rise to the usual term from the conformal anomaly. We thus have
\be 
\begin{tikzpicture}[baseline={([yshift=-.5ex]current bounding box.center)}]
            \draw[very thick, Maroon] (0,-1.2) to (0,-.2);
            \fill[Maroon] (0,-.2) circle (.07);
            \draw[very thick, Maroon, bend left=20] (0,-.2) to (-1,.8);
            \draw[very thick, Maroon, bend right=20] (0,-.2) to (1,.8);
            \node at (-1,1.5) {$\mathbf{L}_{-\nu_\text{L}} \Psi$};
            \node at (1,1.5) {$\mathbf{L}_{-\nu_\text{R}} \Psi$};
            \node at (0,-1.5) {$\Delta_0$};
            \draw[very thick, out=270, in=90] (1.5,1) to (.5,-1);
            \draw[very thick, out=270, in=90] (-1.5,1) to (-.5,-1);
            \draw[very thick, out=270, in=270, looseness=1.5] (-.5,1) to (.5,1);
            \draw[very thick] (-1,1) circle (.5 and .2); 
            \draw[very thick] (1,1) circle (.5 and .2);
            \begin{scope}[shift={(0,-1)}, yscale=.4]
                \draw[very thick] (-.5,0) arc (-180:0:.5);
                \draw[dashed, very thick] (-.5,0) arc (180:0:.5);
        \end{scope}
    \end{tikzpicture}=q^{\Delta+n-\frac{c}{24}} \, \langle \mathbf{L}_{-\nu_\text{L}}\Psi\, |\,  V_{\Delta_0}(1)\, |\,  \mathbf{L}_{-\nu_\text{R}}\Psi   \rangle\ . \label{eq:three-point function once-punctured torus conformal block}
\ee
The three-point function appearing on the right hand side can be reduced to the primary three-point function in a similar way to the procedure in eq.~\eqref{eq:three-point function removing L generator}. 
Plugging it into \eqref{eq:conformal block once-punctured torus abstract expansion} gives the $q$-expansion of the conformal block.

\paragraph{Normalizability.}
What we have discussed so far defines conformal blocks. However, we should also explain that there is an inner product on conformal blocks and the space of conformal blocks can be turned into a Hilbert space.

The inner product was first defined by \cite{Verlinde:1989ua} and made much more concrete in \cite{Collier:2023fwi}. It takes the schematic form
\be 
\langle \mathcal{F}_1 \, | \, \mathcal{F}_2 \rangle=\int_{\mathcal{T}_{g,n}} Z_{bc} \overline{\mathcal{F}}_1 \mathcal{F}_2\ Z_{26-c}\ , \label{eq:inner product Teichmuller space}
\ee
where $Z_{bc}$ is the ghost partition function familiar from string theory and that provides the integration measure for the integral. In fact, this integral is locally precisely the integral that we would compute in string theory. In particular, one needs to combine the conformal blocks with another CFT partition function/correlation function such that the total central charge is $26$ in order to cancel the conformal anomaly. For blocks with external operators, they also have to satisfy the `mass-shell condition' $\Delta_j=1$. We could take $Z_{26-c}$ in principle to be the partition function of an arbitrary CFT with the correct central charge. However, since we are mostly thinking about the regime of large central charge $c$, the corresponding theory is non-unitary. There is essentially a unique candidate given by timelike Liouville theory (also called Liouville theory with $c \le 1$ or imaginary Liouville theory) \cite{Schomerus:2003vv, Zamolodchikov:2005fy, Kostov:2005kk, Ribault:2015sxa}. Thus the inner product in this form is well-defined for central charge $c \ge 25$.

We will make this inner product more concrete below, but for now, we simply want to remark on which conformal blocks are delta-function normalizable and which are not. The result is that normalizable conformal blocks need to have all internal conformal weights above the threshold
\be 
\Delta \ge \frac{c-1}{24}\ .
\ee
These are precisely the conformal blocks that appear in Liouville theory. Thus we will frequently refer to them as Liouville conformal blocks.

The important point for now is that for $c \ge 25$ we can turn the space of Liouville conformal blocks into a Hilbert space with this inner product. Moreover, Liouville conformal blocks constitute a complete delta-function normalizable basis.

The inner product eq.~\eqref{eq:inner product Teichmuller space} is not yet mathematically rigorously defined at the time of writing. There are two mathematical gaps that we formulate for completeness as conjectures.
For physicists, the validity of these conjectures is essentially folklore and we will assume them in the following.
\begin{conj} \label{conj:timelike Liouville exists}
The timelike Liouville correlation function entering \eqref{eq:inner product Teichmuller space} exists.
\end{conj}
\begin{conj} \label{conj:complete basis}
Liouville conformal blocks form a complete basis with respect to the inner product \eqref{eq:inner product Teichmuller space}.
\end{conj}

\subsection{Some Virasoro representation theory and degenerate blocks}
Going forward, it will be essential to recall some basic properties of the representation theory of the Virasoro algebra.
Depending on the situation, it is useful to introduce the following parametrizations of the central charge and the conformal weight in terms of Liouville momentum:
\begin{subequations}
\begin{align} 
c&=1+6Q^2\ , \label{eq:definition Q}\\
Q&=b+b^{-1}\ ,  \label{eq:definition b}\\
\Delta&=\alpha(Q-\alpha)\ , \label{eq:definition alpha}\\
\alpha&=\frac{Q}{2}+i P\ , \label{eq:definition P} \\
p&=i P\ . \label{eq:definition p}%
\end{align}
\end{subequations}
We assume for most of the time that $c \ge 25$ and hence we can choose $0<b\le 1$. Of course, this parametrization is redundant since e.g.\ $\alpha$ and $Q-\alpha$ describe the same state. We refer to this symmetry as reflection symmetry. We also have the redundancy $b \sim -b \sim b^{-1}$, which is visible throughout all the formulas. Notice that $\Delta \ge \frac{c-1}{24}$ corresponds to $P \in \RR$. We will use in this section mainly the capital $P$, but it will be sometimes more convenient further below to use the lowercase $p$ to avoid many cumbersome factors of $i$.

Some Virasoro representations have null vectors. They are known as degenerate representations. Their conformal Liouville momentum takes the form (up to reflection symmetry)
\be 
p_{\langle m,n\rangle }=-\frac{bm}{2}-\frac{n}{2b}\ , \qquad m\, ,\, n \in \ZZ_{\ge 1}\ . \label{eq:Liouville momenta degenerate representations}
\ee
A primary state with such a conformal weight has a null vector at level $mn$.
In particular, the case $m=n=1$ corresponds to the vacuum state, while the case $m=1$, $n=2$ or $m=2$, $n=1$ correspond to the cases in which the Virasoro representation has a null state at level 2. They will play a very important role below. In general, the null state appears at level $mn$. 

We should emphasize that all degenerate representations have conformal weight $\Delta_{\langle m,n \rangle} \le 0$ and do not appear in normalizable Virasoro blocks.

\paragraph{Degenerate fusion rules.} Since degenerate representations are more restrictive, not all three-point functions involving them can be non-vanishing.
For example, say
\be 
\sum_{\nu \in \PP_{mn}} a_\nu \mathbf{L}_{-\nu} \Psi_{\langle m,n \rangle}
\ee
is the null vector in the representation with conformal weight $\Delta_{\langle m,n \rangle}$ and primary state $\Psi_{\langle m,n \rangle}$. Here $a_\nu$ are some explicit numbers (dependent on $b$) that can be worked out for every degenerate state separately. Since we can repeat the same logic as in \eqref{eq:three-point function removing L generator} to evaluate the three-point function involving the null field, we have a constraint of the form
\be 
0=\sum_{\nu \in \PP_{mn}} a_\nu \langle \Psi_1\, |\,  V_{\Delta_2}(1)\, |\, \mathbf{L}_{-\nu} \Psi_{\langle m,n\rangle} \rangle=P_{\langle m,n \rangle}(\Delta_1,\Delta_2)
\ee
for some polynomial $P_{\langle m,n \rangle}$ of degree $mn$. For given $\Delta_1$, there are $mn$ solutions for $\Delta_2$ and thus, there are only $mn$ different allowed fusion channels. 

They can be worked out by evaluating the allowed fusion rules for the field with a null vector at level 2 and then using the associativity of the fusion product. The result is that (up to reflection symmetry)
\be 
\langle  V_{\Delta_1}(z_1)V_{\Delta_2}(z_2) V_{\Delta_{\langle m,n \rangle}}(z_3) \rangle \ne 0 \quad \Longleftrightarrow\quad p_2=p_1 + \frac{rb}{2}+\frac{s}{2b} \label{eq:degenerate fusion rules}
\ee
where $r \in \{-m+1,-m+3,\dots,m-3,m-1\}$ and $s \in \{-n+1,-n+3,\dots,n-3,n-1\}$.

\paragraph{Implications for conformal blocks.} The structure of degenerate representations has several immediate consequences for the conformal blocks.

If is useful to regard the definition \eqref{eq:conformal block four-punctured sphere abstract expansion} as an analytic function in the parameters $(p_1,p_2,p_3,p_4,p)$ (at fixed $z$). Due to the existence of the degenerate representations, we can determine the pole structure explicitly.

Let us consider the limit $p \to \pm p_{\langle m,n \rangle}$. In this case, the Kac-matrix \eqref{eq:Kac matrix definition} is no longer invertible at level $mn$, because there is a null vector in the Verma module. Consequently, the conformal block develops a pole for $p=\pm p_{\langle m,n \rangle}$.
Here and in the following, it is always understood that $m$ and $n$ are positive integers.

We can of course define conformal blocks with a degenerate intermediate field, but for this $p_1$ and $p_2$, as well as $p_3$ and $p_4$ have to satisfy the degenerate fusion rules \eqref{eq:degenerate fusion rules}.
In this case, the pole in the Kac-matrix is compensated by zeros in both three-punctured sphere because the null vector decouples. Since there is a double-zero, it correctly decouples the null vector. For example, for the vacuum block, we have
\be 
\lim_{\Delta \to 0} \hspace{-.8cm} \begin{tikzpicture}[baseline={([yshift=-.5ex]current bounding box.center)}, xscale=.7, yscale=.7]
        \draw[very thick, Maroon, bend left=30] (-1.8,1.2) to (-.6,0);
        \draw[very thick, Maroon, bend right=30] (-1.8,-1.2) to (-.6,0);
        \draw[very thick, Maroon, bend right=30] (2.2,1.2) to (.6,0);
        \draw[very thick, Maroon, bend left=30] (2.2,-1.2) to (.6,0);
        \draw[very thick, Maroon] (-.6,0) to (.6,0);
        \fill[Maroon] (-.6,0) circle (.07);
        \fill[Maroon] (.6,0) circle (.07);
        \draw[very thick] (-2,1.2) circle (.2 and .5);
        \draw[very thick] (-2,-1.2) circle (.2 and .5);
        \begin{scope}[shift={(2,1.2)}, xscale=.4]
            \draw[very thick] (0,-.5) arc (-90:90:.5);
            \draw[very thick, dashed] (0,.5) arc (90:270:.5);
        \end{scope}
        \begin{scope}[shift={(2,-1.2)}, xscale=.4]
            \draw[very thick] (0,-.5) arc (-90:90:.5);
            \draw[very thick, dashed] (0,.5) arc (90:270:.5);
        \end{scope}
        \begin{scope}[xscale=.5]
            \draw[very thick, RoyalBlue] (0,-.9) arc (-90:90:.9);
            \draw[very thick, dashed, RoyalBlue] (0,.9) arc (90:270:.9);
        \end{scope}
        \draw[very thick, out=0, in=180] (-2,1.7) to (0,.9) to (2,1.7);
        \draw[very thick, out=0, in=180] (-2,-1.7) to (0,-.9) to (2,-1.7);
        \draw[very thick, out=0, in=0, looseness=2.5] (-2,-.7) to (-2,.7);
        \draw[very thick, out=180, in=180, looseness=2.5] (2,-.7) to (2,.7);
        \node at (-2.6,1.2) {$\Delta_3$};
        \node at (2.6,1.2) {$\Delta_1$};            
        \node at (2.6,-1.2) {$\Delta_1$};
        \node at (-2.6,-1.2) {$\Delta_3$}; 
        \node at (0,.3) {$\Delta$};
        \end{tikzpicture} \hspace{-.5cm}
        =
        \sum_{n \ge 0} \sum_{\nu_\text{L},\, \nu_\text{R} \in \tilde{\mathbb{P}}_n} \hspace{-.2cm} 
          \begin{tikzpicture}[baseline={([yshift=-.5ex]current bounding box.center)}]
          \begin{scope}[xscale=.7, yscale=.7]
        \draw[very thick, Maroon, bend left=30] (-1.8,1.2) to (-.6,0);
        			\draw[very thick, Maroon, bend right=30] (-1.8,-1.2) to (-.6,0);
        			\draw[very thick, Maroon] (-.6,0) to (.2,0);
        			\fill[Maroon] (-.6,0) circle (.07);
        			\draw[very thick] (-2,1.2) circle (.2 and .5);
        			\draw[very thick] (-2,-1.2) circle (.2 and .5);
        			\begin{scope}[xscale=.4]
            			\draw[very thick] (0,-.5) arc (-90:90:.5);
            			\draw[very thick, dashed] (0,.5) arc (90:270:.5);
        			\end{scope}
        			\draw[very thick, out=0, in=180] (-2,1.7) to (0,.5);
        			\draw[very thick, out=0, in=180] (-2,-1.7) to (0,-.5);
        			\draw[very thick, out=0, in=0, looseness=2.5] (-2,-.7) to (-2,.7);
        \node at (-2.6,1.2) {$\Delta_3$};
        \node at (-2.6,-1.2) {$\Delta_3$}; 
           \node at (.4,1) {$\mathbf{L}_{-\nu_\text{L}}\!\ket{0}$};
		\end{scope}
   \end{tikzpicture} \hspace{-.8cm}
   \times \mathbf{M}_{\nu_\text{L},\nu_\text{R}} \times 
   \hspace{-.8cm}
       \begin{tikzpicture}[baseline={([yshift=-.5ex]current bounding box.center)}]
        \begin{scope}[xscale=.7, yscale=.7]
        			\draw[very thick, Maroon, bend right=30] (2.2,1.2) to (.6,0);
        			\draw[very thick, Maroon, bend left=30] (2.2,-1.2) to (.6,0);
        			\draw[very thick, Maroon] (.2,0) to (.6,0);
        			\fill[Maroon] (.6,0) circle (.07);
        			\begin{scope}[shift={(2,1.2)}, xscale=.4]
            			\draw[very thick] (0,-.5) arc (-90:90:.5);
            			\draw[very thick, dashed] (0,.5) arc (90:270:.5);
        			\end{scope}
        			\begin{scope}[shift={(2,-1.2)}, xscale=.4]
            			\draw[very thick] (0,-.5) arc (-90:90:.5);
            			\draw[very thick, dashed] (0,.5) arc (90:270:.5);
        			\end{scope}
        			\draw[very thick, out=0, in=180] (0,.5) to (2,1.7);
        			\draw[very thick, out=0, in=180] (0,-.5) to (2,-1.7);
        			\draw[very thick, out=180, in=180, looseness=2.5] (2,-.7) to (2,.7);
           \draw[very thick] (0,0) circle (.2 and .5);
        \node at (2.6,1.2) {$\Delta_1$};            
        \node at (2.6,-1.2) {$\Delta_1$};
           \node at (-.5,1) {$\mathbf{L}_{-\nu_\text{R}}\! \ket{0}$};
		\end{scope}
   \end{tikzpicture}\hspace{-.3cm}.
\ee
Here, $\tilde{\mathbb{P}}_n$ denotes all the integer partitions with $\nu_k \ge 2$, i.e.\ we are correctly only summing over descendants of the Verma module.

The same property is more subtle for the torus conformal block. If we want the vacuum to run in the internal channel, we necessarily need to put the external conformal weight $\Delta_0$ to zero to satisfy the degenerate fusion rules. However, when we put $\Delta_0=0$, the three-point function in eq.~\eqref{eq:three-point function once-punctured torus conformal block} becomes just the Kac-Matrix by definition. Thus we have
\be 
 \begin{tikzpicture}[baseline={([yshift=-.5ex]current bounding box.center)}]
        \draw[very thick, Maroon] (0,-.7) to (0,0);
        \fill[Maroon] (0,0) circle (.07);
        \draw[very thick, Maroon, out=140, in=-60] (0,0) to (-1,.8);
        \draw[very thick, Maroon, out=120, in=180] (-1,.8) to (0,1.8);
        \draw[very thick, Maroon, out=40, in=240] (0,0) to (1,.8);
        \draw[very thick, Maroon, out=60, in=0] (1,.8) to (0,1.8);
        \node at (-.4,.6) {$\Delta$};
        \node at (0,-1) {$\Delta_0=0$};
        \draw[very thick, out=270, in=90] (1.5,1) to (.5,-.5);
        \draw[very thick, out=270, in=90] (-1.5,1) to (-.5,-.5);
        \draw[very thick, out=90, in=180] (-1.5,1) to (0,2.2);
        \draw[very thick, out=0,in=90] (0,2.2) to (1.5,1);
        \draw[very thick, bend left=70] (-.7,1) to (.7,1);
        \draw[very thick, bend right=70] (-.8,1.2) to (.8,1.2);
        \begin{scope}[shift={(0,-.5)}, yscale=.4]
            \draw[very thick] (-.5,0) arc (-180:0:.5);
            \draw[dashed, very thick] (-.5,0) arc (180:0:.5);
        \end{scope}
        \begin{scope}[shift={(0,1.8)}, xscale=.4]
            \draw[very thick, RoyalBlue] (0,-.4) arc (-90:90:.4);
            \draw[dashed, very thick, RoyalBlue] (0,.4) arc (90:270:.4);
        \end{scope}
    \end{tikzpicture} =\sum_{n \ge 0} q^{\Delta+n-\frac{c}{24}}\,  |\mathbb{P}_n|=\frac{q^{\Delta-\frac{c-1}{24}}}{\eta(\tau)}\ , \label{eq:conformal block zero-punctured torus limit}
\ee
which is the usual Virasoro character. This is expected since the Virasoro characters are indeed the conformal blocks on the torus without punctures. However, we see that taking $\Delta \to 0$ does not produce the correct Virasoro vacuum character since the null state at level 1 does not decouple. Thus we need to subtract it explicitly to get the correct vacuum conformal block. The reason why this happens is because there is a pole from the Kac-matrix when we take the limit $\Delta \to 0$. This is only compensated by a single zero from the three-point function, whereas we got two zeros from the two three-point functions for the conformal block on the four-punctured sphere.
Similar comments apply when taking $\Delta$ to values corresponding to other degenerate representations. In general, we always have to explicitly decouple null-states that run in a loop of a Riemann surface, while null-states that run through separating curves decouple automatically.

We can also take one of the external representations to a degenerate value. Consider e.g.
\be 
\lim_{\Delta_4 \to 0}\hspace{-.8cm} \begin{tikzpicture}[baseline={([yshift=-.5ex]current bounding box.center)}, xscale=.7, yscale=.7]
        \draw[very thick, Maroon, bend left=30] (-1.8,1.2) to (-.6,0);
        \draw[very thick, Maroon, bend right=30] (-1.8,-1.2) to (-.6,0);
        \draw[very thick, Maroon, bend right=30] (2.2,1.2) to (.6,0);
        \draw[very thick, Maroon, bend left=30] (2.2,-1.2) to (.6,0);
        \draw[very thick, Maroon] (-.6,0) to (.6,0);
        \fill[Maroon] (-.6,0) circle (.07);
        \fill[Maroon] (.6,0) circle (.07);
        \draw[very thick] (-2,1.2) circle (.2 and .5);
        \draw[very thick] (-2,-1.2) circle (.2 and .5);
        \begin{scope}[shift={(2,1.2)}, xscale=.4]
            \draw[very thick] (0,-.5) arc (-90:90:.5);
            \draw[very thick, dashed] (0,.5) arc (90:270:.5);
        \end{scope}
        \begin{scope}[shift={(2,-1.2)}, xscale=.4]
            \draw[very thick] (0,-.5) arc (-90:90:.5);
            \draw[very thick, dashed] (0,.5) arc (90:270:.5);
        \end{scope}
        \begin{scope}[xscale=.5]
            \draw[very thick, RoyalBlue] (0,-.9) arc (-90:90:.9);
            \draw[very thick, dashed, RoyalBlue] (0,.9) arc (90:270:.9);
        \end{scope}
        \draw[very thick, out=0, in=180] (-2,1.7) to (0,.9) to (2,1.7);
        \draw[very thick, out=0, in=180] (-2,-1.7) to (0,-.9) to (2,-1.7);
        \draw[very thick, out=0, in=0, looseness=2.5] (-2,-.7) to (-2,.7);
        \draw[very thick, out=180, in=180, looseness=2.5] (2,-.7) to (2,.7);
        \node at (-2.6,1.2) {$\Delta_3$};
        \node at (2.6,1.2) {$\Delta_2$};            
        \node at (2.6,-1.2) {$\Delta_1$};
        \node at (-2.6,-1.2) {$\Delta_4$}; 
        \node at (0,.3) {$\Delta$};
        \end{tikzpicture}
\ee
Of course, the block does not correspond to a degenerate block unless $\Delta=\Delta_3$. The power-series expansion \eqref{eq:conformal block four-punctured sphere abstract expansion} does not see this and we get a finite value even when $\Delta \ne \Delta_3$. However, when we put $\Delta=\Delta_3$, we simply get the conformal block on the three-punctured sphere by construction.

\paragraph{The explicit form of the inner product.} The inner product \eqref{eq:inner product Teichmuller space} can be made very concrete, as explained in \cite{Collier:2023fwi}. For a four-punctured sphere, we have for example
\be 
\!\scalebox{1}[1.5]{\Bigg\langle} \!\begin{tikzpicture}[baseline={([yshift=-.5ex]current bounding box.center)}, xscale=.65, yscale=.6]
        \draw[very thick, Maroon, bend left=30] (-1.8,1.2) to (-.6,0);
        \draw[very thick, Maroon, bend right=30] (-1.8,-1.2) to (-.6,0);
        \draw[very thick, Maroon, bend right=30] (2.2,1.2) to (.6,0);
        \draw[very thick, Maroon, bend left=30] (2.2,-1.2) to (.6,0);
        \draw[very thick, Maroon] (-.6,0) to (.6,0);
        \fill[Maroon] (-.6,0) circle (.07);
        \fill[Maroon] (.6,0) circle (.07);
        \draw[very thick] (-2,1.2) circle (.2 and .5);
        \draw[very thick] (-2,-1.2) circle (.2 and .5);
        \begin{scope}[shift={(2,1.2)}, xscale=.4]
            \draw[very thick] (0,-.5) arc (-90:90:.5);
            \draw[very thick, dashed] (0,.5) arc (90:270:.5);
        \end{scope}
        \begin{scope}[shift={(2,-1.2)}, xscale=.4]
            \draw[very thick] (0,-.5) arc (-90:90:.5);
            \draw[very thick, dashed] (0,.5) arc (90:270:.5);
        \end{scope}
        \begin{scope}[xscale=.55]
            \draw[very thick, RoyalBlue] (0,-.9) arc (-90:90:.9);
            \draw[very thick, dashed, RoyalBlue] (0,.9) arc (90:270:.9);
        \end{scope}
        \draw[very thick, out=0, in=180] (-2,1.7) to (0,.9) to (2,1.7);
        \draw[very thick, out=0, in=180] (-2,-1.7) to (0,-.9) to (2,-1.7);
        \draw[very thick, out=0, in=0, looseness=2.5] (-2,-.7) to (-2,.7);
        \draw[very thick, out=180, in=180, looseness=2.5] (2,-.7) to (2,.7);
        \node at (-2,1.2) {3};
        \node at (2.4,1.2) {2};            
        \node at (2.4,-1.2) {1};
        \node at (-2,-1.2) {4}; 
        \node at (0,.35) {21};
        \end{tikzpicture} \!\!
        \scalebox{1}[1.5]{\Bigg|} 
\begin{tikzpicture}[baseline={([yshift=-.5ex]current bounding box.center)}, xscale=.65, yscale=.6]
        \draw[very thick, Maroon, bend left=30] (-1.8,1.2) to (-.6,0);
        \draw[very thick, Maroon, bend right=30] (-1.8,-1.2) to (-.6,0);
        \draw[very thick, Maroon, bend right=30] (2.2,1.2) to (.6,0);
        \draw[very thick, Maroon, bend left=30] (2.2,-1.2) to (.6,0);
        \draw[very thick, Maroon] (-.6,0) to (.6,0);
        \fill[Maroon] (-.6,0) circle (.07);
        \fill[Maroon] (.6,0) circle (.07);
        \draw[very thick] (-2,1.2) circle (.2 and .5);
        \draw[very thick] (-2,-1.2) circle (.2 and .5);
        \begin{scope}[shift={(2,1.2)}, xscale=.4]
            \draw[very thick] (0,-.5) arc (-90:90:.5);
            \draw[very thick, dashed] (0,.5) arc (90:270:.5);
        \end{scope}
        \begin{scope}[shift={(2,-1.2)}, xscale=.4]
            \draw[very thick] (0,-.5) arc (-90:90:.5);
            \draw[very thick, dashed] (0,.5) arc (90:270:.5);
        \end{scope}
        \begin{scope}[xscale=.55]
            \draw[very thick, RoyalBlue] (0,-.9) arc (-90:90:.9);
            \draw[very thick, dashed, RoyalBlue] (0,.9) arc (90:270:.9);
        \end{scope}
        \draw[very thick, out=0, in=180] (-2,1.7) to (0,.9) to (2,1.7);
        \draw[very thick, out=0, in=180] (-2,-1.7) to (0,-.9) to (2,-1.7);
        \draw[very thick, out=0, in=0, looseness=2.5] (-2,-.7) to (-2,.7);
        \draw[very thick, out=180, in=180, looseness=2.5] (2,-.7) to (2,.7);
        \node at (-2,1.2) {3};
        \node at (2.4,1.2) {2};            
        \node at (2.4,-1.2) {1};
        \node at (-2,-1.2) {4}; 
        \node at (0,.35) {$21\hspace{-.04cm}'\hspace{.05cm}$};
        \end{tikzpicture} \!\!
        \scalebox{1}[1.5]{\Bigg\rangle}= \frac{\rho_0(P_{21})^{-1}\,\delta(P_{21}-P_{21}')}{ C_0(P_1,P_2,P_{21}) C_0(P_3,P_4,P_{21})}\ . \label{eq:explicit inner product four-punctured sphere}
\ee
We often use the hopefully intuitive naming convention that internal Liouville momenta in the $s$-channel are named $P_{21}$, in the $t$-channel $P_{32}$ etc. 
Here, $C_0(P_1,P_2,P_3)$ is the DOZZ structure constant in a particular normalization and $\rho_0(P)$ is the inverse two-point function normalization in this convention. See eqs.~\eqref{eq:rho0 definition} and \eqref{eq:C0 formula} for their explicit forms. To state this formula, we also had to use a particular normalization of the timelike Liouville correlation functions.
More generally, we get a factor $\rho_0(P)^{-1}$ for every inner curve defining the pair of pants decomposition and a factor $C_0(P_j,P_k,P_\ell)^{-1}$ for every pair of pants in the decomposition. We also get a delta-function setting all the internal Liouville momenta to be equal. Note that we are assuming all Liouville momenta to be positive and are hence not writing the reflected delta-function $\delta(P_{21}+P_{21}')$.

At this point in the discussion \eqref{eq:explicit inner product four-punctured sphere} is far from obvious, but it will follow up to a $P_{21}$-independent normalization once we have derived the explicit form of the crossing kernels. We will not need \eqref{eq:explicit inner product four-punctured sphere} in the actual derivation.

The simple form if the inner product shows in fact that the Hilbert space under consideration is isomorphic to $L^2(\RR_{\ge 0}^{3g-3+n})$ with the standard Lebesgue measure in a simple way.\footnote{We impose Neumann boundary conditions at the boundary. Alternatively, we could have also taken the space $L^2_\text{sym}(\RR^{3g-3+n})$, where every function is required to be symmetric in every coordinate under reflection.} For example, for the four-punctured sphere, the isomorphism is defined by mapping delta-functions to appropriately normalized conformal blocks
\be 
\delta(\mathsf{P}-P_{21} ) \longmapsto \sqrt{\rho_0(P_{21})C_0(P_1,P_2,P_{21}) C_0(P_3,P_4,P_{21})}\, 
\begin{tikzpicture}[baseline={([yshift=-.5ex]current bounding box.center)}, xscale=.65, yscale=.6]
        \draw[very thick, Maroon, bend left=30] (-1.8,1.2) to (-.6,0);
        \draw[very thick, Maroon, bend right=30] (-1.8,-1.2) to (-.6,0);
        \draw[very thick, Maroon, bend right=30] (2.2,1.2) to (.6,0);
        \draw[very thick, Maroon, bend left=30] (2.2,-1.2) to (.6,0);
        \draw[very thick, Maroon] (-.6,0) to (.6,0);
        \fill[Maroon] (-.6,0) circle (.07);
        \fill[Maroon] (.6,0) circle (.07);
        \draw[very thick] (-2,1.2) circle (.2 and .5);
        \draw[very thick] (-2,-1.2) circle (.2 and .5);
        \begin{scope}[shift={(2,1.2)}, xscale=.4]
            \draw[very thick] (0,-.5) arc (-90:90:.5);
            \draw[very thick, dashed] (0,.5) arc (90:270:.5);
        \end{scope}
        \begin{scope}[shift={(2,-1.2)}, xscale=.4]
            \draw[very thick] (0,-.5) arc (-90:90:.5);
            \draw[very thick, dashed] (0,.5) arc (90:270:.5);
        \end{scope}
        \begin{scope}[xscale=.55]
            \draw[very thick, RoyalBlue] (0,-.9) arc (-90:90:.9);
            \draw[very thick, dashed, RoyalBlue] (0,.9) arc (90:270:.9);
        \end{scope}
        \draw[very thick, out=0, in=180] (-2,1.7) to (0,.9) to (2,1.7);
        \draw[very thick, out=0, in=180] (-2,-1.7) to (0,-.9) to (2,-1.7);
        \draw[very thick, out=0, in=0, looseness=2.5] (-2,-.7) to (-2,.7);
        \draw[very thick, out=180, in=180, looseness=2.5] (2,-.7) to (2,.7);
        \node at (-2,1.2) {3};
        \node at (2.4,1.2) {2};            
        \node at (2.4,-1.2) {1};
        \node at (-2,-1.2) {4}; 
        \node at (0,.35) {21};
        \end{tikzpicture}  \ , \label{eq:isomorphism Hilbert spaces}
\ee
where $\mathsf{P}$ is the coordinate on $\RR_{\ge 0}$. The choice of square root is harmless since the expression under the square root is positive for real $P_j$. Alternatively, we could have of course also considered plane waves as a basis for $L^2$-functions in Fourier space. We will explain the meaning of this isomorphism in Section~\ref{subsec:quantum Teichmuller theory}.

\subsection{Crossing transformations} \label{subsec:crossing transformations}
As we have seen, conformal blocks are defined in one channel. Assuming the validity of Conjecture~\ref{conj:complete basis}, they form a complete basis of the space of $L^2$-normalizable holomorphic sections of the line bundle $\mathscr{L}$. As such, the Hilbert space makes no mention of a channel. This means that we can expand conformal blocks as defined in one channel in terms of conformal blocks in the other channel. For example for the four punctured sphere, we can consider the following two channels. We should for example have a schematic identity of the form
\be 
\begin{tikzpicture}[baseline={([yshift=-.5ex]current bounding box.center)}, xscale=.8, yscale=.7]
        \draw[very thick, Maroon, bend left=30] (-1.8,1.2) to (-.6,0);
        \draw[very thick, Maroon, bend right=30] (-1.8,-1.2) to (-.6,0);
        \draw[very thick, Maroon, bend right=30] (2.2,1.2) to (.6,0);
        \draw[very thick, Maroon, bend left=30] (2.2,-1.2) to (.6,0);
        \draw[very thick, Maroon] (-.6,0) to (.6,0);
        \fill[Maroon] (-.6,0) circle (.07);
        \fill[Maroon] (.6,0) circle (.07);
        \draw[very thick] (-2,1.2) circle (.2 and .5);
        \draw[very thick] (-2,-1.2) circle (.2 and .5);
        \begin{scope}[shift={(2,1.2)}, xscale=.4]
            \draw[very thick] (0,-.5) arc (-90:90:.5);
            \draw[very thick, dashed] (0,.5) arc (90:270:.5);
        \end{scope}
        \begin{scope}[shift={(2,-1.2)}, xscale=.4]
            \draw[very thick] (0,-.5) arc (-90:90:.5);
            \draw[very thick, dashed] (0,.5) arc (90:270:.5);
        \end{scope}
        \begin{scope}[xscale=.4]
            \draw[very thick, RoyalBlue] (0,-.9) arc (-90:90:.9);
            \draw[very thick, dashed, RoyalBlue] (0,.9) arc (90:270:.9);
        \end{scope}
        \draw[very thick, out=0, in=180] (-2,1.7) to (0,.9) to (2,1.7);
        \draw[very thick, out=0, in=180] (-2,-1.7) to (0,-.9) to (2,-1.7);
        \draw[very thick, out=0, in=0, looseness=2.5] (-2,-.7) to (-2,.7);
        \draw[very thick, out=180, in=180, looseness=2.5] (2,-.7) to (2,.7);
        \node at (-2,1.2) {3};
        \node at (2.4,1.2) {2};            
        \node at (2.4,-1.2) {1};
        \node at (-2,-1.2) {4}; 
        \node at (0,.3) {21};
        \end{tikzpicture}        
    \hspace{-.3cm} \sim \int\limits_{\ \Delta_{32}} \hspace{-.7cm}\sum\   \mathbb{F}_{\Delta_{21},\Delta_{32}} \!\begin{bmatrix}
                \Delta_3 & \Delta_2 \\
                \Delta_4 & \Delta_1
    \end{bmatrix}
    \ 
    \begin{tikzpicture}[baseline={([yshift=-.5ex]current bounding box.center)}, xscale=.8, yscale=.7]
        \draw[very thick, Maroon, bend left=10] (-1.8,1.2) to (0,.5);
        \draw[very thick, Maroon, bend right=10] (-1.8,-1.2) to (0,-.5);
        \draw[very thick, Maroon, bend right=10] (2.2,1.2) to (0,.5);
        \draw[very thick, Maroon, bend left=10] (2.2,-1.2) to (0,-.5);
        \draw[very thick, Maroon] (0,-.5) to (0,.5);
        \fill[Maroon] (0,.5) circle (.07);
        \fill[Maroon] (0,-.5) circle (.07);
        \draw[very thick] (-2,1.2) circle (.2 and .5);
        \draw[very thick] (-2,-1.2) circle (.2 and .5);
        \begin{scope}[shift={(2,1.2)}, xscale=.4]
            \draw[very thick] (0,-.5) arc (-90:90:.5);
            \draw[very thick, dashed] (0,.5) arc (90:270:.5);
        \end{scope}
        \begin{scope}[shift={(2,-1.2)}, xscale=.4]
            \draw[very thick] (0,-.5) arc (-90:90:.5);
            \draw[very thick, dashed] (0,.5) arc (90:270:.5);
        \end{scope}
        \begin{scope}[yscale=.33]
            \draw[very thick, RoyalBlue] (-.97,0) arc (-180:0:.97);
            \draw[very thick, dashed, RoyalBlue] (.97,0) arc (0:180:.97);
        \end{scope}
        \draw[very thick, out=0, in=180] (-2,1.7) to (0,.9) to (2,1.7);
        \draw[very thick, out=0, in=180] (-2,-1.7) to (0,-.9) to (2,-1.7);
        \draw[very thick, out=0, in=0, looseness=2.5] (-2,-.7) to (-2,.7);
        \draw[very thick, out=180, in=180, looseness=2.5] (2,-.7) to (2,.7);
        \node at (-2,1.2) {3};
        \node at (2.4,1.2) {2};            
        \node at (2.4,-1.2) {1};
        \node at (-2,-1.2) {4};   
        \node at (.3,0) {32};
    \end{tikzpicture} .
\ee
for some object $\mathbb{F}$. $\mathbb{F}$ goes under different names in the literature: Virasoro crossing kernel, Virasoro fusion kernel, or Virasoro fusion matrix. We will call it the spherical crossing kernel, since the label Virasoro should not be necessary and we want to distinguish it from the modular crossing kernel introduced below. We will drop the label spherical if no confusion is possible. $\mathbb{F}$ of course also depends implicitly on the central charge $c$, which we assume to be fixed. We should also mention that we follow the standard convention of the literature and write the indices of $\mathbb{F}$ in an opposite way from the usual convention for matrix multiplication.

\paragraph{Definition of $\mathbb{F}$.} At this point, it is not clear what conformal weights $\Delta_{32}$ appear on the right-hand side. We will make this formula now precise. 
Thanks to the existence of the inner product on the space of conformal blocks \eqref{eq:inner product Teichmuller space}, we know that it is sufficient to expand the right-hand side only into normalizable blocks. For this reason, it is often more convenient to parametrize the blocks and the fusion matrix $\mathbb{F}$ via their Liouville momentum $P$ as defined in \eqref{eq:definition P}. We can then write
\be 
\begin{tikzpicture}[baseline={([yshift=-.5ex]current bounding box.center)}, xscale=.8, yscale=.7]
        \draw[very thick, Maroon, bend left=30] (-1.8,1.2) to (-.6,0);
        \draw[very thick, Maroon, bend right=30] (-1.8,-1.2) to (-.6,0);
        \draw[very thick, Maroon, bend right=30] (2.2,1.2) to (.6,0);
        \draw[very thick, Maroon, bend left=30] (2.2,-1.2) to (.6,0);
        \draw[very thick, Maroon] (-.6,0) to (.6,0);
        \fill[Maroon] (-.6,0) circle (.07);
        \fill[Maroon] (.6,0) circle (.07);
        \draw[very thick] (-2,1.2) circle (.2 and .5);
        \draw[very thick] (-2,-1.2) circle (.2 and .5);
        \begin{scope}[shift={(2,1.2)}, xscale=.4]
            \draw[very thick] (0,-.5) arc (-90:90:.5);
            \draw[very thick, dashed] (0,.5) arc (90:270:.5);
        \end{scope}
        \begin{scope}[shift={(2,-1.2)}, xscale=.4]
            \draw[very thick] (0,-.5) arc (-90:90:.5);
            \draw[very thick, dashed] (0,.5) arc (90:270:.5);
        \end{scope}
        \begin{scope}[xscale=.4]
            \draw[very thick, RoyalBlue] (0,-.9) arc (-90:90:.9);
            \draw[very thick, dashed, RoyalBlue] (0,.9) arc (90:270:.9);
        \end{scope}
        \draw[very thick, out=0, in=180] (-2,1.7) to (0,.9) to (2,1.7);
        \draw[very thick, out=0, in=180] (-2,-1.7) to (0,-.9) to (2,-1.7);
        \draw[very thick, out=0, in=0, looseness=2.5] (-2,-.7) to (-2,.7);
        \draw[very thick, out=180, in=180, looseness=2.5] (2,-.7) to (2,.7);
        \node at (-2,1.2) {3};
        \node at (2.4,1.2) {2};            
        \node at (2.4,-1.2) {1};
        \node at (-2,-1.2) {4}; 
        \node at (0,.3) {21};
        \end{tikzpicture}        
    \hspace{-.3cm} = \int_0^\infty\!\!  \d P_{32}\    \mathbb{F}_{P_{21},P_{32}} \!\begin{bmatrix}
                P_3 & P_2 \\
                P_4 & P_1
    \end{bmatrix}
    \ 
    \begin{tikzpicture}[baseline={([yshift=-.5ex]current bounding box.center)}, xscale=.8, yscale=.7]
        \draw[very thick, Maroon, bend left=10] (-1.8,1.2) to (0,.5);
        \draw[very thick, Maroon, bend right=10] (-1.8,-1.2) to (0,-.5);
        \draw[very thick, Maroon, bend right=10] (2.2,1.2) to (0,.5);
        \draw[very thick, Maroon, bend left=10] (2.2,-1.2) to (0,-.5);
        \draw[very thick, Maroon] (0,-.5) to (0,.5);
        \fill[Maroon] (0,.5) circle (.07);
        \fill[Maroon] (0,-.5) circle (.07);
        \draw[very thick] (-2,1.2) circle (.2 and .5);
        \draw[very thick] (-2,-1.2) circle (.2 and .5);
        \begin{scope}[shift={(2,1.2)}, xscale=.4]
            \draw[very thick] (0,-.5) arc (-90:90:.5);
            \draw[very thick, dashed] (0,.5) arc (90:270:.5);
        \end{scope}
        \begin{scope}[shift={(2,-1.2)}, xscale=.4]
            \draw[very thick] (0,-.5) arc (-90:90:.5);
            \draw[very thick, dashed] (0,.5) arc (90:270:.5);
        \end{scope}
        \begin{scope}[yscale=.33]
            \draw[very thick, RoyalBlue] (-.97,0) arc (-180:0:.97);
            \draw[very thick, dashed, RoyalBlue] (.97,0) arc (0:180:.97);
        \end{scope}
        \draw[very thick, out=0, in=180] (-2,1.7) to (0,.9) to (2,1.7);
        \draw[very thick, out=0, in=180] (-2,-1.7) to (0,-.9) to (2,-1.7);
        \draw[very thick, out=0, in=0, looseness=2.5] (-2,-.7) to (-2,.7);
        \draw[very thick, out=180, in=180, looseness=2.5] (2,-.7) to (2,.7);
        \node at (-2,1.2) {3};
        \node at (2.4,1.2) {2};            
        \node at (2.4,-1.2) {1};
        \node at (-2,-1.2) {4};   
        \node at (.3,0) {32};
    \end{tikzpicture} . \label{eq:definition F}
\ee
This is the defining relation for the spherical crossing kernel $\mathbb{F}$.

\paragraph{Definition of $\mathbb{S}$.} There is a similar object one can define for the crossing transformation of the conformal blocks of the once-punctured torus. We define the modular crossing kernel (also known as torus crossing kernel, modular fusion kernel or modular kernel) as
\be 
 \begin{tikzpicture}[baseline={([yshift=-.5ex]current bounding box.center)}]
        \draw[very thick, Maroon] (0,-.7) to (0,0);
        \fill[Maroon] (0,0) circle (.07);
        \draw[very thick, Maroon, out=140, in=-60] (0,0) to (-1,.8);
        \draw[very thick, Maroon, out=120, in=180] (-1,.8) to (0,1.8);
        \draw[very thick, Maroon, out=40, in=240] (0,0) to (1,.8);
        \draw[very thick, Maroon, out=60, in=0] (1,.8) to (0,1.8);
        \node at (-.5,.6) {$1$};
        \node at (0,-1) {$0$};
        \draw[very thick, out=270, in=90] (1.5,1) to (.5,-.5);
        \draw[very thick, out=270, in=90] (-1.5,1) to (-.5,-.5);
        \draw[very thick, out=90, in=180] (-1.5,1) to (0,2.2);
        \draw[very thick, out=0,in=90] (0,2.2) to (1.5,1);
        \draw[very thick, bend left=70] (-.7,1) to (.7,1);
        \draw[very thick, bend right=70] (-.8,1.2) to (.8,1.2);
        \begin{scope}[shift={(0,-.5)}, yscale=.4]
            \draw[very thick] (-.5,0) arc (-180:0:.5);
            \draw[dashed, very thick] (-.5,0) arc (180:0:.5);
        \end{scope}
        \begin{scope}[shift={(0,1.8)}, xscale=.4]
            \draw[very thick, RoyalBlue] (0,-.4) arc (-90:90:.4);
            \draw[dashed, very thick, RoyalBlue] (0,.4) arc (90:270:.4);
        \end{scope}
    \end{tikzpicture} =
    \int_0^\infty\!\! \d P_2\  \mathbb{S}_{P_1,P_2}[P_0]\ 
    \begin{tikzpicture}[baseline={([yshift=-.5ex]current bounding box.center)}]
        \draw[very thick, Maroon] (0,-.7) to (0,0);
        \fill[Maroon] (0,0) circle (.07);
        \draw[very thick, Maroon, bend left=20] (0,-.2) to (.3,.8);
        \draw[very thick, Maroon, dashed, bend left=50] (.3,.8) to (.85,.15);
        \draw[very thick, Maroon, bend left=20] (.85,.15) to (0,0);
        \node at (-.1,.5) {2};
        \node at (0,-1) {0};
        \draw[very thick, out=270, in=90] (1.5,1) to (.5,-.5);
        \draw[very thick, out=270, in=90] (-1.5,1) to (-.5,-.5);
        \draw[very thick, out=90, in=180] (-1.5,1) to (0,2.2);
        \draw[very thick, out=0,in=90] (0,2.2) to (1.5,1);
        \draw[very thick, bend left=70] (-.7,1) to (.7,1);
        \draw[very thick, bend right=70] (-.8,1.2) to (.8,1.2);
        \begin{scope}[shift={(0,-.5)}, yscale=.4]
            \draw[very thick] (-.5,0) arc (-180:0:.5);
            \draw[dashed, very thick] (-.5,0) arc (180:0:.5);
        \end{scope}
        \draw[very thick, RoyalBlue] (0,1) circle (1.1 and .8);
    \end{tikzpicture} \label{eq:definition S}
\ee

\paragraph{Definition of $\mathbb{B}$ and $\mathbb{T}$.} Two more objects are useful to define. The braiding move is defined as
\be 
\begin{tikzpicture}[baseline={([yshift=-.5ex]current bounding box.center)}, scale=.8]
        \draw[very thick, Maroon] (0,-1.2) to (0,-.2);
        \fill[Maroon] (0,-.2) circle (.07);
        \draw[very thick, Maroon, bend left=20] (0,-.2) to (-1,.8);
        \draw[very thick, Maroon, bend right=20] (0,-.2) to (1,.8);
        \node at (-.4,.4) {1};
        \node at (.4,.4) {2};
        \node at (.17,-.55) {3};
        \draw[very thick, out=270, in=90] (1.5,1) to (.5,-1);
        \draw[very thick, out=270, in=90] (-1.5,1) to (-.5,-1);
        \draw[very thick, out=270, in=270, looseness=1.5] (-.5,1) to (.5,1);
        \draw[very thick] (-1,1) circle (.5 and .2); 
        \draw[very thick] (1,1) circle (.5 and .2);
        \begin{scope}[shift={(0,-1)}, yscale=.4]
            \draw[very thick] (-.5,0) arc (-180:0:.5);
            \draw[dashed, very thick] (-.5,0) arc (180:0:.5);
        \end{scope}
    \end{tikzpicture}
    \ = \mathbb{B}^{P_1,P_2}_{P_3} \ 
    \begin{tikzpicture}[baseline={([yshift=-.5ex]current bounding box.center)}, scale=.8]
        \draw[very thick, Maroon] (0,-1.2) to (0,-.2);
        \fill[Maroon] (0,-.2) circle (.07);
        \draw[very thick, Maroon, bend left=20] (0,-.2) to (-1,.8);
        \draw[very thick, Maroon, bend left=20] (0,-.2) to (-1.1,.1);
        \draw[very thick, dashed, Maroon, bend left=30] (-1.1,.1) to (-.4,.7);
        \draw[very thick, Maroon] (-.4,.7) .. controls (0,.1) and (.5,.1) .. (1,.8);
        \node at (-.27,.22) {1};
        \node at (.5,.08) {2};
        \node at (.17,-.55) {3};
        \draw[very thick, out=270, in=90] (1.5,1) to (.5,-1);
        \draw[very thick, out=270, in=90] (-1.5,1) to (-.5,-1);
        \draw[very thick, out=270, in=270, looseness=1.5] (-.5,1) to (.5,1);
        \draw[very thick] (-1,1) circle (.5 and .2); 
        \draw[very thick] (1,1) circle (.5 and .2);
        \begin{scope}[shift={(0,-1)}, yscale=.4]
            \draw[very thick] (-.5,0) arc (-180:0:.5);
            \draw[dashed, very thick] (-.5,0) arc (180:0:.5);
        \end{scope}
    \end{tikzpicture} \label{eq:definition B}
\ee
In this case, there is a single block on the three-punctured sphere and thus we do not have an integral on the right-hand side. In fact, it is elementary to determine $\mathbb{B}_{P_1,P_2}^{P_3}$ from the general structure of the three-point function on the sphere given in eq.~\eqref{eq:conformal block three-punctured sphere}.
Performing the braiding move twice corresponds to taking $z_2$ clockwise around $z_1$. Thus we have
\begin{align} 
\begin{tikzpicture}[baseline={([yshift=-.5ex]current bounding box.center)}, scale=.8]
        \draw[very thick, Maroon] (0,-1.2) to (0,-.2);
        \fill[Maroon] (0,-.2) circle (.07);
        \draw[very thick, Maroon, bend left=20] (0,-.2) to (-1,.8);
        \draw[very thick, Maroon, bend right=20] (0,-.2) to (1,.8);
        \node at (-.4,.4) {1};
        \node at (.4,.4) {2};
        \node at (.17,-.55) {3};
        \draw[very thick, out=270, in=90] (1.5,1) to (.5,-1);
        \draw[very thick, out=270, in=90] (-1.5,1) to (-.5,-1);
        \draw[very thick, out=270, in=270, looseness=1.5] (-.5,1) to (.5,1);
        \draw[very thick] (-1,1) circle (.5 and .2); 
        \draw[very thick] (1,1) circle (.5 and .2);
        \begin{scope}[shift={(0,-1)}, yscale=.4]
            \draw[very thick] (-.5,0) arc (-180:0:.5);
            \draw[dashed, very thick] (-.5,0) arc (180:0:.5);
        \end{scope}
    \end{tikzpicture}
    &=
    \mathbb{B}^{P_1,P_2}_{P_3}\, \mathbb{B}^{P_2,P_1}_{P_3}
    \begin{tikzpicture}[baseline={([yshift=-.5ex]current bounding box.center)}, scale=.8]
        \draw[very thick, Maroon] (0,-1.2) to (0,-.2);
        \fill[Maroon] (0,-.2) circle (.07);
        \draw[very thick, Maroon, bend right=10] (0,-.2) to (-.63,-.5);
        \draw[very thick, Maroon, dashed, bend right=20]
        (-.63,-.5) to (.63,-.5);
        \draw[very thick, Maroon] (.63,-.5) .. controls  (.5,0) and (-.5,0) .. (-.4,.7);
        \draw[very thick, Maroon, dashed, out=200, in=-45] (-.4,.7) to (-1.3,.4);
        \draw[very thick, Maroon, bend right=50] (-1.3,.4) to (-1,.8);
        \draw[very thick, Maroon, bend left=10] (0,-.2) to (-.8,-.2);
        \draw[very thick, dashed, Maroon, bend right=10] (-.8,-.2) to (.4,.7);
        \draw[very thick, Maroon, bend right=50] (.4,.7) to (1,.8);
        \node at (-.5,.38) {1};
        \node at (1,.5) {2};
        \node at (0,-1.45) {3};
        \draw[very thick, out=270, in=90] (1.5,1) to (.5,-1);
        \draw[very thick, out=270, in=90] (-1.5,1) to (-.5,-1);
        \draw[very thick, out=270, in=270, looseness=1.5] (-.5,1) to (.5,1);
        \draw[very thick] (-1,1) circle (.5 and .2); 
        \draw[very thick] (1,1) circle (.5 and .2);
        \begin{scope}[shift={(0,-1)}, yscale=.4]
            \draw[very thick] (-.5,0) arc (-180:0:.5);
            \draw[dashed, very thick] (-.5,0) arc (180:0:.5);
        \end{scope}
    \end{tikzpicture} \\
    &=
    \mathbb{B}^{P_1,P_2}_{P_3}\, \mathbb{B}^{P_2,P_1}_{P_3}\, 
    \mathrm{e}^{2\pi i (\Delta_1+\Delta_2-\Delta_3)}
    \begin{tikzpicture}[baseline={([yshift=-.5ex]current bounding box.center)}, scale=.8]
        \draw[very thick, Maroon] (0,-1.2) to (0,-.2);
        \fill[Maroon] (0,-.2) circle (.07);
        \draw[very thick, Maroon, bend left=20] (0,-.2) to (-1,.8);
        \draw[very thick, Maroon, bend right=20] (0,-.2) to (1,.8);
        \node at (-.4,.4) {1};
        \node at (.4,.4) {2};
        \node at (.17,-.55) {3};
        \draw[very thick, out=270, in=90] (1.5,1) to (.5,-1);
        \draw[very thick, out=270, in=90] (-1.5,1) to (-.5,-1);
        \draw[very thick, out=270, in=270, looseness=1.5] (-.5,1) to (.5,1);
        \draw[very thick] (-1,1) circle (.5 and .2); 
        \draw[very thick] (1,1) circle (.5 and .2);
        \begin{scope}[shift={(0,-1)}, yscale=.4]
            \draw[very thick] (-.5,0) arc (-180:0:.5);
            \draw[dashed, very thick] (-.5,0) arc (180:0:.5);
        \end{scope}
    \end{tikzpicture}\ ,
\end{align}
since the twisted picture is related to the untwisted one by moving $z_2$ around $z_1$. $\mathbb{B}_{P_3}^{P_1,P_2}$ is also symmetric in $P_1 \leftrightarrow P_2$ since we could have reflected the picture.
Thus we have
\be 
    \mathbb{B}^{P_1,P_2}_{P_3}= \pm\,  \mathrm{e}^{\pi i (-\Delta_1-\Delta_2+\Delta_3)}\ . \label{eq:braiding phase}
\ee
The sign has to be the $+$ sign by continuity. Indeed, it has to depend continuously on $\Delta_j$ and has to be trivial for $\Delta_2=0$, since braiding does nothing for the identity.

Another important move is a Dehn twist, which is obtained by moving a line around a puncture as follows:
\be 
    \begin{tikzpicture}[baseline={([yshift=-.5ex]current bounding box.center)}, scale=.8]
        \begin{scope}
            \draw[very thick, Maroon] (0,-1.2) to (0,-.2);
            \fill[Maroon] (0,-.2) circle (.07);
            \draw[very thick, Maroon, bend left=20] (0,-.2) to (-1,.8);
            \draw[very thick, Maroon, bend right=20] (0,-.2) to (1,.8);
            \node at (-.4,.4) {1};
            \node at (.4,.4) {2};
            \node at (.17,-.55) {3};
            \draw[very thick, out=270, in=90] (1.5,1) to (.5,-1);
            \draw[very thick, out=270, in=90] (-1.5,1) to (-.5,-1);
            \draw[very thick, out=270, in=270, looseness=1.5] (-.5,1) to (.5,1);
            \draw[very thick] (-1,1) circle (.5 and .2); 
            \draw[very thick] (1,1) circle (.5 and .2);
            \begin{scope}[shift={(0,-1)}, yscale=.4]
                \draw[very thick] (-.5,0) arc (-180:0:.5);
                \draw[dashed, very thick] (-.5,0) arc (180:0:.5);
            \end{scope}
        \end{scope}
        \end{tikzpicture}
        = \mathbb{T}_{P_2}
        \begin{tikzpicture}[baseline={([yshift=-.5ex]current bounding box.center)}, scale=.8]
        \begin{scope}
            \draw[very thick, Maroon] (0,-1.2) to (0,-.2);
            \fill[Maroon] (0,-.2) circle (.07);
            \draw[very thick, Maroon, bend left=20] (0,-.2) to (-1,.8);
            \draw[very thick, Maroon, bend right=10] (0,-.2) to (.9,-.1);
            \draw[very thick, dashed, Maroon] (.9,-.1) .. controls (.7,.5) and (.6,.1) .. (.4,.7);
            \draw[very thick, Maroon, out=-30, in=-90] (.4,.7) to (1,.8);
            \node at (-.9,.3) {1};
            \node at (.5,.1) {2};
            \node at (-.17,-.55) {3};
            \draw[very thick, out=270, in=90] (1.5,1) to (.5,-1);
            \draw[very thick, out=270, in=90] (-1.5,1) to (-.5,-1);
            \draw[very thick, out=270, in=270, looseness=1.5] (-.5,1) to (.5,1);
            \draw[very thick] (-1,1) circle (.5 and .2); 
            \draw[very thick] (1,1) circle (.5 and .2);
            \begin{scope}[shift={(0,-1)}, yscale=.4]
                \draw[very thick] (-.5,0) arc (-180:0:.5);
                \draw[dashed, very thick] (-.5,0) arc (180:0:.5);
            \end{scope}
        \end{scope}
    \end{tikzpicture}\ . \label{eq:definition T}
\ee
$\mathbb{T}$ is related in a simple way to the braiding move, since we have
\be 
    \begin{tikzpicture}[baseline={([yshift=-.5ex]current bounding box.center)}, scale=.8]
        \begin{scope}
            \draw[very thick, Maroon] (0,-1.2) to (0,-.2);
            \fill[Maroon] (0,-.2) circle (.07);
            \draw[very thick, Maroon, bend left=20] (0,-.2) to (-1,.8);
            \draw[very thick, Maroon, bend right=20] (0,-.2) to (1,.8);
            \node at (-.4,.4) {1};
            \node at (.4,.4) {2};
            \node at (.17,-.55) {3};
            \draw[very thick, out=270, in=90] (1.5,1) to (.5,-1);
            \draw[very thick, out=270, in=90] (-1.5,1) to (-.5,-1);
            \draw[very thick, out=270, in=270, looseness=1.5] (-.5,1) to (.5,1);
            \draw[very thick] (-1,1) circle (.5 and .2); 
            \draw[very thick] (1,1) circle (.5 and .2);
            \begin{scope}[shift={(0,-1)}, yscale=.4]
                \draw[very thick] (-.5,0) arc (-180:0:.5);
                \draw[dashed, very thick] (-.5,0) arc (180:0:.5);
            \end{scope}
        \end{scope}
    \end{tikzpicture}
    =\mathbb{B}_{P_1,P_2}^{P_3}
    \begin{tikzpicture}[baseline={([yshift=-.5ex]current bounding box.center)}, scale=.8]
        \begin{scope}
            \draw[very thick, Maroon] (0,-1.2) to (0,-.2);
            \fill[Maroon] (0,-.2) circle (.07);
            \draw[very thick, Maroon, bend left=20] (0,-.2) to (-1,.8);
            \draw[very thick, Maroon, bend left=20] (0,-.2) to (-1.1,.1);
            \draw[very thick, dashed, Maroon, bend left=30] (-1.1,.1) to (-.4,.7);
            \draw[very thick, Maroon] (-.4,.7) .. controls (0,.1) and (.5,.1) .. (1,.8);
            \node at (-.27,.2) {1};
            \node at (.5,.05) {2};
            \node at (.17,-.55) {3};
            \draw[very thick, out=270, in=90] (1.5,1) to (.5,-1);
            \draw[very thick, out=270, in=90] (-1.5,1) to (-.5,-1);
            \draw[very thick, out=270, in=270, looseness=1.5] (-.5,1) to (.5,1);
            \draw[very thick] (-1,1) circle (.5 and .2); 
            \draw[very thick] (1,1) circle (.5 and .2);
            \begin{scope}[shift={(0,-1)}, yscale=.4]
                \draw[very thick] (-.5,0) arc (-180:0:.5);
                \draw[dashed, very thick] (-.5,0) arc (180:0:.5);
            \end{scope}
        \end{scope}
        \end{tikzpicture}
        =
\mathbb{B}_{P_1,P_2}^{P_3}\mathbb{B}_{P_2,P_3}^{P_1}
        \begin{tikzpicture}[baseline={([yshift=-.5ex]current bounding box.center)}, scale=.8]
        \begin{scope}
            \draw[very thick, Maroon] (0,-1.2) to (0,-.2);
            \fill[Maroon] (0,-.2) circle (.07);
            \draw[very thick, Maroon, bend left=20] (0,-.2) to (-1,.8);
            \draw[very thick, Maroon, bend left=20] (0,-.2) to (.6,-.5);
            \draw[very thick, dashed, Maroon, bend left=70] (.6,-.5) to (-.4,.7);
            \draw[very thick, Maroon] (-.4,.7) .. controls (0,.1) and (.5,.1) .. (1,.8);
            \node at (-.9,.3) {1};
            \node at (.5,.05) {2};
            \node at (-.17,-.55) {3};
            \draw[very thick, out=270, in=90] (1.5,1) to (.5,-1);
            \draw[very thick, out=270, in=90] (-1.5,1) to (-.5,-1);
            \draw[very thick, out=270, in=270, looseness=1.5] (-.5,1) to (.5,1);
            \draw[very thick] (-1,1) circle (.5 and .2); 
            \draw[very thick] (1,1) circle (.5 and .2);
            \begin{scope}[shift={(0,-1)}, yscale=.4]
                \draw[very thick] (-.5,0) arc (-180:0:.5);
                \draw[dashed, very thick] (-.5,0) arc (180:0:.5);
            \end{scope}
        \end{scope}
    \end{tikzpicture}\ . \label{eq:B twice relation to T}
\ee
Comparing \eqref{eq:definition T} and \eqref{eq:B twice relation to T} gives
\be 
\mathbb{T}_{P_2}=\mathbb{B}_{P_1,P_2}^{P_3}\mathbb{B}_{P_2,P_3}^{P_1}=\mathrm{e}^{-2\pi i \Delta_2}\ . \label{eq:Dehn twist phase}
\ee
This follows also directly from the fact that $\mathrm{e}^{2\pi i L_0}$ is the operator implementing the Dehn twist.

We discuss these `trivial' crossing transformations in more detail for the four-punctured sphere in Appendix~\ref{app:four-punctured sphere}, where we make them completely explicit.

\paragraph{General crossing transformations.} Starting from these building blocks, one can study the crossing transformations of conformal blocks on arbitrary Riemann surfaces. To do so, let us fix a pair of pants decomposition on the surface $\Sigma_{g,n}$. Consider then the union of two adjacent three-punctured spheres appearing in the pair of pants decomposition. They form together a subsurface which is either a four-punctured sphere or a once-punctured torus. We can then consider a crossing transformation on this subsurface. The crossing transformations are described by the same objects $\mathbb{F}$, $\mathbb{S}$, $\mathbb{B}$ or $\mathbb{T}$ that we already defined.
For example, we have on the two-punctured torus
\be 
    \!\!\!\begin{tikzpicture}[baseline={([yshift=-.5ex]current bounding box.center)}]
        \begin{scope}[xscale=.7, yscale=.7]
        			\draw[very thick, Maroon, bend right=30] (2.2,1.2) to (.6,0);
        			\draw[very thick, Maroon, bend left=30] (2.2,-1.2) to (.6,0);
        			\draw[very thick, Maroon] (-.6,0) to (.6,0);
        			\fill[Maroon] (-.6,0) circle (.07);
        			\fill[Maroon] (.6,0) circle (.07);
                    \draw[very thick, Maroon, out=135, in=90, looseness=1.2] (-.6,0) to (-3.15,0);
                    \draw[very thick, Maroon, out=-135, in=-90, looseness=1.2] (-.6,0) to (-3.15,0);
                    \draw[very thick, out=180, in=180, looseness=2] (-2,1.7) to (-2,-1.7);
                    \draw[very thick, bend right=40] (-2.9,0.1) to (-1.1,.1);
        		    \draw[very thick, bend left=40] (-2.75,0) to (-1.25,0);
        			\begin{scope}[shift={(2,1.2)}, xscale=.4]
            			\draw[very thick] (0,-.5) arc (-90:90:.5);
            			\draw[very thick, dashed] (0,.5) arc (90:270:.5);
        			\end{scope}
        			\begin{scope}[shift={(2,-1.2)}, xscale=.4]
            			\draw[very thick] (0,-.5) arc (-90:90:.5);
            			\draw[very thick, dashed] (0,.5) arc (90:270:.5);
        			\end{scope}
        			\begin{scope}[xscale=.4]
            			\draw[very thick, RoyalBlue] (0,-.9) arc (-90:90:.9);
            			\draw[very thick, dashed, RoyalBlue] (0,.9) arc (90:270:.9);
        			\end{scope}
                    \draw[very thick, RoyalBlue] (-2.75,0) arc (0:180:.61 and .3);
                    \draw[very thick, dashed, RoyalBlue] (-2.75,0) arc (0:-180:.61 and .3);
        			\draw[very thick, out=0, in=180] (-2,1.7) to (0,.9) to (2,1.7);
        			\draw[very thick, out=0, in=180] (-2,-1.7) to (0,-.9) to (2,-1.7);
        			\draw[very thick, out=180, in=180, looseness=2.5] (2,-.7) to (2,.7);
           \node at (2.45,-1.2) {$1$};
           \node at (2.45,1.2) {$2$};
           \node at (0,.3) {$\Delta$};
           \node at (-3.45,.68) {$\Delta_1'$};
		\end{scope}
   \end{tikzpicture} \!\!\!\! =\int_0^\infty\!\! \d P \ \mathbb{S}_{P_1',P_2'}[P] \! \!\! 
       \begin{tikzpicture}[baseline={([yshift=-.5ex]current bounding box.center)}]
        \begin{scope}[xscale=.7, yscale=.7]
        			\draw[very thick, Maroon, bend right=30] (2.2,1.2) to (.6,0);
        			\draw[very thick, Maroon, bend left=30] (2.2,-1.2) to (.6,0);
        			\draw[very thick, Maroon] (-.6,0) to (.6,0);
                    \draw[very thick, Maroon, bend right=20] (-1.25,0) to (-.6,0);
        			\fill[Maroon] (-.6,0) circle (.07);
        			\fill[Maroon] (.6,0) circle (.07);
                    \draw[very thick, Maroon, bend right=30] (-.6,0) to (-1,1.3);
                    \draw[very thick, Maroon, dashed, bend right=30] (-1,1.3) to (-1.25,0);
                    \draw[very thick, out=180, in=180, looseness=2] (-2,1.7) to (-2,-1.7);
                    \draw[very thick, bend right=40] (-2.9,0.1) to (-1.1,.1);
        		    \draw[very thick, bend left=40] (-2.75,0) to (-1.25,0);
        			\begin{scope}[shift={(2,1.2)}, xscale=.4]
            			\draw[very thick] (0,-.5) arc (-90:90:.5);
            			\draw[very thick, dashed] (0,.5) arc (90:270:.5);
        			\end{scope}
        			\begin{scope}[shift={(2,-1.2)}, xscale=.4]
            			\draw[very thick] (0,-.5) arc (-90:90:.5);
            			\draw[very thick, dashed] (0,.5) arc (90:270:.5);
        			\end{scope}
        			\begin{scope}[xscale=.4]
            			\draw[very thick, RoyalBlue] (0,-.9) arc (-90:90:.9);
            			\draw[very thick, dashed, RoyalBlue] (0,.9) arc (90:270:.9);
        			\end{scope}
                    \draw[very thick, RoyalBlue] (-2,0) circle (1.2 and .6);
        			\draw[very thick, out=0, in=180] (-2,1.7) to (0,.9) to (2,1.7);
        			\draw[very thick, out=0, in=180] (-2,-1.7) to (0,-.9) to (2,-1.7);
        			\draw[very thick, out=180, in=180, looseness=2.5] (2,-.7) to (2,.7);
           \node at (2.45,-1.2) {$1$};
           \node at (2.45,1.2) {$2$};
           \node at (0,.3) {$\Delta$};
           \node at (-1.7,1) {$\Delta_2'$};
		\end{scope}
   \end{tikzpicture}\ .
\ee
The fact that the same crossing kernel appears follows directly from the power series definition of conformal blocks \eqref{eq:general conformal block power series expansion} since we can always reduce the conformal blocks to their basic building blocks. We can also always remove descendants by using Virasoro symmetry. Since this is independent of the conformal block channel, conformal blocks of descendants have the same crossing properties as the conformal blocks of their primaries.

\paragraph{Unitarity.} The crossing transformations actually define \emph{unitary} operators on the Hilbert space of Liouville conformal blocks endowed with the inner product \eqref{eq:inner product Teichmuller space}. Indeed, this follows directly from the fact that the inner product makes no reference to a conformal block decomposition and we can just change the integration variable in the integral over Teichm\"uller space to compensate for the crossing transformation.

\paragraph{Analyticity.} Initially, the crossing kernels $\mathbb{F}$ and $\mathbb{S}$ are defined for real values of all the Liouville momenta. However, as we discussed above, conformal blocks are meromorphic functions in all the momenta $P_j$. This immediately implies that for fixed $P_{32}$, the spherical crossing kernel
\be 
\mathbb{F}_{P_{21},P_{32}}\! \begin{bmatrix}
    P_3 & P_2 \\ P_4 & P_1
\end{bmatrix}
\ee
can be extended to a meromorphic function in the complex plane in all the other arguments. The only poles are located where the conformal block itself has a pole, which happens when the internal conformal weight coincides with a degenerate value, i.e.\ $P_{21}= \pm P_{\langle m,n \rangle}$. This then also implies the same property for $P_{32}$, since the spherical crossing kernel has a certain tetrahedral symmetry exchanging the roles of $P_{21}$ and $P_{32}$, see eq.~\eqref{eq:tetrahedrally symmetric combination} below. Similar arguments show that the modular crossing kernel can be extended to a meromorphic function in all three Liouville momenta.

\paragraph{Existence.}
Let us comment on one very important difference compared to the approaches that have appeared in the literature \cite{Ponsot:1999uf, Ponsot:2000mt}. We exhibited the space of Liouville conformal blocks as a Hilbert space with a well-defined inner product.
This makes it obvious that the crossing kernels $\mathbb{F}$ and $\mathbb{S}$ need to exist since we are guaranteed that a state can be expanded in terms of the complete basis.\footnote{Modulo the conjectures~\ref{conj:timelike Liouville exists} and \ref{conj:complete basis}.} Without the inner product, it is not obvious why we should restrict our attention only to Liouville conformal blocks.
In particular this also means that once we determine explicit formulas for the crossing kernels in Section~\ref{sec:bootstrapping the crossing kernels}, they are guaranteed to satisfy all the consistency conditions that we will spell out in the next Subsection~\ref{subsec:Moore Seiberg consistency conditions}. In other approaches it is usually very hard to demonstrate these identities and the only complete proof in the literature proceeds by translating them to a different basis of the Hilbert space where the relations are automatic \cite{Teschner:2013tqy}. For genus 0 blocks, consistency follows also from the representation theory of the quantum group $\mathcal{U}_q(\mathfrak{sl}(2,\mathbb{R}))$ \cite{Ponsot:2000mt,Nidaiev:2013bda}.

\subsection{General consistency conditions} \label{subsec:Moore Seiberg consistency conditions}
The objects $\mathbb{F}$ and $\mathbb{S}$ (as well as the simple phases $\mathbb{B}$ and $\mathbb{T}$) satisfy a number of general consistency conditions that were first described by Moore and Seiberg \cite{Moore:1988qv}. The resulting structure is often called the Moore-Seiberg groupoid. Mathematically speaking, it is a groupoid since the mapping class group acts faithfully on the set of decorated pair of pants decomposition as in Figure~\ref{fig:four-punctured sphere pair of pants decomposition}. The different orbits are described by different trivalent graphs. We will however not make use of this terminology in the following. 

There are in total six consistency conditions, besides the explicit formulas for the braiding $\mathbb{B}$ given in eq.~\eqref{eq:braiding phase} and for the Dehn twist $\mathbb{T}$ given in eq.~\eqref{eq:Dehn twist phase} that we already explained how to derive. These consistency conditions arise by considering the basic transformations on various surfaces of low genus. There are two consistency conditions coming from the four-punctured sphere, one from the five-punctured sphere, two from the once-punctured torus and one from the two-punctured torus. We explain them in turn. They are given in eqs.~\eqref{eq:g=0, n=4 idempotency F}, \eqref{eq:g=0, n=4 hexagon equation}, \eqref{eq:g=0, n=5 pentagon equation}, \eqref{eq:g=1, n=1 idempotency S}, \eqref{eq:g=1, n=1 (ST)^3} and \eqref{eq:g=1, n=2 relation}.
It was argued in \cite{Moore:1988qv} and mathematically proven in \cite{Bakalov:2000} that these consistency conditions are complete. We will now explain them in turn.

\paragraph{Symmetries of $\mathbb{F}$.} There are a couple of obvious symmetries of $\mathbb{F}$ that follow directly from the definition in eq.~\eqref{eq:definition F}, since we can rotate and reflect the picture. They are
\be 
\mathbb{F}_{P_{21},P_{32}}\!\begin{bmatrix}
    P_3 & P_2 \\ P_4 & P_1
\end{bmatrix}=
\mathbb{F}_{P_{21},P_{32}}\!\begin{bmatrix}
    P_2 & P_3 \\ P_1 & P_4
\end{bmatrix}=
\mathbb{F}_{P_{21},P_{32}}\!\begin{bmatrix}
    P_4 & P_1 \\ P_3 & P_2
\end{bmatrix} \label{eq:obvious symmetries}
\ee
We should also recall that due to reflection symmetry, any quantity is invariant under the replacement $P_j \to -P_j$. We hence restrict to positive Liouville momenta.

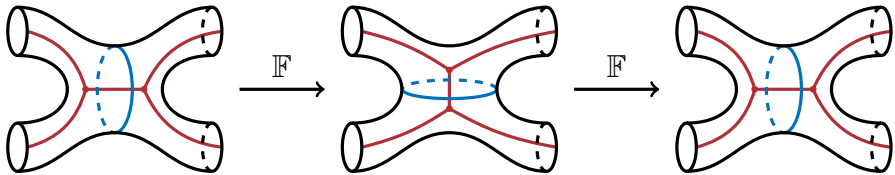
\begin{figure}[ht]
	\centering
	    \begin{tikzpicture}[scale=.8]
	    		\begin{scope}[scale=.8]
        			\draw[very thick, Maroon, bend left=30] (-1.8,1.2) to (-.6,0);
        			\draw[very thick, Maroon, bend right=30] (-1.8,-1.2) to (-.6,0);
        			\draw[very thick, Maroon, bend right=30] (2.2,1.2) to (.6,0);
        			\draw[very thick, Maroon, bend left=30] (2.2,-1.2) to (.6,0);
        			\draw[very thick, Maroon] (-.6,0) to (.6,0);
        			\fill[Maroon] (-.6,0) circle (.07);
        			\fill[Maroon] (.6,0) circle (.07);
        			\draw[very thick] (-2,1.2) circle (.2 and .5);
        			\draw[very thick] (-2,-1.2) circle (.2 and .5);
        			\begin{scope}[shift={(2,1.2)}, xscale=.4]
            			\draw[very thick] (0,-.5) arc (-90:90:.5);
            			\draw[very thick, dashed] (0,.5) arc (90:270:.5);
        			\end{scope}
        			\begin{scope}[shift={(2,-1.2)}, xscale=.4]
            			\draw[very thick] (0,-.5) arc (-90:90:.5);
            			\draw[very thick, dashed] (0,.5) arc (90:270:.5);
        			\end{scope}
        			\begin{scope}[xscale=.4]
            			\draw[very thick, RoyalBlue] (0,-.9) arc (-90:90:.9);
            			\draw[very thick, RoyalBlue, dashed] (0,.9) arc (90:270:.9);
        			\end{scope}
        			\draw[very thick, out=0, in=180] (-2,1.7) to (0,.9) to (2,1.7);
        			\draw[very thick, out=0, in=180] (-2,-1.7) to (0,-.9) to (2,-1.7);
        			\draw[very thick, out=0, in=0, looseness=2.5] (-2,-.7) to (-2,.7);
        			\draw[very thick, out=180, in=180, looseness=2.5] (2,-.7) to (2,.7);
			\end{scope}
			\begin{scope}[shift={(5.5,0)}, scale=.8]
        			\draw[very thick, Maroon, bend left=10] (-1.8,1.2) to (0,.4);
        			\draw[very thick, Maroon, bend right=10] (-1.8,-1.2) to (0,-.4);
        			\draw[very thick, Maroon, bend right=10] (2.2,1.2) to (0,.4);
        			\draw[very thick, Maroon, bend left=10] (2.2,-1.2) to (0,-.4);
        			\draw[very thick, Maroon] (0,-.4) to (0,.4);
        			\fill[Maroon] (0,.4) circle (.07);
        			\fill[Maroon] (0,-.4) circle (.07);
        			\draw[very thick] (-2,1.2) circle (.2 and .5);
        			\draw[very thick] (-2,-1.2) circle (.2 and .5);
        			\begin{scope}[shift={(2,1.2)}, xscale=.4]
        			    \draw[very thick] (0,-.5) arc (-90:90:.5);
        			    \draw[very thick, dashed] (0,.5) arc (90:270:.5);
        			\end{scope}
        			\begin{scope}[shift={(2,-1.2)}, xscale=.4]
        			    \draw[very thick] (0,-.5) arc (-90:90:.5);
        			    \draw[very thick, dashed] (0,.5) arc (90:270:.5);
        			\end{scope}
        			\begin{scope}[yscale=.2]
        			    \draw[very thick, RoyalBlue] (-.97,0) arc (-180:0:.97);
        			    \draw[very thick, dashed, RoyalBlue] (.97,0) arc (0:180:.97);
        			\end{scope}
        			\draw[very thick, out=0, in=180] (-2,1.7) to (0,.9) to (2,1.7);
        			\draw[very thick, out=0, in=180] (-2,-1.7) to (0,-.9) to (2,-1.7);
        			\draw[very thick, out=0, in=0, looseness=2.5] (-2,-.7) to (-2,.7);
        			\draw[very thick, out=180, in=180, looseness=2.5] (2,-.7) to (2,.7);   
        		\end{scope}
	    		\begin{scope}[shift={(11,0)}, scale=.8]
        			\draw[very thick, Maroon, bend left=30] (-1.8,1.2) to (-.6,0);
        			\draw[very thick, Maroon, bend right=30] (-1.8,-1.2) to (-.6,0);
        			\draw[very thick, Maroon, bend right=30] (2.2,1.2) to (.6,0);
        			\draw[very thick, Maroon, bend left=30] (2.2,-1.2) to (.6,0);
        			\draw[very thick, Maroon] (-.6,0) to (.6,0);
        			\fill[Maroon] (-.6,0) circle (.07);
        			\fill[Maroon] (.6,0) circle (.07);
        			\draw[very thick] (-2,1.2) circle (.2 and .5);
        			\draw[very thick] (-2,-1.2) circle (.2 and .5);
        			\begin{scope}[shift={(2,1.2)}, xscale=.4]
            			\draw[very thick] (0,-.5) arc (-90:90:.5);
            			\draw[very thick, dashed] (0,.5) arc (90:270:.5);
        			\end{scope}
        			\begin{scope}[shift={(2,-1.2)}, xscale=.4]
            			\draw[very thick] (0,-.5) arc (-90:90:.5);
            			\draw[very thick, dashed] (0,.5) arc (90:270:.5);
        			\end{scope}
        			\begin{scope}[xscale=.4]
            			\draw[very thick, RoyalBlue] (0,-.9) arc (-90:90:.9);
            			\draw[very thick, RoyalBlue, dashed] (0,.9) arc (90:270:.9);
        			\end{scope}
        			\draw[very thick, out=0, in=180] (-2,1.7) to (0,.9) to (2,1.7);
        			\draw[very thick, out=0, in=180] (-2,-1.7) to (0,-.9) to (2,-1.7);
        			\draw[very thick, out=0, in=0, looseness=2.5] (-2,-.7) to (-2,.7);
        			\draw[very thick, out=180, in=180, looseness=2.5] (2,-.7) to (2,.7);
			\end{scope}
			\begin{scope}[shift={(2.75,0)}]
					\draw[very thick, ->] (-.7,0) to node[above] {$\mathbb{F}$} (.7,0);
			\end{scope}
			\begin{scope}[shift={(8.25,0)}]
					\draw[very thick, ->] (-.7,0) to node[above] {$\mathbb{F}$} (.7,0);
			\end{scope}
    \end{tikzpicture}
    \caption{Idempotency of $\mathbb{F}$.} \label{fig:idempotency F}
\end{figure}

\paragraph{Idempotency of $\mathbb{F}$.}
There is a simple equation from the fact that applying crossing symmetry on the four-punctured sphere twice gives the original block back. We display the series of moves in Figure~\ref{fig:idempotency F}. To make the figure less crammed, we suppressed the labels of the internal and external lines, as well as the labels on $\mathbb{F}$.

\begin{figure}[ht]
	\centering
	\begin{tikzpicture}[scale=.8]
		\begin{scope}[scale=.8]
			\draw[very thick, Maroon, bend left=30] (-1.8,1.2) to (-.6,0);
        		\draw[very thick, Maroon, bend right=30] (-1.8,-1.2) to (-.6,0);
        		\draw[very thick, Maroon, bend right=30] (2.2,1.2) to (.6,0);
        		\draw[very thick, Maroon, bend left=30] (2.2,-1.2) to (.6,0);
        		\draw[very thick, Maroon] (-.6,0) to (.6,0);
        		\fill[Maroon] (-.6,0) circle (.07);
        		\fill[Maroon] (.6,0) circle (.07);
        		\draw[very thick] (-2,1.2) circle (.2 and .5);
        		\draw[very thick] (-2,-1.2) circle (.2 and .5);
        		\begin{scope}[shift={(2,1.2)}, xscale=.4]
            		\draw[very thick] (0,-.5) arc (-90:90:.5);
            		\draw[very thick, dashed] (0,.5) arc (90:270:.5);
        		\end{scope}
        		\begin{scope}[shift={(2,-1.2)}, xscale=.4]
            		\draw[very thick] (0,-.5) arc (-90:90:.5);
            		\draw[very thick, dashed] (0,.5) arc (90:270:.5);
        		\end{scope}
        		\begin{scope}[xscale=.4]
            		\draw[very thick, RoyalBlue] (0,-.9) arc (-90:90:.9);
            		\draw[very thick, dashed, RoyalBlue] (0,.9) arc (90:270:.9);
        		\end{scope}
        		\draw[very thick, out=0, in=180] (-2,1.7) to (0,.9) to (2,1.7);
        		\draw[very thick, out=0, in=180] (-2,-1.7) to (0,-.9) to (2,-1.7);
        		\draw[very thick, out=0, in=0, looseness=2.5] (-2,-.7) to (-2,.7);
        		\draw[very thick, out=180, in=180, looseness=2.5] (2,-.7) to (2,.7);
		\end{scope}
		\begin{scope}[shift={(3,0)}]
			\draw[very thick,->] (-1,0) to node[above] {$\mathbb{B}$} (1,0);
		\end{scope}
		\begin{scope}[shift={(6,0)}, scale=.8]
			\draw[very thick, Maroon, bend left=30] (-1.8,1.2) to (-.6,0);
        		\draw[very thick, Maroon, bend right=30] (-1.8,-1.2) to (-.6,0);
        		\draw[very thick, Maroon, bend right=30] (2.2,1.2) to (.6,0);
        		\draw[very thick, Maroon, bend left=40] (2.2,-1.2) to (.97,-.1);
        		\draw[very thick, Maroon, dashed, bend right=30] (.97,-.1) to (.6,1.05);
        		\draw[very thick, Maroon, bend right=10] (.6,1.05) to (.6,0);
        		\draw[very thick, Maroon] (-.6,0) to (.6,0);
        		\fill[Maroon] (-.6,0) circle (.07);
        		\fill[Maroon] (.6,0) circle (.07);
        		\draw[very thick] (-2,1.2) circle (.2 and .5);
        		\draw[very thick] (-2,-1.2) circle (.2 and .5);
        		\begin{scope}[shift={(2,1.2)}, xscale=.4]
            		\draw[very thick] (0,-.5) arc (-90:90:.5);
            		\draw[very thick, dashed] (0,.5) arc (90:270:.5);
        		\end{scope}
        		\begin{scope}[shift={(2,-1.2)}, xscale=.4]
            		\draw[very thick] (0,-.5) arc (-90:90:.5);
            		\draw[very thick, dashed] (0,.5) arc (90:270:.5);
        		\end{scope}
        		\begin{scope}[xscale=.4]
            		\draw[very thick, RoyalBlue] (0,-.9) arc (-90:90:.9);
            		\draw[very thick, dashed, RoyalBlue] (0,.9) arc (90:270:.9);
        		\end{scope}
        		\draw[very thick, out=0, in=180] (-2,1.7) to (0,.9) to (2,1.7);
        		\draw[very thick, out=0, in=180] (-2,-1.7) to (0,-.9) to (2,-1.7);
        		\draw[very thick, out=0, in=0, looseness=2.5] (-2,-.7) to (-2,.7);
        		\draw[very thick, out=180, in=180, looseness=2.5] (2,-.7) to (2,.7);
		\end{scope}
		\begin{scope}[shift={(8,-2)}]
			\draw[very thick,->] (0,.7) to node[above, xshift=2] {$\mathbb{F}$} (1,0);
		\end{scope}
		\begin{scope}[shift={(9,-3.5)}, scale=.8]
			\draw[very thick, Maroon, bend right=10] (-1.8,-1.2) to (.1,.2);
			\draw[very thick, Maroon, bend left=20] (.1,.2) to (-.8,.7);
			\draw[very thick, Maroon, bend right=20] (-.8,.7) to (-1.8,1.2);
			\draw[very thick, Maroon, bend left=20] (-.8,.7) to (-.4,.97);
			\draw[very thick, Maroon, bend left=20, dashed] (-.4,.97) to (1,-.2);
			\draw[very thick, Maroon] (1,-.2) .. controls (.6,-.5) .. (2.2,-1.2);
			\draw[very thick, Maroon, bend left=20] (.1,.2) to (2.2,1.2);
			\fill[Maroon] (.1,.2) circle (.06);
			\fill[Maroon] (-.8,.7) circle (.06);
        		\draw[very thick] (-2,1.2) circle (.2 and .5);
        		\draw[very thick] (-2,-1.2) circle (.2 and .5);
        		\begin{scope}[shift={(2,1.2)}, xscale=.4]
            		\draw[very thick] (0,-.5) arc (-90:90:.5);
            		\draw[very thick, dashed] (0,.5) arc (90:270:.5);
        		\end{scope}
        		\begin{scope}[shift={(2,-1.2)}, xscale=.4]
            		\draw[very thick] (0,-.5) arc (-90:90:.5);
            		\draw[very thick, dashed] (0,.5) arc (90:270:.5);
        		\end{scope}
        		\draw[very thick, RoyalBlue, bend left=10] (-1,-.2) to (.5,1);
        		\draw[very thick, RoyalBlue, bend left=10] (1,.2) to (-.5,-1);
        		\draw[very thick, RoyalBlue, bend right=30, dashed] (1,.2) to (.5,1);
        		\draw[very thick, RoyalBlue, bend right=30, dashed] (-1,-.2) to (-.5,-1);
        		\draw[very thick, out=0, in=180] (-2,1.7) to (0,.9) to (2,1.7);
        		\draw[very thick, out=0, in=180] (-2,-1.7) to (0,-.9) to (2,-1.7);
        		\draw[very thick, out=0, in=0, looseness=2.5] (-2,-.7) to (-2,.7);
        		\draw[very thick, out=180, in=180, looseness=2.5] (2,-.7) to (2,.7);
		\end{scope}
		\begin{scope}[shift={(8,-5)}]
			\draw[very thick,<-] (0,-.7) to node[below, xshift=2] {$\mathbb{B}$} (1,0);
		\end{scope}
		\begin{scope}[shift={(6,-7)}, scale=.8]
			\draw[very thick, Maroon, bend right=10] (-1.8,-1.2) to (.1,.2);
			\draw[very thick, Maroon, bend left=20] (.1,.2) to (-.8,.7);
			\draw[very thick, Maroon, bend right=20] (-.8,.7) to (-1.8,1.2);
			\draw[very thick, Maroon, bend right=20] (-.8,.7) to (-1,.2);
			\draw[very thick, Maroon, bend right=20, dashed] (-1,.2) to (1,-.2);
			\draw[very thick, Maroon] (1,-.2) .. controls (.6,-.5) .. (2.2,-1.2);
			\draw[very thick, Maroon, bend left=20] (.1,.2) to (2.2,1.2);
			\fill[Maroon] (.1,.2) circle (.06);
			\fill[Maroon] (-.8,.7) circle (.06);
        		\draw[very thick] (-2,1.2) circle (.2 and .5);
        		\draw[very thick] (-2,-1.2) circle (.2 and .5);
        		\begin{scope}[shift={(2,1.2)}, xscale=.4]
            		\draw[very thick] (0,-.5) arc (-90:90:.5);
            		\draw[very thick, dashed] (0,.5) arc (90:270:.5);
        		\end{scope}
        		\begin{scope}[shift={(2,-1.2)}, xscale=.4]
            		\draw[very thick] (0,-.5) arc (-90:90:.5);
            		\draw[very thick, dashed] (0,.5) arc (90:270:.5);
        		\end{scope}
        		\draw[very thick, RoyalBlue,  bend left=10] (-1,-.2) to (.5,1);
        		\draw[very thick, RoyalBlue, bend left=10] (1,.2) to (-.5,-1);
        		\draw[very thick, RoyalBlue, bend right=30, dashed] (1,.2) to (.5,1);
        		\draw[very thick, RoyalBlue, bend right=30, dashed] (-1,-.2) to (-.5,-1);
        		\draw[very thick, out=0, in=180] (-2,1.7) to (0,.9) to (2,1.7);
        		\draw[very thick, out=0, in=180] (-2,-1.7) to (0,-.9) to (2,-1.7);
        		\draw[very thick, out=0, in=0, looseness=2.5] (-2,-.7) to (-2,.7);
        		\draw[very thick, out=180, in=180, looseness=2.5] (2,-.7) to (2,.7);
		\end{scope}
		\begin{scope}[shift={(-2,-2)}]
			\draw[very thick,->] (0,.7) to node[above, xshift=-1] {$\mathbb{F}$} (-1,0);
		\end{scope}		
		\begin{scope}[shift={(-3,-3.5)}, scale=.8]
        		\draw[very thick, Maroon, bend left=10] (-1.8,1.2) to (0,.4);
        		\draw[very thick, Maroon, bend right=10] (-1.8,-1.2) to (0,-.4);
        		\draw[very thick, Maroon, bend right=10] (2.2,1.2) to (0,.4);
        		\draw[very thick, Maroon, bend left=10] (2.2,-1.2) to (0,-.4);
        		\draw[very thick, Maroon] (0,-.4) to (0,.4);
        		\fill[Maroon] (0,.4) circle (.07);
        		\fill[Maroon] (0,-.4) circle (.07);
        		\draw[very thick] (-2,1.2) circle (.2 and .5);
        		\draw[very thick] (-2,-1.2) circle (.2 and .5);
        		\begin{scope}[shift={(2,1.2)}, xscale=.4]
        		    \draw[very thick] (0,-.5) arc (-90:90:.5);
        		    \draw[very thick, dashed] (0,.5) arc (90:270:.5);
        		\end{scope}
        		\begin{scope}[shift={(2,-1.2)}, xscale=.4]
        		    \draw[very thick] (0,-.5) arc (-90:90:.5);
        		    \draw[very thick, dashed] (0,.5) arc (90:270:.5);
        		\end{scope}
        		\begin{scope}[yscale=.2]
        		    \draw[very thick, RoyalBlue] (-.97,0) arc (-180:0:.97);
        		    \draw[very thick, RoyalBlue, dashed] (.97,0) arc (0:180:.97);
        		\end{scope}
        		\draw[very thick, out=0, in=180] (-2,1.7) to (0,.9) to (2,1.7);
        		\draw[very thick, out=0, in=180] (-2,-1.7) to (0,-.9) to (2,-1.7);
        		\draw[very thick, out=0, in=0, looseness=2.5] (-2,-.7) to (-2,.7);
        		\draw[very thick, out=180, in=180, looseness=2.5] (2,-.7) to (2,.7);
        	\end{scope}  
		\begin{scope}[shift={(-2,-5)}]
			\draw[very thick,->] (0,-.7) to node[below, xshift=-1] {$\mathbb{B}$} (-1,0);
		\end{scope}		
		\begin{scope}[shift={(0,-7)}, scale=.8]
        		\draw[very thick, Maroon, bend left=10] (-1.8,1.2) to (0,.4);
        		\draw[very thick, Maroon, bend right=10] (-1.8,-1.2) to (0,-.4);
        		\draw[very thick, Maroon, bend right=10] (2.2,1.2) to (0,.4);
        		\draw[very thick, Maroon] (0,-.4) to (0,.4);
        		\draw[very thick, Maroon, bend right=20, dashed] (-1.2,-.4) to (1.2,-.4);
			\draw[very thick, Maroon] (1.2,-.4) .. controls (1.2,-.8) .. (2.2,-1.2);
			\draw[very thick, Maroon, bend left=10] (-1.2,-.4) to (0,-.4);
        		\fill[Maroon] (0,.4) circle (.07);
        		\fill[Maroon] (0,-.4) circle (.07);
        		\draw[very thick] (-2,1.2) circle (.2 and .5);
        		\draw[very thick] (-2,-1.2) circle (.2 and .5);
        		\begin{scope}[shift={(2,1.2)}, xscale=.4]
        		    \draw[very thick] (0,-.5) arc (-90:90:.5);
        		    \draw[very thick, dashed] (0,.5) arc (90:270:.5);
        		\end{scope}
        		\begin{scope}[shift={(2,-1.2)}, xscale=.4]
        		    \draw[very thick] (0,-.5) arc (-90:90:.5);
        		    \draw[very thick, dashed] (0,.5) arc (90:270:.5);
        		\end{scope}
        		\begin{scope}[yscale=.2]
        		    \draw[very thick, RoyalBlue] (-.97,0) arc (-180:0:.97);
        		    \draw[very thick, RoyalBlue, dashed] (.97,0) arc (0:180:.97);
        		\end{scope}
        		\draw[very thick, out=0, in=180] (-2,1.7) to (0,.9) to (2,1.7);
        		\draw[very thick, out=0, in=180] (-2,-1.7) to (0,-.9) to (2,-1.7);
        		\draw[very thick, out=0, in=0, looseness=2.5] (-2,-.7) to (-2,.7);
        		\draw[very thick, out=180, in=180, looseness=2.5] (2,-.7) to (2,.7);
        	\end{scope}        
		\begin{scope}[shift={(3,-7)}]
			\draw[very thick,->] (-1,0) to node[above] {$\mathbb{F}$} (1,0);
		\end{scope}	
	\end{tikzpicture}
    	\caption{The hexagon equation.}
        \label{fig:hexagon equation}
\end{figure}
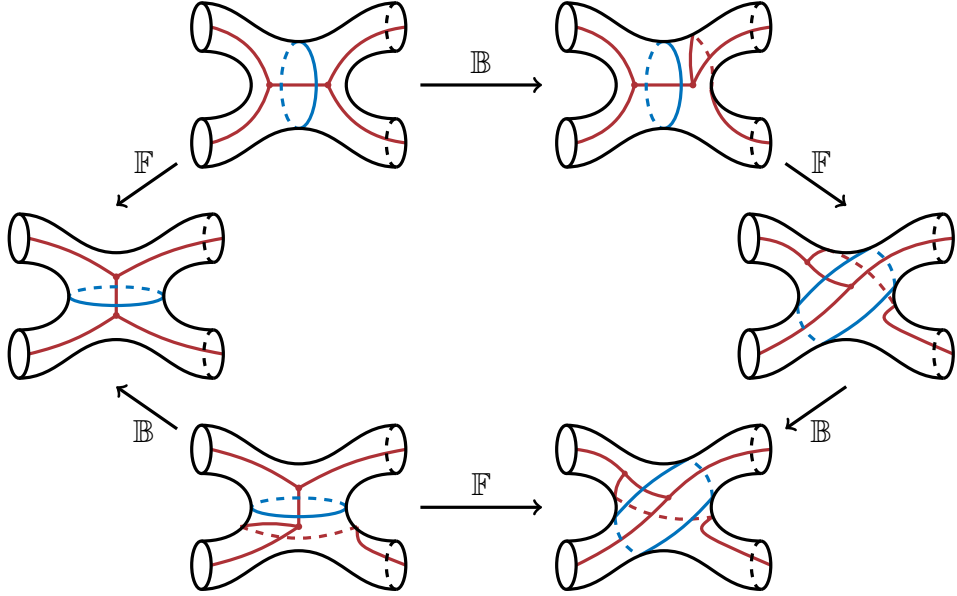

By comparing the left-most picture and the right-most picture in Figure~\ref{fig:idempotency F} and using that conformal blocks are a complete basis for the $L^2$-sections of the line bundle over Teichm\"uller space, we conclude that
\be 
        \int_0^\infty\!\! \d P_{32}\ \mathbb{F}_{P_{21},P_{32}} \!\begin{bmatrix}
            P_3 & P_2 \\
            P_4 & P_1
        \end{bmatrix}
        \mathbb{F}_{P_{32},P_{21}'} \!\begin{bmatrix}
            P_4 & P_3 \\
            P_1 & P_2
        \end{bmatrix}=\delta(P_{21}-P_{21}')\ . \label{eq:g=0, n=4 idempotency F}
    \ee
\paragraph{Hexagon equation.} This is the second consistency condition appearing for the four-punctured sphere. It can be obtained by applying the series of moves displayed in Figure~\ref{fig:hexagon equation}. The form of the figure should explain the name of the equation. Comparing the two different paths through the figure gives the equation
\begin{multline}
            \int_0^\infty\!\! \d P_{32}\ \mathrm{e}^{\pi i (\sum_{j=1}^4\Delta_j-\Delta_{21}-\Delta_{32}-\Delta_{31})} \, \mathbb{F}_{P_{21},P_{32}} \!\begin{bmatrix}
            P_3 & P_2 \\
            P_4 & P_1
        \end{bmatrix}
        \mathbb{F}_{P_{32},P_{31}} \!\begin{bmatrix}
            P_1 & P_3 \\
            P_4 & P_2
        \end{bmatrix} \\
        =
        \mathbb{F}_{P_{21},P_{31}} \!\begin{bmatrix}
            P_3 & P_1 \\
            P_4 & P_2
        \end{bmatrix}\ , \label{eq:g=0, n=4 hexagon equation}
        \end{multline}
where we collected all phases on one side of the equation. One can also apply the orientation reversed series of moves which leads to the complex conjugated equation of \eqref{eq:g=0, n=4 hexagon equation}, where the phases on the LHS are opposite.

\begin{figure}[ht]
    \centering
	\begin{tikzpicture}[scale=.8]
		\begin{scope}
			\draw[Maroon, very thick, bend left=100, looseness=1.85] (-1.8,1) to (-1.8,-1);
			\draw[Maroon, very thick, bend right=100, looseness=2.55] (2.2,1) to (2.2,-1);
			\draw[Maroon, very thick] (-.73,-.1) to (0,-.3) to (.73,-.1);
			\draw[Maroon, very thick] (0,-.3) to (0,-1.9);
			\fill[Maroon] (-.73,-.1) circle (0.07);
			\fill[Maroon] (.73,-.1) circle (0.07);
			\fill[Maroon] (0,-.3) circle (0.07);
			\begin{scope}[shift={(2,1)}, xscale=.4]
        		    \draw[very thick] (0,-.5) arc (-90:90:.5);
        		    \draw[very thick, dashed] (0,.5) arc (90:270:.5);
        		\end{scope}	
			\begin{scope}[shift={(2,-1)}, xscale=.4]
        		    \draw[very thick] (0,-.5) arc (-90:90:.5);
        		    \draw[very thick, dashed] (0,.5) arc (90:270:.5);
        		\end{scope}	
        		\draw[very thick] (-2,1) circle (.2 and .5);
        		\draw[very thick] (-2,-1) circle (.2 and .5);
        		\begin{scope}[shift={(0,-1.7)}, yscale=.4]
        		    \draw[very thick] (-.5,0) arc (-180:0:.5);
        		    \draw[dashed, very thick] (-.5,0) arc (180:0:.5);
        		\end{scope}        
        		\draw[very thick] (-2,-1.5) .. controls (-1,-1) and (-.5,-1)  .. (-.5,-1.7);	
        		\draw[very thick] (2,-1.5) .. controls (1,-1) and (.5,-1)  .. (.5,-1.7);		
        		\draw[very thick, bend right=100, looseness=4] (-2,-.5) to (-2,.5);
        		\draw[very thick, bend left=100, looseness=4] (2,-.5) to (2,.5);
        		\draw[very thick, out=0, in=180] (-2,1.5) to (0,.6) to (2,1.5);
        		
        		\draw[very thick, RoyalBlue, bend right=15] (-.7,-1.15) to (-.3,.65);
        		\draw[very thick, RoyalBlue, dashed, bend left=15] (-.7,-1.15) to (-.3,.65);
        		\draw[very thick, RoyalBlue, bend right=15] (.7,-1.15) to (.3,.65);
        		\draw[very thick, RoyalBlue, dashed, bend left=15] (.7,.-1.15) to (.3,.65);
		\end{scope}
		\begin{scope}[shift={(-6.5,-1.5)}]
			\draw[Maroon, very thick, bend left=100, looseness=1.85] (-1.8,1) to (-1.8,-1);
			\fill[Maroon] (-.73,-.1) circle (.07);
			\fill[Maroon] (.2,-.1) circle (.07);
			\fill[Maroon] (.6,-.9) circle (.07);
			\draw[very thick, Maroon] (-.73,-.1) to (.2,-.1) to (.6,-.9);
			\draw[very thick, Maroon, bend left=10] (.6,-.9) to (2.2,-1);
			\draw[very thick, Maroon, bend right=20] (.6,-.9) to (0,-1.9);
			\draw[very thick, Maroon, bend left=20] (.2,-.1) to (2.2,1);
			\begin{scope}[shift={(2,1)}, xscale=.4]
        		    \draw[very thick] (0,-.5) arc (-90:90:.5);
        		    \draw[very thick, dashed] (0,.5) arc (90:270:.5);
        		\end{scope}	
			\begin{scope}[shift={(2,-1)}, xscale=.4]
        		    \draw[very thick] (0,-.5) arc (-90:90:.5);
        		    \draw[very thick, dashed] (0,.5) arc (90:270:.5);
        		\end{scope}	
        		\draw[very thick] (-2,1) circle (.2 and .5);
        		\draw[very thick] (-2,-1) circle (.2 and .5);
        		\begin{scope}[shift={(0,-1.7)}, yscale=.4]
        		    \draw[very thick] (-.5,0) arc (-180:0:.5);
        		    \draw[dashed, very thick] (-.5,0) arc (180:0:.5);
        		\end{scope}        
        		\draw[very thick] (-2,-1.5) .. controls (-1,-1) and (-.5,-1)  .. (-.5,-1.7);	
        		\draw[very thick] (2,-1.5) .. controls (1,-1) and (.5,-1)  .. (.5,-1.7);		
        		\draw[very thick, bend right=100, looseness=4] (-2,-.5) to (-2,.5);
        		\draw[very thick, bend left=100, looseness=4] (2,-.5) to (2,.5);
        		\draw[very thick, out=0, in=180] (-2,1.5) to (0,.6) to (2,1.5);
        		
        		\draw[very thick, bend right=20, RoyalBlue] (-.7,-1.2) to (.9,-.2);
        		\draw[very thick, dashed, bend left=20, RoyalBlue] (-.7,-1.2) to (.9,-.2);
        		\draw[very thick, bend right=20, RoyalBlue] (-.9,-1.13) to (.2,.6);
        		\draw[very thick, dashed, bend left=20, RoyalBlue] (-.9,-1.13) to (.2,.6);
		\end{scope}
		\begin{scope}[shift={(-3.5,-5.5)}]
			\fill[Maroon] (0,.4) circle (.07);
			\fill[Maroon] (0,-.3) circle (.07);
			\fill[Maroon] (.6,-.9) circle (.07);
			\draw[very thick, Maroon] (0,.4) to (0,-.3) to (.6,-.9);
			\draw[very thick, Maroon, bend left=10] (.6,-.9) to (2.2,-1);
			\draw[very thick, Maroon, bend right=20] (.6,-.9) to (0,-1.9);
			\draw[very thick, Maroon, bend left=10] (0,-.3) to (-1.8,-1);
			\draw[very thick, Maroon] (0,.4) .. controls (.7,.4) and (1.5,.9) .. (2.2,1);
			\draw[very thick, Maroon, bend right=10] (0,.4) .. controls (-.7,.4) and (-1.3,.8) .. (-1.8,1);
			\begin{scope}[shift={(2,1)}, xscale=.4]
        		    \draw[very thick] (0,-.5) arc (-90:90:.5);
        		    \draw[very thick, dashed] (0,.5) arc (90:270:.5);
        		\end{scope}	
			\begin{scope}[shift={(2,-1)}, xscale=.4]
        		    \draw[very thick] (0,-.5) arc (-90:90:.5);
        		    \draw[very thick, dashed] (0,.5) arc (90:270:.5);
        		\end{scope}	
        		\draw[very thick] (-2,1) circle (.2 and .5);
        		\draw[very thick] (-2,-1) circle (.2 and .5);
        		\begin{scope}[shift={(0,-1.7)}, yscale=.4]
        		    \draw[very thick] (-.5,0) arc (-180:0:.5);
        		    \draw[dashed, very thick] (-.5,0) arc (180:0:.5);
        		\end{scope}        
        		\draw[very thick] (-2,-1.5) .. controls (-1,-1) and (-.5,-1)  .. (-.5,-1.7);	
        		\draw[very thick] (2,-1.5) .. controls (1,-1) and (.5,-1)  .. (.5,-1.7);		
        		\draw[very thick, bend right=100, looseness=4] (-2,-.5) to (-2,.5);
        		\draw[very thick, bend left=100, looseness=4] (2,-.5) to (2,.5);
        		\draw[very thick, out=0, in=180] (-2,1.5) to (0,.6) to (2,1.5);
        		
        		\draw[very thick, RoyalBlue, bend right=20] (-.7,-1.2) to (.9,-.2);
        		\draw[very thick, RoyalBlue, dashed, bend left=20] (-.7,-1.2) to (.9,-.2);
        		\draw[very thick, RoyalBlue, bend right=20] (-.85,.1) to (.85,.1);
        		\draw[very thick, RoyalBlue, dashed, bend left=20] (-.85,.1) to (.85,.1);
		\end{scope}
		\begin{scope}[shift={(6.5,-1.5)}, xscale=-1]
			\draw[Maroon, very thick, bend left=100, looseness=2.55] (-2.2,1) to (-2.2,-1);
			\fill[Maroon] (-.73,-.1) circle (.07);
			\fill[Maroon] (.2,-.1) circle (.07);
			\fill[Maroon] (.6,-.9) circle (.07);
			\draw[very thick, Maroon] (-.73,-.1) to (.2,-.1) to (.6,-.9);
			\draw[very thick, Maroon, bend left=10] (.6,-.9) to (1.8,-1);
			\draw[very thick, Maroon, bend right=20] (.6,-.9) to (0,-1.9);
			\draw[very thick, Maroon, bend left=20] (.2,-.1) to (1.8,1);
			\begin{scope}[shift={(-2,1)}, xscale=.4]
        		    \draw[very thick, dashed] (0,-.5) arc (-90:90:.5);
        		    \draw[very thick] (0,.5) arc (90:270:.5);
        		\end{scope}	
			\begin{scope}[shift={(-2,-1)}, xscale=.4]
        		    \draw[very thick, dashed] (0,-.5) arc (-90:90:.5);
        		    \draw[very thick] (0,.5) arc (90:270:.5);
        		\end{scope}	
        		\draw[very thick] (2,1) circle (.2 and .5);
        		\draw[very thick] (2,-1) circle (.2 and .5);
        		\begin{scope}[shift={(0,-1.7)}, yscale=.4]
        		    \draw[very thick] (-.5,0) arc (-180:0:.5);
        		    \draw[dashed, very thick] (-.5,0) arc (180:0:.5);
        		\end{scope}        
        		\draw[very thick] (-2,-1.5) .. controls (-1,-1) and (-.5,-1)  .. (-.5,-1.7);	
        		\draw[very thick] (2,-1.5) .. controls (1,-1) and (.5,-1)  .. (.5,-1.7);		
        		\draw[very thick, bend right=100, looseness=4] (-2,-.5) to (-2,.5);
        		\draw[very thick, bend left=100, looseness=4] (2,-.5) to (2,.5);
        		\draw[very thick, out=0, in=180] (-2,1.5) to (0,.6) to (2,1.5);
        		
        		\draw[very thick, bend right=20, RoyalBlue] (-.7,-1.2) to (.9,-.2);
        		\draw[very thick, dashed, bend left=20, RoyalBlue] (-.7,-1.2) to (.9,-.2);
        		\draw[very thick, bend right=20, RoyalBlue] (-.9,-1.13) to (.2,.6);
        		\draw[very thick, dashed, bend left=20, RoyalBlue] (-.9,-1.13) to (.2,.6);
		\end{scope}
		\begin{scope}[shift={(3.5,-5.5)}, xscale=-1]
			\fill[Maroon] (0,.4) circle (.07);
			\fill[Maroon] (0,-.3) circle (.07);
			\fill[Maroon] (.6,-.9) circle (.07);
			\draw[very thick, Maroon] (0,.4) to (0,-.3) to (.6,-.9);
			\draw[very thick, Maroon, bend left=10] (.6,-.9) to (1.8,-1);
			\draw[very thick, Maroon, bend right=20] (.6,-.9) to (0,-1.9);
			\draw[very thick, Maroon, bend left=10] (0,-.3) to (-2.2,-1);
			\draw[very thick, Maroon] (0,.4) .. controls (.7,.4) and (1.5,.9) .. (1.8,1);
			\draw[very thick, Maroon, bend right=10] (0,.4) .. controls (-.7,.4) and (-1.3,.8) .. (-2.2,1);
			\begin{scope}[shift={(-2,1)}, xscale=.4]
        		    \draw[very thick, dashed] (0,-.5) arc (-90:90:.5);
        		    \draw[very thick] (0,.5) arc (90:270:.5);
        		\end{scope}	
			\begin{scope}[shift={(-2,-1)}, xscale=.4]
        		    \draw[very thick, dashed] (0,-.5) arc (-90:90:.5);
        		    \draw[very thick] (0,.5) arc (90:270:.5);
        		\end{scope}	
        		\draw[very thick] (2,1) circle (.2 and .5);
        		\draw[very thick] (2,-1) circle (.2 and .5);
        		\begin{scope}[shift={(0,-1.7)}, yscale=.4]
        		    \draw[very thick] (-.5,0) arc (-180:0:.5);
        		    \draw[dashed, very thick] (-.5,0) arc (180:0:.5);
        		\end{scope}        
        		\draw[very thick] (-2,-1.5) .. controls (-1,-1) and (-.5,-1)  .. (-.5,-1.7);	
        		\draw[very thick] (2,-1.5) .. controls (1,-1) and (.5,-1)  .. (.5,-1.7);		
        		\draw[very thick, bend right=100, looseness=4] (-2,-.5) to (-2,.5);
        		\draw[very thick, bend left=100, looseness=4] (2,-.5) to (2,.5);
        		\draw[very thick, out=0, in=180] (-2,1.5) to (0,.6) to (2,1.5);
        		
        		\draw[very thick, bend right=20, RoyalBlue] (-.7,-1.2) to (.9,-.2);
        		\draw[very thick, dashed, bend left=20, RoyalBlue] (-.7,-1.2) to (.9,-.2);
        		\draw[very thick, bend right=20, RoyalBlue] (-.85,.1) to (.85,.1);
        		\draw[very thick, dashed, bend left=20, RoyalBlue] (-.85,.1) to (.85,.1);
		\end{scope}
		\begin{scope}[shift={(3.3,-.5)}]
			\draw[very thick,->] (-.7,.2) to node[above, xshift=2] {$\mathbb{F}$} (.7,-.7);
		\end{scope}
		\begin{scope}[shift={(-3.3,-.5)}]
			\draw[very thick,->] (.7,.2) to node[above, xshift=-2] {$\mathbb{F}$} (-.7,-.7);
		\end{scope}
		\begin{scope}[shift={(5.3,-3.4)}]
			\draw[very thick,->] (-.3,0) to node[above, xshift=-2] {$\mathbb{F}$} (-1.5,-1);
		\end{scope}
		\begin{scope}[shift={(-5.3,-3.4)}]
			\draw[very thick,->] (.3,0) to node[above, xshift=2] {$\mathbb{F}$} (1.5,-1);
		\end{scope}
		\begin{scope}[shift={(0,-5.5)}]
			\draw[very thick, ->] (-1,0) to node[above] {$\mathbb{F}$} (1,0);
		\end{scope}
	\end{tikzpicture}
    	\caption{The pentagon equation.}
      \label{fig:pentagon equation}
\end{figure}
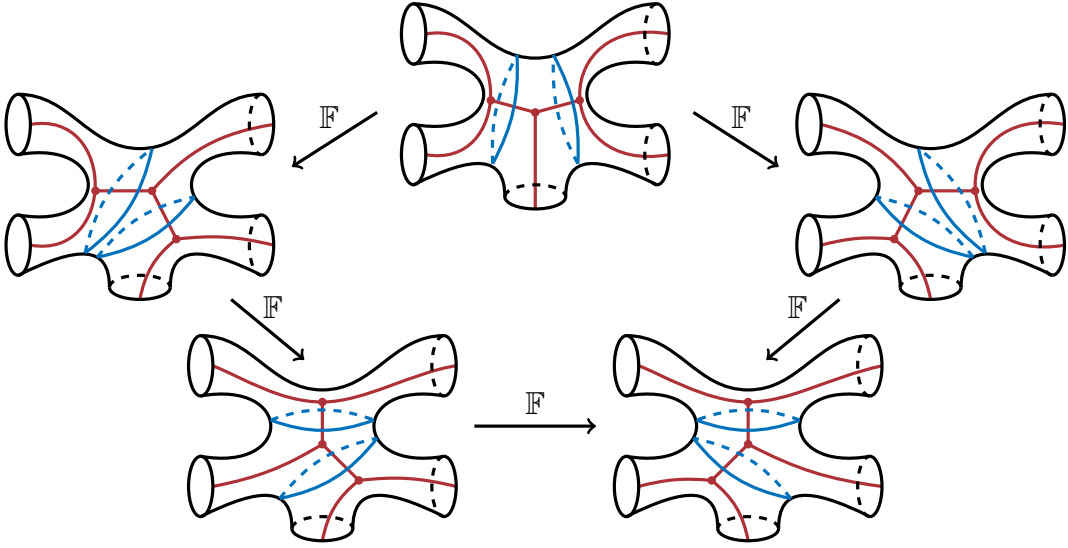

\paragraph{Pentagon equation.}
The pentagon equation is a consistency condition involving only $\mathbb{F}$ that arises from the five-punctured sphere. It is obtained by comparing the two different paths in Figure~\ref{fig:pentagon equation}. The form of the picture again gives rise to the name of the equation. Using again the fact that normalizable conformal blocks form a complete basis gives the pentagon equation
        \begin{multline}
        \int_0^\infty\!\!  \d P_{32}\ \mathbb{F}_{P_{21},P_{32}} \!\begin{bmatrix}
            P_3 & P_2 \\
            P_{54} & P_1
        \end{bmatrix}
        \mathbb{F}_{P_{54},P_{51}} \!\begin{bmatrix}
            P_4 & P_{32} \\
            P_5 & P_1
        \end{bmatrix}
        \mathbb{F}_{P_{32},P_{43}} \!\begin{bmatrix}
            P_4 & P_3 \\
            P_{51} & P_2
        \end{bmatrix} \\
        =
        \mathbb{F}_{P_{21},P_{51}} \!\begin{bmatrix}
            P_{43} & P_2 \\
            P_5 & P_1
        \end{bmatrix}
        \mathbb{F}_{P_{54},P_{43}} \!\begin{bmatrix}
            P_4 & P_3 \\
            P_5 & P_{21}
        \end{bmatrix}\ . \label{eq:g=0, n=5 pentagon equation}
        \end{multline}

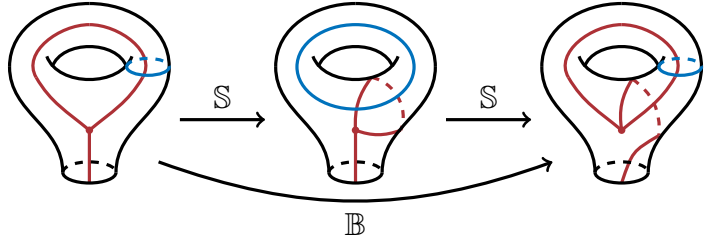
\begin{figure}[ht]
	\centering
	\begin{tikzpicture}[scale=.7]
		\begin{scope}
        		\draw[very thick, Maroon] (0,-1.2) to (0,-.2);
        		\fill[Maroon] (0,-.2) circle (.07);
        		\draw[very thick, Maroon, out=140, in=-60] (0,-.2) to (-1,.8);
        		\draw[very thick, Maroon, out=120, in=180] (-1,.8) to (0,1.8);
        		\draw[very thick, Maroon, out=40, in=240] (0,-.2) to (1,.8);
        		\draw[very thick, Maroon, out=60, in=0] (1,.8) to (0,1.8);
        		\draw[very thick, out=270, in=90] (1.5,1) to (.5,-1);
        		\draw[very thick, out=270, in=90] (-1.5,1) to (-.5,-1);
        		\draw[very thick, out=90, in=180] (-1.5,1) to (0,2.2);
        		\draw[very thick, out=0,in=90] (0,2.2) to (1.5,1);
        		\draw[very thick, bend left=70] (-.7,1) to (.7,1);
        		\draw[very thick, bend right=70] (-.8,1.2) to (.8,1.2);
        		\begin{scope}[shift={(0,-1)}, yscale=.4]
        		    \draw[very thick] (-.5,0) arc (-180:0:.5);
        		    \draw[dashed, very thick] (-.5,0) arc (180:0:.5);
        		\end{scope}
        		\begin{scope}[shift={(1.1,1)}, yscale=.4]
        		    \draw[very thick, RoyalBlue] (-.4,0) arc (-180:0:.4);
        		    \draw[dashed, very thick, RoyalBlue] (-.4,0) arc (180:0:.4);
        		\end{scope}
    		\end{scope} 
    		\begin{scope}[shift={(2.5,0)}]
    			\draw[very thick, ->] (-.8,0) to node[above] {$\mathbb{S}$} (.8,0);
    		\end{scope}
    		\begin{scope}[shift={(5,0)}]
        		\draw[very thick, Maroon] (0,-1.2) to (0,-.2);
        		\fill[Maroon] (0,-.2) circle (.07);
        		\draw[very thick, Maroon, bend left=20] (0,-.2) to (.3,.8);
        		\draw[very thick, Maroon, dashed, bend left=50] (.3,.8) to (.85,-.2);
        		\draw[very thick, Maroon, bend left=20] (.85,-.2) to (0,-.2);
        		\draw[very thick, out=270, in=90] (1.5,1) to (.5,-1);
        		\draw[very thick, out=270, in=90] (-1.5,1) to (-.5,-1);
        		\draw[very thick, out=90, in=180] (-1.5,1) to (0,2.2);
        		\draw[very thick, out=0,in=90] (0,2.2) to (1.5,1);
        		\draw[very thick, bend left=70] (-.7,1) to (.7,1);
        		\draw[very thick, bend right=70] (-.8,1.2) to (.8,1.2);
        		\begin{scope}[shift={(0,-1)}, yscale=.4]
        		    \draw[very thick] (-.5,0) arc (-180:0:.5);
        		    \draw[dashed, very thick] (-.5,0) arc (180:0:.5);
        		\end{scope}
        		\draw[very thick, RoyalBlue] (0,1) circle (1.1 and .8);
    		\end{scope}
    		\begin{scope}[shift={(7.5,0)}]
    			\draw[very thick, ->] (-.8,0) to node[above] {$\mathbb{S}$} (.8,0);
    		\end{scope}
    		\begin{scope}[shift={(10,0)}]
        		\draw[very thick, Maroon, bend left=20] (0,-.2) to (.2,.77);
        		\draw[very thick, Maroon, dashed, bend left=30] (.2,.77) to (.7,-.3);
        		\draw[very thick, Maroon] (.7,-.3) .. controls (.2,-.6) .. (0,-1.2);
        		\fill[Maroon] (0,-.2) circle (.07);
        		\draw[very thick, Maroon, out=140, in=-60] (0,-.2) to (-1,.8);
        		\draw[very thick, Maroon, out=120, in=180] (-1,.8) to (0,1.8);
        		\draw[very thick, Maroon, out=40, in=240] (0,-.2) to (1,.8);
        		\draw[very thick, Maroon, out=60, in=0] (1,.8) to (0,1.8);
        		\draw[very thick, out=270, in=90] (1.5,1) to (.5,-1);
        		\draw[very thick, out=270, in=90] (-1.5,1) to (-.5,-1);
        		\draw[very thick, out=90, in=180] (-1.5,1) to (0,2.2);
        		\draw[very thick, out=0,in=90] (0,2.2) to (1.5,1);
        		\draw[very thick, bend left=70] (-.7,1) to (.7,1);
        		\draw[very thick, bend right=70] (-.8,1.2) to (.8,1.2);
        		\begin{scope}[shift={(0,-1)}, yscale=.4]
        		    \draw[very thick] (-.5,0) arc (-180:0:.5);
        		    \draw[dashed, very thick] (-.5,0) arc (180:0:.5);
        		\end{scope}
        		\begin{scope}[shift={(1.1,1)}, yscale=.4]
        		    \draw[very thick, RoyalBlue] (-.4,0) arc (-180:0:.4);
        		    \draw[dashed, very thick, RoyalBlue] (-.4,0) arc (180:0:.4);
        		\end{scope}
    		\end{scope} 
    		\begin{scope}[shift={(5,-1)}]
    			\draw[very thick, ->, bend right=20] (-3.7,.2) to node[below] {$\mathbb{B}$} (3.7,.2);
    		\end{scope}
    	\end{tikzpicture} 
    	\caption{Idempotency of $\mathbb{S}$.}
     \label{fig:idempotency S}
\end{figure}

\paragraph{Idempotency of $\mathbb{S}$.} Let us now move on to consistency conditions satisfied by the modular crossing kernel $\mathbb{S}$. $\mathbb{S}$ is also essentially idempotent up to a phase. Indeed from Figure~\ref{fig:idempotency S}, we deduce the relation
\be 
            \int_0^\infty\!\!  \d P_2\  \mathbb{S}_{P_1,P_2}[P_0]\, \mathbb{S}_{P_2,P_3}[P_0] =\mathrm{e}^{\pi i \Delta_0}\delta(P_1-P_3) \ .  \label{eq:g=1, n=1 idempotency S}
\ee

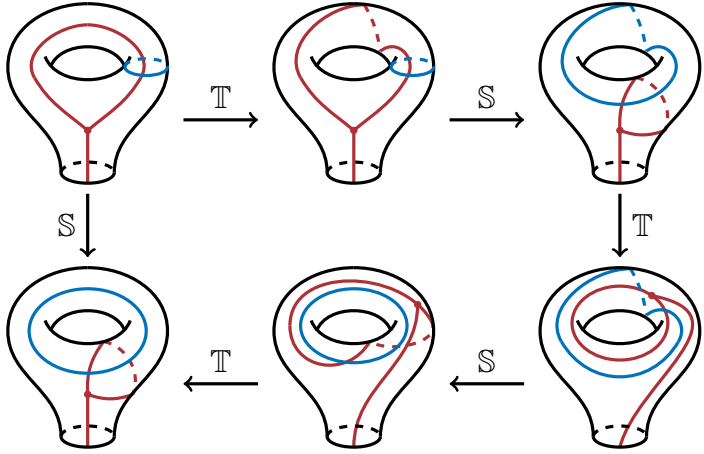
\begin{figure}[ht]
	\centering
	\begin{tikzpicture}[scale=.7]
		\begin{scope}
        		\draw[very thick, Maroon] (0,-1.2) to (0,-.2);
        		\fill[Maroon] (0,-.2) circle (.07);
        		\draw[very thick, Maroon, out=140, in=-60] (0,-.2) to (-1,.8);
        		\draw[very thick, Maroon, out=120, in=180] (-1,.8) to (0,1.8);
        		\draw[very thick, Maroon, out=40, in=240] (0,-.2) to (1,.8);
        		\draw[very thick, Maroon, out=60, in=0] (1,.8) to (0,1.8);
        		\draw[very thick, out=270, in=90] (1.5,1) to (.5,-1);
        		\draw[very thick, out=270, in=90] (-1.5,1) to (-.5,-1);
        		\draw[very thick, out=90, in=180] (-1.5,1) to (0,2.2);
        		\draw[very thick, out=0,in=90] (0,2.2) to (1.5,1);
        		\draw[very thick, bend left=70] (-.7,1) to (.7,1);
        		\draw[very thick, bend right=70] (-.8,1.2) to (.8,1.2);
        		\begin{scope}[shift={(0,-1)}, yscale=.4]
        		    \draw[very thick] (-.5,0) arc (-180:0:.5);
        		    \draw[dashed, very thick] (-.5,0) arc (180:0:.5);
        		\end{scope}
        		\begin{scope}[shift={(1.1,1)}, yscale=.4]
        		    \draw[very thick, RoyalBlue] (-.4,0) arc (-180:0:.4);
        		    \draw[dashed, very thick, RoyalBlue] (-.4,0) arc (180:0:.4);
        		\end{scope}
    		\end{scope} 
    		\begin{scope}[shift={(0,-2)}]
    			\draw[very thick, ->] (0,.6) to node[left] {$\mathbb{S}$} (0,-.6);
    		\end{scope}
    		\begin{scope}[shift={(0,-5)}]
        		\draw[very thick, Maroon] (0,-1.2) to (0,-.2);
        		\fill[Maroon] (0,-.2) circle (.07);
        		\draw[very thick, Maroon, bend left=20] (0,-.2) to (.3,.8);
        		\draw[very thick, Maroon, dashed, bend left=50] (.3,.8) to (.85,-.2);
        		\draw[very thick, Maroon, bend left=20] (.85,-.2) to (0,-.2);
        		\draw[very thick, out=270, in=90] (1.5,1) to (.5,-1);
        		\draw[very thick, out=270, in=90] (-1.5,1) to (-.5,-1);
        		\draw[very thick, out=90, in=180] (-1.5,1) to (0,2.2);
        		\draw[very thick, out=0,in=90] (0,2.2) to (1.5,1);
        		\draw[very thick, bend left=70] (-.7,1) to (.7,1);
        		\draw[very thick, bend right=70] (-.8,1.2) to (.8,1.2);
        		\begin{scope}[shift={(0,-1)}, yscale=.4]
        		    \draw[very thick] (-.5,0) arc (-180:0:.5);
        		    \draw[dashed, very thick] (-.5,0) arc (180:0:.5);
        		\end{scope}
        		\draw[very thick, RoyalBlue] (0,1) circle (1.1 and .8);
    		\end{scope}
    		\begin{scope}[shift={(2.5,0)}]
    			\draw[very thick,->] (-.7,0) to node[above] {$\mathbb{T}$} (.7,0);
    		\end{scope}
    		\begin{scope}[shift={(5,0)}]
        		\draw[very thick, Maroon] (0,-1.2) to (0,-.2);
        		\fill[Maroon] (0,-.2) circle (.07);
        		\draw[very thick, Maroon, out=140, in=-60] (0,-.2) to (-1,.8);
        		\draw[very thick, Maroon, out=120, in=180] (-1,.8) to (.2,2.18);
        		\draw[very thick, Maroon, out=40, in=240] (0,-.2) to (1,.8);
        		\draw[very thick, Maroon, dashed, bend left=10] (.2,2.18) to (.5,1.3);
        		\draw[very thick, Maroon, out=60, in=40, looseness=1.5] (1,.8) to (.5,1.3);
        		\draw[very thick, out=270, in=90] (1.5,1) to (.5,-1);
        		\draw[very thick, out=270, in=90] (-1.5,1) to (-.5,-1);
        		\draw[very thick, out=90, in=180] (-1.5,1) to (0,2.2);
        		\draw[very thick, out=0,in=90] (0,2.2) to (1.5,1);
        		\draw[very thick, bend left=70] (-.7,1) to (.7,1);
        		\draw[very thick, bend right=70] (-.8,1.2) to (.8,1.2);
        		\begin{scope}[shift={(0,-1)}, yscale=.4]
        		    \draw[very thick] (-.5,0) arc (-180:0:.5);
        		    \draw[dashed, very thick] (-.5,0) arc (180:0:.5);
        		\end{scope}
        		\begin{scope}[shift={(1.1,1)}, yscale=.4]
        		    \draw[very thick, RoyalBlue] (-.4,0) arc (-180:0:.4);
        		    \draw[dashed, very thick, RoyalBlue] (-.4,0) arc (180:0:.4);
        		\end{scope}
    		\end{scope} 
    		\begin{scope}[shift={(7.5,0)}]
    			\draw[very thick,->] (-.7,0) to node[above] {$\mathbb{S}$} (.7,0);
    		\end{scope}
    		\begin{scope}[shift={(10,0)}]
        		\draw[very thick, Maroon] (0,-1.2) to (0,-.2);
        		\fill[Maroon] (0,-.2) circle (.07);
        		\draw[very thick, Maroon, bend left=20] (0,-.2) to (.3,.8);
        		\draw[very thick, Maroon, dashed, bend left=50] (.3,.8) to (.85,-.2);
        		\draw[very thick, Maroon, bend left=20] (.85,-.2) to (0,-.2);
        		\draw[very thick,out=-120, in=-60, RoyalBlue] (1,.8) to (-1,.8);
        		\draw[very thick, out=120, in=180, RoyalBlue] (-1,.8) to (.2,2.18);
        		\draw[very thick, dashed, bend left=10, RoyalBlue] (.2,2.18) to (.5,1.3);
        		\draw[very thick, out=60, in=40, looseness=1.5, RoyalBlue] (1,.8) to (.5,1.3);
        		\draw[very thick, out=270, in=90] (1.5,1) to (.5,-1);
        		\draw[very thick, out=270, in=90] (-1.5,1) to (-.5,-1);
        		\draw[very thick, out=90, in=180] (-1.5,1) to (0,2.2);
        		\draw[very thick, out=0,in=90] (0,2.2) to (1.5,1);
        		\draw[very thick, bend left=70] (-.7,1) to (.7,1);
        		\draw[very thick, bend right=70] (-.8,1.2) to (.8,1.2);
        		\begin{scope}[shift={(0,-1)}, yscale=.4]
        		    \draw[very thick] (-.5,0) arc (-180:0:.5);
        		    \draw[dashed, very thick] (-.5,0) arc (180:0:.5);
        		\end{scope}
    		\end{scope} 
    		\begin{scope}[shift={(10,-2)}]
    			\draw[very thick, ->] (0,.6) to node[right] {$\mathbb{T}$} (0,-.6);    		
    		\end{scope}
		\begin{scope}[shift={(10,-5)}]
			\fill[Maroon] (.6,1.67) circle (.07);
        		\draw[very thick, Maroon] (0,1.15) circle (.9 and .7);
        		\draw[very thick, Maroon] (.6,1.67) .. controls (2.5,1.3) and (.3,0) .. (0,-1.2);
        		\draw[very thick,out=-120, in=-60, looseness=1.2, RoyalBlue] (1.1,.8) to (-1.1,.8);
        		\draw[very thick, out=120, in=180, RoyalBlue] (-1.1,.8) to (.2,2.18);
        		\draw[very thick, dashed, bend left=10, RoyalBlue] (.2,2.18) to (.5,1.3);
        		\draw[very thick, out=60, in=40, looseness=1.5, RoyalBlue] (1.1,.8) to (.5,1.3);
        		\draw[very thick, out=270, in=90] (1.5,1) to (.5,-1);
        		\draw[very thick, out=270, in=90] (-1.5,1) to (-.5,-1);
        		\draw[very thick, out=90, in=180] (-1.5,1) to (0,2.2);
        		\draw[very thick, out=0,in=90] (0,2.2) to (1.5,1);
        		\draw[very thick, bend left=70] (-.7,1) to (.7,1);
        		\draw[very thick, bend right=70] (-.8,1.2) to (.8,1.2);
        		\begin{scope}[shift={(0,-1)}, yscale=.4]
        		    \draw[very thick] (-.5,0) arc (-180:0:.5);
        		    \draw[dashed, very thick] (-.5,0) arc (180:0:.5);
        		\end{scope}
    		\end{scope} 
    		\begin{scope}[shift={(7.5,-5)}]
    			\draw[very thick, ->] (.7,0) to node[above] {$\mathbb{S}$} (-.7,0);    		
    		\end{scope}
		\begin{scope}[shift={(5,-5)}]
			\draw[very thick, Maroon, in=90, out=120, looseness=1.1] (1.45,1.2) to (-1.2,1.1);
			\draw[very thick, Maroon, dashed, out=-60, in=-30] (1.45,1.2) to (.3,.8);
			\draw[very thick, Maroon, out=240,in=-90, looseness=1.4] (.3,.8) to (-1.2,1.1);
			\fill[Maroon] (1.2,1.5) circle (0.07);
			\draw[very thick, Maroon, in=90,out=-90] (1.2,1.5) to (0,-1.2);
        		\draw[very thick, RoyalBlue] (0,1.1) circle (1 and .7);
        		\draw[very thick, out=270, in=90] (1.5,1) to (.5,-1);
        		\draw[very thick, out=270, in=90] (-1.5,1) to (-.5,-1);
        		\draw[very thick, out=90, in=180] (-1.5,1) to (0,2.2);
        		\draw[very thick, out=0,in=90] (0,2.2) to (1.5,1);
        		\draw[very thick, bend left=70] (-.7,1) to (.7,1);
        		\draw[very thick, bend right=70] (-.8,1.2) to (.8,1.2);
        		\begin{scope}[shift={(0,-1)}, yscale=.4]
        		    \draw[very thick] (-.5,0) arc (-180:0:.5);
        		    \draw[dashed, very thick] (-.5,0) arc (180:0:.5);
        		\end{scope}
    		\end{scope} 
    		\begin{scope}[shift={(2.5,-5)}]
    			\draw[very thick, ->] (.7,0) to node[above] {$\mathbb{T}$} (-.7,0);    
    		\end{scope}
	\end{tikzpicture}
	\caption{The $\PSL(2,\ZZ)$ relation.} \label{fig:SL2Z relation}
\end{figure}

\paragraph{$\PSL(2,\ZZ)$ relation.}
There is another relation deduced from the once-punctured torus. It is the projective analogue of the familiar relation $(ST)^3=1$ in $\SL(2,\ZZ)$. It can be deduced from the pictures in Figure~\ref{fig:SL2Z relation}. Equality of the two paths through the picture yields
\be 
\int_0^\infty\!\! \d P_2 \ \mathbb{S}_{P_1,P_2}[P_0] \, \mathrm{e}^{-2\pi i \sum_{j=1}^3(\Delta_j-\frac{c}{24})} \, \mathbb{S}_{P_2,P_3}[P_0]=  \mathbb{S}_{P_1,P_3}[P_0]\ .  \label{eq:g=1, n=1 (ST)^3}
\ee
Notably, this relation involves also the central charge $c$ explicitly. This is because we are applying Dehn twists around a non-separating curve, meaning that the surface remains connected when we pinch the blue curve in Figure~\ref{fig:SL2Z relation}. In this case, there is a Casimir energy and the action of the Dehn twist gets shifted to $\mathrm{e}^{-2\pi i(\Delta-\frac{c}{24})}$. More generally, this is a choice that we are making and the phases are related to the projectiveness of the action of crossing transformations. We will discuss this point in more detail in Section~\ref{subsec:representation mapping class group}.

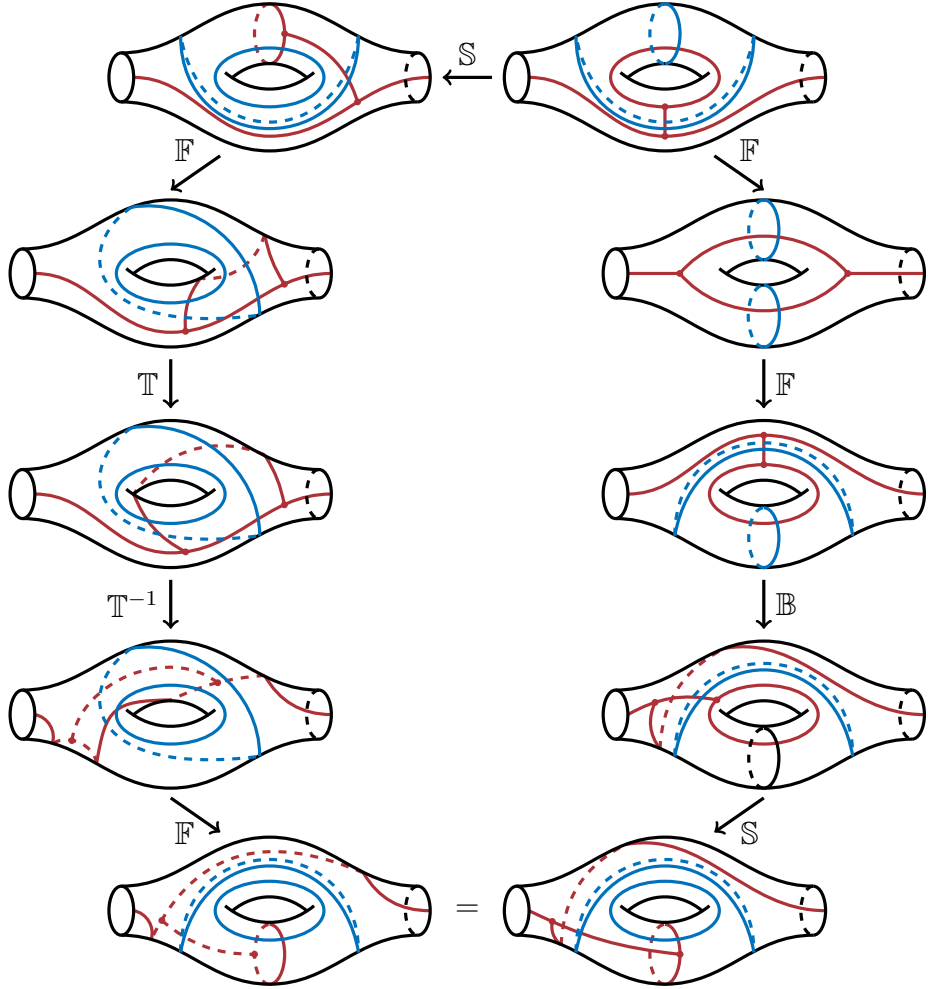
\begin{figure}[ht]
	\centering
		\begin{tikzpicture}[scale=.65]
    		\begin{scope}[shift={(4,0)}]
        		\draw[very thick, Maroon, in=180, out=0] (-2.75,0) to (0,-1.2) to (3.25,0);
        		\draw[very thick, Maroon] (0,0) circle (1.1 and .6);
        		\fill[Maroon] (0,-.6) circle (.07);
        		\fill[Maroon] (0,-1.2) circle (.07);
        		\draw[very thick, Maroon] (0,-.6) to (0,-1.2);
        		\draw[very thick, in=180, out=0] (-3,.5) to (0,1.5) to (3,.5);
        		\draw[very thick, in=180, out=0] (-3,-.5) to (0,-1.5) to (3,-.5);
        		\draw[very thick, bend right=40] (-.9,0.1) to (.9,.1);
        		\draw[very thick, bend left=40] (-.75,0) to (.75,0);
        		\begin{scope}[xscale=.5]
        		    \draw[very thick] (-6,-.5)  arc (-90:270:.5);
        		\end{scope}
        		\begin{scope}[xscale=.5]
        		    \draw[very thick] (6,-.5)  arc (-90:90:.5);
        		    \draw[dashed, very thick] (6,.5)  arc (90:270:.5);
        		\end{scope}
        		\begin{scope}[xscale=.5]
        		    \draw[very thick, RoyalBlue] (0,1.5)  arc (90:-90:.61);
        		    \draw[dashed, RoyalBlue, very thick] (0,1.5)  arc (90:270:.61);
        		\end{scope}
        		\draw[very thick, RoyalBlue, dashed, bend left=80, looseness=1.7] (1.8,.85) to (-1.8,.85);
        		\draw[very thick, RoyalBlue, bend left=90, looseness=1.8] (1.8,.85) to (-1.8,.85);
    		\end{scope}
    		\begin{scope}[shift={(6,-4)}]
        		\fill[Maroon] (-1.7,0) circle (.07);
        		\fill[Maroon] (1.7,0) circle (.07);
        		\draw[very thick, Maroon] (-1.7,0) to (-2.75,0);
        		\draw[very thick, Maroon, bend left=50] (-1.7,0) to (1.7,0);
        		\draw[very thick, Maroon, bend right=50] (-1.7,0) to (1.7,0);
        		\draw[very thick, Maroon] (1.7,0) to (3.25,0);
        		\draw[very thick, in=180, out=0] (-3,.5) to (0,1.5) to (3,.5);
        		\draw[very thick, in=180, out=0] (-3,-.5) to (0,-1.5) to (3,-.5);
        		\draw[very thick, bend right=40] (-.9,0.1) to (.9,.1);
        		\draw[very thick, bend left=40] (-.75,0) to (.75,0);
        		\begin{scope}[xscale=.5]
        		    \draw[very thick] (-6,-.5)  arc (-90:270:.5);
        		\end{scope}
        		\begin{scope}[xscale=.5]
        		    \draw[very thick] (6,-.5)  arc (-90:90:.5);
        		    \draw[dashed, very thick] (6,.5)  arc (90:270:.5);
        		\end{scope}
        		\begin{scope}[xscale=.5]
        		    \draw[very thick, RoyalBlue] (0,-1.5)  arc (-90:90:.63);
        		    \draw[dashed, very thick, RoyalBlue] (0,-1.5)  arc (-90:-270:.63);
        		\end{scope}
        		\begin{scope}[xscale=.5]
        		    \draw[very thick, RoyalBlue] (0,1.5)  arc (90:-90:.61);
        		    \draw[dashed, RoyalBlue, very thick] (0,1.5)  arc (90:270:.61);
        		\end{scope}
    		\end{scope}
		\begin{scope}[shift={(6,-8.5)}]
        		\draw[very thick, Maroon, in=180, out=0] (-2.75,0) to (0,1.2) to (3.25,0);
        		\draw[very thick, Maroon] (0,0) circle (1.1 and .6);
        		\fill[Maroon] (0,.6) circle (.07);
        		\fill[Maroon] (0,1.2) circle (.07);
        		\draw[very thick, Maroon] (0,.6) to (0,1.2);
        		\draw[very thick, in=180, out=0] (-3,-.5) to (0,-1.5) to (3,-.5);
        		\draw[very thick, in=180, out=0] (-3,.5) to (0,1.5) to (3,.5);
        		\draw[very thick, bend right=40] (-.9,0.1) to (.9,.1);
        		\draw[very thick, bend left=40] (-.75,0) to (.75,0);
        		\begin{scope}[xscale=.5]
        		    \draw[very thick] (-6,-.5)  arc (-90:270:.5);
        		\end{scope}
        		\begin{scope}[xscale=.5]
        		    \draw[very thick] (6,-.5)  arc (-90:90:.5);
        		    \draw[dashed, very thick] (6,.5)  arc (90:270:.5);
        		\end{scope}
        		\begin{scope}[xscale=.5]
        		    \draw[very thick, RoyalBlue] (0,-1.5)  arc (-90:90:.61);
        		    \draw[dashed, very thick, RoyalBlue] (0,-1.5)  arc (-90:-270:.61);
        		\end{scope}
        		\draw[very thick, RoyalBlue, bend left=80, looseness=1.7] (-1.8,-.85) to (1.8,-.85);
        		\draw[very thick, RoyalBlue, dashed, bend left=90, looseness=1.8] (-1.8,-.85) to (1.8,-.85);
    		\end{scope}
    		\begin{scope}[shift={(-4,0)}]
        		\draw[very thick, Maroon, in=180, out=0] (-2.75,0) to (0,-1.2) to (3.25,0);
        		\begin{scope}[xscale=.5]
        		    \draw[very thick, Maroon] (0,1.5)  arc (90:-90:.61);
        		    \draw[dashed, very thick, Maroon] (0,1.5)  arc (90:270:.61);
        		\end{scope}
        		\draw[very thick, Maroon, bend left=30] (.305,.89) to (1.78,-.5);        
        		\fill[Maroon] (.305,.89) circle (.07);		
        		\fill[Maroon] (1.78,-.5) circle (.07);	
        		\draw[very thick, in=180, out=0] (-3,.5) to (0,1.5) to (3,.5);
        		\draw[very thick, in=180, out=0] (-3,-.5) to (0,-1.5) to (3,-.5);
        		\draw[very thick, bend right=40] (-.9,0.1) to (.9,.1);
        		\draw[very thick, bend left=40] (-.75,0) to (.75,0);
        		\begin{scope}[xscale=.5]
        		    \draw[very thick] (-6,-.5)  arc (-90:270:.5);
        		\end{scope}
        		\begin{scope}[xscale=.5]
        		    \draw[very thick] (6,-.5)  arc (-90:90:.5);
        		    \draw[dashed, very thick] (6,.5)  arc (90:270:.5);
        		\end{scope}
        		\draw[very thick, RoyalBlue] (0,0) circle (1.1 and .6);
        		\draw[very thick, RoyalBlue, dashed, bend left=80, looseness=1.7] (1.8,.85) to (-1.8,.85);
        		\draw[very thick, RoyalBlue, bend left=90, looseness=1.8] (1.8,.85) to (-1.8,.85);
    		\end{scope}

		\begin{scope}[shift={(6,-13)}]
        		\draw[very thick, Maroon, in=180, out=10] (-.8,1.35) to (3.25,0);
        		\draw[very thick, Maroon] (0,0) circle (1.1 and .6);
        		\draw[very thick, Maroon, bend left=20] (-2.75,0) to (-.95,.3); 
        		\fill[Maroon] (-2.2,.25) circle (0.07);
        		\fill[Maroon] (-.95,.3) circle (0.07);
        		\draw[very thick, Maroon, bend left=30, dashed] (-2.1,-.7) to (-.8,1.35);
        		\draw[very thick, Maroon, bend left=30] (-2.1,-.7) to (-2.2,.25);
        		\draw[very thick, in=180, out=0] (-3,-.5) to (0,-1.5) to (3,-.5);
        		\draw[very thick, in=180, out=0] (-3,.5) to (0,1.5) to (3,.5);
        		\draw[very thick, bend right=40] (-.9,0.1) to (.9,.1);
        		\draw[very thick, bend left=40] (-.75,0) to (.75,0);
        		\begin{scope}[xscale=.5]
        		    \draw[very thick] (-6,-.5)  arc (-90:270:.5);
        		\end{scope}
        		\begin{scope}[xscale=.5]
        		    \draw[very thick] (6,-.5)  arc (-90:90:.5);
        		    \draw[dashed, very thick] (6,.5)  arc (90:270:.5);
        		\end{scope}
        		\begin{scope}[xscale=.5]
        		    \draw[very thick] (0,-1.5)  arc (-90:90:.61);
        		    \draw[dashed, very thick] (0,-1.5)  arc (-90:-270:.61);
        		\end{scope}
        		\draw[very thick, RoyalBlue, bend left=80, looseness=1.7] (-1.8,-.85) to (1.8,-.85);
        		\draw[very thick, RoyalBlue, dashed, bend left=90, looseness=1.8] (-1.8,-.85) to (1.8,-.85);
    		\end{scope}
   		\begin{scope}[shift={(-6,-4)}]
        		\draw[very thick, Maroon, in=180, out=0] (-2.75,0) to (0,-1.2) to (3.25,0);	
        		\fill[Maroon] (2.3,-.22) circle (.07);	
        		\fill[Maroon] (.3,-1.18) circle (.07);
        		\draw[very thick, Maroon, bend left=10] (2.3,-.22) to (1.9,.8);	
        		\draw[very thick, Maroon, dashed, bend left=40] (1.9,.8) to (.6,-.1);
        		\draw[very thick, Maroon, bend right=20] (.6,-.1) to (.3,-1.18);
        		\draw[very thick, in=180, out=0] (-3,.5) to (0,1.5) to (3,.5);
        		\draw[very thick, in=180, out=0] (-3,-.5) to (0,-1.5) to (3,-.5);
        		\draw[very thick, bend right=40] (-.9,0.1) to (.9,.1);
        		\draw[very thick, bend left=40] (-.75,0) to (.75,0);
        		\begin{scope}[xscale=.5]
        		    \draw[very thick] (-6,-.5)  arc (-90:270:.5);
        		\end{scope}
        		\begin{scope}[xscale=.5]
        		    \draw[very thick] (6,-.5)  arc (-90:90:.5);
        		    \draw[dashed, very thick] (6,.5)  arc (90:270:.5);
        		\end{scope}
        		\draw[very thick, RoyalBlue] (0,0) circle (1.1 and .6);
			\draw[very thick, bend left=50, RoyalBlue] (-.8,1.35) to (1.8,-.85);
			\draw[very thick, dashed, RoyalBlue] (-.8,1.35) .. controls (-2,.5) and (-1.7,-1.3) .. (1.8,-.85);
    		\end{scope}
   		\begin{scope}[shift={(-6,-8.5)}]
        		\draw[very thick, Maroon, in=180, out=0] (-2.75,0) to (0,-1.2) to (3.25,0);	
        		\fill[Maroon] (2.3,-.22) circle (.07);	
        		\fill[Maroon] (.3,-1.18) circle (.07);
        		\draw[very thick, Maroon, bend right=10] (2.3,-.22) to (1.9,.8);	
        		\draw[very thick, Maroon, dashed, bend right=40] (1.9,.8) to (-.75,0);
        		\draw[very thick, Maroon, bend right=20] (-.75,0) to (.3,-1.18);
        		\draw[very thick, in=180, out=0] (-3,.5) to (0,1.5) to (3,.5);
        		\draw[very thick, in=180, out=0] (-3,-.5) to (0,-1.5) to (3,-.5);
        		\draw[very thick, bend right=40] (-.9,0.1) to (.9,.1);
        		\draw[very thick, bend left=40] (-.75,0) to (.75,0);
        		\begin{scope}[xscale=.5]
        		    \draw[very thick] (-6,-.5)  arc (-90:270:.5);
        		\end{scope}
        		\begin{scope}[xscale=.5]
        		    \draw[very thick] (6,-.5)  arc (-90:90:.5);
        		    \draw[dashed, very thick] (6,.5)  arc (90:270:.5);
        		\end{scope}
        		\draw[very thick, RoyalBlue] (0,0) circle (1.1 and .6);
			\draw[very thick, bend left=50, RoyalBlue] (-.8,1.35) to (1.8,-.85);
			\draw[very thick, dashed, RoyalBlue] (-.8,1.35) .. controls (-2,.5) and (-1.7,-1.3) .. (1.8,-.85);
    		\end{scope}
		\begin{scope}[shift={(4,-17)}]
        		\draw[very thick, Maroon, in=180, out=10] (-.8,1.35) to (3.25,0);
        		\draw[very thick, Maroon, bend right=10] (-2.75,0) to (.305,-.89); 
        		\fill[Maroon] (.305,-.89) circle (0.07);
        		\fill[Maroon] (-2.28,-.22) circle (0.07);
        		\begin{scope}[xscale=.5]
        		    \draw[very thick, Maroon] (0,-1.5)  arc (-90:90:.61);
        		    \draw[dashed, very thick, Maroon] (0,-1.5)  arc (-90:-270:.61);
        		\end{scope}
        		\draw[very thick, Maroon, bend left=30, dashed] (-2.1,-.7) to (-.8,1.35);
        		\draw[very thick, Maroon, bend left=30] (-2.1,-.7) to (-2.28,-.22);
        		\draw[very thick, in=180, out=0] (-3,-.5) to (0,-1.5) to (3,-.5);
        		\draw[very thick, in=180, out=0] (-3,.5) to (0,1.5) to (3,.5);
        		\draw[very thick, bend right=40] (-.9,0.1) to (.9,.1);
        		\draw[very thick, bend left=40] (-.75,0) to (.75,0);
        		\begin{scope}[xscale=.5]
        		    \draw[very thick] (-6,-.5)  arc (-90:270:.5);
        		\end{scope}
        		\begin{scope}[xscale=.5]
        		    \draw[very thick] (6,-.5)  arc (-90:90:.5);
        		    \draw[dashed, very thick] (6,.5)  arc (90:270:.5);
        		\end{scope}
        		\draw[very thick, RoyalBlue] (0,0) circle (1.1 and .6);
        		\draw[very thick, RoyalBlue, bend left=80, looseness=1.7] (-1.8,-.85) to (1.8,-.85);
        		\draw[very thick, RoyalBlue, dashed, bend left=90, looseness=1.8] (-1.8,-.85) to (1.8,-.85);
    		\end{scope}
   		\begin{scope}[shift={(-6,-13)}]
        		\draw[very thick, Maroon, bend left=30] (3.25,0) to (1.9,.8);
        		\draw[very thick, Maroon, dashed, bend right=10] (1.9,.8) to (0,.3);
        		\draw[very thick, Maroon, bend right=40, looseness=1.5] (0,.3) to (-1.5,-1);
        		\draw[very thick, Maroon, dashed, bend right=60] (-1.5,-1) to (-2.4,-.6);
        		\draw[very thick, Maroon, bend right=50] (-2.4,-.6) to (-2.75,0);
        		\draw[very thick, Maroon, dashed, bend left=60] (-2,-.53) to (.95,.65);
        		\fill[Maroon] (-2,-.53) circle (.07);
        		\fill[Maroon] (.95,.65) circle (.07);
        		\draw[very thick, in=180, out=0] (-3,.5) to (0,1.5) to (3,.5);
        		\draw[very thick, in=180, out=0] (-3,-.5) to (0,-1.5) to (3,-.5);
        		\draw[very thick, bend right=40] (-.9,0.1) to (.9,.1);
        		\draw[very thick, bend left=40] (-.75,0) to (.75,0);
        		\begin{scope}[xscale=.5]
        		    \draw[very thick] (-6,-.5)  arc (-90:270:.5);
        		\end{scope}
        		\begin{scope}[xscale=.5]
        		    \draw[very thick] (6,-.5)  arc (-90:90:.5);
        		    \draw[dashed, very thick] (6,.5)  arc (90:270:.5);
        		\end{scope}
        		\draw[very thick, RoyalBlue] (0,0) circle (1.1 and .6);
			\draw[very thick, bend left=50, RoyalBlue] (-.8,1.35) to (1.8,-.85);
			\draw[very thick, dashed, RoyalBlue] (-.8,1.35) .. controls (-2,.5) and (-1.7,-1.3) .. (1.8,-.85);
    		\end{scope}
   		\begin{scope}[shift={(-4,-17)}]
			 \begin{scope}[xscale=.5]
        		    \draw[very thick, Maroon] (0,-1.5)  arc (-90:90:.61);
        		    \draw[dashed, very thick, Maroon] (0,-1.5)  arc (-90:-270:.61);
        		\end{scope}
        		\draw[very thick, Maroon, bend left=30] (3.25,0) to (1.9,.77);
        		\draw[very thick, Maroon, dashed] (1.9,.77) ..controls (0,1.6) and (-1.5,1.3) .. (-2.4,-.6);
        		\draw[very thick, Maroon, bend right=50] (-2.4,-.6) to (-2.75,0);
        		\draw[very thick, Maroon, dashed, bend right=20] (-2.18,-.2) to (-.305,-.89);
        		\fill[Maroon] (-2.18,-.2) circle (.07);
        		\fill[Maroon] (-.305,-.89) circle (.07);
        		\draw[very thick, in=180, out=0] (-3,.5) to (0,1.5) to (3,.5);
        		\draw[very thick, in=180, out=0] (-3,-.5) to (0,-1.5) to (3,-.5);
        		\draw[very thick, bend right=40] (-.9,0.1) to (.9,.1);
        		\draw[very thick, bend left=40] (-.75,0) to (.75,0);
        		\begin{scope}[xscale=.5]
        		    \draw[very thick] (-6,-.5)  arc (-90:270:.5);
        		\end{scope}
        		\begin{scope}[xscale=.5]
        		    \draw[very thick] (6,-.5)  arc (-90:90:.5);
        		    \draw[dashed, very thick] (6,.5)  arc (90:270:.5);
        		\end{scope}
        		\draw[very thick, RoyalBlue] (0,0) circle (1.1 and .6);
        		\draw[very thick, RoyalBlue, bend left=80, looseness=1.7] (-1.8,-.85) to (1.8,-.85);
        		\draw[very thick, RoyalBlue, dashed, bend left=90, looseness=1.8] (-1.8,-.85) to (1.8,-.85);
    		\end{scope}
  		\begin{scope}[shift={(0,0)}]
    			\draw[very thick,->] (.5,0) to node[above] {$\mathbb{S}$} (-.5,0);
    		\end{scope}
       		\begin{scope}[shift={(6,-10.75)}]
    			\draw[very thick,->] (0,.5) to node[right] {$\mathbb{B}$} (0,-.5);
    		\end{scope}
    		\begin{scope}[shift={(-6,-2.3)}]
    			\draw[very thick,->] (1,.7) to node[above, xshift=-4] {$\mathbb{F}$} (0,0);
    		\end{scope}
        		\begin{scope}[shift={(6,-6.25)}]
    			\draw[very thick,->] (0,.5) to node[right] {$\mathbb{F}$} (0,-.5);
    		\end{scope}
          		\begin{scope}[shift={(-6,-6.25)}]
    			\draw[very thick,->] (0,.5) to node[left] {$\mathbb{T}$} (0,-.5);
    		\end{scope}
    		\begin{scope}[shift={(-6,-10.75)}]
    			\draw[very thick,->] (0,.5) to node[left] {$\mathbb{T}^{-1}$} (0,-.5);
    		\end{scope}
    		\begin{scope}[shift={(6,-14.7)}]
    			\draw[very thick,<-] (-1,-.7) to node[below, yshift=2, xshift=4] {$\mathbb{S}$} (0,0);
    		\end{scope}
  		\begin{scope}[shift={(-6,-14.7)}]
    			\draw[very thick,<-] (1,-.7) to node[below, yshift=2, xshift=-4] {$\mathbb{F}$} (0,0);
    		\end{scope}
   		\begin{scope}[shift={(6,-2.3)}]
    			\draw[very thick,->] (-1,.7) to node[above, xshift=4] {$\mathbb{F}$} (0,0);
    		\end{scope}
    		\begin{scope}[shift={(0,-17)}]
    			\node at (0,0) {$=$};
    		\end{scope}
	\end{tikzpicture}
	\caption{The relation at $g=1$, $n=2$.} \label{fig:relation g=1, n=2}
\end{figure}

\paragraph{Relation at $g=1$, $n=2$.} There is one last consistency relation that can be derived from the consistency on the two-punctured torus. The sequence of moves for this relation is quite complicated and is depicted in Figure~\ref{fig:relation g=1, n=2}.
In particular, this is the only relation that relates the data of $\mathbb{F}$ and $\mathbb{S}$. Carefully following the two different paths through the picture gives the following identity:
\begin{multline}
            \mathbb{S}_{P_1,P_2}[P_3] \int_0^\infty\!\! \d P_4 \ \mathbb{F}_{P_3,P_4}\!\begin{bmatrix}
                P_2 & P_0' \\
                P_2 & P_0
            \end{bmatrix}
            \mathrm{e}^{2\pi i  (\Delta_4-\Delta_2)}\, 
        \mathbb{F}_{P_4,P_5}\!\begin{bmatrix}
                P_0 & P_0' \\
                P_2 & P_2
            \end{bmatrix} \\
        =
        \int_0^\infty\!\! \d P_6 \ \mathbb{F}_{P_3,P_6}\!\begin{bmatrix}
            P_1 & P_0 \\
            P_1 & P_0'
        \end{bmatrix}
        \mathbb{F}_{P_1,P_5}\!\begin{bmatrix}
            P_0 & P_0' \\
            P_6 & P_6
        \end{bmatrix} \mathrm{e}^{\pi i (\Delta_0+\Delta_0'-\Delta_5)} \, \mathbb{S}_{P_6,P_2}[P_5]\ .\label{eq:g=1, n=2 relation}
        \end{multline}

\subsection{Vacuum blocks} \label{subsec:vacuum blocks}
An important special case is the crossing transformation of conformal blocks where some of the internal or external lines are the vacuum. In this case the blocks are not normalizable, but their crossing transformation can still be expressed in terms of the basis of conformal blocks. Concretely, this means that
\be 
\mathbb{F}_{P_3,P_{32}}\!\begin{bmatrix}
    P_3 & P_2 \\ \id & P_1
\end{bmatrix}\ , \qquad \mathbb{F}_{\id,P_{32}}\!\begin{bmatrix}
    P_3 & P_2 \\
    P_3 & P_2
\end{bmatrix}\ , \qquad 
\mathbb{S}_{P_1,P_2}[\id]\ , \qquad
\mathbb{S}_{\id,P}[\id] \label{eq:special vacuum F and S}
\ee
are well-defined objects. Here and in the following, we often simply write $\id$ when the corresponding state is the identity. This corresponds to the value $P=\frac{iQ}{2}$, although taking the limit $P \to \frac{iQ}{2}$ doesn't necessarily always give the right result since we may need to decouple the null-state by hand as discussed after eq.~\eqref{eq:conformal block zero-punctured torus limit}.
\paragraph{Spherical crossing kernel with an external identity.}
The first case in \eqref{eq:special vacuum F and S} is trivial to determine. The four-punctured block with one trivial external operator is just a three-point block given by \eqref{eq:conformal block three-punctured sphere}. It in particular coincides in the two channels and thus we simply have
\be 
\mathbb{F}_{P_3,P_{32}}\!\begin{bmatrix}
    P_3 & P_2 \\ \id & P_1
\end{bmatrix}=\delta(P_{32}-P_1)\ . \label{eq:F symbol identity delta function}
\ee
\paragraph{Modular crossing kernel with external identity.}
Similarly, $\mathbb{S}_{P_1,P_2}[\id]$ can be computed directly, since it is just the usual modular S-matrix on the space of Virasoro characters. The Virasoro characters are given by
\be 
\frac{q^{P^2}}{\eta(\tau)}\ ,
\ee
where as usual $q=\mathrm{e}^{2\pi i \tau}$.
For the modular transformation, we have
\be 
\frac{\mathrm{e}^{-\frac{2\pi i \tau}{P_1^2}}}{\eta(-\frac{1}{\tau})}
=\int_0^\infty\!\! \d P_2\ \mathbb{S}_{P_1,P_2}[\id] \, \frac{\mathrm{e}^{2\pi i \tau P_2^2}}{\eta(\tau)}\ ,
\ee
where we used the modular transformation of $\eta(\tau)$. We read off
\be 
\mathbb{S}_{P_1,P_2}[\id]=2 \sqrt{2}\, \cos(4\pi P_1 P_2)\ . \label{eq:SP1,P2[1]}
\ee
For $\mathbb{S}_{\id,P}[\id]$, we have
\be 
\mathbb{S}_{\id,P}[\id]=\mathbb{S}_{P_1=\frac{i}{2}(b+b^{-1}),P}[\id]-\mathbb{S}_{P_1=\frac{i}{2}(b-b^{-1}),P}[\id]=4 \sqrt{2}\, \sinh(2\pi b P) \sinh(2\pi b^{-1} P)\ , \label{eq:S1P[1] computation}
\ee
since we recall from the discussion after eq.~\eqref{eq:conformal block zero-punctured torus limit} that we need to decouple the null-state in the vacuum representation explicitly. 
Finally, the second case in \eqref{eq:special vacuum F and S} also admits a relatively simple formula, but it is not yet obvious from what we discussed so far. We will see below that it is related to the DOZZ formula.

\subsection{Special cases of the consistency conditions} \label{subsec:special cases consistency conditions}
It is useful to consider a number of special cases of the Moore-Seiberg consistency conditions. 

\paragraph{Tetrahedral symmetry of $\mathbb{F}$.} The crossing kernel $\mathbb{F}$ has in fact a bigger symmetry group then the basic invariances \eqref{eq:obvious symmetries}. Consider the pentagon identity \eqref{eq:g=0, n=5 pentagon equation}. We can specialize $P_4=P_5$, $P_{21}=P_3$ and $P_{54}=\id$. Even though $P_{54}$ is not above threshold, this is allowed by analyticity of the crossing kernel as discussed above.
Evaluating the pentagon identity on this starting conformal block and using \eqref{eq:F symbol identity delta function} leads to
\be 
\mathbb{F}_{\id,P_{51}}\!\begin{bmatrix}
    P_4 & P_1 \\P_4 & P_1
\end{bmatrix} 
\mathbb{F}_{P_1,P_{43}}\!\begin{bmatrix}
    P_4 & P_3 \\P_{51} & P_2
\end{bmatrix}=
\mathbb{F}_{P_3,P_{51}}\!\begin{bmatrix}
    P_{43} & P_2 \\ P_4 & P_1
\end{bmatrix}
\mathbb{F}_{\id,P_{43}}\!\begin{bmatrix}
    P_4 & P_3 \\
    P_4 & P_3
\end{bmatrix} \label{eq:exchange symmetry F degenerated pentagon}
\ee
Thus, this identity allows us to exchange labels of $\mathbb{F}$ at the cost of the simpler factors with the crossing kernel of the identity block.

We can further specialize this by putting $P_1=P_4$ and solve for the last crossing kernel $\mathbb{F}$. This gives
\be 
\mathbb{F}_{\id,P_{43}}\!\begin{bmatrix}
    P_4 & P_3 \\
    P_4 & P_3
\end{bmatrix}=\frac{\mathbb{F}_{\id,P_{51}}\!\begin{bmatrix}
    P_4 & P_4 \\P_4 & P_4
\end{bmatrix} 
\mathbb{F}_{P_4,P_{43}}\!\begin{bmatrix}
    P_4 & P_3 \\P_{51} & P_2
\end{bmatrix}}{\mathbb{F}_{P_3,P_{51}}\!\begin{bmatrix}
    P_{43} & P_2 \\ P_4 & P_4
\end{bmatrix}}
\ee
This identity for $\mathbb{F}$ on the left-hand side holds for any choice of $P_{51}$ and $P_2$. We will now consider the limit $P_{51} \to \id$, i.e.\ $P_{51} \to \frac{iQ}{2}$. Even though the appearing $\mathbb{F}$ symbols appearing on the right hand side do no longer have a clear meaning in terms of crossing, this allows us to deduce a certain symmetry property of the left-hand side. The last factor in the numerator is simple, it tends to a $\delta$-function imposing $P_{43}=P_2$. Except for the first factor in the numerator, everything on the right-hand side is then symmetric under exchange of $P_4$ and $P_{43}$. Thus we conclude that for the ratio of the left-hand side with $P_4$ and $P_{43}$ exchanged, we have
\be 
\frac{\mathbb{F}_{\id,P_{43}}\!\begin{bmatrix}
    P_4 & P_3 \\
    P_4 & P_3
\end{bmatrix}}{\mathbb{F}_{\id,P_{4}}\!\begin{bmatrix}
    P_{43} & P_3 \\
    P_{43} & P_3
\end{bmatrix}}=\lim_{P_{51} \to \id } \frac{\mathbb{F}_{\id,P_{51}}\!\begin{bmatrix}
    P_4 & P_4 \\P_4 & P_4
\end{bmatrix} }{\mathbb{F}_{\id,P_{51}}\!\begin{bmatrix}
    P_{43} & P_{43} \\P_{43} & P_{43}
\end{bmatrix} } \ . \label{eq:ratio F exchange rho0}
\ee
This means that we can write
\be 
\mathbb{F}_{\id,P_{43}}\!\begin{bmatrix}
    P_4 & P_3 \\
    P_4 & P_3
\end{bmatrix}=C_0(P_3,P_4,P_{43}) \rho_0(P_{43}) \ , \label{eq:F1 in terms of C0 rho0}
\ee
for some totally symmetric function $C_0(P_3,P_4,P_{43})$ and some function $\rho_0(P)$. The right hand side of eq.\eqref{eq:ratio F exchange rho0} is then the ratio $\frac{\rho_0(P_{43})}{\rho_0(P_4)}$. Given that $\mathbb{F}$ behaves in a simple way under $P_3 \to \id$, see eq.~\eqref{eq:F symbol identity delta function}, we have
\be 
\lim_{P_1 \to \id} C_0(P_1,P_2,P_3)=\rho_0(P_2)^{-1} \delta(P_2-P_3)\ . \label{eq:limit C0 two-point function}
\ee
These definitions define both $C_0(P_1,P_2,P_3)$ and $\rho_0(P)$ only up to a constant that cancels out of the combination \eqref{eq:F1 in terms of C0 rho0}. We will fix a convenient choice for the constant below.
The equation \eqref{eq:exchange symmetry F degenerated pentagon} as well as the trivial symmetries \eqref{eq:obvious symmetries} can be summarized by saying that the combination
\be 
    \begin{tikzpicture}[baseline={([yshift=-.5ex]current bounding box.center)}, scale=.7]
    \draw[very thick] (0,0) to node[below] {$P_{32}$} (4,0);
    \draw[very thick, dashed] (0,0) to node[right, shift={(0,-.1)}] {$P_4$} (2.8,1.5) to node[left] {$P_3$} (4,0);
    \draw[very thick, dashed] (2.8,1.5) to node[left, shift={(.1,-.2)}] {$P_{21}$} (2.5,3.5);
    \draw[very thick] (0,0) to node[left] {$P_1$} (2.5,3.5) to node[right] {$P_2$} (4,0);
\end{tikzpicture}\longleftrightarrow \frac{C_0(P_1,P_2,P_{21}) C_0(P_3,P_4,P_{21})}{\rho_0(P_{32})} \, \mathbb{F}_{P_{21},P_{32}} \!\begin{bmatrix}
        P_3 & P_2 \\
        P_4 & P_1
    \end{bmatrix} \label{eq:tetrahedrally symmetric combination}
\ee
has the tetrahedral symmetry implied by the picture.

\paragraph{Consequences for $\mathbb{S}$.} We now work out simple consequences of the Moore-Seiberg relations for the modular crossing kernel $\mathbb{S}$. The most important consequence originates from the last identity of the two-punctured torus \eqref{eq:g=1, n=2 relation}. Let us specialize $P_0'=P_0$, which gives
\begin{multline}
            \mathbb{S}_{P_1,P_2}[P_3] \int_0^\infty\!\!\d P_4 \ \mathbb{F}_{P_3,P_4}\!\begin{bmatrix}
                P_2 & P_0 \\
                P_2 & P_0
            \end{bmatrix}
            \mathrm{e}^{2\pi i  (\Delta_4-\Delta_2)}\, 
        \mathbb{F}_{P_4,P_5}\!\begin{bmatrix}
                P_0 & P_0 \\
                P_2 & P_2
            \end{bmatrix} \\
        =
        \int_0^\infty\!\! \d P_6 \ \mathbb{F}_{P_3,P_6}\!\begin{bmatrix}
            P_1 & P_0 \\
            P_1 & P_0
        \end{bmatrix}
        \mathbb{F}_{P_1,P_5}\!\begin{bmatrix}
            P_0 & P_0 \\
            P_6 & P_6
        \end{bmatrix} \mathrm{e}^{\pi i (2\Delta_0-\Delta_5)} \, \mathbb{S}_{P_6,P_2}[P_5]\ .
\end{multline}
We can then set $P_1=P_3=\id$. This is allowed since it means that we are evaluating the identity on a conformal block with identities as internal lines. The first $\mathbb{F}$-symbol in the second line gives a delta-function $\delta(P_6-P_1)$, which localizes the integral. We hence obtain
\begin{multline} 
\mathbb{S}_{\id,P_2}[\id]\int_0^\infty\!\! \d P_4 \ \mathbb{F}_{\id,P_4}\!\begin{bmatrix}
    P_2 & P_0 \\
    P_2 & P_0
\end{bmatrix}\, \mathrm{e}^{2\pi i(\Delta_4-\Delta_2)} \, \mathbb{F}_{P_4,P_5} \!\begin{bmatrix}
    P_0 & P_0 \\
    P_2 & P_2
\end{bmatrix}\\
=\mathbb{F}_{\id,P_5} \!\begin{bmatrix}
    P_0 & P_0 \\ P_0 & P_0
\end{bmatrix} \, \mathrm{e}^{\pi i(2 \Delta_0-\Delta_5)} \, \mathbb{S}_{P_0,P_2}[P_5]\ .
\end{multline}
We can rename $P_0 \to P_1$, $P_5 \to P_0$ and $P_4 \to P_{21}$ plug in \eqref{eq:F1 in terms of C0 rho0} to obtain
\begin{multline}
    \mathbb{S}_{P_1,P_2}[P_0]=\mathbb{S}_{\id,P_2}[\id]\int_0^\infty\!\! \d P_{21}\ \frac{\rho_0(P_{21}) C_0(P_1,P_2,P_{21})}{\rho_0(P_0) C_0(P_1,P_1,P_0)} \, \mathrm{e}^{\pi i (2 \Delta_{21}-2\Delta_2-2\Delta_1+\Delta_0)} \\ \times  \mathbb{F}_{P_{21},P_0} \!\begin{bmatrix}
        P_1 & P_1 \\P_2 & P_2 
    \end{bmatrix}\ , \label{eq:S F relation}
\end{multline}
where $\mathbb{S}_{\id,P_2}[\id]$ is given explicitly by \eqref{eq:S1P[1] computation}.
Notice that this is almost symmetric under exchange of $P_1$ and $P_2$. In fact, we see that the combination
\be 
\frac{\mathbb{S}_{P_1,P_2}[P_0]}{\mathbb{S}_{\id,P_2}[\id] }\, C_0(P_1,P_1,P_0) \label{eq:S symmetric under exchange combination}
\ee
is invariant under $P_1 \leftrightarrow P_2$ exchange. 

Consider finally the limit $P_0 \to \id$, which should recover $\mathbb{S}_{P_1,P_2}[\id]$ as computed in eq.~\eqref{eq:SP1,P2[1]}. In particular, $\mathbb{S}_{P_1,P_2}[\id]$ is symmetric in $P_1$ and $P_2$. Using the limit \eqref{eq:limit C0 two-point function}, symmetry of \eqref{eq:S symmetric under exchange combination} implies that the functions $\mathbb{S}_{1,P}[\id]$ and $\rho_0(P)$ must agree up to a constant. Since the relation eq.~\eqref{eq:F1 in terms of C0 rho0} defined $\rho_0(P)$ only up to a free overall constant, we can define
\be 
\rho_0(P) \equiv \mathbb{S}_{\id,P}[\id]=4 \sqrt{2}\, \sinh(2\pi b P) \sinh(2\pi b^{-1} P)\ , \label{eq:rho0 definition}
\ee
where we used eq.~\eqref{eq:S1P[1] computation}.
Eq.~\eqref{eq:S F relation} hence expresses $\mathbb{S}$ fully in terms of $\mathbb{F}$ and thus the main task in the following will be the determination of $\mathbb{F}$, which satisfies the pentagon equation \eqref{eq:g=0, n=5 pentagon equation} and hexagon equation \eqref{eq:g=0, n=4 hexagon equation}.

\paragraph{Relation to the DOZZ formula.} $C_0(P_1,P_2, P_3)$ is the DOZZ-formula for the three-point function of Liouville theory \cite{Dorn:1994xn, Zamolodchikov:1995aa}, albeit in a slightly unusual normalization where the two-point function is given by $\rho_0(P)^{-1}$, see eq.~\eqref{eq:limit C0 two-point function}.\footnote{From a bootstrap point of view, this normalization is actually much nicer than the convention normalization used in the literature, since it is in particular reflection symmetric.} This can be seen by noticing that the pentagon identity implies that $C_0(P_1,P_2,P_3)$ gives a solution to the crossing equations for generic central charges $c \ge 25$ and is thus uniquely fixed. 
Start with \eqref{eq:g=0, n=4 idempotency F} (after a relabelling) and apply the exchange formula \eqref{eq:exchange symmetry F degenerated pentagon} twice, yielding
\begin{align}
    \delta(P_{32}-P_{32}')&=\int_0^\infty\!\! \mathrm{d}P_{21}\ \mathbb{F}_{P_{32},P_{21}}\!\begin{bmatrix}
P_2 & P_1 \\
P_3 & P_4
\end{bmatrix}
\mathbb{F}_{P_{21},P_{32}'}\!\begin{bmatrix}
P_3 & P_2 \\
P_4 & P_1
\end{bmatrix} \\
&=\int_0^\infty\!\! \mathrm{d}P_{21}\ \frac{\rho_0(P_{21}) C_0(P_1,P_2,P_{21})}{\rho_0(P_4)C_0(P_1,P_4,P_{32})}\, \mathbb{F}_{P_2,P_4}\!\begin{bmatrix}
P_{21} & P_3 \\
P_1 & P_{32}
\end{bmatrix}
\mathbb{F}_{P_{21},P_{32}'}\!\begin{bmatrix}
P_3 & P_2 \\
P_4 & P_1
\end{bmatrix} \\
&=\int_0^\infty\!\! \mathrm{d}P_{21}\ \frac{\rho_0(P_{21})\rho_0(P_4) C_0(P_1,P_2,P_{21}) C_0(P_3,P_4,P_{21})}{\rho_0(P_4)\rho_0(P_{32})C_0(P_1,P_4,P_{32})C_0(P_2,P_3,P_{32})}\nonumber\\
&\qquad\qquad\qquad\qquad\qquad\qquad\times \mathbb{F}_{P_{21},P_{32}}\!\begin{bmatrix}
P_2 & P_3 \\
P_1 & P_4
\end{bmatrix}
\mathbb{F}_{P_{21},P_{32}'}\!\begin{bmatrix}
P_3 & P_2 \\
P_4 & P_1
\end{bmatrix} \ .
\end{align}
We can recast this as
\begin{multline}
    \int_0^\infty\!\! \mathrm{d}P_{21}\ \rho_0(P_{21}) C_0(P_1,P_2,P_{21}) C_0(P_3,P_4,P_{21})\mathbb{F}_{P_{21},P_{32}}\!\begin{bmatrix}
P_2 & P_3 \\
P_1 & P_4
\end{bmatrix}
\mathbb{F}_{P_{21},P_{32}'}\!\begin{bmatrix}
P_3 & P_2 \\
P_4 & P_1
\end{bmatrix} 
\\
=\rho_0(P_{32})C_0(P_1,P_4,P_{32})C_0(P_2,P_3,P_{32})\delta(P_{32}-P_{32}')\ . \label{eq:crossing symmetry Liouville theory}
\end{multline}
This equality expresses precisely the crossing symmetry of Liouville theory. Indeed, if we multiply it with the left- and right-moving $t$-channel blocks, integrate over $P_{32}$ and $P_{32}'$ and use the definition of the crossing kernel, we get
\begin{multline}  
\int_0^\infty\!\! \mathrm{d}P_{21}\ \rho_0(P_{21}) C_0(P_1,P_2,P_{21}) C_0(P_4,P_3,P_{21}) \left| \!\!
\begin{tikzpicture}[baseline={([yshift=-.5ex]current bounding box.center)}, xscale=.6, yscale=.6]
        \draw[very thick, Maroon, bend left=30] (-1.8,1.2) to (-.7,0);
        \draw[very thick, Maroon, bend right=30] (-1.8,-1.2) to (-.7,0);
        \draw[very thick, Maroon, bend right=30] (2.2,1.2) to (.7,0);
        \draw[very thick, Maroon, bend left=30] (2.2,-1.2) to (.7,0);
        \draw[very thick, Maroon] (-.7,0) to (.7,0);
        \fill[Maroon] (-.7,0) circle (.07);
        \fill[Maroon] (.7,0) circle (.07);
        \draw[very thick] (-2,1.2) circle (.2 and .5);
        \draw[very thick] (-2,-1.2) circle (.2 and .5);
        \begin{scope}[shift={(2,1.2)}, xscale=.4]
            \draw[very thick] (0,-.5) arc (-90:90:.5);
            \draw[very thick, dashed] (0,.5) arc (90:270:.5);
        \end{scope}
        \begin{scope}[shift={(2,-1.2)}, xscale=.4]
            \draw[very thick] (0,-.5) arc (-90:90:.5);
            \draw[very thick, dashed] (0,.5) arc (90:270:.5);
        \end{scope}
        \begin{scope}[xscale=.6]
            \draw[very thick, RoyalBlue] (0,-.9) arc (-90:90:.9);
            \draw[very thick, dashed, RoyalBlue] (0,.9) arc (90:270:.9);
        \end{scope}
        \draw[very thick, out=0, in=180] (-2,1.7) to (0,.9) to (2,1.7);
        \draw[very thick, out=0, in=180] (-2,-1.7) to (0,-.9) to (2,-1.7);
        \draw[very thick, out=0, in=0, looseness=2.5] (-2,-.7) to (-2,.7);
        \draw[very thick, out=180, in=180, looseness=2.5] (2,-.7) to (2,.7);
        \node at (-2.55,1.2) {$P_3$};
        \node at (2.55,1.2) {$P_2$};            
        \node at (2.55,-1.2) {$P_1$};
        \node at (-2.55,-1.2) {$P_4$}; 
        \node at (0.02,.36) {$P_{21}$};
        \end{tikzpicture}  
        \!\!\right|^2 \\
=\int_0^\infty\!\! \mathrm{d}P_{32}\ \rho_0(P_{32})C_0(P_4,P_1,P_{32})C_0(P_3,P_2,P_{32})
\left|\!\!
\begin{tikzpicture}[baseline={([yshift=-.5ex]current bounding box.center)}, xscale=.6, yscale=.6]
        \draw[very thick, Maroon, bend left=10] (-1.8,1.2) to (0,.6);
        \draw[very thick, Maroon, bend right=10] (-1.8,-1.2) to (0,-.6);
        \draw[very thick, Maroon, bend right=10] (2.2,1.2) to (0,.6);
        \draw[very thick, Maroon, bend left=10] (2.2,-1.2) to (0,-.6);
        \draw[very thick, Maroon] (0,-.6) to (0,.6);
        \fill[Maroon] (0,.6) circle (.07);
        \fill[Maroon] (0,-.6) circle (.07);
        \draw[very thick] (-2,1.2) circle (.2 and .5);
        \draw[very thick] (-2,-1.2) circle (.2 and .5);
        \begin{scope}[shift={(2,1.2)}, xscale=.4]
            \draw[very thick] (0,-.5) arc (-90:90:.5);
            \draw[very thick, dashed] (0,.5) arc (90:270:.5);
        \end{scope}
        \begin{scope}[shift={(2,-1.2)}, xscale=.4]
            \draw[very thick] (0,-.5) arc (-90:90:.5);
            \draw[very thick, dashed] (0,.5) arc (90:270:.5);
        \end{scope}
        \begin{scope}[yscale=.5]
            \draw[very thick, RoyalBlue] (-.97,0) arc (-180:0:.97);
            \draw[very thick, dashed, RoyalBlue] (.97,0) arc (0:180:.97);
        \end{scope}
        \draw[very thick, out=0, in=180] (-2,1.7) to (0,.9) to (2,1.7);
        \draw[very thick, out=0, in=180] (-2,-1.7) to (0,-.9) to (2,-1.7);
        \draw[very thick, out=0, in=0, looseness=2.5] (-2,-.7) to (-2,.7);
        \draw[very thick, out=180, in=180, looseness=2.5] (2,-.7) to (2,.7);
        \node at (-2.55,1.2) {$P_3$};
        \node at (2.55,1.2) {$P_2$};            
        \node at (2.55,-1.2) {$P_1$};
        \node at (-2.55,-1.2) {$P_4$};   
        \node at (.45,.08) {$P_{32}$};
    \end{tikzpicture} \!\!\right|^2\ .
\end{multline}
We should emphasize here that only the crossratio is complex conjugated for the right-moving block, but the Liouville momenta that enter the crossing kernel are not complex conjugated.
Thus, we will see that $C_0(P_1,P_2,P_3)$ coincides with the DOZZ structure constants in a normalization in which the two-point function is $\rho_0(P)^{-1}$. Note also that contrary to the usual normalization of the DOZZ formula, $C_0(P_1,P_2,P_3)$ is reflection symmetric.

Imposing the consequences of the pentagon equation for $C_0(P_1,P_2,P_3)$ recovers Teschner's argument for the uniqueness of the DOZZ formula \cite{Teschner:1995yf}. We spell it out in Section~\ref{subsec:C0 bootstrap}.

\subsection{Degenerate conformal blocks and the shift relation} \label{subsec:degenerate conformal blocks}
So far, we have only explained very general constraints, but have not made much use of Virasoro symmetry, ecxept in our explicit computation of $\mathbb{S}_{P_1,P_2}[\id]$, see eq.~\eqref{eq:SP1,P2[1]}. Virasoro symmetry enters if one considers degenerate Virasoro representations whose Verma module possesses a null vector. After the vacuum representation, the next simplest such degenerate representation occurs for $p=p_{\langle 2,1 \rangle}=-b-\frac{1}{2b}$. There is a similar representation with $b \to b^{-1}$, but we will focus on the first one. We will in the following also focus on the four-punctured sphere, since as we explained we only need to determine $\mathbb{F}$.
\paragraph{Imaginary momenta.} From this point onward, it will be much more convention to use the imaginary Liouville momenta $p$ as defined in eq.~\eqref{eq:definition p}. By abuse of notation, we write
\be 
\mathbb{F}_{p_{21},p_{32}}\!\begin{bmatrix}
    p_3 & p_2 \\ p_4 & p_1
\end{bmatrix} \equiv \mathbb{F}_{P_{21}=i p_{21},P_{32}=i p_{32}}\!\begin{bmatrix}
    P_3=i p_3 & P_2=i p_2 \\ P_4=i p_4 & P_1=i p_1
\end{bmatrix}\ ,
\ee
which should not lead to any confusion in the following.

\paragraph{The BPZ differential equation.} Letting $\Psi$ denote the highest weight state of the Verma module, the null vector in question is 
\be  
\big(L_{-2}+b^{-2} L_{-1}^2\big)\Psi\ .
\ee
Consider now the possible conformal blocks appearing in the correlation function
\be  
\big\langle V_{p_1}(1) V_{p_2}(\infty) V_{p_3}(0) V_{p_{\langle 2,1 \rangle}}(z) \big\rangle
\ee
The vertex operators denote Virasoro primary vertex operators in any unitary CFT -- not necessarily Liouville theory. Here we assume that $0<z<1$, where the conformal block is single-valued. Since $L_{-1}$ acts as a derivative and $L_{-2}$ appears in the regular term of the OPE expansion, we can derive the following constraint on the correlation function
\begin{align}
    0&=-z(z-1)b^{-2} \partial_z^2 \big\langle V_{p_1}(1) V_{p_2}(\infty) V_{p_3}(0) V_{p_{\langle 2,1 \rangle}}(z) \big\rangle\nonumber\\
    &\qquad-z(z-1)\Res_{x=z} \frac{x(x-1)}{x-z}\big\langle V_{p_1}(1) V_{p_2}(\infty) V_{p_3}(0) T(x) V_{p_{\langle 2,1 \rangle}}(z) \big\rangle\nonumber\\
    &\qquad+\left((2z-1) \partial_z +\Delta_4\right)\big\langle V_{p_1}(1) V_{p_2}(\infty) V_{p_3}(0) V_{p_{\langle 2,1 \rangle}}(z) \big\rangle
\end{align}
Here the residue pick out the $L_{-2}$ term acting on the degenerate field, but there are also contributions from the first order pole and second order pole for $x \to z$, which we subtract in the third line. One then rewrites the residues a sum of the three residues at $x=0$, $x=1$ and $x=\infty$ which can be explicitly evaluated. The result is a second order differential equation known as BPZ equation \cite{Belavin:1984vu},
\begin{multline}
    0=\bigg(\!-b^{-2}z(z-1)\partial_z^2+(2z-1)\partial_z+\frac{\Delta_3}{z}+\frac{\Delta_1}{1-z} -\Delta_2+\Delta_{\langle 2,1 \rangle}\bigg)\\
    \big\langle V_{p_1}(1) V_{p_2}(\infty) V_{p_3}(0) V_{p_{\langle 2,1 \rangle}}(z) \big\rangle\ .  
\end{multline}
Here, $\Delta_{\langle 2,1 \rangle}=-\frac{1}{2}-\frac{3b^2}{4}$.
This is a Fuchsian differential equation with three regular singular points. Its solution can hence be expressed in terms of the Gauss hypergeometric function. 

\paragraph{Degenerate conformal blocks.}
We write the two linearly independent solutions as
\vspace{-.3cm}
\begin{multline}
\!\!\!\!\begin{tikzpicture}[baseline={([yshift=-1.1ex]current bounding box.center)}, xscale=.6, yscale=.6]
        \draw[very thick, Maroon, bend left=30] (-1.8,1.2) to (-.6,0);
        \draw[very thick, Maroon, bend right=30] (-1.8,-1.2) to (-.6,0);
        \draw[very thick, Maroon, bend right=30] (2.2,1.2) to (.6,0);
        \draw[very thick, Maroon, bend left=30] (2.2,-1.2) to (.6,0);
        \draw[very thick, Maroon] (-.6,0) to (.6,0);
        \fill[Maroon] (-.6,0) circle (.07);
        \fill[Maroon] (.6,0) circle (.07);
        \draw[very thick] (-2,1.2) circle (.2 and .5);
        \draw[very thick] (-2,-1.2) circle (.2 and .5);
        \begin{scope}[shift={(2,1.2)}, xscale=.4]
            \draw[very thick] (0,-.5) arc (-90:90:.5);
            \draw[very thick, dashed] (0,.5) arc (90:270:.5);
        \end{scope}
        \begin{scope}[shift={(2,-1.2)}, xscale=.4]
            \draw[very thick] (0,-.5) arc (-90:90:.5);
            \draw[very thick, dashed] (0,.5) arc (90:270:.5);
        \end{scope}
        \begin{scope}[xscale=.5]
            \draw[very thick, RoyalBlue] (0,-.9) arc (-90:90:.9);
            \draw[very thick, dashed, RoyalBlue] (0,.9) arc (90:270:.9);
        \end{scope}
        \draw[very thick, out=0, in=180] (-2,1.7) to (0,.9) to (2,1.7);
        \draw[very thick, out=0, in=180] (-2,-1.7) to (0,-.9) to (2,-1.7);
        \draw[very thick, out=0, in=0, looseness=2.5] (-2,-.7) to (-2,.7);
        \draw[very thick, out=180, in=180, looseness=2.5] (2,-.7) to (2,.7);
        \node at (-2.6,1.2) {$p_3$};
        \node at (2.6,1.2) {$p_2$};            
        \node at (2.6,-1.2) {$p_1$};
        \node at (-2.9,-1.2) {$p_{\langle 2,1 \rangle}$}; 
        \node at (0,1.75) {$p_{3}+\frac{\varepsilon b}{2}$};
        \end{tikzpicture}\!\!\!
=z^{\frac{bQ}{2}-\varepsilon b p_3}(1-z)^{\frac{bQ}{2}- b p_1}   \\
\times {}_2F_1 \left[\begin{array}{c}
     \frac{1}{2}-b p_1+ \varepsilon b(p_2-p_3)\quad \frac{1}{2}-b p_1+ \varepsilon b(-p_2-p_3)  \\
     1-2\varepsilon b p_3
\end{array}; z\right] \label{eq:degenerate s channel conformal blocks}
\end{multline}
for $\varepsilon=+,-$.
Here, we can identify the internal Liouville momentum  by looking at the leading term as $z \to 0$ and comparing with eqs.~\eqref{eq:conformal block four-punctured sphere abstract expansion}, \eqref{eq:right three punctured sphere conformal block} and \eqref{eq:left three punctured sphere conformal block}, which gives the following leading order behaviour
\be  
z^{-\Delta_3-\Delta_{\langle 2,1 \rangle}+\Delta(p_3+\frac{\varepsilon b}{2})} \left(1+\mathcal{O}(z)\right)\ .
\ee
Similarly, we can write down solutions in the $t$-channel basis,
\begin{multline}
\begin{tikzpicture}[baseline={([yshift=-.5ex]current bounding box.center)}, xscale=.6, yscale=.6]
        \draw[very thick, Maroon, bend left=10] (-1.8,1.2) to (0,.5);
        \draw[very thick, Maroon, bend right=10] (-1.8,-1.2) to (0,-.5);
        \draw[very thick, Maroon, bend right=10] (2.2,1.2) to (0,.5);
        \draw[very thick, Maroon, bend left=10] (2.2,-1.2) to (0,-.5);
        \draw[very thick, Maroon] (0,-.5) to (0,.5);
        \fill[Maroon] (0,.5) circle (.07);
        \fill[Maroon] (0,-.5) circle (.07);
        \draw[very thick] (-2,1.2) circle (.2 and .5);
        \draw[very thick] (-2,-1.2) circle (.2 and .5);
        \begin{scope}[shift={(2,1.2)}, xscale=.4]
            \draw[very thick] (0,-.5) arc (-90:90:.5);
            \draw[very thick, dashed] (0,.5) arc (90:270:.5);
        \end{scope}
        \begin{scope}[shift={(2,-1.2)}, xscale=.4]
            \draw[very thick] (0,-.5) arc (-90:90:.5);
            \draw[very thick, dashed] (0,.5) arc (90:270:.5);
        \end{scope}
        \begin{scope}[yscale=.4]
            \draw[very thick, RoyalBlue] (-.97,0) arc (-180:0:.97);
            \draw[very thick, dashed, RoyalBlue] (.97,0) arc (0:180:.97);
        \end{scope}
        \draw[very thick, out=0, in=180] (-2,1.7) to (0,.9) to (2,1.7);
        \draw[very thick, out=0, in=180] (-2,-1.7) to (0,-.9) to (2,-1.7);
        \draw[very thick, out=0, in=0, looseness=2.5] (-2,-.7) to (-2,.7);
        \draw[very thick, out=180, in=180, looseness=2.5] (2,-.7) to (2,.7);
        \node at (-2.6,1.2) {$p_3$};
        \node at (2.6,1.2) {$p_2$};            
        \node at (2.6,-1.2) {$p_1$};
        \node at (-2.9,-1.2) {$p_{\langle 2,1 \rangle}$};   
        \node at (-2.2,0) {$p_{1}+\frac{\eta b}{2}$};
    \end{tikzpicture}\hspace{-.5cm}=z^{\frac{bQ}{2}+b p_3}(1-z)^{\frac{bQ}{2}-\eta b p_1} \\
\times {}_2F_1 \left[\begin{array}{c}
     \frac{1}{2}+b p_3+\eta b(-p_1+p_2)\quad \frac{1}{2}+b p_3+\eta b(-p_1-p_2)  \\
     1-2\eta bp_1 
\end{array}; 1-z\right] \label{eq:degenerate t channel conformal blocks}
\end{multline}
for $\eta=+,-$. 

\paragraph{Degenerate crossing matrix.} We have the connection identity (see e.g.\ \cite[eq.~(15.3.6)]{Abramowitz})
\begin{multline}  
{}_2F_1\left[\begin{array}{c}
     a \quad b\\ c
\end{array}; z\right]=\frac{\Gamma(c)\Gamma(c-a-b)}{\Gamma(c-a)\Gamma(c-b)}\,  {}_2F_1 \left[\begin{array}{c}
     a \quad b\\ a+b-c+1
\end{array}; 1-z\right]\\
+\frac{\Gamma(c)\Gamma(a+b-c)}{\Gamma(a)\Gamma(b)} (1-z)^{c-a-b} \, {}_2 F_1 \left[\begin{array}{c}
     c-a \quad c-b\\ c-a-b+1
\end{array}; 1-z\right]
\end{multline}
for the Gauss hypergeometric function. Applying this to the $s$-channel conformal block \eqref{eq:degenerate s channel conformal blocks} lets us rewrite them in terms of $t$-channel conformal blocks. Using the definition of the $\mathbb{F}$ in eq.~\eqref{eq:definition F} lets us read off
\be   
\mathbb{F}_{p_3+\frac{\varepsilon b}{2},p_{32}}\!\begin{bmatrix}
    p_3 & p_2 \\ 
    p_{\langle 2,1 \rangle} & p_1
\end{bmatrix}=\sum_{\eta=\pm} \frac{\Gamma(1-2b\varepsilon p_3)\Gamma(2b\eta p_1)}{\Gamma(\frac{1}{2}+\eta b p_1\pm bp_2-\varepsilon b p_3)} \delta\big(p_{32}-(p_1+\tfrac{\eta b}{2})\big)\ . \label{eq:degenerate crossing kernel}
\ee
Here we used a convenient notation that we will often employ in the following. We will often encounter products of functions with different signs and mean by the $\pm$,
\be 
\Gamma(a\pm b) \equiv \Gamma(a+b)\Gamma(a-b)\ .
\ee
If there are more than one $\pm$ sign in a function in the following, we will take the product over all possibilities.
Compare this equation with the trivial case of $p_4=\id=p_{\langle 1,1 \rangle}$, which is given by \eqref{eq:F symbol identity delta function}. 
Of course, we could have derived the corresponding formula for the other degenerate field $p_{\langle 1,2\rangle}=-\frac{b}{2}-\frac{1}{b}$ by replacing $b \to b^{-1}$.
\paragraph{Shift relation.} Knowing this degenerate crossing kernel leads to a shift relation for the full crossing kernel. Let us put $p_4=p_{\langle 2,1 \rangle}$ in the pentagon identity \eqref{eq:g=0, n=5 pentagon equation}. We then also have to put $p_{54}=p_5+\frac{b\varepsilon}{2}$ to satisfy the degenerate fusion rules. The integral on the left-hand side 
becomes a sum in this case, since $p_{32}$ can only take the values $p_{51}+\frac{b \eta}{2}$ for $\eta=\pm$ due to the second crossing kernel. Finally, the last crossing kernel on the left-hand side of \eqref{eq:g=0, n=5 pentagon equation} is also degenerate and leads to a delta-function setting $p_{43}=p_3+\frac{b \zeta}{2}$ with $\zeta=\pm$. Comparing the coefficients of the delta-functions on the left- and right-hand side of the equation leads to the following special case of the pentagon equation:
\begin{multline} 
\sum_{\eta=\pm} \frac{ \Gamma(-b^2-2b \eta p_{32})\Gamma(1-2 b \eta p_{32})}{\Gamma(\frac{1}{2}\pm b p_2+b \zeta p_3-b \eta p_{32})\Gamma(\frac{1-b^2}{2} \pm b p_1- b \varepsilon p_4-b \eta p_{32})}\, \mathbb{F}_{p_{21},p_{32}+\frac{b \eta}{2}}\!\begin{bmatrix}
    p_3 & p_2 \\
    p_4+\frac{b \varepsilon}{2} & p_1
\end{bmatrix}   \\
=\frac{1}{\Gamma(\frac{1}{2} \pm b p_{21}+ b \zeta p_3-b \varepsilon p_4)} \, \mathbb{F}_{p_{21},p_{32}} \!\begin{bmatrix}
    p_3+\frac{b \zeta}{2} & p_2 \\
    p_4 & p_1
\end{bmatrix}\ . \label{eq:shift relation epsilon zeta}
\end{multline}
We replaced $p_5 \to p_4$ and $p_{51} \to p_{32}$ to put the generic crossing kernels in the standard form.
We also cancelled common factors of the left- and right-hand side. This equation holds for any of the four cases $\epsilon,\,\zeta=\pm$. However all of these four cases contain the same information since we already know that the crossing kernel $\mathbb{F}$ is reflection symmetric in all entries and thus e.g.\ the cases $\varepsilon=\pm$ are related by sending $p_4 \to -p_4$.  This is confirmed by the Gamma functions in this equation which only contains the combinations $\zeta p_3$ and $\varepsilon p_4$. We can thus set $\varepsilon=\zeta=+$ without loss of generality. We also replace $p_4 \to p_4-\frac{b}{2}$ to bring it into a slightly nicer form. This gives the final form of the equation to which we refer to as the shift relation
\begin{multline} 
\sum_{\eta=\pm} \frac{ \Gamma(-b^2-2b \eta p_{32})\Gamma(1-2b \eta p_{32})}{\Gamma(\frac{1}{2}\pm b p_2+b p_3-b \eta p_{32})\Gamma(\frac{1}{2} \pm b p_1-b p_4- b \eta p_{32})} \, \mathbb{F}_{p_{21},p_{32}+\frac{b \eta}{2}}\!\begin{bmatrix}
    p_3 & p_2 \\
    p_4 & p_1
\end{bmatrix}  \\
=\frac{1}{\Gamma(\frac{1+b^2}{2} \pm b p_{21}+b p_3-b p_4)} \, \mathbb{F}_{p_{21},p_{32}} \!\begin{bmatrix}
    p_3+\frac{b}{2} & p_2 \\
    p_4-\frac{b}{2} & p_1
\end{bmatrix} \ .\label{eq:shift relation}
\end{multline}
Of course, a similar shift relation also holds when we replace $b \to b^{-1}$ since we could have taken the degenerate field $p_{\langle 1,2 \rangle}$ instead of $p_{\langle 2,1 \rangle}$.

This equation constitutes a starting point to bootstrap the answer for the crossing kernel. As we shall see in the next Section, this shift equation fixes the answer in fact uniquely in terms of explicit special functions. Let us note that the existence of an explicit solution is perhaps unexpected: after all, it is hard to compute conformal blocks directly, but we can still work out their transformation behaviour under crossing!

\section{Bootstrapping the crossing kernels} \label{sec:bootstrapping the crossing kernels}
We will now explain how the various constraints derived in the previous Section fix the result for the crossing kernels uniquely.
\subsection{Special functions} \label{subsec:special functions}
As can be anticipated from the DOZZ formula, the solution will involve various special functions. Let us recall the definition of the double Gamma function, which is the main special function that is needed. It is a meromorphic function on the plane with the following functional equations
\be  
\Gamma_b(z+b)=\frac{\sqrt{2\pi}\,  b^{bz-\frac{1}{2}}}{\Gamma(bz)}\, \Gamma_b(z)\ , \qquad \Gamma_b(z+b^{-1})=\frac{\sqrt{2\pi}\,  b^{-b^{-1}z+\frac{1}{2}}}{\Gamma(b^{-1}z)}\, \Gamma_b(z) \label{eq:Gammab functional equations}
\ee
and depends continuously on $b$ (which we assume to be real and positive).
These functional equations determine the function up to an overall normalization. Indeed, the standard argument for this goes as follows. Assume first that $b^2 \not \in \QQ$. Then $b$ and $b^{-1}$ are incommensurable, which means that $b \ZZ \oplus b^{-1} \ZZ$ is a dense set in $\RR$ and thus determines $\Gamma_b(z)$ on a dense subset of the real line up to overall normalization. This is enough by analyticity to extend the definition to the whole complex plane. For rational $b^2$, we can approximate it arbitrarily well by an irrational $b^2$ for which the definition is unique and use the continuity assumption on $b$.

Since $\Gamma(z)$ does not have any zeros, it follows that $\Gamma_b(z)$ has no zeros as well. The pole structure follows from the poles of the Gamma function. It implies that $\Gamma_b(z)$ has simple poles at $z=-m b-n b^{-1}$, $m,\, n \in \ZZ_{\ge 0}$. For $b^2 \in \QQ$ some of these poles may coincide and form higher order poles.
Notice that the shift relations are highly overconstrained and it is non-trivial that they are compatible. For example, one can check the two orders to apply the functional equation twice to $\Gamma_b(z+b+b^{-1})$ give the same result.

Such a function does exist and can be explicitly constructed by the integral 
\be  
\log \Gamma_b(z)=\int_0^\infty \frac{\mathrm{d}t}{t}\left(\frac{\mathrm{e}^{\frac{t}{2}(Q-2z)}-1}{4 \sinh(\frac{bt}{2})\sinh(\frac{t}{2b})}-\frac{1}{8}\left(Q-2z\right)^2 \mathrm{e}^{-t}-\frac{Q-2z}{2t}\right)\ . \label{eq:Gammab integral representation main text}
\ee
The integral representation converges for $\Re(z)>0$, i.e.\ in the half-plane where the function has no poles. Its values on the half-plane $\Re(z)<0$ following then by repeately applying the functional equation \eqref{eq:Gammab functional equations}.
This function is called the double Gamma function. Since it plays a major role in the theory, we have written an extensive Appendix~\ref{app:special functions}. Besides listing standard properties of these functions, we also explain their miraculous integral identities that we will need later on.

We also define the combination
\be 
S_b(z)\equiv \frac{\Gamma_b(z)}{\Gamma_b(Q-z)}\ , \label{eq:Sb definition}
\ee
which satisfies the functional equation
\be 
    S_b(z+b)=2 \sin(\pi b z) S_b(z)\ ,
\ee
thanks to the reflection formula of the Gamma function.
This function is called the double sine function.\footnote{$S_b(z)$ is also called the hyperbolic gamma function and is closely related to Faddeev's quantum dilogarithm \cite{Faddeev:1995nb}.} Many more properties of the functions are explained in Appendix~\ref{app:special functions}.

\subsection{Bootstrapping \texorpdfstring{$C_0$}{C0}}\label{subsec:C0 bootstrap}
We start by showing that the shift relation \eqref{eq:shift relation} fixes the special case of the crossing kernel $\mathbb{F}$ related to $C_0$ as defined by \eqref{eq:F1 in terms of C0 rho0}. We will use the same abuse of notation as before,
\begin{align} 
C_0(p_1,p_2,p_3)&=C_0(P_1=i p_1,P_2=i p_2,P_3=i p_3)\ , \\
\rho_0(p) &= \rho_0(P=i p)\ .
\end{align}
\paragraph{Solving the shift relation.} For this purpose, we specialize \eqref{eq:shift relation} further by setting $p_1=p_2$, $p_4=p_3$ and $p_{21}=\id$. The crossing kernel on the right hand side of \eqref{eq:shift relation} vanishes since the corresponding conformal block is incompatible with the degenerate fusion rules. Thus we obtain a two term shift relation instead of a three term shift relation. This simplification drastically reduces the complexity of the problem. We get the shift relation
\be 
\sum_{\eta=\pm} \frac{ \Gamma(-b^2-2 b \eta p_{32})\Gamma(1-2 b \eta p_{32})}{\Gamma(\frac{1}{2}\pm b p_2\pm b p_3-b \eta p_{32})} \, \mathbb{F}_{\id,p_{32}+\frac{b \eta}{2}}\!\begin{bmatrix}
    p_3 & p_2 \\
    p_3 & p_2
\end{bmatrix}=0\ .
\ee
This implies the following shift relation for $C_0(p_1,p_2,p_3)$:
\be 
\frac{C_0(p_1+b,p_2,p_3)}{C_0(p_1,p_2,p_3)}=\frac{\Gamma(1+b^2+2b p_1)\Gamma(1+2b^2+2b p_1)\Gamma(\frac{1-b^2}{2}-b p_1\pm b p_2 \pm b p_3)}{\Gamma(1-2b p_1)\Gamma(1-b^2-2b p_1)\Gamma(\frac{1+b^2}{2}+b p_1 \pm bp_2 \pm b p_2)}\ , \label{eq:C0 shift relation}
\ee
as well as the same shift relation with $b \to b^{-1}$. Here we renamed $p_{32}=p_1+\frac{b}{2}$. By the same basic reasoning as explained after eq.~\eqref{eq:Gammab functional equations}, these shift relations determine $C_0(p_1,p_2,p_3)$ uniquely for $b^2 \not\in \QQ$, up to an overall constant. 
The solution is given up to an overall constant by
\be 
C_0(P_1,P_2,P_3)= \frac{\Gamma_b(2Q)\Gamma_b(\frac{Q}{2} \pm i P_1\pm i P_2 \pm i P_3)}{\sqrt{2}\,\Gamma_b(Q)^3\prod_{j=1}^3\Gamma_b(Q\pm 2i P_j)}\ . \label{eq:C0 formula}
\ee
We reverted back to using the real Liouville momenta $P_j$ in this formula for reference purposes.
It straightforward to verify using the functional equations of the double Gamma function \eqref{eq:Gammab functional equations} that this formula indeed solves the shift relation \eqref{eq:C0 shift relation}. We already put the correct overall constant in \eqref{eq:C0 formula}. To check its correctness, we have to show that the two-point function \eqref{eq:limit C0 two-point function} is correctly reproduced. We note as a non-trivial consistency check that $C_0(P_1,P_2,P_3)$ is manifestly symmetric in all three entries and symmetric under $P_j \to -P_j$. 

\paragraph{Reduction to the two-point function.} So let us consider the limit $P_3 \to \frac{iQ}{2}$, which we denote as before by $P_3 \to \id$. We also assume as usual that $P_1,\, P_2>0$ are above threshold. In the limit $P_3 \to \frac{iQ}{2}$, $C_0(P_1,P_2,P_3)$ naively vanishes since the factor $\Gamma_b(Q+2i P_3)$ in the denominator has a pole. The only way to potentially get a finite answer is by compensating it with another pole in the numerator. Let us set $P_3=\frac{iQ}{2}-i \varepsilon$ for small $\varepsilon>0$. Then the terms in the numerator become
\be 
\Gamma_b(Q-\varepsilon \pm i P_1 \pm i P_2)\ , \qquad \Gamma_b(\varepsilon \pm i P_1 \pm i P_2)\ .
\ee
Since $\Gamma_b(z)$ has poles for $z=-mb-n b^{-1}$ with $m,\, n \in \ZZ_{\ge 0}$ only the second type of terms can produce a pole when the argument is close to 0. We assumed $P_1,\, P_2 \ge 0$ and thus this only happens fo $P_1$ close to $P_2$. 
When taking the limit, we can now set $P_2=P_1$ and $\varepsilon=0$ in all regular factors. For the singular factors, we use that
\be 
\Res_{z=0} \Gamma_b(z)=\frac{\Gamma_b(Q)}{2\pi}\ ,
\ee
which follows from the functional equation \eqref{eq:Gammab functional equations}, see also eq.~\eqref{eq:Gammab residue 0}. We thus have
\begin{align}
    \lim_{P_3 \to \id} C_0(P_1,P_2,P_3)&=\frac{\Gamma_b(\pm 2i P_1)}{\sqrt{2}\, \Gamma_b(Q) \Gamma_b(Q \pm 2i P_1)} \times \frac{\Gamma_b(\varepsilon \pm i (P_1-P_2))}{\Gamma_b(2 \varepsilon)} \\
    &=\frac{\Gamma(1 \pm 2i b P_1)\Gamma(\pm 2i b^{-1} P_1)}{4\sqrt{2}\, b^2\pi^3}\times  \frac{\varepsilon}{\varepsilon^2+(P_1-P_2)^2} \\
    &=\frac{\delta(P_1-P_2)}{4\sqrt{2}\, \sinh(2\pi b P_1) \sinh(2\pi b^{-1} P_1)}=\frac{1}{\rho_0(P)}\, \delta(P_1-P_2)\ .
\end{align}
Here we used that $\frac{\varepsilon}{\pi(x^2+\varepsilon^2)}$ tends to the delta function for $\varepsilon \to 0$ (in the distributional sense). This confirms the normalization in eq.~\eqref{eq:C0 formula}.

\subsection{The degenerate crossing kernels}
We will take a small detour from our main goal to determine the crossing kernel $\mathbb{F}$ and discuss the case of degenerate crossing kernels separately. 
By degenerate crossing kernel we mean that we take one of the external operators, say $p_4$, to the degenerate value $p_{\langle m,n \rangle}$. In this case, there are only finitely many blocks and we obtain the generalization of \eqref{eq:degenerate crossing kernel}, which is the special case with $p_4=p_{\langle 2,1 \rangle}$.
These are the crossing kernels that appear in (generalized) minimal models \cite{Zamolodchikov:2005fy}.

It is straightforward to see that they are recursively determined by the shift relation \eqref{eq:shift relation}, see \cite{Nemkov:2021huu} for a similar discussion in the special case $\langle 3,1 \rangle$. Indeed, consider \eqref{eq:shift relation} as a shift relation in $p_4$. Take \eqref{eq:shift relation}, set $p_3 \to p_3-\frac{b}{2}$, $p_4=p_{\langle m,n \rangle}$ and solve for the right-hand side. We obtain
\begin{multline}
    \mathbb{F}_{p_{21},p_{32}} \!\begin{bmatrix}
    p_3 & p_2 \\
    p_{\langle m+1,n\rangle} & p_1
\end{bmatrix} =\sum_{\eta=\pm} \frac{ \Gamma(-b^2-2b \eta p_{32})\Gamma(1-2b \eta p_{32})}{\Gamma(\frac{1-b^2}{2}\pm b p_2+ \zeta b p_3-b \eta p_{32})} \\
\times \frac{\Gamma(\frac{1}{2} \pm b p_{21}+\zeta b  p_3-b p_{\langle m,n \rangle})}{\Gamma(\frac{1}{2} \pm b p_1-b p_{\langle m,n \rangle}- b \eta p_{32})}\, \mathbb{F}_{p_{21},p_{32}+\frac{b \eta}{2}}\!\begin{bmatrix}
    p_3-\frac{\zeta b}{2} & p_2 \\
    p_{\langle m,n \rangle} & p_1
\end{bmatrix} \ . \label{eq:degenerate shift relation}
\end{multline}
Here we reintroduced $\zeta$ as in eq.~\eqref{eq:shift relation epsilon zeta} since it will be useful in the following.
For this equation to make sense, we need to set $p_{21}=p_3+\frac{rb}{2}+\frac{s}{2b}$, where $r \in \{-m,-m+2,\dots,m\}$ and $s \in \{-n+1,-n+3,\dots,n-1\}$, see eq.~\eqref{eq:degenerate fusion rules}. The left-hand side will then produce a number of delta-functions that will set similarly $p_{32}=p_1+\frac{r'b}{2}+\frac{s'}{2b}$. The same delta functions can be found on the right-hand side and we can compare their coefficients.

On the right hand side, we have then $p_{21}=p_3'+\frac{\zeta b}{2}$, where $p_3'=p_3-\frac{\zeta b}{2}$ is the third argument of the crossing kernel appearing on the right hand side. For this to satisfy the degenerate fusion rules, we choose $\zeta=+$ when $r=-m$ and $\zeta=-$ when $r=m$; in all other cases the choice does not matter. The right-hand side then only involves the degenerate crossing kernels. We should also note that the Gamma function prefactor appearing on the right hand side are non-singular for these choices. Thus \eqref{eq:degenerate shift relation} together with its $b \to b^{-1}$ analogue fully determines the degenerate crossing kernels recursively.

\subsection{Rewriting the shift relations in a symmetric normalization}
Since we have now determined $C_0(P_1,P_2,P_3)$, it will be more convenient from now on to use a normalization for which the full tetrahedral symmetry of $\mathbb{F}$ becomes manifest. We already discussed such a normalization in eq.~\eqref{eq:tetrahedrally symmetric combination}. It will be actually slightly more convenient to use a different normalization convention that still preserves tetrahedral symmetry, but breaks reflection symmetry.

We set
\begin{multline}
    \mathbb{F}_{p_{21},p_{32}}\!\begin{bmatrix}
        p_3 & p_2 \\ p_4 & p_1
    \end{bmatrix}=\prod_{\pm_1 \pm_2 \pm_3=+} \frac{\Gamma_b(\frac{Q}{2} \pm_1 p_{2}\pm_2p_3\pm_3 p_{32})\Gamma_b(\frac{Q}{2} \pm_1 p_1 \pm_2 p_4 \pm_3 p_{32})}{\Gamma_b(\frac{Q}{2} \pm_1p_1\pm_2p_2\mp_3p_{21})\Gamma_b(\frac{Q}{2} \pm_1p_3\pm_2p_4\mp_3p_{21})}\\
    \times 
     \frac{\Gamma_b(Q\pm 2p_{21})}{\Gamma_b(\pm 2 p_{32})}\times  \sixjnorm{p_1}{p_2}{p_{21}}{p_3}{p_4}{p_{32}}\ . \label{eq:sixjnorm definition}
\end{multline}
The notation means that we are taking the product over the three choices of signs by with the constraint that the product of the signs is positive. In fact we have
\begin{multline} 
\sixjnorm{p_1}{p_2}{p_{21}}{p_3}{p_4}{p_{32}}=\frac{2\sqrt{2}\, \Gamma_b(Q)^6}{\Gamma_b(2Q)^2} \prod_{(p^{(1)},p^{(2)},p^{(3)}) \in \mathcal{V}} V(p^{(1)},p^{(2)},p^{(3)}) \prod_{p \in \mathcal{E}} E(p) \\
\times \rho_0(p_{32})^{-1} C_0(p_1,p_2,p_{21}) C_0(p_3,p_4,p_{21}) \, \mathbb{F}_{p_{21},p_{32}}\!\begin{bmatrix}
    p_3 & p_2 \\
    p_4 & p_1
\end{bmatrix}\ .
\end{multline}
Here, $\mathcal{V}$ is the set of the four vertices of the tetrahedron in eq.~\eqref{eq:tetrahedrally symmetric combination} and $\mathcal{E}$ the set of six edges. The vertex and edge factors take the form
\begin{subequations}
    \begin{align}
        V(p_1,p_2,p_3)&=\prod_{\pm_1 \pm_2 \pm_3=+} \Gamma_b(\tfrac{Q}{2} \pm_1 p_1\pm_2p_2\pm_3 p_{3})\ , \\
        E(p)&=\Gamma_b(Q \pm 2p)\ .
    \end{align}
\end{subequations}
Given that eq.~\eqref{eq:tetrahedrally symmetric combination} is tetrahedrally symmetric, it then also follows that \eqref{eq:sixjnorm definition} defines a tetrahedrally symmetric combination. Thus it will be more convenient to rename $p_{21} \to p_3, \, p_3 \to p_4, p_4 \to p_5,\, p_{32} \to p_6$ to reflect the extended symmetry. The labels of the tetrahedron are then as follows,
\be 
\sixjnorm{p_1}{p_2}{p_3}{p_4}{p_5}{p_6} \longleftrightarrow \begin{tikzpicture}[baseline={([yshift=-.5ex]current bounding box.center)}, scale=.7]
    \draw[very thick] (0,0) to node[below] {$p_6$} (4,0);
    \draw[very thick, dashed] (0,0) to node[right, shift={(0,-.1)}] {$p_5$} (2.8,1.5) to node[left] {$p_4$} (4,0);
    \draw[very thick, dashed] (2.8,1.5) to node[left, shift={(.1,-.2)}] {$p_3$} (2.5,3.5);
    \draw[very thick] (0,0) to node[left] {$p_1$} (2.5,3.5) to node[right] {$p_2$} (4,0);
\end{tikzpicture}\ .
\ee
Let us also note that in this notation, the symmetry group of the tetrahedron is generated by permuting columns and by permuting pairs of entries of the first row with their partners in the second row, i.e.
\be 
\sixjnorm{p_1}{p_2}{p_3}{p_4}{p_5}{p_6}=\sixjnorm{p_2}{p_1}{p_3}{p_5}{p_4}{p_6}=\sixjnorm{p_1}{p_3}{p_2}{p_4}{p_6}{p_5}=\sixjnorm{p_4}{p_5}{p_3}{p_1}{p_2}{p_6}\ .
\ee 
We refer to this object as the normalized crossing kernel.
As a word of caution, this normalization does \emph{not} coincide with the normalization of the Racah-Wigner $6j$ symbols that was used e.g.\ in \cite{Teschner:2012em}. We explain the relation to that normalization in Section~\ref{subsec:unitarity}.

Let us also note that under reflection, it satisfies the reflection equation
\begin{multline} 
\sixjnorm{-p_1}{p_2}{p_3}{p_4}{p_5}{p_6}=\sixjnorm{p_1}{p_2}{p_3}{p_4}{p_5}{p_6} \\
\times \prod_{\pm_1 \pm_2 \pm_3=+} S_b(\tfrac{Q}{2} \pm_1 p_1\pm_2p_2\pm_3 p_3)S_b(\tfrac{Q}{2} \pm_1 p_1\pm_2p_5\pm_3 p_6)\ , \label{eq:sixjnorm reflection symmetry}
\end{multline}
where we used the double sine function defined in eq.~\eqref{eq:Sb definition}.

It will thus be most convenient to rewrite the shift relations in the symmetric normalization.
When plugging in \eqref{eq:sixjnorm definition} into \eqref{eq:shift relation}, we obtain
\begin{multline}
    \sum_{\eta=\pm} \eta \cos\big(\pi b(p_1+\eta p_5+p_6)\big)\cos\big(\pi b(p_2+\eta p_4-p_6)\big) \sixjnorm{p_1}{p_2}{p_3}{p_4}{p_5}{p_6 \pm \frac{ \eta b}{2}} \\
    =\sin(2\pi b p_6) \cos\big(\tfrac{\pi b^2}{2}+\pi b(-p_3+p_4-p_5)\big) \sixjnorm{p_1}{p_2}{p_3}{p_4+\frac{b}{2}}{p_5-\frac{b}{2}}{p_6}\ . \label{eq:normalized shift relation}
\end{multline}
Besides having tetrahedral symmetry, this normalization of the crossing kernel features only trigonometric functions in the shift relation. Of course any shift relation related by tetrahedral symmetry is also valid. Moreover, the same shift relation holds if we replace $b \to b^{-1}$.

\subsection{Uniqueness of the solution} \label{subsec:uniqueness}
Now that we have derived the shift relation \eqref{eq:normalized shift relation}, the next step is to show that its solution is unique up to an overall constant. Contrary to the case of a two-term shift equation such as for $C_0$, showing uniqueness is actually rather difficult in this case and the arguments for this that have been given in the literature are incomplete, e.g. \cite{Ponsot:1999uf, Ponsot:2001ng,  Teschner:2003en, Remy:2020suk, Roussillon:2020lyc}.\footnote{Let us note that the previous proofs contain at least two gaps. The proof given in \cite[Appendix C]{Ponsot:2001ng} that there are at most two solutions to such a shift equation is incorrect. Since the Wronskian $\mathcal{W}(f,g)$ defined in eq.~(65) depends on $b$, it only satisfies a shift equation in $b$, but not in $b^{-1}$. This does not suffice to deduce uniqueness. Second, it was claimed in \cite{Ponsot:2001ng} that the correct solution out of the two potential solutions can be selected by imposing reflection symmetry. This can fail since both linearly independent solutions could be reflection symmetric. A toy example where this happens is e.g.\ the two-term shift equation
\be 
\sin(4\pi bp) f(p)-\sin(\pi b(2p+b)) f(p-b)-\sin(\pi b(2p-b)) f(p+b)=0
\ee
and the corresponding equation with $b \leftrightarrow b^{-1}$. It has the two linearly independent solutions $f(p)=\mathrm{e}^{\pi i p^2}$ and $f(p)=\mathrm{e}^{-\pi i p^2}$, which both have the same parity under reflection $p \to -p$. In \cite{Roussillon:2020lyc} the argument is completely missing and it just asserted that the shift equations have a unique solution without proof. \label{footnote:error}
}
Thus we will be quite careful in explaining the argument. Our argument is inspired by \cite{Rain:determinants}.\footnote{I thank Guillaume Remy for discussions about this argument.}

Let us clearly state what we exactly need for this proof. We assume the solution to be a meromorphic (single-valued) function in all the Liouville momenta, as was justified in Section~\ref{subsec:crossing transformations}. We also assume that the solution has tetrahedral symmetry as was discussed in Section~\ref{subsec:special cases consistency conditions}.
We will then show that for each $b^2 \in \RR \setminus \QQ$, the shift equations \eqref{eq:normalized shift relation} have a unique such solution, up to an overall normalization constant. The normalization constant will be fixed by requiring the correct trivial crossing kernel \eqref{eq:F symbol identity delta function}. By continuity in $b$, the solution can then be extended to values $b^2 \in \QQ$. This proof does \emph{not} need the hexagon equation \eqref{eq:g=0, n=4 hexagon equation}, nor does it need reflection symmetry in the Liouville momenta.

\paragraph{A shift relation in one variable.} A useful step for establishing uniqueness is to reduce the number of variables in the shift relation essentially to one. This can be done by taking a suitable linear combination of the shift relation \eqref{eq:normalized shift relation}. Let us denote by $\text{Sh}(p_1,p_2,p_3,p_4,p_5,p_6)$ the shift relation in eq.~\eqref{eq:normalized shift relation}. Consider now the following three shift relations:
\be 
\text{Sh}(p_1,p_2,p_3,p_4-\tfrac{b}{2},p_5+\tfrac{b}{2},p_6)\ , \qquad \text{Sh}(p_2,p_1,p_3,p_5,p_4,p_6+\tfrac{ \varepsilon b}{2})\ , \quad \varepsilon=\pm\ .
\ee
They constitute three shift relations for the normalized crossing kernel with parameters
\be 
\sixjnorm{p_1}{p_2}{p_3}{p_4}{p_5}{p_6}\ , \qquad \sixjnorm{p_1}{p_2}{p_3}{p_4}{p_5}{p_6+ \varepsilon b}\ , \qquad \sixjnorm{p_1}{p_2}{p_3}{p_4-\frac{b}{2}}{p_5+\frac{b}{2}}{p_6+\frac{\varepsilon b}{2}}\ ,
\ee
for $\varepsilon=\pm$.
The latter two can then be eliminated to derive a shift relation only on the former three. In particular, we may think of $(p_1,p_2,p_3,p_4,p_5)$ of being constant in the following.
This single-variable shift relation explicitly takes the form
\begin{align}
    \sixjnorm{p_1}{p_2}{p_3}{p_4}{p_5}{p_6+b}+\alpha_b(p_6)\sixjnorm{p_1}{p_2}{p_3}{p_4}{p_5}{p_6}+\beta_b(p_6) \sixjnorm{p_1}{p_2}{p_3}{p_4}{p_5}{p_6-b}=0\ . \label{eq:single variable shift relation}
\end{align}
Here the functions $\alpha_b(p_6)$ and $\beta_b(p_6)$ take the explicit form
\begin{subequations}
\begin{align}
    \alpha_b(p_6)&= -\frac{s(2 p_6+b) }{c (p_{15|6}-\frac{b}{2}) c  (p_{246}+\frac{b}{2}) }\bigg[\frac{c(p_{26|4}+\frac{b}{2}) c(p_{56|1}+\frac{b}{2})}{s(2p_6+b)} \nonumber\\
    &\qquad+\frac{1}{c(p_{24|6}-\frac{b}{2})c(p_{156}+\frac{b}{2})} \bigg(s(2  p_6) c(p_{5|34} +\tfrac{b}{2}) c (p_{35|4}+\tfrac{b}{2})\nonumber\\
    &\qquad\qquad\qquad+\frac{c (p_{2|46}+\frac{b}{2}) c(p_{15|6}+\frac{b}{2}) c(p_{246}-\frac{b}{2}) c (p_{5|16}+\frac{b}{2})}{s(2 p_6-b)}\bigg)\bigg] \ ,\\
    \beta_b(p_6)&=-\frac{\sin(\pi b(2p_6+b))\cos(\pi b(p_{1|5} \pm (p_6-\frac{b}{2})) \cos(\pi b(p_{2|4} \pm (p_6-\frac{b}{2})))}{\sin(\pi b(2p_6-b))\cos(\pi b(p_{15} \pm (p_6+\frac{b}{2}))\cos(\pi b(p_{24} \pm (p_6+\frac{b}{2})))}\ .
\end{align} \label{eq:alpha and beta}%
\end{subequations}
In writing $\alpha_b(p_6)$ we used the shortcuts $s(x) \equiv \sin(\pi b x)$ and $c(x)=\cos(\pi b x)$. We also used the short-hand notation
\be 
p_{I|J} \equiv \sum_{i \in I} p_i-\sum_{j \in J} p_j\ , \label{eq:definition pI|J}
\ee
which we will frequently employ below.
Of course $\alpha_b$ and $\beta_b$ also depend on the other Liouville momenta, but we consider them as external parameters in the following. As usual the same shift with $b \to b^{-1}$ also holds. We will only need the explicit form of $\beta_b(p_6)$ below. 

\paragraph{A matrix shift relation.} It is useful to repackage these shift relations in a matrix. Let us write
\be 
f(x)=\sixjnorm{p_1}{p_2}{p_3}{p_4}{p_5}{x}
\ee
and think of $p_1,\dots,p_5$ as fixed numbers in the following. We can then form the matrix-valued function
\be 
\mathbf{F}(x)=\begin{pmatrix}
    f(x) & f(x-b^{-1}) \\ f(x-b) & f(x-Q)
\end{pmatrix}
\ee
$\mathbf{F}(x)$ satisfies a first-order difference equation
\be 
\mathbf{F}(x+b)=\mathbf{M}_b(x) \mathbf{F}(x)\ , \label{eq:matrix shift relation b}
\ee
where
\be 
\mathbf{M}_b(x)=\begin{pmatrix}
    -\alpha_b(x) & -\beta_b(x) \\ 1 & 0
\end{pmatrix}\ . \label{eq:definition matrix M}
\ee
For this step, we had to use the fact that $\alpha_b(x)$ and $\beta_b(x)$ are $b^{-1}$-periodic, which in turn relies on the judicious choice of normalization in \eqref{eq:sixjnorm definition}.
There is a similar equation
\be 
\mathbf{F}(x+b^{-1})=\mathbf{F}(x) \mathbf{M}_{b^{-1}}(x)^{\mathsf{T}}\ .\label{eq:matrix shift relation b^-1}
\ee
\paragraph{A shift equation for the determinant.} Before diving into the full matrix shift equation, we note that it implies a first order shift relation for the determinant of $\mathbf{F}(x)$. The determinant is a discrete version of the Wronskian of $f(x)$, also known as Casoratian. We have
\be 
\det \mathbf{F}_b(x+b)=\beta_b(x) \det \mathbf{F}_b(x) \label{eq:difference equation determinant}
\ee
and the corresponding shift for $b \to b^{-1}$. This is a scalar first order shift relation and thus has a unique solution for $b^2 \not \in \QQ$, up to a function depending only on the remaining parameters $(p_1,\dots,p_5)$.
In the following, it will only be important that there exists a solution of the shift relations such that $\det \mathbf{F}(x)$ does not vanish identically. We will see below that this is the case.

\paragraph{A shift equation in the adjoint representation.} Let us now consider two solutions $\mathbf{F}(x)$ and $\mathbf{G}(x)$ of the matrix-valued shift relation. We will assume that $\det \mathbf{G}(x)$ does not vanish identically. Thus the proof relies on the existence of a solution with non-vanishing determinant. We exhibit a solution below that indeed satisfies this property, see the computation after eq.~\eqref{eq:sixjnorm determinant}. Thus we can form the combination
\be 
\mathbf{N}(x) \equiv \mathbf{F}(x) \mathbf{G}(x)^{-1}\ . \label{eq:N FG^-1}
\ee
$\mathbf{N}(x)$ satisfies the following simple two shift relations,
\be 
\mathbf{N}(x+b)=\mathbf{M}_b(x) \mathbf{N}(x) \mathbf{M}_b(x)^{-1}\ , \qquad \mathbf{N}(x+b^{-1})=\mathbf{N}(x)\ . \label{eq:shift relation N}
\ee
Thus the entries of $\mathbf{N}(x)$ are $b^{-1}$-periodic and $\mathbf{N}(x)$ satisfies a difference equation in $b$ and transforms in the adjoint representation. We now show that the only such matrix-valued function is $\mathbf{N}(x)$ proportional to the identity, which will finish the proof of uniqueness. The remaining proof relies on `difference Galois theory', a version of Galois theory for difference equations. We keep the present discussion elementary and assume no prior knowledge. For a lot of background, see \cite{Van2012Galois}.

\paragraph{Difference Galois theory.} The main ingredient of difference Galois theory is a difference field, which is a field together with an automorphism corresponding to the shift operator. We take the field to be
\be 
\KK=\text{$b^{-1}$-periodic meromorphic functions}\ . \label{eq:definition field K}
\ee
The shift operator $\tau$ acts by shifting a function $f(x)$ to $f(x+b)$. The field of constants $\KK^\tau$ is the subfield of $\KK$ fixed by the translation. In case of $b^2 \not \in \QQ$, the only meromorphic function that is both $b$ and $b^{-1}$-periodic is the constant function. Thus
\be 
\KK^\tau\cong \CC\ .
\ee
Let us now consider a general difference equation (we are ultimately interested in $v(x)=\mathbf{N}(x)$ and $A(x)=\text{Ad}_{\mathbf{M}_b(x)}$, the action of $\mathbf{M}_b(x)$ by conjugation)
\be 
v(x+b)=A(x) v(x)\ . \label{eq:formal difference equation}
\ee
Here, $A(x)$ is an $n \times n$ matrix in the field $\KK$, i.e.\ it satisfies $A(x+b^{-1})=A(x)$. $v(x)$ is an $n$-vector. It is important that the field of constants is trivial since we can multiply a solution $v(x)$ by an element in $\KK^\tau$ to get another solution. Thus the solution space naturally forms a vector space over $\KK^\tau$.

Just like in ordinary Galois theory for polynomials, one then constructs a \emph{ring}-extension by formally adjoining the solutions of the general difference equation \eqref{eq:formal difference equation} to the base field. Let us summarize the $n$-solutions into a matrix-valued function $V(x)$. The resulting ring is given by
\be 
\tilde{R}=(\KK[V,(\det V)^{-1}],\tau)\ .
\ee
This ring is the polynomial ring generated by all the entries of the matrix-valued function $V(x)$, as well as by $(\det V(x))^{-1}$. $\tau$ acts on $V$ by the defining difference equation \eqref{eq:formal difference equation}, i.e.
\be 
\tau(V)=AV\ .
\ee
Sometimes it is necessary to divide out an ideal in order to get the proper notion of a field extension in the difference setting, i.e.
\be 
R=\tilde{R}/I\ .
\ee
The price to pay is that $R$ is not necessarily an integral domain and thus we cannot simply pass to its field of fraction. In particular, it is more convenient to consider a \emph{ring} extension in the difference setting. The ring $R$ is then called a Picard-Vessiot ring. It is a $\KK$-algebra.

The Galois group of the difference equation is now simply the automorphism group of the ring extension, meaning
\be 
\Gal(R/\KK)=\{\phi:R \to R\, | \, \phi|_\KK=\id \} \subset \GL(n,\CC)\ . \label{eq:Galois group definition}
\ee
The Galois group is specified by its action on $V$ and is thus naturally a (closed algebraic) subgroup of $\GL(n,\CC)$.

\paragraph{Irreducibility implies uniqueness.} Let us go back to the case of interest. We want to find solutions of the shift relation \eqref{eq:matrix shift relation b} that actually lie in the base field $\KK$. By definition of the Galois group \eqref{eq:Galois group definition}, this means that it is fixed by the action of the Galois group. Even though we don't need it, the converse statement also holds -- if the Galois group fixes a solution $v$, then $v$ lies actually in $\KK$.

The Galois group only depends on $A(x)$. In our case, $A(x)=\Ad_{\mathbf{M}_b(x)}$ is given by the conjugation of the matrix $\mathbf{M}_b(x)$.  In such a situation, the Galois group is the image under the adjoint map of the Galois group of the simpler system for the fundamental representation
\be 
v(x+b)=\mathbf{M}_b(x) v(x)\ . \label{eq:shift equation fundamental representation}
\ee
See e.g.\ \cite{Van2012Galois} for an extensive discussion in the differential setting.
It will thus be sufficient to show that the Galois group of this difference equation \eqref{eq:shift equation fundamental representation} is irreducible. Since the three-dimensional representation of the Galois group is then still irreducible, this ensures that the only element of the ring extension $R$ fixed under the Galois group is a multiple of the identity matrix. We do this in the following two paragraphs which will finish the proof.

\paragraph{The Riccati equation.} We now consider the concrete difference equation \eqref{eq:shift equation fundamental representation}
over the field of $b^{-1}$-periodic meromorphic functions. We recall that the matrix $\mathbf{M}_b(x)$ was given in \eqref{eq:definition matrix M}. If the Galois group of this difference system is not irreducible, there must be a vector $v_0(x)$ whose direction (in the two-dimensional space $R/\KK$) is fixed under the action of the Galois group. Clearly the vector $(\begin{smallmatrix}
    1 \\ 0
\end{smallmatrix})$ cannot be fixed since $\mathbf{M}_b(x)$ maps it to $(\begin{smallmatrix}
    -\alpha_b(x) \\ 1
\end{smallmatrix})$. Thus we can parametrize $v_0(x)$ as
\be 
v_0(x)=\begin{pmatrix}
    u(x) \\ 1
\end{pmatrix}\ .
\ee
Then we have
\be 
\begin{pmatrix}
    u(x+b) \\ 1
\end{pmatrix}=\begin{pmatrix}
    -\alpha_b(x) & -\beta_b(x) \\ 1 & 0
\end{pmatrix} \begin{pmatrix}
    u(x) \\ 1
\end{pmatrix}=\begin{pmatrix}
    -\alpha_b(x) u(x)-\beta_b(x) \\ u(x)
\end{pmatrix}\ .
\ee
Thus for the direction of the line to be fixed, we must have that 
\be 
u(x) u(x+b)+\alpha_b(x) u(x)+\beta_b(x)=0 \label{eq:Riccati equation}
\ee
has a solution in the field $\KK$. This follows the same logic as in the last paragraph of Section~\ref{subsec:uniqueness}. This equation is called the Riccati equation in the literature, see e.g.\ \cite{Van2012Galois}. Thus in order to have an irreducible Galois group, it simply remains to show that \eqref{eq:Riccati equation} does not have any solutions in $\KK$, i.e.\ solutions that also satisfy $u(x+b^{-1})=u(x)$.
\paragraph{No solutions of the Riccati equation.} Finally, we have to show that the Riccati equation \emph{generically} has no solution. For this, we may choose a special value of $b$. Indeed, irreducibility of the Galois group of one value of $b$ will show irreducibility generically, since the Galois group can only decrease under specialization.
Let us first note that for $b^2 \not\in \QQ$, there is a one-parameter family of solutions, by the by now familiar argument that uses that $\ZZ b\oplus \ZZ b^{-1}$ is  dense in $\mathbb{R}$, see e.g.\ the explanation after eq.~\eqref{eq:Gammab functional equations}.
We will make the choice $b=1+\varepsilon$ and take the limit $\varepsilon \to 0$ and show that no solution exists in this case. We could have chosen $b=1$ directly, but the issue with that is that the field of constants becomes bigger if $b^2 \in \QQ$. However, everything is continuous and thus we simply have to ask about solutions to the difference equation
\be 
u(x+1)u(x)+\alpha_1(x)u(x)+\beta_1(x)=0\ , \qquad u(x+1)=u(x)\ . 
\ee
We can hence replace $u(x+1)$ by $u(x)$ and solve the quadratic equation, giving
\be 
u(x)=\frac{1}{2}\left(-\alpha_1(x)\pm\sqrt{\alpha_1(x)^2-4\beta_1(x)}\right)\ . \label{eq:u candidate solution}
\ee
For this solution to be acceptable, it cannot have any branch cuts since we were searching for \emph{meromorphic} solutions. However, it is easy to see that $\alpha_1(x)^2-4\beta_1(x)$ can have (simple) zeros, which means that the candidate solution \eqref{eq:u candidate solution} is in fact not meromorphic. We also remark that the same argument also works when $b^2=n \in \NN$. When $b^2=\frac{m}{n}$, we can also suitably adapt it by plugging the Riccati equation several times into itself, which shows that $u(x)$ is the solution of a higher-order polynomial.
We thus conclude that the Galois group is irreducible.

\subsection{Ponsot and Teschner's solution for \texorpdfstring{$\mathbb{F}$}{F}} \label{subsec:Ponsot Teschner solution}
Now that we have shown that the normalized crossing kernel is uniquely fixed by the shift relation \eqref{eq:normalized shift relation} up to overall normalization, we will give the general solution. It was first found by Ponsot \& Teschner in \cite{Ponsot:1999uf, Ponsot:2000mt} by using an analogy to the quantum group $\mathcal{U}_g(\mathfrak{sl}(2,\mathbb{R}))$, called the modular double \cite{Faddeev:1999fe}. We use the form provided by Teschner \& Vartanov \cite{Teschner:2012em} and discuss alternative forms later. We claim that 
\begin{multline}  
\sixjnorm{p_1}{p_2}{p_3}{p_4}{p_5}{p_6}=\int_{\frac{Q}{4}+i \RR}\! \frac{\mathrm{d}p}{i}\ S_b(p-p_{123})S_b(p-p_{156})S_b(p-p_{246})S_b(p-p_{345})\\
\times S_b(\tfrac{Q}{2}+p_{1245}-p) S_b(\tfrac{Q}{2}+p_{1346}-p)S_b(\tfrac{Q}{2}+p_{2356}-p)S_b(\tfrac{Q}{2}-p)\ . \label{eq:sixjnorm solution J form}
\end{multline}
We denoted $p_I=\sum_{i \in I} p_i$ for some set $I$. The integration contour over $p$ is the positively oriented imaginary axis shifted by $\frac{Q}{4}$ so that all the singularities are avoided.
Notice that the factors in the first line are associated to the vertices of the tetrahedron while the first three factors in the second line are associated to pairs of opposing edges. Thus tetrahedral symmetry of \eqref{eq:sixjnorm solution J form} is obvious.

\begin{figure}[ht]
    \centering
    \begin{tikzpicture}[scale=.7]
    \begin{scope}[shift={(2,0)}]
        \draw[very thick, rounded corners=20, RoyalBlue, fill=RoyalBlue!20!white] (3.7,-1) to (-.7,0) to (3.7,1);
        \foreach \x\y in {0/0,1/0,2/0,3/0,1/1,2/1,3/1,2/2,3/2,3/3,4/4,4/3}{
            \fill[RoyalBlue] ({\x-.2*\y},{.4*\y-.23*\x}) circle (.07);     
        }
    \end{scope}
    \begin{scope}[shift={(2,3)}]
        \draw[very thick, rounded corners=20, RoyalBlue, fill=RoyalBlue!20!white] (3.7,-1) to (-.7,0) to (3.7,1);
        \foreach \x\y in {0/0,1/0,2/0,3/0,1/1,2/1,3/1,2/2,3/2,3/3,4/4,4/3}{
            \fill[RoyalBlue] ({\x-.2*\y},{.4*\y-.23*\x}) circle (.07);     
        }
    \end{scope}
    \begin{scope}[shift={(2,5.5)}]
        \draw[very thick, rounded corners=20, RoyalBlue, fill=RoyalBlue!20!white] (3.7,-1) to (-.7,0) to (3.7,1);
        \foreach \x\y in {0/0,1/0,2/0,3/0,1/1,2/1,3/1,2/2,3/2,3/3,4/4,4/3}{
            \fill[RoyalBlue] ({\x-.2*\y},{.4*\y-.23*\x}) circle (.07);     
        }
    \end{scope}
    \begin{scope}[shift={(2,8)}]
        \draw[very thick, rounded corners=20, RoyalBlue, fill=RoyalBlue!20!white] (3.7,-1) to (-.7,0) to (3.7,1);
        \foreach \x\y in {0/0,1/0,2/0,3/0,1/1,2/1,3/1,2/2,3/2,3/3,4/4,4/3}{
            \fill[RoyalBlue] ({\x-.2*\y},{.4*\y-.23*\x}) circle (.07);     
        }
    \end{scope}
    \begin{scope}[rotate=180, shift={(2,.5)}]
        \draw[very thick, rounded corners=20, RoyalBlue, fill=RoyalBlue!20!white] (3.7,-1) to (-.7,0) to (3.7,1);
        \foreach \x\y in {0/0,1/0,2/0,3/0,1/1,2/1,3/1,2/2,3/2,3/3,4/4,4/3}{
            \fill[RoyalBlue] ({\x-.2*\y},{.4*\y-.23*\x}) circle (.07);     
        }
    \end{scope}
    \begin{scope}[rotate=180, shift={(2,-1.8)}]
        \draw[very thick, rounded corners=20, RoyalBlue, fill=RoyalBlue!20!white] (3.7,-1) to (-.7,0) to (3.7,1);
        \foreach \x\y in {0/0,1/0,2/0,3/0,1/1,2/1,3/1,2/2,3/2,3/3,4/4,4/3}{
            \fill[RoyalBlue] ({\x-.2*\y},{.4*\y-.23*\x}) circle (.07);     
        }
    \end{scope}
    \begin{scope}[rotate=180, shift={(2,-4.3)}]
        \draw[very thick, rounded corners=20, RoyalBlue, fill=RoyalBlue!20!white] (3.7,-1) to (-.7,0) to (3.7,1);
        \foreach \x\y in {0/0,1/0,2/0,3/0,1/1,2/1,3/1,2/2,3/2,3/3,4/4,4/3}{
            \fill[RoyalBlue] ({\x-.2*\y},{.4*\y-.23*\x}) circle (.07);     
        }
    \end{scope}
    \begin{scope}[rotate=180, shift={(2,-6.8)}]
        \draw[very thick, rounded corners=20, RoyalBlue, fill=RoyalBlue!20!white] (3.7,-1) to (-.7,0) to (3.7,1);
        \foreach \x\y in {0/0,1/0,2/0,3/0,1/1,2/1,3/1,2/2,3/2,3/3,4/4,4/3}{
            \fill[RoyalBlue] ({\x-.2*\y},{.4*\y-.23*\x}) circle (.07);     
        }
    \end{scope}
    \draw[very thick, Maroon,->] (0,-1) .. controls (-.5,2) and (.5,5) .. (0,8.5);
    \end{tikzpicture}
    \caption{The $p$-plane and the contour for the integral \eqref{eq:sixjnorm solution J form}. We drew the picture for complex $b$ with $\Re(b)>0$ to make the pole structure clearer. The picture is drawn for $p_j \in i \RR$. In other cases, the contour is deformed smoothly as the poles move around.}
    \label{fig:p plane J}
\end{figure}
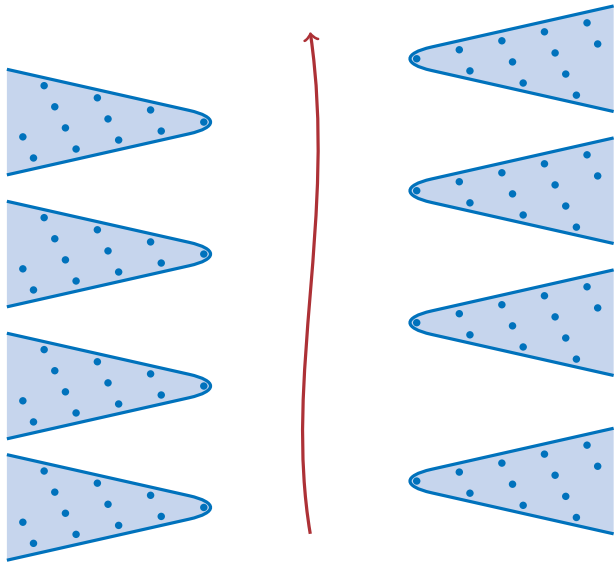

Recall that $S_b(z)$ has poles for $z=-m b-n b^{-1}$ and $m,\, n \in \ZZ_{\ge 0}$. Thus the pole structure of the integrand in the $p$-plane takes the form as displayed in Figure~\ref{fig:p plane J}. There are four `wedges' of singularities to left and right and the $p$-contour threads in between them. Such types of integrals are of Mellin-Barnes type. We discuss them in Appendix~\ref{app:hypergeometric functions} in more detail.

\paragraph{Convergence of the integral.} We first remark that the integral in \eqref{eq:sixjnorm solution J form} converges. Indeed, this follows from the asymptotics of the double sine function, see eq.~\eqref{eq:S_b asymptotics}. For $p=\frac{Q}{4}+i x$ and large $\Re(x)$, the integrand behaves as
\be 
\mathrm{e}^{2\pi i (p_1p_4+p_2p_5+p_3p_6)-\pi i Q\sum_{j=1}^6 p_j-2\pi Q x}\ .
\ee
The first terms are all phases, the only important part is the exponential suppression from the last term. We find a similar exponential suppression for $\Re (x) \to -\infty$ and thus the integral converges.
\paragraph{Check of the shift relation.} We will now demonstrate that this formula indeed solves the shift relation \eqref{eq:normalized shift relation}. We have
\begin{align}
    &\!\!\sum_{\eta=\pm} \eta \cos\big(\pi b(p_1+\eta p_5+p_6)\big)\cos\big(\pi b(p_2+\eta p_4-p_6)\big) \sixjnorm{p_1}{p_2}{p_3}{p_4}{p_5}{p_6 \pm \frac{ \eta b}{2}} \nonumber \\
    &\qquad -\sin(2\pi b p_6) \cos\big(\tfrac{\pi b^2}{2}+\pi b(-p_3+p_4-p_5)\big) \sixjnorm{p_1}{p_2}{p_3}{p_4+\frac{b}{2}}{p_5-\frac{b}{2}}{p_6} \\
    &=\int \frac{\d p}{i}\bigg[ \frac{\cos(\pi bp_{24|6})\cos(\pi b p_{156})\cos(\pi b(p-p_{2356}))}{\cos(\pi b(\frac{Q}{2}+p_{156}-p))}+\sin(2\pi b p_6)\sin(\pi b(\tfrac{Q}{2}+p_{4|35}))\nonumber\\
    &\qquad-\frac{\cos(\pi bp_{2|46})\cos(\pi bp_{16|5})\cos(\pi b(\frac{Q}{2}+p_{246}-p))}{\cos(\pi b(p-p_{1346})}\bigg] \sixjnorm{p_1}{p_2}{p_3}{p_4+\frac{b}{2}}{p_5-\frac{b}{2}}{p_6}_\text{int}\!\!\!\!\!(p) \\
    &=\int \frac{\d p}{i} \bigg[ \frac{\sin(\pi b(\frac{Q}{2}-p))\sin(\pi b(\frac{Q}{2}+p_{1245}-p))\cos(\pi b(p_{2356}-p))}{\cos(\pi b(\frac{Q}{2}+p_{156}-p))}\nonumber\\
    &\qquad-\frac{\sin(\pi b(p-p_{123}))\sin(\pi b(p-p_{345}))\cos(\pi b(\frac{Q}{2}+p_{246}-p))}{\cos(\pi b(p-p_{1346}))}\bigg] \nonumber\\
    &\qquad\qquad\times\sixjnorm{p_1}{p_2}{p_3}{p_4+\frac{b}{2}}{p_5-\frac{b}{2}}{p_6}_\text{int}\!\!\!\!\!(p)\ .
\end{align}
Here, the subscript `int' on the normalized crossing kernel denotes the integrand in eq.~\eqref{eq:sixjnorm solution J form}, which also depends on $p$. We also used the notation introduced in eq.~\eqref{eq:definition pI|J} for brevity.
In passing to the last line, we used a trigonometric identity that is straightforward to check.
Using the functional equation of the double sine function, the integrand can now be written in the form
\be 
f(p+b)-f(p)\ ,
\ee
where
\begin{multline}
    f(p)=-\frac{1}{4}\, S_b(p-p_{123})S_b(p-\tfrac{b}{2}-p_{156})S_b(p-p_{345}) S_b(p-\tfrac{b}{2}-p_{246})S_b(\tfrac{Q}{2}+b+p_{1245}-p) \\ \times S_b(\tfrac{Q}{2}+\tfrac{b}{2}+p_{1346}-p) S_b(\tfrac{Q}{2}+\tfrac{b}{2}+p_{2356}-p)S_b(\tfrac{Q}{2}+b-p)\ .
\end{multline}
By definition, the contour avoids all poles and thus we can shift the contour of the first term back, $p+b \to p$, without crossing any poles. We conclude 
\be  
\int \frac{\mathrm{d}p}{i} \ \big(f(p+b)-f(p)\big)=0
\ee
and thus the shift relation \eqref{eq:normalized shift relation} is satisfied by \eqref{eq:sixjnorm solution J form}.
\paragraph{Symmetry properties.} We also have to demonstrate that \eqref{eq:sixjnorm solution J form} has the required symmetry properties. It has manifest tetrahedral symmetry from the definition as we already noticed below eq.~\eqref{eq:sixjnorm solution J form}. We should notice that in order to prove the uniqueness of the solution, we did not make use of reflection symmetry and thus reflection symmetry of the solution \eqref{eq:sixjnorm solution J form} should be automatic. However, this is not obvious from the formula \eqref{eq:sixjnorm solution J form} and we now explain how to see reflection symmetry more explicitly.

In order to do so, we take a short detour into the theory of $b$-deformed hypergeometric functions. This theory was developed in the context of the elliptic Gamma function of which the double sine function is a special limit. We have provided more background in Appendix~\ref{app:hypergeometric functions}. The $b$-deformed hypergeometric function is defined as
\be 
V_b(a_1,\dots,a_8)=\int \frac{\d z}{2i}\, \frac{\prod_{j=1}^8 S_b(a_j \pm z)}{S_b(\pm 2 z)}\ . \label{eq:b-deformed hypergeometric function main text}
\ee
Here, the eight parameters $(a_1,\dots,a_8)$ satisfy the single `balancing condition' $\sum_i a_i=2Q$. The definition has an obvious $\S_8$ symmetry permuting the eight parameters $a_i$. However, as was shown by \cite{Spridonov:theta, Rains:Transformations}, the hypergeometric function behaves covariantly under the extended symmetry group given by the Weyl group $W(\mathrm{E}_7)$. This is manifested in the identity
\be 
V_b(a_1,\dots,a_8)=\prod_{1 \le j<k \le 4} S_b(a_j+a_k)\prod_{5 \le j<k \le 8} S_b(a_{j}+a_{k}) V_b(\tilde{a}_1,\dots,\tilde{a}_8) \label{eq:hypergeometric function E7 symmetry transformation main text}
\ee
with 
\be 
\nu_j=\begin{cases}
    \frac{1}{2}(Q-a_{1234})+a_j\ , \qquad j=1,\dots,4\ , \\
    \frac{1}{2}(Q-a_{5678})+a_j\ , \qquad j=5,\dots,8\ .
\end{cases}
\ee
We give a proof of this identity in Appendix~\ref{app:hypergeometric functions}.
The type of integrand that we encounter in the proposed solution of the normalized crossing kernel in eq.~\eqref{eq:sixjnorm solution J form} is a special case of this hypergeometric function. 
The hypergeometric function $V_b(a_1,\dots,a_8)$ reduces to the form \eqref{eq:sixjnorm solution J form} by replacing $a_j \mapsto a_j+x$ for $j=1,2,3,4$ and $a_j \to a_j-x$ for $j=5,6,7,8$. One then considers the limit $x \to +i\infty$. See \eqref{eq:hypergeometric function D6 degeneration asymmetric} for the precise formula.
This limit breaks some of the symmetries of the hypergeometric function, but some non-trivial remnant of eq.~\eqref{eq:hypergeometric function E7 symmetry transformation main text} remains. In particular, the symmetry group is extended from the trivial $\S_4 \times \S_4$ symmetry to the Weyl group $W(\mathrm{D}_6)\cong \mathrm{S}_6 \ltimes \ZZ_2^5$. It is now a simple exercise to trace through the definitions and the transformation property \eqref{eq:hypergeometric function E7 symmetry transformation main text} and confirm that the solution \eqref{eq:sixjnorm solution J form} indeed exhibits the correct behaviour under reflection symmetry.

\paragraph{Pole structure.} Let us briefly discuss the pole structure of the solution \eqref{eq:sixjnorm solution J form}.
A pole appears whenever the contour in $p$ gets pinched in-between two of the wedges. Let us write $U_1,\dots,U_4$ for the sums of $p_j$'s appearing in the first four terms and $V_1,\dots,V_4$ for those appearing in the second four terms. We hence write
\be  
\sixjnorm{p_1}{p_2}{p_3}{p_4}{p_5}{p_6}=\frac{1}{i} \int \mathrm{d}p\ \prod_{j=1}^4 S_b(p-U_j)S_b(V_j-p)\ .
\ee
Then the integrand has poles for $p-U_j=-m b-n b^{-1}$ and $V_j-p=-m b-n b^{-1}$. Consequently, the contour gets pinched whenever $V_k-U_j=-m b-n b^{-1}$ for some $j$ and $k$. There are hence 16 groups of singularities. They are of the form
\be  
\frac{Q}{2}\pm_1 p_1 \pm_2 p_2 \mp_3 p_3=-m b-n b^{-1} \label{eq:sixjnorm poles}
\ee
under the constraint $\pm_1 \pm_2 \pm_3=+$. We get a similar set of poles for the other vertices of the tetrahedron with momenta $(p_1,p_5,p_6)$, $(p_3,p_4,p_5)$ and $(p_2, p_4, p_6)$.

\paragraph{Determinant formula.} We recall that we needed that 
\be 
\det\!{}_6 \sixjnorm{p_1}{p_2}{p_3}{p_4}{p_5}{p_6}\equiv \det\begin{pmatrix}
\sixjnorm{p_1}{p_2}{p_3}{p_4}{p_5}{p_6} & \sixjnorm{p_1}{p_2}{p_3}{p_4}{p_5}{p_6-b^{-1}} \\
\sixjnorm{p_1}{p_2}{p_3}{p_4}{p_5}{p_6-b } & \sixjnorm{p_1}{p_2}{p_3}{p_4}{p_5}{p_6-Q}
\end{pmatrix} \label{eq:sixjnorm determinant}
\ee
to not vanish identically in the uniqueness proof of Section~\ref{subsec:uniqueness}, see the discussion around eq.~\eqref{eq:N FG^-1}. We now compute its value. In particular, it will not be zero. From the difference equation \eqref{eq:difference equation determinant} for the determinant the $p_6$-dependence is fully fixed. We get
\begin{multline} 
\det\!{}_6 \sixjnorm{p_1}{p_2}{p_3}{p_4}{p_5}{p_6}=f(p_1,\dots,p_5) S_b(Q \pm (2p_6-Q)) \\
\times \frac{S_b(-\frac{Q}{2}+p_6 \pm p_{1|5})S_b(-\frac{Q}{2}+p_6 \pm p_{2|4})}{S_b(\frac{Q}{2}+p_6 \pm p_{15})S_b(\frac{Q}{2}+p_6 \pm p_{24})}\ .
\end{multline}
To evaluate the prefactor $f(p_1,\dots,p_5)$, we can choose a special value of $p_6$. We take 
\be 
p_6=\frac{Q}{2}-p_{15}\ .
\ee
This is chosen such that the first entry in eq.~\eqref{eq:sixjnorm determinant} has a pole, but not the other entries, compare with eq.~\eqref{eq:sixjnorm poles}. Thus we can take residues on both sides. We get
\begin{multline} 
f(p_1,\dots,p_5) \frac{S_b(-2p_1)S_b(-2p_5)}{S_b(-2p_{15})} S_b(-p_{15} \pm p_{2|4})S_b(p_{15} \pm p_{24}))\\
=-2\pi \Res_{p_6=\frac{Q}{2}-p_{15}} \sixjnorm{p_1}{p_2}{p_3}{p_4}{p_5}{p_6} \sixjnorm{p_1}{p_2}{p_3}{p_4}{p_5}{-\frac{Q}{2}-p_{15}}\ . \label{eq:residue compute f}
\end{multline}
Let us first compute the last normalized crossing kernel,
\begin{align}
    \sixjnorm{p_1}{p_2}{p_3}{p_4}{p_5}{-\frac{Q}{2}-p_{15}}&=\int_{\frac{Q}{4}+i \RR} \frac{\d p}{i}\  S_b(p-p_{123}) S_b(\tfrac{Q}{2}+p+p_{15|24}) S_b(p-p_{345}) \nonumber \\
    &\qquad\times S_b(\tfrac{Q}{2}+p_{1245}-p) S_b(p_{34|5}-p)S_b(p_{23|1}-p)\\
    &=\frac{S_b(-2p_1)S_b(-2p_5)}{S_b(-2p_{15})} S_b(\tfrac{Q}{2}+p_{45|3}) S_b(-p_{15} \pm p_{2|4}) \nonumber \\&\qquad\times S_b(\tfrac{Q}{2}+p_{13|2}) S_b(\tfrac{Q}{2}+p_{35|4})S_b(\tfrac{Q}{2}+p_{12|3})\ . \label{eq:value sixjnorm p6=-p15-Q/2}
\end{align}
Here we used the integral identity \eqref{eq:S_b beta integral identity asymmetric} and we refer to Appendix~\ref{app:hypergeometric functions} for further explanations about its origin.

Computing the residue of the other normalized crossing kernel is actually simpler. Setting $p_6=\frac{Q}{2}-p_{15}-\varepsilon$ for small $\varepsilon$ leads to the factors
\be 
S_b(p-\tfrac{Q}{2}+\varepsilon)S_b(\tfrac{Q}{2}-p)
\ee
in the integrand. Thus the contour gets pinched at $p=\frac{Q}{2}$. We can put $p=\frac{Q}{2}$ in all other regular factors. For the singular factor, we can use $S_b(z) \sim \frac{1}{2\pi z}$ for small $z$ and evaluate the residue explicitly. This leads to
\be 
\Res_{p_6=\frac{Q}{2}-p_{15}} \sixjnorm{p_1}{p_2}{p_3}{p_4}{p_5}{p_6}=\frac{1}{2\pi} S_b(p_{15} \pm p_{24}) S_b(\tfrac{Q}{2}-p_1 \pm p_{23}) S_b(\tfrac{Q}{2}-p_5 \pm p_{34})\ .  \label{eq:value residue sixjnorm p6=-p15+Q/2}
\ee
We can then insert \eqref{eq:value sixjnorm p6=-p15-Q/2} and \eqref{eq:value residue sixjnorm p6=-p15+Q/2} into \eqref{eq:residue compute f} to obtain
\be 
f(p_1,\dots,p_5)=-\prod_{\pm_1 \pm_2 \pm_3=+} S_b(\tfrac{Q}{2} \pm_1 p_1 \pm_2 p_2 \mp_3 p_3) S_b(\tfrac{Q}{2} \pm_1 p_3 \pm_2 p_4 \mp_3 p_5)\ .
\ee
As a consistency check, we notice that this preserves all remaining tetrahedral symmetries. Most importantly, we remark that it does not vanish identically, which is what we needed in the proof of the uniqueness.

\paragraph{Normalization.} The last property that we need to check in order to confirm that \eqref{eq:sixjnorm solution J form} indeed constitutes a solution of the bootstrap problem is the normalization of eq.~\eqref{eq:sixjnorm solution J form}. We do this by showing that it correctly reproduces the initial case \eqref{eq:F symbol identity delta function} where we send on of the external momenta to $\id$.

Thus let us put $p_{21}=p_3$ and set $p_4=\varepsilon-\frac{Q}{2}$. We expect to get a delta function setting $p_{32}$ to $p_1$ or $p_{32}$ to $-p_1$. Since we have already demonstrate the invariance under reflection symmetry, it suffices to focus on one of those. The latter one turns out to be simpler.

Note that because of the prefactors in \eqref{eq:sixjnorm definition}, this limit naively vanishes, because the last factor in the denominator $\Gamma_b(\frac{Q}{2}\pm_1 p_3\pm_2 p_4 \mp_3 p_{21})$ diverges for $\pm_1=\pm_3$ and $\pm_2=+$. Thus this factor vanishes to second order in $\varepsilon \to 0$. In order to get a finite result, we need to compensate this behaviour by another diverging factor. One can easily see that without further specifying parameters, the normalized crossing kernel is regular and thus the result can have at most delta-function support in the parameter space. As we mentioned below, we will focus on the delta-function of the form $\delta(p_{32}+p_1)$, supported on the imaginary line $i \RR$.
Then we get a further divergence in the prefactor from the last factors in the numerator of eq.~\eqref{eq:sixjnorm definition}. They read
\be  
\prod_{\pm_1 \pm_2 \pm_3=\pm} \Gamma_b\big(\tfrac{Q}{2}\pm_1p_1\pm_2p_4\pm_3p_{32})\ .
\ee
When $p_{32} +p_1 \sim 0$, we can make the argument of the double Gamma function to be close to a pole when $\pm_1=\pm_3$ and $\pm_2=+$. Using that
\be  
\frac{\varepsilon}{\varepsilon^2+P^2} \longrightarrow \pi \delta(P)
\ee
is an approximation of a delta-function this divergence appears for $p_{32}=-p_1$.\footnote{Since the delta-function is supported on the imaginary line $p \in i \RR$, we should use the real Liouville momentum $P$ here.} It remains to compute
\begin{multline} 
\lim_{\varepsilon \to 0} \varepsilon \sixjnorm{p_1}{p_2}{p_3}{p_3}{\varepsilon-\frac{Q}{2}}{-p_1}\\
=\lim_{\varepsilon \to 0} \, \varepsilon \int \frac{\mathrm{d}p}{i} \ S_b(p-p_{123})S_b(p_{123}+\varepsilon-p)S_b(p-p_{23|1})S_b(p_{23|1}+\varepsilon-p)  \\
\times S_b(\tfrac{Q}{2}+p-\varepsilon)S_b(\tfrac{Q}{2}-p) S_b(\tfrac{Q}{2}+p-2p_3-\varepsilon)S_b(\tfrac{Q}{2}+2p_3-p)  \ .
\end{multline}
The first two factors pinch the contour at $p\sim p_{123}$, which leads to the desired pole in the integral. Similarly, the third and fourth factor pinch the contour at $p=p_{32|1}$ and this leads to a further contribution to the pole. For both contributions, we may set $p$ to these values and $\varepsilon=0$ in all other regular factors. In particular upon setting $\varepsilon=0$, the factors in the last line equal 1.
This gives
\begin{align}
\lim_{\varepsilon \to 0} \varepsilon \sixjnorm{p_1}{p_2}{p_3}{p_3}{\varepsilon-\frac{Q}{2}}{-p_1}&=\frac{1}{\pi} \, S_b(\pm 2 p_1)\ ,
\end{align}
where a factor 2 accounts for the two different contributions. When we combine the computation for the prefactor and the normalized 6j symbol, we get
\be  
\lim_{p_4 \to 0} \mathbb{F}_{p_{3},p_{32}}\!\begin{bmatrix}
    p_3 & p_2 \\ p_4 & p_1
\end{bmatrix}\overset{p_{32}+p_1 \sim 0}{=}\delta(P_{32}+P_1)\ .
\ee
The other delta function term $\delta(P_{32}-P_1)$ can also be obtained directly from a similar limiting procedure, but reflection symmetry of the crossing kernel guarantees the presence of the other term.

\subsection{More symmetries and alternative forms}
We have demonstrated at this point that the crossing kernel given by eq.~\eqref{eq:sixjnorm definition} and \eqref{eq:sixjnorm solution J form} is the unique solution to the bootstrap problem. We now mention various further properties. We in particular need the alternative form \eqref{eq:alternative integral representation of the normalized crossing kernel} to derive the modular crossing kernel below.

\paragraph{Regge symmetries.} The crossing kernel $\mathbb{F}$ has actually more symmetries besides the tetrahedral symmetry and reflection symmetry. There is an analogue of the Regge symmetries that one encounters in 6j symbols of $\mathrm{SU}(2)$ representations \cite{Regge:1959}. For $\mathrm{SU}(2)$, this symmetry is somewhat mysterious, as it does not have an obvious representation-theoretic explanation. However, for Virasoro blocks there is a natural explanation in terms of the AGT correspondence \cite{Alday:2009aq}.
Indeed, as we already mentioned, the symmetry group of the $b$-deformed hypergeometric function \eqref{eq:sixjnorm solution J form} is the Weyl group of $\mathrm{D}_6$, which has $6! \cdot 2^5=\num{23040}$ elements. Tetrahedral symmetry and reflection symmetry only account for $4! \cdot 2^6=\num{1536}$ of these symmetries. 

Consider the solution \eqref{eq:sixjnorm solution J form}. We have
\be 
\sixjnorm{\frac{1}{2}p_{1245}-p_1}{\frac{1}{2}p_{1245}-p_2}{p_3}{\frac{1}{2}p_{1245}-p_4}{\frac{1}{2}p_{1245}-p_5}{p_6}=\sixjnorm{\frac{1}{2}p_{1245}-p_4}{\frac{1}{2}p_{1245}-p_5}{p_3}{\frac{1}{2}p_{1245}-p_1}{\frac{1}{2}p_{1245}-p_2}{p_6}=\sixjnorm{p_1}{p_2}{p_3}{p_4}{p_5}{p_6}\ ,
\ee
where we used tetrahedral symmetry in the first equality. The second equality follows directly from the solution \eqref{eq:sixjnorm solution J form}, since under this replacement, we merely exchange two of the factors in the integrand. The prefactor also respects this symmetry and we thus have
\be 
\mathbb{F}_{p_{21},p_{32}}\!\begin{bmatrix}
    p_3 & p_2 \\
    p_4 & p_1
\end{bmatrix}=\mathbb{F}_{p_{21},p_{32}}\!\begin{bmatrix}
    \frac{1}{2}p_{1234}-p_3 & \frac{1}{2}p_{1234}-p_2 \\
    \frac{1}{2}p_{1234}-p_4 & \frac{1}{2}p_{1234}-p_1
\end{bmatrix}\ . \label{eq:Regge symmetry}
\ee
This Regge symmetry, together with tetrahedral and reflection symmetry generates now the full symmetry group of the crossing kernel. Eq.~\eqref{eq:Regge symmetry} can be proven much more directly by noting that the shift relation \eqref{eq:shift relation} exhibits this symmetry. Since the solution to the shift relation is unique up to an overall constant which can easily be fixed from special cases, the symmetry follows. 
The Regge symmetry follows from a similar symmetry of the four-point block on the sphere, see eq.~\eqref{eq:conformal block four-punctured sphere Regge symmetry}. One explanation of this symmetry can be given in terms of the AGT correspondence. The four-point block on the sphere is related to the Nekrasov instanton partition function \cite{Nekrasov:2002qd} for $\mathrm{SU}(2)$ $\mathcal{N}=2$ gauge theory in four dimensions with four flavours. This theory has an extended $\mathrm{SO}(8)$ flavour symmetry and the Regge symmetry is a manifestation of the $\mathrm{SO}(8)$ Weyl group action \cite{Giribet:2009hm}. We included further details of this statement in Appendix~\ref{app:four-punctured sphere}.

\paragraph{Alternative forms.} There are alternative integral representations that make other symmetries of the crossing kernel manifest. For example, we may apply any symmetry transformation of the $W(\mathrm{D}_6)$ symmetry group to eq.~\eqref{eq:sixjnorm solution J form}. Since the formula only makes $\mathrm{S}_4 \times \mathrm{S}_4$ symmetry manifest, this leads to many similar expressions, but with different parameters. In particular, the original expression given by Ponsot and Teschner \cite{Ponsot:1999uf} corresponds to a choice that does not make tetrahedral symmetry manifest.

More interestingly, we can also give a different looking integral representation. It is obtained from the $W(\mathrm{E}_7)$ symmetry of the underlying $b$-deformed hypergeometric function \eqref{eq:b-deformed hypergeometric function main text}. We obtain in this way
\begin{multline}
    \sixjnorm{p_1}{p_2}{p_3}{p_4}{p_5}{p_6}=\prod_{j=1}^3S_b(\tfrac{Q}{2}-p_j \pm (p_{j+3}-p_{456}))\\
    \times \int \frac{\mathrm{d}p}{2i}\ \frac{\prod_{j=1}^3 S_b\big(\frac{1}{2}(\frac{Q}{2}+p_{123})-p_j \pm p_{j+3} \pm p\big)}{S_b( \pm 2p)} \ . \label{eq:alternative integral representation of the normalized crossing kernel}
\end{multline}
This expression makes an $\mathrm{S}_3$ subgroup of the tetrahedral symmetry as well as reflection symmetry in $p_4$, $p_5$ and $p_6$ manifest.

\paragraph{Pole structure.}
We already analyzed the pole structure of the normalized crossing kernels in Section~\ref{subsec:Ponsot Teschner solution}, see eq.~\eqref{eq:sixjnorm poles}. Let us now extend this to the full crossing kernel, including the prefactor in eq.~\eqref{eq:sixjnorm definition}. 

The prefactor cancels all the poles associated to the vertices $(p_1,p_2,p_{21})$ and $(p_3,p_4,p_{21})$ of the tetrahedron. The other double Gamma functions have the effect of inserting the missing poles for the other two vertices that are predicted by reflection symmetry. Overall, we get the following poles
\begin{subequations}
\begin{align}
    \frac{Q}{2} \pm p_2\pm p_3\pm p_{32}&=-m b-n b^{-1}\ , \\
    \frac{Q}{2} \pm p_1\pm p_4\pm p_{32}&=-m b-n b^{-1}\ .
\end{align}
\end{subequations}
Additionally, we also get a series of poles for $\pm p_{21}=-\frac{(m+1) b}{2}-\frac{(n+1) }{2b}$. Finally, we get a series of zeros for $\pm p_{32}=-\frac{m b}{2}-\frac{n }{2b}$.\footnote{For this argument to work, we need the statement that for generic choices of momenta the hypergeometric function does not have zeros at these locations. This follows for example from the same fact for the ordinary hypergeometric function ${}_2F_1$, which can be obtained as a special limit from the $b$-deformed hypergeometric function.}

The poles in $p_{21}$ are exactly what we expected, see the discussion about analyticity in Section~\ref{subsec:crossing transformations}.
All other poles and zeros can be deduced by making use of tetrahedral symmetry \eqref{eq:tetrahedrally symmetric combination}.
More physically, these other poles were interpreted in \cite{Collier:2018exn} in terms of double-twist operators.

\subsection{The modular crossing kernel \texorpdfstring{$\mathbb{S}$}{S}}
We already expressed the modular crossing kernel in terms of the crossing kernel on the sphere, see eqs.~\eqref{eq:S F relation} and \eqref{eq:rho0 definition}. We now evaluate this relation explicitly to obtain a solution for $\mathbb{S}$. Our computation is very similar to the derivation in \cite{Teschner:2013tqy}. It is convenient to reflect one of the Liouville momenta in the crossing kernel on the right hand side in \eqref{eq:S F relation}. Taking one $p_2 \to -p_2$ and plugging in the definitions leads to 
\begin{multline}
    \mathbb{S}_{p_1,p_2}[p_0]=\frac{\rho_0(p_2) \mathrm{e}^{\pi i(\Delta_0-2\Delta_1-2\Delta_2+\frac{Q^2}{2})}\, \Gamma_b(Q \pm 2 p_1) \Gamma_b(\frac{Q}{2}-p_0 \pm 2p_2)}{2\sqrt{2}\, \Gamma_b(Q \pm 2 p_2) \Gamma_b(\frac{Q}{2}- p_0 \pm 2 p_1)} \\
    \times \int \frac{\d p}{i} \ \rho_0(p)\,  \mathrm{e}^{-2\pi i p^2} \sixjnorm{p_1}{p_2}{p}{-p_2}{p_1}{p_0}\ .
\end{multline}
The contour over $p$ extends from $-i\infty$ to $+i\infty$. This accounts for a factor of 2, since the contour in \eqref{eq:S F relation} only runs from $0$ to $\infty$.
Let us denote the integral in the second line as $\mathcal{I}$. 
We will now explain how to simplify it. We will use the alternative expression \eqref{eq:alternative integral representation of the normalized crossing kernel} for the crossing kernel since it makes reflection symmetry in $p$ manifest. We have
\begin{multline}
    \sixjnorm{p_1}{p_2}{p}{-p_2}{p_1}{p_0}=\sixjnorm{p_2}{-p_2}{p_0}{p_1}{p_1}{p}=S_b(\tfrac{Q}{2}-p_0 \pm 2p_1) \\ \times \int \frac{\d q}{2i} \ \frac{S_b(\frac{Q}{4}+\frac{p_0}{2} \pm p_1 \pm p_2 \pm q) S_b(\frac{Q}{4}-\frac{p_0}{2} \pm p \pm q)}{S_b( \pm 2 q)}\ .
\end{multline}
We now recall that $\frac{1}{\sqrt{2}}\rho_0(p)= S_b(\pm 2p)^{-1}$. We can thus write the integral $\mathcal{I}$ as follows,
\begin{multline}
    \mathcal{I}=2\sqrt{2} S_b(\tfrac{Q}{2}-p_0 \pm 2p_1)\int\frac{\d q}{2i} \ \frac{S_b(\frac{Q}{4}+\frac{p_0}{2} \pm p_1 \pm p_2 \pm q)}{S_b(\pm 2q)}\\
    \times \int \frac{\d p}{2i} \ \mathrm{e}^{-2\pi i p^2} \frac{S_b(\frac{Q}{4}-\frac{p_0}{2} \pm q \pm p)}{S_b(\pm 2p)}\ .
\end{multline}
We interchanged the integral over $p$ and $q$. The integral over $p$ can now be performed with the help of the degenerated Euler beta-integral identity \eqref{eq:S_b beta integral identity 2 a}. We obtain
\begin{multline}
    \mathcal{I}=2\sqrt{2} S_b(\tfrac{Q}{2}-p_0)S_b(\tfrac{Q}{2}-p_0 \pm 2p_1) \mathrm{e}^{\frac{\pi i}{4}(Q-2p_0)Q} \\
    \times \int \frac{\d q}{2i}\ \mathrm{e}^{-2\pi i q^2} \frac{S_b(\frac{Q}{4}+\frac{p_0}{2} \pm p_1 \pm p_2 \pm q)}{S_b(\pm 2q)}\ . 
\end{multline}
We thus obtain the formula
\begin{multline}
    \mathbb{S}_{p_1,p_2}[p_0]=S_b(\tfrac{Q}{2}-p_0) \rho_0(p_2) \frac{\Gamma_b(Q \pm 2 p_1) \Gamma_b(\frac{Q}{2}-p_0 \pm 2p_2)}{\Gamma_b(Q \pm 2 p_2) \Gamma_b(\frac{Q}{2}+ p_0 \pm 2 p_1)}\\
    \times \mathrm{e}^{\frac{\pi i}{2}(-p_0 Q-2p_0^2)+2\pi i(p_1^2+p_2^2)} \int \frac{\d p}{2i}\ \mathrm{e}^{-2\pi i p^2} \frac{S_b(\frac{Q}{4}+\frac{p_0}{2} \pm p_1 \pm p_2 \pm p)}{S_b(\pm 2p)}\ . \label{eq:modular crossing kernel first form}
\end{multline}
We renamed $q \to p$. The required integral needs a contour deformation to converge absolutely. The asymptotics $p \to \pm i \infty$ are $\mathrm{e}^{-2\pi i p(p \pm 2p_0)}$. One can make it convergent by rotating the contour slightly $p \to p \mathrm{e}^{i 0^+}$ in a positive direction.

\paragraph{Symmetries and hypergeometric function.} The formula \eqref{eq:modular crossing kernel first form} exhibits some of the expected symmetries. It is manifestly reflection symmetric under $p_1 \to -p_1$ and under $p_2 \to -p_2$. It also behaves in a simple way when exchanging $p_1$ and $p_2$. It is indeed easy to confirm that the combination \eqref{eq:S symmetric under exchange combination} is invariant under exchange of $p_1$ and $p_2$. Finally, the reflection symmetry in $p_0$ is less obvious. 
To see it, we notice that \eqref{eq:modular crossing kernel first form} is a special case of the $b$-deformed hypergeometric function that we already encountered for the crossing kernel on the sphere. This limit is described further in Appendix~\ref{subapp:degenerations}. The degenerated hypergeometric function inherits some of the $W(\mathrm{E}_7)$ symmetries of the parent hypergeometric function.
The degenerated hypergeometric function has $\mathrm{S}_4 \times \ZZ_2$ symmetry, but only a $\ZZ_4 \times \ZZ_2 \times \ZZ_2$ symmetry acts non-trivially here because the arguments are further specialized.
One $\ZZ_2$ symmetry is non-trivial and is the integral identity \eqref{eq:S4Z2 hypergeometric function Z2 symmetry}. It demonstrates invariance under $p_0 \to -p_0$.

We should also notice that \eqref{eq:modular crossing kernel first form} is almost invariant when we change the signs of the exponentials. In fact, we can apply eq.~\eqref{eq:S4Z2 hypergeometric function opposite sign exponent} followed by sending $p_0 \to -p_0$, which gives the alternative expression
\begin{multline}
    \mathbb{S}_{p_1,p_2}[p_0]=S_b(\tfrac{Q}{2}-p_0) \rho_0(p_2) \frac{\Gamma_b(Q \pm 2 p_1) \Gamma_b(\frac{Q}{2}-p_0 \pm 2p_2)}{\Gamma_b(Q \pm 2 p_2) \Gamma_b(\frac{Q}{2}+ p_0 \pm 2 p_1)}\\
    \times \mathrm{e}^{\frac{\pi i}{4}(Q-2p_0)Q-2\pi i(p_1^2+p_2^2)} \int \frac{\d p}{2i}\ \mathrm{e}^{2\pi i p^2} \frac{S_b(\frac{Q}{4}+\frac{p_0}{2} \pm p_1 \pm p_2 \pm p)}{S_b(\pm 2p)}\ . 
\end{multline}
In particular, we learn that for the choice $p_0,\, p_1 ,\, p_2 \in i \RR$, the combination
\be 
\mathrm{e}^{-\frac{\pi i \Delta_0}{2}} \mathbb{S}_{p_1,p_2}[p_0]
\ee
is real, as expected from the Moore-Seiberg relation \eqref{eq:g=1, n=1 idempotency S}.

\paragraph{Alternative forms.} This not the form one usually encounters in the literature \cite{Teschner:2003at, Teschner:2013tqy}. It can be brought into a more standard form by using the $W(\mathrm{E}_7)$ symmetry of the underlying hypergeometric function, which leads to two essentially different other integral representations. We describe the details in Appendix~\ref{subapp:degenerations}. Using eq.~\eqref{eq:S4Z2 hypergeometric function relation first form second form}, we obtain the representation
\begin{multline}
\mathbb{S}_{p_1,p_2}[p_0]=\frac{\mathrm{e}^{\frac{\pi i}{2} \Delta_0} \rho_0(p_2)\Gamma_b(Q \pm 2p_1) \Gamma_b(\frac{Q}{2}-p_0 \pm 2p_2)}{S_b(\frac{Q}{2}+p_0)\Gamma_b(Q \pm 2p_2) \Gamma_b(\frac{Q}{2}-p_0 \pm 2p_1)} \\
\times \int \frac{\d p}{i} \ \mathrm{e}^{4\pi i p p_1} S_b(\tfrac{Q}{4}+\tfrac{p_0}{2}  \pm p_2 \pm p)\ . \label{eq:modular crossing kernel second form}
\end{multline}
This form is arguably a bit simpler, but makes less symmetry manifest. It is the standard form known from the literature. Convergence is again a little bit subtle, since the integrand asymptotes to an oscillatory function as $p \to \pm i \infty$. Convergence can be restored by rotating the contour slightly. There is a third form that is obtained from eq.~\eqref{eq:S4Z2 hypergeometric function relation first form third form}, which we also state for completeness. It reads
\begin{align}
    \mathbb{S}_{p_1,p_2}[p_0]&=\rho_0(p_2) \mathrm{e}^{\frac{\pi i}{2}(3p_1^2-2p_1p_2+3p_2^2-2p_0^2-\frac{Q^2}{4})} \frac{\Gamma_b(Q \pm 2p_1)\Gamma_b(\frac{Q}{2} \pm p_0-2p_2)}{\Gamma_b(Q \pm 2p_2) \Gamma_b(\frac{Q}{2} \pm p_0+2p_1)} \nonumber\\
    &\qquad\times \int \frac{\d p}{i}\ \mathrm{e}^{\frac{\pi i p}{2}(2Q+2p_1+2p_2-p)} S_b(\tfrac{Q}{2}- p_1-p_2-p)S_b(\tfrac{Q}{2}+ p_1-p_2-p)\nonumber\\
    &\qquad\qquad\times S_b(\tfrac{Q}{2}- p_1+p_2-p)S_b(\pm p_0+p_1+p_2+p)\ .\label{eq:modular crossing kernel third form}
\end{align}

\paragraph{Reduction for $p_0 \to \id$.} Let us check for consistency that the modular crossing kernel reduces back to $2 \sqrt{2} \cos(4\pi p_1 p_2)$ when we take the limit $p_0 \to \id$, see eq.~\eqref{eq:SP1,P2[1]}. We can take the limit using either of the three forms. Let us take \eqref{eq:modular crossing kernel second form} and set $p_0=-\frac{Q}{2}+2 \varepsilon$. The prefactor $S_b(\frac{Q}{2}+p_0)^{-1}$ goes to zero in the limit, but this is compensated by the pinching of the contour at $p=\pm p_2$. We thus get
\begin{align}
    \lim_{\varepsilon \to 0} \mathbb{S}_{p_1,p_2}[-\tfrac{Q}{2}+2 \varepsilon]&=4\pi \rho_0(p_2) \lim_{\varepsilon \to 0} \varepsilon\int \frac{\d p}{i} \ \mathrm{e}^{4\pi i p p_1} S_b(\varepsilon  \pm p_2 \pm p) \\
    &=\rho_0(p_2) S_b(\pm 2p_2) (\mathrm{e}^{4\pi i p_1 p_2}+\mathrm{e}^{-4\pi i p_1 p_2}) \\
    &=2 \sqrt{2} \cos(4\pi p_1p_2)\ ,
\end{align}
as expected.

\paragraph{Pole structure.} Let us finally also mention the analytic structure of the modular crossing kernel. It is easiest to discuss from the form \eqref{eq:modular crossing kernel first form}, since we do not have to worry about convergence issues related to the exponentials underneath the integral. As we discussed, the integral always converges with the appropriate contour.
The modular crossing kernel has then the following sets of poles
\be 
Q \pm 2p_1,\, \ \frac{Q}{2} \pm p_0 \pm 2p_2 = -m b-n b^{-1}\ .
\ee
Additionally, there is a series of zeros for $ \pm 2p_2=-m b-n b^{-1}$.
The first set of poles corresponds to the expected pole when $p_1$ belongs to a degenerate representation where the Kac-Matrix fails to be invertible. The second series of poles as well as the series of zeros follows from the symmetry \eqref{eq:S symmetric under exchange combination}.

\section{Further properties and applications} \label{sec:further properties and applications}
We have now fully derived the kernels on the four-punctured sphere and the once-punctured torus. In this section, we will go back and explain the crossing transformations on arbitrary surfaces and various applications. We will switch back to using the real Liouville momenta $P$.

\subsection{Unitarity} \label{subsec:unitarity}

\paragraph{Extending to arbitrary genus.} We remind the reader that even though we mostly discussed the case of the four-punctured sphere and the once-punctured torus, the derived crossing kernels are the building block for arbitrary crossing transformations on any genus surfaces, see the discussion in Section~\ref{subsec:crossing transformations}.

\paragraph{Moore-Seiberg consistency conditions.} Even though we only imposed a subset of the Moore-Seiberg consistency conditions on the crossing kernels, we are guaranteed that they all hold from abstract reasoning. Indeed, we started from the Hilbert space of normalizable blocks with respect to the inner product \eqref{eq:inner product Teichmuller space}. We then identified a particular basis of blocks in one particular conformal block channel. Changing of the pair of pants decomposition is a purely geometric operation that satisfies a set of relations which were proven to be necessary and complete in \cite{Bakalov:2000}. 
Since we have shown that the solution of $\mathbb{F}$ and $\mathbb{S}$ to the bootstrap equations is unique, the full set of relations necessarily has to hold on the space of conformal blocks. As we have also mentioned already in Section~\ref{subsec:crossing transformations}, these transformations are unitarity by virtue of the definition of the inner product \eqref{eq:inner product Teichmuller space}.

\paragraph{Check of unitarity.} We can explicitly check that the crossing kernel on the sphere and the once-punctured torus preserve the inner product \eqref{eq:explicit inner product four-punctured sphere}. Let us do so for the crossing kernel on the four-punctured sphere. Inserting the definition of the crossing kernel gives
\begin{align}
    \delta(P_{21}-P_{21}')&=C_0(P_1,P_2,P_{21}) C_0(P_3,P_4,P_{21}) \rho_0(P_{21}) \int_0^\infty\!\! \d P_{32}\, \d P_{32}'\ \overline{\mathbb{F}_{P_{21},P_{32}}\!\begin{bmatrix}
        P_3 & P_2 \\ P_4 & P_1
    \end{bmatrix}}\nonumber\\
    &\qquad\times \mathbb{F}_{P_{21}',P_{32}'}\!\begin{bmatrix}
        P_3 & P_2 \\ P_4 & P_1
    \end{bmatrix}
\!\scalebox{1}[1.5]{\Bigg\langle} \!\begin{tikzpicture}[baseline={([yshift=-.5ex]current bounding box.center)}, xscale=.65, yscale=.6]
        \draw[very thick, Maroon, bend left=10] (-1.8,1.2) to (0,.6);
        \draw[very thick, Maroon, bend right=10] (-1.8,-1.2) to (0,-.6);
        \draw[very thick, Maroon, bend right=10] (2.2,1.2) to (0,.6);
        \draw[very thick, Maroon, bend left=10] (2.2,-1.2) to (0,-.6);
        \draw[very thick, Maroon] (0,-.6) to (0,.6);
        \fill[Maroon] (0,.6) circle (.07);
        \fill[Maroon] (0,-.6) circle (.07);
        \draw[very thick] (-2,1.2) circle (.2 and .5);
        \draw[very thick] (-2,-1.2) circle (.2 and .5);
        \begin{scope}[shift={(2,1.2)}, xscale=.4]
            \draw[very thick] (0,-.5) arc (-90:90:.5);
            \draw[very thick, dashed] (0,.5) arc (90:270:.5);
        \end{scope}
        \begin{scope}[shift={(2,-1.2)}, xscale=.4]
            \draw[very thick] (0,-.5) arc (-90:90:.5);
            \draw[very thick, dashed] (0,.5) arc (90:270:.5);
        \end{scope}
        \begin{scope}[yscale=.5]
            \draw[very thick, RoyalBlue] (-.97,0) arc (-180:0:.97);
            \draw[very thick, dashed, RoyalBlue] (.97,0) arc (0:180:.97);
        \end{scope}
        \draw[very thick, out=0, in=180] (-2,1.7) to (0,.9) to (2,1.7);
        \draw[very thick, out=0, in=180] (-2,-1.7) to (0,-.9) to (2,-1.7);
        \draw[very thick, out=0, in=0, looseness=2.5] (-2,-.7) to (-2,.7);
        \draw[very thick, out=180, in=180, looseness=2.5] (2,-.7) to (2,.7);
        \node at (-2,1.2) {$3$};
        \node at (2.4,1.2) {$2$};            
        \node at (2.4,-1.2) {$1$};
        \node at (-2,-1.2) {$4$};   
        \node at (.4,-.03) {$32$};
    \end{tikzpicture} \!\!
        \scalebox{1}[1.5]{\Bigg|} \begin{tikzpicture}[baseline={([yshift=-.5ex]current bounding box.center)}, xscale=.65, yscale=.6]
        \draw[very thick, Maroon, bend left=10] (-1.8,1.2) to (0,.6);
        \draw[very thick, Maroon, bend right=10] (-1.8,-1.2) to (0,-.6);
        \draw[very thick, Maroon, bend right=10] (2.2,1.2) to (0,.6);
        \draw[very thick, Maroon, bend left=10] (2.2,-1.2) to (0,-.6);
        \draw[very thick, Maroon] (0,-.6) to (0,.6);
        \fill[Maroon] (0,.6) circle (.07);
        \fill[Maroon] (0,-.6) circle (.07);
        \draw[very thick] (-2,1.2) circle (.2 and .5);
        \draw[very thick] (-2,-1.2) circle (.2 and .5);
        \begin{scope}[shift={(2,1.2)}, xscale=.4]
            \draw[very thick] (0,-.5) arc (-90:90:.5);
            \draw[very thick, dashed] (0,.5) arc (90:270:.5);
        \end{scope}
        \begin{scope}[shift={(2,-1.2)}, xscale=.4]
            \draw[very thick] (0,-.5) arc (-90:90:.5);
            \draw[very thick, dashed] (0,.5) arc (90:270:.5);
        \end{scope}
        \begin{scope}[yscale=.5]
            \draw[very thick, RoyalBlue] (-.97,0) arc (-180:0:.97);
            \draw[very thick, dashed, RoyalBlue] (.97,0) arc (0:180:.97);
        \end{scope}
        \draw[very thick, out=0, in=180] (-2,1.7) to (0,.9) to (2,1.7);
        \draw[very thick, out=0, in=180] (-2,-1.7) to (0,-.9) to (2,-1.7);
        \draw[very thick, out=0, in=0, looseness=2.5] (-2,-.7) to (-2,.7);
        \draw[very thick, out=180, in=180, looseness=2.5] (2,-.7) to (2,.7);
        \node at (-2,1.2) {$3$};
        \node at (2.4,1.2) {$2$};            
        \node at (2.4,-1.2) {$1$};
        \node at (-2,-1.2) {$4$};   
        \node at (.45,-.03) {$32'$};
    \end{tikzpicture}
 \!\!
        \scalebox{1}[1.5]{\Bigg\rangle} \\
        &=\int_0^\infty\!\! \d P_{32} \ \frac{C_0(P_1,P_2,P_{21}) C_0(P_3,P_4,P_{21}) \rho_0(P_{21})}{C_0(P_1,P_4,P_{32}) C_0(P_2,P_3,P_{32}) \rho_0(P_{32})}\nonumber \\
        &\qquad\qquad\times \mathbb{F}_{P_{21},P_{32}}\!\begin{bmatrix}
        P_3 & P_2 \\ P_4 & P_1
    \end{bmatrix}\mathbb{F}_{P_{21}',P_{32}} \!\begin{bmatrix}
        P_3 & P_2 \\ P_4 & P_1
    \end{bmatrix} \\
    &=\int_0^\infty\!\! \d P_{32}\ \mathbb{F}_{P_{21},P_{32}}\!\begin{bmatrix}
        P_3 & P_2 \\ P_4 & P_1
    \end{bmatrix}\mathbb{F}_{P_{32},P_{21}'}\!\begin{bmatrix}
        P_4 & P_3 \\ P_1 & P_2
    \end{bmatrix}\ .
\end{align}
Here we used that the crossing kernel is real and the tetrahedral symmetry of $\mathbb{F}$ to exchange the arguments in the second crossing kernel \eqref{eq:tetrahedrally symmetric combination}. The identity then indeed holds -- it is simply the idempotency of the crossing operation \eqref{eq:g=0, n=4 idempotency F}. We should also mention that this is essentially the inverse computation that shows crossing symmetry of the Liouville correlator that we explained in Section~\ref{subsec:vacuum blocks}. A very similar argument shows unitarity of the modular crossing kernel, while unitarity of $\mathbb{B}$ is of course obvious since it is just a phase.

\paragraph{Uniqueness of the inner product.} We can go further and show that \eqref{eq:explicit inner product four-punctured sphere} is the \emph{unique} inner product with respect to which the crossing kernels are unitary, up to a symmetric overall function $F(P_1,\dots,P_n)$. This proves in particular the equality of \eqref{eq:inner product Teichmuller space} and \eqref{eq:explicit inner product four-punctured sphere}, up to an overall factor depending only on the external Liouville momenta.

Unitarity of the Dehn twists implies immediately that a compatible inner product must be diagonal in the sense that
\be 
\!\scalebox{1}[1.5]{\Bigg\langle} \!\begin{tikzpicture}[baseline={([yshift=-.5ex]current bounding box.center)}, xscale=.65, yscale=.6]
        \draw[very thick, Maroon, bend left=30] (-1.8,1.2) to (-.6,0);
        \draw[very thick, Maroon, bend right=30] (-1.8,-1.2) to (-.6,0);
        \draw[very thick, Maroon, bend right=30] (2.2,1.2) to (.6,0);
        \draw[very thick, Maroon, bend left=30] (2.2,-1.2) to (.6,0);
        \draw[very thick, Maroon] (-.6,0) to (.6,0);
        \fill[Maroon] (-.6,0) circle (.07);
        \fill[Maroon] (.6,0) circle (.07);
        \draw[very thick] (-2,1.2) circle (.2 and .5);
        \draw[very thick] (-2,-1.2) circle (.2 and .5);
        \begin{scope}[shift={(2,1.2)}, xscale=.4]
            \draw[very thick] (0,-.5) arc (-90:90:.5);
            \draw[very thick, dashed] (0,.5) arc (90:270:.5);
        \end{scope}
        \begin{scope}[shift={(2,-1.2)}, xscale=.4]
            \draw[very thick] (0,-.5) arc (-90:90:.5);
            \draw[very thick, dashed] (0,.5) arc (90:270:.5);
        \end{scope}
        \begin{scope}[xscale=.55]
            \draw[very thick, RoyalBlue] (0,-.9) arc (-90:90:.9);
            \draw[very thick, dashed, RoyalBlue] (0,.9) arc (90:270:.9);
        \end{scope}
        \draw[very thick, out=0, in=180] (-2,1.7) to (0,.9) to (2,1.7);
        \draw[very thick, out=0, in=180] (-2,-1.7) to (0,-.9) to (2,-1.7);
        \draw[very thick, out=0, in=0, looseness=2.5] (-2,-.7) to (-2,.7);
        \draw[very thick, out=180, in=180, looseness=2.5] (2,-.7) to (2,.7);
        \node at (-2,1.2) {3};
        \node at (2.4,1.2) {2};            
        \node at (2.4,-1.2) {1};
        \node at (-2,-1.2) {4}; 
        \node at (0,.35) {21};
        \end{tikzpicture} \!\!
        \scalebox{1}[1.5]{\Bigg|} 
\begin{tikzpicture}[baseline={([yshift=-.5ex]current bounding box.center)}, xscale=.65, yscale=.6]
        \draw[very thick, Maroon, bend left=30] (-1.8,1.2) to (-.6,0);
        \draw[very thick, Maroon, bend right=30] (-1.8,-1.2) to (-.6,0);
        \draw[very thick, Maroon, bend right=30] (2.2,1.2) to (.6,0);
        \draw[very thick, Maroon, bend left=30] (2.2,-1.2) to (.6,0);
        \draw[very thick, Maroon] (-.6,0) to (.6,0);
        \fill[Maroon] (-.6,0) circle (.07);
        \fill[Maroon] (.6,0) circle (.07);
        \draw[very thick] (-2,1.2) circle (.2 and .5);
        \draw[very thick] (-2,-1.2) circle (.2 and .5);
        \begin{scope}[shift={(2,1.2)}, xscale=.4]
            \draw[very thick] (0,-.5) arc (-90:90:.5);
            \draw[very thick, dashed] (0,.5) arc (90:270:.5);
        \end{scope}
        \begin{scope}[shift={(2,-1.2)}, xscale=.4]
            \draw[very thick] (0,-.5) arc (-90:90:.5);
            \draw[very thick, dashed] (0,.5) arc (90:270:.5);
        \end{scope}
        \begin{scope}[xscale=.55]
            \draw[very thick, RoyalBlue] (0,-.9) arc (-90:90:.9);
            \draw[very thick, dashed, RoyalBlue] (0,.9) arc (90:270:.9);
        \end{scope}
        \draw[very thick, out=0, in=180] (-2,1.7) to (0,.9) to (2,1.7);
        \draw[very thick, out=0, in=180] (-2,-1.7) to (0,-.9) to (2,-1.7);
        \draw[very thick, out=0, in=0, looseness=2.5] (-2,-.7) to (-2,.7);
        \draw[very thick, out=180, in=180, looseness=2.5] (2,-.7) to (2,.7);
        \node at (-2,1.2) {3};
        \node at (2.4,1.2) {2};            
        \node at (2.4,-1.2) {1};
        \node at (-2,-1.2) {4}; 
        \node at (0,.35) {$21\hspace{-.04cm}'\hspace{.05cm}$};
        \end{tikzpicture} \!\!
        \scalebox{1}[1.5]{\Bigg\rangle}= \frac{F(P_{21};P_1,P_2,P_3,P_4)\,\delta(P_{21}-P_{21}')}{ \rho_0(P_{21})C_0(P_1,P_2,P_{21}) C_0(P_3,P_4,P_{21})}
\ee
for some function $F$. We restricted to the four-punctured sphere for simplicity, the argument for different values of $g$ and $n$ is analogous. The function $F$ has to satisfy the properties
\be 
F(P_{21};P_1,P_2,P_3,P_4)=F(P_{21};P_2,P_1,P_3,P_4)=F(P_{21};P_3,P_4,P_1,P_2)
\ee
implied by unitarity of braiding and rotating the picture. Unitarity of $\mathbb{F}$ together with the idempotency of $\mathbb{F}$ now imposes
\be 
F(P_{21};P_1,P_2,P_3,P_4)=F(P_{32};P_1,P_2,P_3,P_4)\ .
\ee
Since $P_{21}$ and $P_{32}$ is arbitrary, $F$ cannot depend on its first argument and moreover has to be a symmetric function in all its remaining arguments, which demonstrates the claim.

\paragraph{Boundedness.} As a corollary of unitarity, we also note that the operators $\mathbb{F}$ and $\mathbb{S}$ are in particular bounded operators on the space of normalizable conformal blocks. This implies that any composition of them will be well-defined and we do not have to worry about convergence of integrals etc.\ when composing them.

\paragraph{Extension to $c \in \CC \setminus (-\infty,1]$.} Our logic to establish the Moore-Seiberg relations relied on the existence of an inner product on the space of conformal blocks \eqref{eq:inner product Teichmuller space}, which we abstractly only defined for $c \ge 25$, i.e.\ $b \in \RR$. However, it is now trivial to analytically continue in $b$. The crossing kernel can be defined for any $b \in \CC \setminus \{i \RR\}$. For $b \in i \RR$, the solution presented here breaks down since for example the double Gamma function $\Gamma_b$ does not exist.

\paragraph{Racah-Wigner normalization.} The form of the inner product suggests to use a different normalization for the crossing kernel, which was used in \cite{Teschner:2012em, Teschner:2013tqy}. It corresponds to the Racah-Wigner normalization of 6j symbols. We set
\be 
\sixj{P_1}{P_2}{P_{21}}{P_3}{P_4}{P_{32}}=\rho_0(P_{32})^{-1}\sqrt{\frac{C_0(P_1,P_2,P_{21})C_0(P_3,P_4,P_{21})}{C_0(P_2,P_3,P_{32})C_0(P_1,P_4,P_{32})}} \ \mathbb{F}_{P_{21},P_{32}} \! \begin{bmatrix}
    P_3 & P_2 \\ P_4 & P_1
\end{bmatrix}\ .
\ee
In view of \eqref{eq:tetrahedrally symmetric combination}, this normalization is also tetrahedrally symmetric, as well as reflection symmetric. The downside of using this normalization is of course the appearance of non-analytic factors, but since $C_0(P_1,P_2,P_3)>0$ for real Liouville momenta, the choice of branch is unambiguous. Define then the measure
\be 
\d \mu(P)=\d P\, \rho_0(P)\ ,
\ee
which plays the role of the Plancherel measure on the Virasoro group, i.e.\ the natural measure on the space of all representations. The crossing transformations are then described by
\be 
\begin{tikzpicture}[baseline={([yshift=-.5ex]current bounding box.center)}, xscale=.8, yscale=.7]
        \draw[very thick, Maroon, bend left=30] (-1.8,1.2) to (-.6,0);
        \draw[very thick, Maroon, bend right=30] (-1.8,-1.2) to (-.6,0);
        \draw[very thick, Maroon, bend right=30] (2.2,1.2) to (.6,0);
        \draw[very thick, Maroon, bend left=30] (2.2,-1.2) to (.6,0);
        \draw[very thick, Maroon] (-.6,0) to (.6,0);
        \draw[very thick, Maroon] (-.73,.13) to (-.47,-.13);
        \draw[very thick, Maroon] (-.73,-.13) to (-.47,.13);
        \draw[very thick, Maroon] (.73,.13) to (.47,-.13);
        \draw[very thick, Maroon] (.73,-.13) to (.47,.13);
        \draw[very thick] (-2,1.2) circle (.2 and .5);
        \draw[very thick] (-2,-1.2) circle (.2 and .5);
        \begin{scope}[shift={(2,1.2)}, xscale=.4]
            \draw[very thick] (0,-.5) arc (-90:90:.5);
            \draw[very thick, dashed] (0,.5) arc (90:270:.5);
        \end{scope}
        \begin{scope}[shift={(2,-1.2)}, xscale=.4]
            \draw[very thick] (0,-.5) arc (-90:90:.5);
            \draw[very thick, dashed] (0,.5) arc (90:270:.5);
        \end{scope}
        \begin{scope}[xscale=.4]
            \draw[very thick, RoyalBlue] (0,-.9) arc (-90:90:.9);
            \draw[very thick, dashed, RoyalBlue] (0,.9) arc (90:270:.9);
        \end{scope}
        \draw[very thick, out=0, in=180] (-2,1.7) to (0,.9) to (2,1.7);
        \draw[very thick, out=0, in=180] (-2,-1.7) to (0,-.9) to (2,-1.7);
        \draw[very thick, out=0, in=0, looseness=2.5] (-2,-.7) to (-2,.7);
        \draw[very thick, out=180, in=180, looseness=2.5] (2,-.7) to (2,.7);
        \node at (-2,1.2) {3};
        \node at (2.4,1.2) {2};            
        \node at (2.4,-1.2) {1};
        \node at (-2,-1.2) {4}; 
        \node at (0,.3) {21};
        \end{tikzpicture}        
    \hspace{-.3cm} = \int_0^\infty\!\!  \d \mu(P_{32})\  \sixj{P_1}{P_2}{P_{21}}{P_3}{P_4}{P_{32}} 
    \ 
    \begin{tikzpicture}[baseline={([yshift=-.5ex]current bounding box.center)}, xscale=.8, yscale=.7]
        \draw[very thick, Maroon, bend left=10] (-1.8,1.2) to (0,.5);
        \draw[very thick, Maroon, bend right=10] (-1.8,-1.2) to (0,-.5);
        \draw[very thick, Maroon, bend right=10] (2.2,1.2) to (0,.5);
        \draw[very thick, Maroon, bend left=10] (2.2,-1.2) to (0,-.5);
        \draw[very thick, Maroon] (0,-.5) to (0,.5);
        \draw[very thick, Maroon] (.13,-.37) to (-.13,-.63);
        \draw[very thick, Maroon] (-.13,-.37) to (.13,-.63);
        \draw[very thick, Maroon] (.13,.37) to (-.13,.63);
        \draw[very thick, Maroon] (-.13,.37) to (.13,.63);
        \draw[very thick] (-2,1.2) circle (.2 and .5);
        \draw[very thick] (-2,-1.2) circle (.2 and .5);
        \begin{scope}[shift={(2,1.2)}, xscale=.4]
            \draw[very thick] (0,-.5) arc (-90:90:.5);
            \draw[very thick, dashed] (0,.5) arc (90:270:.5);
        \end{scope}
        \begin{scope}[shift={(2,-1.2)}, xscale=.4]
            \draw[very thick] (0,-.5) arc (-90:90:.5);
            \draw[very thick, dashed] (0,.5) arc (90:270:.5);
        \end{scope}
        \begin{scope}[yscale=.33]
            \draw[very thick, RoyalBlue] (-.97,0) arc (-180:0:.97);
            \draw[very thick, dashed, RoyalBlue] (.97,0) arc (0:180:.97);
        \end{scope}
        \draw[very thick, out=0, in=180] (-2,1.7) to (0,.9) to (2,1.7);
        \draw[very thick, out=0, in=180] (-2,-1.7) to (0,-.9) to (2,-1.7);
        \draw[very thick, out=0, in=0, looseness=2.5] (-2,-.7) to (-2,.7);
        \draw[very thick, out=180, in=180, looseness=2.5] (2,-.7) to (2,.7);
        \node at (-2,1.2) {3};
        \node at (2.4,1.2) {2};            
        \node at (2.4,-1.2) {1};
        \node at (-2,-1.2) {4};   
        \node at (.3,0) {32};
    \end{tikzpicture} ,
\ee
where the appearing conformal blocks should now be normalized with additional factors of $\sqrt{C_0}$ for every pair of pants appearing in the decomposition. We indicated this by a cross at the junction of the dual graph.

\subsection{The line bundle of conformal blocks} \label{subsec:line bundle conformal blocks}
We will now discuss the line bundle $\mathscr{L}$ of conformal blocks in more detail. Recall that we defined conformal blocks in Section~\ref{subsec:definition of conformal blocks} as the sections of a line bundle $\mathscr{L}$ over Teichm\"uller space, whose curvature coincides with the Weil-Petersson symplectic form \eqref{eq:line bundle conformal blocks curvature}.
\paragraph{From the line bundle on $\mathcal{T}_{g,n}$ to a projective line bundle on $\mathcal{M}_{g,n}$.} It is more common in the literature to talk about conformal blocks in terms of a projective line bundle $\mathscr{L}_\text{proj}$ on $\mathcal{M}_{g,n}$. Let us recall that a projective line bundle is a line bundle, whose overlap functions $g_{\alpha\beta}:U_\alpha \cap U_\beta \longrightarrow \mathbb{C}^*$ on a covering $(U_\alpha)_{\alpha \in \mathcal{A}}$ do not satisfy the consistency conditions $g_{\alpha \beta}g_{\beta \gamma} g_{\gamma\alpha}=1$ on the triple overlap $U_\alpha \cap U_\beta \cap U_\gamma$, but rather 
\be 
g_{\alpha \beta}g_{\beta \gamma} g_{\gamma\alpha}=\sigma_{\alpha\beta\gamma}\ . \label{eq:projective line bundle triple overlap}
\ee
The right hand side defines a class in $\mathrm{H}^2(\mathcal{M}_{g,n},\mathrm{U}(1))$ (via \v{C}ech cohomology), which is the obstruction of a projective line bundle to lift to a usual line bundle. Let us also recall that one can define a first Chern class of a projective line bundle taking values in $\mathrm{H}^2(\mathcal{M}_{g,n},\mathbb{R})$. The projective line bundle is an ordinary line bundle when its first Chern class can be lifted to an integer cohomology class.

The projective line bundle $\mathscr{L}_\text{proj}$ arises as follows. Take the collection of \emph{all} normalizable conformal blocks indexed by the collection of $3g-3+n$ internal Liouville momenta. As such the collection of all normalizable Liouville conformal blocks defines an infinite-dimensional holomorphic vector bundle $\mathscr{V}$ over Teichm\"uller space $\mathcal{T}_{g,n}$. We can then restrict $\mathscr{V}$ to moduli space $\mathcal{M}_{g,n}$, where the fiber at every point in $\mathcal{M}_{g,n}$ is still defined as the space spanned by all normalizable conformal blocks.
However, $\mathscr{V}$ when viewed on moduli space might fail to be a vectorbundle. Since we took a quotient in passing from $\mathcal{T}_{g,n}$ to $\mathcal{M}_{g,n}$, there are new triple overlaps $U_\alpha \cap U_\beta \cap U_\gamma$ which might only satisfy the weaker consistency condition \eqref{eq:projective line bundle triple overlap}. This in fact happens and we only get a projective vector bundle $\mathscr{V}_\text{proj}$ on $\mathcal{M}_{g,n}$. Up to isomorphism, we can canonically separate out the projective part and write
\be 
\mathscr{V}_\text{proj}=\mathscr{L}_\text{proj} \otimes \mathscr{V}'\ ,
\ee
where $\mathscr{V}'$ is an actual vector bundle (with vanishing first Chern class) and $\mathscr{L}_\text{proj}$ a projective line bundle on $\mathcal{M}_{g,n}$.

\paragraph{The Monster theory.}
One can say this in a more explicit and down-to-earth way as follows. Recall that the monster CFT $\mathbb{M}$ is a chiral CFT of central charge 24 with $\num{196883}$ Virasoro primaries with $\Delta=2$ (and no Virasoro primaries with $\Delta=1$) \cite{Frenkel:1988xz}. Let $Z_{\mathbb{M}}(\mathbf{m})$ and $Z_{\mathbb{M}}^{(1)}(z,\mathbf{m})$ denote the partition function on $\Sigma_g$ and the one-point function of the primary with $\Delta=2$ on $\Sigma_{g,1}$, respectively.\footnote{It does not matter which one of the $\num{196883}$ primaries we pick since they are all equal by Monster symmetry.} Then instead of considering a conformal block, we can consider the combination
\be 
Z_{\mathbb{M}}(\mathbf{m})^{-\frac{c}{24}}\prod_{j=1}^n \left(\frac{Z_{\mathbb{M}}^{(1)}(z_j,\mathbf{m})}{Z_{\mathbb{M}}(\mathbf{m})}\right)^{-\frac{1}{2}\Delta_j} \begin{tikzpicture}[baseline={([yshift=-.5ex]current bounding box.center)}, xscale=.65, yscale=.55]
        \draw[very thick, Maroon, bend left=30] (-1.7,1.2) to (-.6,0);
        \draw[very thick, Maroon, bend right=30] (-1.7,-1.2) to (-.6,0);
        \draw[very thick, Maroon, bend right=30] (2.3,1.2) to (.6,0);
        \draw[very thick, Maroon, bend left=30] (2.3,-1.2) to (.6,0);
        \draw[very thick, Maroon] (-.6,0) to (.6,0);
        \fill[Maroon] (-.6,0) circle (.07);
        \fill[Maroon] (.6,0) circle (.07);
        \draw[very thick] (-2,1.2) circle (.3 and .5);
        \draw[very thick] (-2,-1.2) circle (.3 and .5);
        \begin{scope}[shift={(2,1.2)}, xscale=.6]
            \draw[very thick] (0,-.5) arc (-90:90:.5);
            \draw[very thick, dashed] (0,.5) arc (90:270:.5);
        \end{scope}
        \begin{scope}[shift={(2,-1.2)}, xscale=.6]
            \draw[very thick] (0,-.5) arc (-90:90:.5);
            \draw[very thick, dashed] (0,.5) arc (90:270:.5);
        \end{scope}
        \begin{scope}[xscale=.4]
            \draw[very thick, RoyalBlue] (0,-.9) arc (-90:90:.9);
            \draw[very thick, dashed, RoyalBlue] (0,.9) arc (90:270:.9);
        \end{scope}
        \draw[very thick, out=0, in=180] (-2,1.7) to (0,.9) to (2,1.7);
        \draw[very thick, out=0, in=180] (-2,-1.7) to (0,-.9) to (2,-1.7);
        \draw[very thick, out=0, in=0, looseness=2.5] (-2,-.7) to (-2,.7);
        \draw[very thick, out=180, in=180, looseness=2.5] (2,-.7) to (2,.7);
        \node at (-2,1.2) {3};
        \node at (2.55,1.2) {2};            
        \node at (2.55,-1.2) {1};
        \node at (-2,-1.2) {4}; 
        \end{tikzpicture}\ ,
\ee
where we drew a four-punctured sphere for concreteness. The prefactor is chosen such that the conformal anomaly cancels and this combination behaves like a function under change of coordinates and Weil transformations. Thus this combination is a \emph{function} on $\mathcal{T}_{g,n}$. Notice that the branch of the prefactor can be chosen continuously since Teichm\"uller space is simply connected. These renormalized conformal blocks now form again an infinite-dimensional vectorbundle. It can be restricted to moduli space since the curvature of $\mathscr{L}$ is cancelled and thus the first Chern class is trivial. In particular, 
\be 
Z_{\mathbb{M}}(\mathbf{m})^{\frac{c}{24}}\prod_{j=1}^n \left(\frac{Z_{\mathbb{M}}^{(1)}(z_j,\mathbf{m})}{Z_{\mathbb{M}}(\mathbf{m})}\right)^{\frac{1}{2}\Delta_j}
\ee
is a holomorphic section of $\mathscr{L}_\text{proj}$. On $\mathcal{M}_{g,n}$, the phases of the prefactors can no longer necessarily be chosen consistently which reflects the projectiveness of $\mathscr{L}_\text{proj}$.

\paragraph{Curvature without punctures.} We now explain the origin of the curvature of $\mathscr{L}$ and thus $\mathscr{L}_\text{proj}$, a calculation due to Friedan and Shenker \cite{Friedan:1986ua}. Let us first consider a surface without any punctures.
To do so, we can use that the curvature of a line bundle is given by
\be 
F=\partial \bar{\partial} \log s\ ,
\ee
where $s$ is a section of the line bundle. In our context, a derivative is a functional derivative acting as follows:
\be 
\partial_\mu  = \int \d^2 z \  \mu(z)  T(z) \ ,
\ee
where $\mu$ is a Beltrami differential.\footnote{This means that it is a $(-1,1)$-differential and is holomorphic, $\bar{\partial} \mu=0$.} 
Thus the curvature is controlled by the contact terms appearing between $T(z)$ and $\overline{T}(w)$. By dimensional analysis, the only possible such contact terms are
\be 
T(z) \overline{T}(w)=\alpha c R \delta^2(z-w)+\beta c \partial_z \partial_{\partial w} \delta^2(z-w)+\text{reg}\ . \label{eq:TTbar contact terms}
\ee
The first term does not contribute to the curvature since we can use that $\bar{\partial} \mu=0$ and integration by parts to eliminate it. 
The numerical constants $\alpha$ and $\beta$ can in principle be worked out by using the $T(z) T(w)$ OPE and relating it to $T(z) \overline{T}(w)$ via conservation of energy-momentum. We will simply quote the result of Friedan and Shenker below.
Thus the curvature of $\mathscr{L}$ is given by
\be 
F(\mu,\mu')=\beta c \int \d^2z \sqrt{g} R\, \overline{\mu} \mu' \ .
\ee
This is the Weil-Petersson form on moduli space. It is usually written after specializing to a constant negative curvature metric where $R$ is a constant.
After tracing through the definitions carefully, we have
\be 
\frac{F(\mathscr{L})}{2\pi}=\frac{c\, \omega_\text{WP}}{48\pi^2}\ .
\ee
For the projective line bundle $\mathscr{L}_\text{proj}$ on $\mathcal{M}_{g,n}$, it also makes sense to ask about the first Chern class which is
\be 
c_1(\mathscr{L}_\text{proj})=\frac{c\, [\omega_\text{WP}]}{48\pi^2}=\frac{c\, \kappa_1}{24}
\ee
Here we used the standard notation $\kappa_1=\frac{1}{2\pi^2} [\omega_\text{WP}] \in \H^2(\overline{\mathcal{M}}_g,\RR)$.

Friedan and Shenker write $c_1(\mathscr{L}_\text{proj})=\frac{c}{2} \lambda_1$, where $\lambda_1$ is the first Chern-class of the Hodge bundle over $\mathcal{M}_{g}$. It is known that $\lambda_1=\frac{1}{12}\kappa_1$ on $\mathcal{M}_g$ and thus the two formulas are equivalent.\footnote{Notice that we do not discuss the compactification $\overline{\mathcal{M}}_{g}$ and thus contributions from boundary divisors are omitted.}

\paragraph{Curvature with punctures.}
In the presence of punctures, the curvature is modified as follows. The contact terms of $T(z)$ and $\overline{T}(w)$ now also involve contributions from the insertions. One can quickly work out the contribution to $c_1(\mathscr{L}_\text{proj})$ as follows.
The contributions will be proportional to the classes $\psi_i$ on $\mathcal{M}_{g,n}$ which are defined as follows. Consider the line-bundle $\mathbb{L}_j$ over $\mathcal{M}_{g,n}$ whose fiber at the point $\Sigma_{g,n} \in \mathcal{M}_{g,n}$ consists of the cotangent bundle at the $j$-th marked point. Thus a holomorphic section of $\mathbb{L}_1^{\Delta_1} \otimes \cdots \otimes\mathbb{L}_n^{\Delta_n}$ behaves like a conformal block under coordinate changes, since a field of conformal weight $1$ transforms like a differential $\d z$ on the surface. Conventionally, the first Chern class of $\mathbb{L}_j$ is called $\psi_j$.

Hence the curvature of the projective line bundle $\mathscr{L}_\text{proj}$ is given by
\be 
c_1(\mathscr{L}_\text{proj})=\frac{c}{24}\, \pi^*\kappa_1+\sum_{j=1}^n \Delta_j \psi_j\ .
\ee
Here $\pi:\mathcal{M}_{g,n} \longrightarrow \mathcal{M}_g$ is the map that forgets the locations of the $n$ marked points on the surface. The conformal anomaly does not depend on the locations of the marked points and is thus simply given by the pullback of the curvature form from $\mathcal{M}_g$ to $\mathcal{M}_{g,n}$.
We have \cite{Witten:1990hr}
\be 
\pi^* \kappa_1=\kappa_1-\sum_{j=1}^n \psi_j\ ,
\ee
where the cohomology on the right hand side are in $\H^2(\mathcal{M}_{g,n},\RR)$.
Thus in the presence of punctures the curvature takes the form
\be 
c_1(\mathscr{L}_\text{proj})=\frac{c}{24}\, \kappa_1+\sum_{j=1}^n \left(\Delta_j-\frac{c}{24}\right) \psi_j\ .
\ee
We can compare this with the formula of the cohomology class of the Weil-Petersson form \cite{Mirzakhani:2006eta},
\be 
\frac{1}{2\pi^2}\, [\omega_\text{WP}]=\kappa_1+\frac{1}{4\pi^2} \sum_{j=1}^n \ell_j^2 \psi_j\ . \label{eq:first Chern class line bundle kappa psi}
\ee
This leads to 
\be 
c_1(\mathscr{L}_\text{proj})=\frac{c \, [\omega_\text{WP}(\ell_1,\dots,\ell_n)]}{48\pi^2}\ ,
\ee
with the relation between $\Delta$ and $\ell$ given by eq.~\eqref{eq:conformal weight geodesic length relation}.
The fact that the curvature of the projective line bundle of conformal blocks is given by the Weil-Petersson form leads to a fascinating interplay between conformal blocks and hyperbolic geometry.

\paragraph{Relation to string theory.} Let us mention how this formula is related to the familiar statement of the Weyl anomaly cancellation in string theory \cite{Polyakov:1981rd}. As is well-known, the matter theory in bosonic string theory has to have central charge $c=26$ and vertex operators must have conformal weight $\Delta_j=1$ in order to get sensible integrals over moduli space. In view of \eqref{eq:first Chern class line bundle kappa psi}, this means that the holomorphic part of the integrand in string theory is a section of a line bundle with first Chern class
\be 
\frac{1}{12}\bigg(13\kappa_1-\sum_{i=1}^n \psi_i \bigg)\ .
\ee
This in fact coincides with the first Chern class $c_1(\mathscr{K})$ of the canonical bundle\footnote{The canonical bundle is the top exterior power of the holomorphic tangent bundle.} $\mathscr{K}$ on moduli space \cite{HarrisMumford} and thus can be viewed as a top exterior form on moduli space. The projectiveness is irrelevant since we are combining left- and right-moving parts which cancels all the phases. Together with the right-movers it then makes sense to integrate it over $\mathcal{M}_{g,n}$.

\subsection{The representation of the mapping class group} \label{subsec:representation mapping class group}
We can now assemble the pieces and discuss the set of crossing transformations as a whole.

\paragraph{Mapping class group.} Understanding the crossing transformations of conformal blocks in particular leads to a (projective) representation of the mapping class of the surface. Let us recall some details about the mapping class group $\Map(\Sigma_{g,n})$. It is defined as the group of `large diffeomorphisms' of the surface\footnote{We also assume these diffeomorphisms to be orientation-preserving.}
\be 
\Map(\Sigma_{g,n})=\Diff(\Sigma_{g,n})/\Diff_0(\Sigma_{g,n})\ ,
\ee
where $\Diff_0(\Sigma_{g,n})$ is the group of diffeomorphisms that are homotopic to the identity. We also require the mapping class group to treat punctures as distinguishable and every puncture has to be preserved individually. More precisely, this is called the pure mapping class group and in the following `mapping class group' always means `pure mapping class group'.
It is generated by the Dehn twists around all the simple closed curves of the surface.

We can also obtain a representation of the actual mapping class group that allow the punctures to get permuted, but in this case all punctures need to be indistinguishable and thus carry the same label, $P_j \equiv P$. We will not consider this further.

\paragraph{Central extension.} Since the vector bundle of conformal blocks over $\mathcal{M}_{g,n}$ is projective, we get a representation of $\Map(\Sigma_{g,n})$ which is also projective, i.e.\ a representation of the central extension. Central extensions are classified by the cohomology group of the mapping class group, which in turn is simply the cohomology of $\mathcal{M}_{g,n}$. For sufficiently high genus ($g \ge 4$ is enough),\footnote{It should be noted that there are some accidents at low genus that make the cohomology group smaller.}
\be 
\H^2(\Map(\Sigma_{g,n}),\mathrm{U}(1)) \cong \H^2(\mathcal{M}_{g,n},\mathrm{U}(1)) \cong \mathrm{U}(1)^{n+1}\ .
\ee
The isomorphism is a difficult theorem due to Harer \cite{HarerH2}. Thus one can add $n+1$ additional central generators to the mapping class group. These $n+1$ generators are represented by the $\mathrm{U}(1)$ reduction of the cohomology classes $\kappa_1$ and $\psi_1,\dots,\psi_n$ that we discussed in Section~\ref{subsec:line bundle conformal blocks}. Let us now explain what they mean on the level of the central extension.

\begin{figure}[ht]
    \centering
    \begin{tikzpicture}
    \begin{scope}
        \draw[very thick] (0,0) circle (.5);
        \draw[very thick, Maroon] (.5,0) to (2,0);
        \draw[very thick,<-] (30:1) arc (30:150:1);
    \end{scope}
    \begin{scope}[shift={(3,0)}]
        \node at (0,0) {$\Longrightarrow$};
    \end{scope}
    \begin{scope}[shift={(5,0)}]
        \draw[very thick] (0,0) circle (.5);
        \draw[very thick, Maroon, out=0, in=0, looseness=1.5] (.5,0) to (0,1);
        \draw[very thick, Maroon, out=180, in=90] (0,1) to (-1,0);        
        \draw[very thick, Maroon, out=-90, in=180] (-1,0) to (0,-1);
        \draw[very thick, Maroon, out=0, in=180] (0,-1) to (2,0);
    \end{scope}
    \end{tikzpicture}
    \caption{A transformation of the centrally extended mapping class group of the surface.}
    \label{fig:mapping class group puncture}
\end{figure}
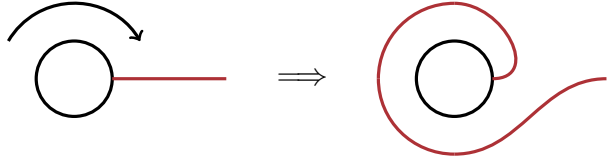

The $n$ factors corresponding to $\psi_1,\dots,\psi_n$ are easy to understand. They just arise by treating punctures as holes in the surface. This means in particular that the centrally extended mapping class is also allowed to rotate boundary punctures as in Figure~\ref{fig:mapping class group puncture}. 
There is one more central generator known as the Euler class corresponding to the class $\kappa_1$.

The normalizations of $\kappa_1$ and $\psi_1,\dots,\psi_n$ are such that they make sense as an element of the cohomology groups with integer coefficients. Put differently, there are line bundles (and not just projective line bundles) over moduli spaces such that these classes are realized as Chern classes. This means that the central extension of the representation is characterized by the reduction of $c_1(\mathscr{L}_\text{proj})$ from $\mathrm{H}^2(\mathcal{M}_{g,n},\mathbb{R})$ to $\mathrm{H}^2(\mathcal{M}_{g,n},\mathrm{U}(1))$,
\be 
\left\{\frac{c}{24} \right \} \kappa_1+\sum_{j=1}^n \left\{\Delta_j-\frac{c}{24} \right \} \psi_j \in \H^2(\mathcal{M}_{g,n},\mathrm{U}(1))\ .
\ee
where $\{x\}$ denotes the fractional part. 

The phases $\mathrm{e}^{2\pi i \Delta_j}$ and $\mathrm{e}^{\frac{\pi ic}{12}}$ are of course the familiar framing anomalies from Chern-Simons theory, see e.g.\ \cite{Witten:1988hf}. 
We should also note that we can always redefine generators of the representation according to $\rho(g) \mapsto \eta(g) \rho(g)$, where $\eta(g)$ are arbitrary phases. The choice of phases in the Moore-Seiberg consistency conditions \eqref{eq:g=0, n=4 idempotency F} -- \eqref{eq:g=1, n=2 relation} is just one consistent choice of phases.

\paragraph{Faithfulness.} 
The unitary projective representation of the mapping class group is actually faithful. This can be proven as follows.

Let us assume for the sake of contradiction that the mapping class group does not act faithfully. Hence there is a mapping class group element $\gamma \ne \id$ such that $U(\gamma)$ fixes every conformal block (or acts by a pure overall phase), where $U(\gamma)$ is the representation on the Hilbert space of normalizable conformal blocks. Since the kernels and their products are analytic, we also conclude that every possibly unnormalizable section of the line bundle of conformal blocks $\mathscr{L}$ is invariant under $\gamma$. Thus to reach a contradiction, we simply have to construct a holomorphic section of $\mathscr{L}$ that is not invariant under $\gamma$.

Fix two distinct points $x$ and $y$ in Teichm\"uller space related by $\gamma$, i.e.\ $y=\gamma(x)$. The key to construct such a section is holomorphic separability of Teichm\"uller space. This means that for two distinct points $x$ and $y$ there is always a holomorphic function $f$ such that $f(x) \ne f(y)$, see e.g.\ \cite{Bers1961}. This is a simple consequence of the fact that Teichm\"uller space may be realized as an open ball of $\CC^{3g-3+n}$. Let us denote points in Teichm\"uller space by $(z_1,\dots,z_n,\mathbf{m})$, where $z_j$ are the locations of the punctures and $\mathbf{m}$ are all the remaining moduli. Thus we write such a function as $f(z_1,\dots,z_n,\mathbf{m})$ and the two special points $x$ and $y$ as $(z_1^{(j)},\dots,z_n^{(j)},\mathbf{m}^{(j)})$ for $j=1,\, 2$. This is not yet quite what we want since $f$ is a function and not a section of the line bundle $\mathscr{L}$. However, we can produce a section of $\mathscr{L}$ from $f$ for example by combing $f$ with a suitable combination of Monster CFT partition functions and one-point functions as in Section~\ref{subsec:line bundle conformal blocks}.\footnote{This construction does not work for genus 0 since the one-point function vanishes on the sphere. However, proving faithfulness of the mapping class group representation for higher genus immediately implies the same at genus 0, since we can embed the genus 0 mapping class group into a genus 1 mapping class group.} 
Then
\be 
Z_{\mathbb{M}}(\mathbf{m})^{\frac{c}{24}}\prod_{j=1}^n \left(\frac{Z_{\mathbb{M}}^{(1)}(z_j,\mathbf{m})}{Z_{\mathbb{M}}(\mathbf{m})}\right)^{\frac{1}{2}\Delta_j} f(z_1,\dots,z_n,\mathbf{m})
\ee
is a holomorphic section of the line bundle $\mathscr{L}$. Since we are on Teichm\"uller space, the fractional exponents can be defined consistently.\footnote{Since the partition function and the one-point function of the monster CFT can have singularities on codimension 1 subloci of $\mathcal{T}_{g,n}$, we should choose the two points $(z_1^{(j)},\dots,z_n^{(j)},\mathbf{m}^{(j)})$ such that we avoid those.} This section is by construction \emph{not} invariant under $\gamma$ (not even up to projective phases). Thus we see that the representation is indeed faithful.

\subsection{Quantum Teichm\"uller theory} \label{subsec:quantum Teichmuller theory}
We now explain a different perspective on conformal blocks via the quantization of Teichm\"uller space. 

\paragraph{Conformal blocks from quantizing Teichm\"uller space.}
As we have explained, normalizable Virasoro conformal blocks are holomorphic sections of a line bundle over Teichm\"uller space. The curvature of this line bundle is precisely the Weil-Petersson symplectic form on moduli space, see eq.~\eqref{eq:line bundle conformal blocks curvature}.
We should also note that Teichm\"uller space is actually a K\"ahler manifold. 
This is precisely the situation one encounters in geometric quantization. When quantizing a phase space, one needs to pick a Hilbert space of wavefunctions which depend on ``half of the coordinates''. For K\"ahler manifolds, there is a natural way to do so. One takes the Hilbert space to be holomorphic sections of a line bundle whose curvature is given by the symplectic form. 

In the language of geometric quantization, conformal blocks are thus the Hilbert space associated to the quantization of Teichm\"uller space, where the symplectic form is given by the Weil-Petersson form.
Quantization of Teichm\"uller space was studied from a variety of perspectives in the literature and goes under the name of quantum Teichm\"uller theory. In the mathematics literature, typically a different quantization scheme is employed that makes no reference to conformal blocks \cite{Kashaev:1998fc, Chekhov:1999tn}. The connection between the two quantizations was made in \cite{Teschner:2005bz, Teschner:2013tqy}.

\paragraph{A 3d TQFT.} One can go further and define a three-dimensional TQFT whose Hilbert space is given by the space of conformal blocks. This TQFT was called ``Virasoro TQFT'' in \cite{Collier:2023fwi}, but appeared earlier in different forms under the name of ``Teichm\"uller TQFT'' (although the equivalence of these theories is not obvious) \cite{EllegaardAndersen:2011vps}, see also \cite{Teschner:2005bz}. In fact, three-dimensional quantum gravity may be obtained by combining two copies of Virasoro TQFT \cite{Collier:2023fwi}.

This perspective is useful, since it immediately gives us new operators to consider. Every vertex operator with Liouville momentum $P$ gives rise to a Wilson line operator in the three-dimensional theory. Performing the path integral of the 3d TQFT on a handlebody geometry with an appropriate set of Wilson lines inserted gives the corresponding conformal block in the boundary. We have e.g.
\be 
\begin{tikzpicture}[baseline={([yshift=-.5ex]current bounding box.center)}, xscale=.65, yscale=.6]
        \draw[very thick, Maroon, bend left=30] (-1.8,1.2) to (-.6,0);
        \draw[very thick, Maroon, bend right=30] (-1.8,-1.2) to (-.6,0);
        \draw[very thick, Maroon, bend right=30] (2.2,1.2) to (.6,0);
        \draw[very thick, Maroon, bend left=30] (2.2,-1.2) to (.6,0);
        \draw[very thick, Maroon] (-.6,0) to (.6,0);
        \fill[Maroon] (-.6,0) circle (.07);
        \fill[Maroon] (.6,0) circle (.07);
        \draw[very thick] (-2,1.2) circle (.2 and .5);
        \draw[very thick] (-2,-1.2) circle (.2 and .5);
        \begin{scope}[shift={(2,1.2)}, xscale=.4]
            \draw[very thick] (0,-.5) arc (-90:90:.5);
            \draw[very thick, dashed] (0,.5) arc (90:270:.5);
        \end{scope}
        \begin{scope}[shift={(2,-1.2)}, xscale=.4]
            \draw[very thick] (0,-.5) arc (-90:90:.5);
            \draw[very thick, dashed] (0,.5) arc (90:270:.5);
        \end{scope}
        \begin{scope}[xscale=.55]
            \draw[very thick, RoyalBlue] (0,-.9) arc (-90:90:.9);
            \draw[very thick, dashed, RoyalBlue] (0,.9) arc (90:270:.9);
        \end{scope}
        \draw[very thick, out=0, in=180] (-2,1.7) to (0,.9) to (2,1.7);
        \draw[very thick, out=0, in=180] (-2,-1.7) to (0,-.9) to (2,-1.7);
        \draw[very thick, out=0, in=0, looseness=2.5] (-2,-.7) to (-2,.7);
        \draw[very thick, out=180, in=180, looseness=2.5] (2,-.7) to (2,.7);
        \node at (-2,1.2) {3};
        \node at (2.4,1.2) {2};            
        \node at (2.4,-1.2) {1};
        \node at (-2,-1.2) {4}; 
        \node at (0,.35) {21};
        \end{tikzpicture}=Z_\text{Vir} \! \left(\begin{tikzpicture}[baseline={([yshift=-.5ex]current bounding box.center)}, xscale=.65, yscale=.6]
        \draw[very thick, Maroon, bend left=30] (-1.8,1.2) to (-.8,0);
        \draw[very thick, Maroon, bend right=30] (-1.8,-1.2) to (-.8,0);
        \draw[very thick, Maroon, bend right=30] (1.8,1.2) to (.8,0);
        \draw[very thick, Maroon, bend left=30] (1.8,-1.2) to (.8,0);
        \draw[very thick, Maroon] (-.8,0) to (.8,0);
        \fill[Maroon] (-.8,0) circle (.07);
        \fill[Maroon] (.8,0) circle (.07);
        \draw[very thick] (0,0) circle (2.3 and 1.9);
        \node at (0,.35) {21};
        \node at (-1.6,.7) {3};
        \node at (-1.6,-.7) {4};
        \node at (1.6,.7) {2};
        \node at (1.6,-.7) {1};
        \end{tikzpicture}\right)\ ,
\ee
where $Z_\text{Vir}$ denotes the partition function of the Virasoro TQFT and the red lines now denote Wilson lines extending into the bulk of the ball. This statement requires a particular normalization of the junction of three Wilson lines in the bulk, but it will not play a role in the following.
This is in parallel to the corresponding statement in Chern-Simons theory/WZW models.

\paragraph{Verlinde lines.} 
As in any 2d CFT, we can also consider Verlinde lines \cite{Verlinde:1988sn}.
The 3d perspective makes it trivial to define them. The action of Verlinde lines on a conformal block is given by inserting the Wilson line parallel and close to the boundary in the three-dimensional path integral. For example, the Verlinde line encircling punctures 1 and 2 on the four-punctured sphere acts as follows on a conformal block:
\be 
Z_\text{Vir} \! \left(\begin{tikzpicture}[baseline={([yshift=-.5ex]current bounding box.center)}, xscale=.65, yscale=.6]
        \draw[very thick, Maroon, bend left=30] (-1.8,1.2) to (-.8,0);
        \draw[very thick, Maroon, bend right=30] (-1.8,-1.2) to (-.8,0);
        \draw[very thick, Maroon, bend right=30] (1.8,1.2) to (.8,0);
        \draw[very thick, Maroon, bend left=30] (1.8,-1.2) to (.8,0);
        \draw[very thick, Maroon] (-.8,0) to (.8,0);
        \fill[Maroon] (-.8,0) circle (.07);
        \fill[Maroon] (.8,0) circle (.07);
        \draw[very thick] (0,0) circle (2.3 and 1.9);
        \node at (0,.35) {21};
        \node at (-1.6,.7) {3};
        \node at (-1.6,-.7) {4};
        \node at (1.6,.7) {2};
        \node at (1.6,-.7) {1};
        \end{tikzpicture}\right) \longmapsto
Z_\text{Vir} \! \left(\begin{tikzpicture}[baseline={([yshift=-.5ex]current bounding box.center)}, xscale=.65, yscale=.6]
        \draw[very thick, Maroon, bend left=30] (-1.8,1.2) to (-.8,0);
        \draw[very thick, Maroon, bend right=30] (-1.8,-1.2) to (-.8,0);
        \draw[very thick, Maroon, bend right=30] (1.8,1.2) to (.8,0);
        \draw[very thick, Maroon, bend left=30] (1.8,-1.2) to (.8,0);
        \draw[very thick, Maroon] (-.8,0) to (0,0);
        \fill[Maroon] (-.8,0) circle (.07);
        \fill[Maroon] (.8,0) circle (.07);
        \fill[white] (-.5,0) circle (.1);
        \draw[Maroon, very thick] (0,0) circle (.5 and 1);
        \fill[white] (.5,0) circle (.1);
        \draw[Maroon, very thick] (0,0) to (.8,0);
        \draw[very thick] (0,0) circle (2.3 and 1.9);]
        \node at (0,.35) {21};
        \node at (-1.6,.7) {3};
        \node at (-1.6,-.7) {4};
        \node at (1.6,.7) {2};
        \node at (1.6,-.7) {1};
        \node at (0,1.38) {$P_\text{V}$};
        \end{tikzpicture}\right)\ ,
\ee
where $P_\text{V}$ is the Liouville momentum of the Verlinde line. This trick was explained in \cite{Gaiotto:2014lma}.

\paragraph{Hyperbolic parametrization.} We could have considered Teichm\"uller space from a hyperbolic viewpoint, not a holomorphic one as we have done so far. By the uniformization theorem, we can give any Riemann surface (of negative Euler characteristic) a unique hyperbolic structure. We can then consider a pair of pants decomposition defined by simple closed geodesics. Teichm\"uller space may then be parametrized by Fenchel-Nielsen coordinates -- the lengths $\ell_i$ and twists $\theta_i$  of the simple closed geodesics defining the pair of pants decomposition. The Weil-Petersson symplectic form takes a canonical form in these coordinates \cite{Wolpert:Fenchel, Wolpert:Symplectic}:
\be 
\omega_\text{WP}=\sum_{i=1}^{3g-3+n} \d \ell_i \wedge \d \theta_i\ . \label{eq:Weil-Petersson form Fenchel-Nielsen coordinates}
\ee
We will now explain that conformal blocks are precisely the intertwiners between the two perspectives.

\paragraph{Relating $P$ to the hyperbolic lengths.}
We can use Verlinde line operators to identify a relation of the internal Liouville momentum $P_j$ with the hyperbolic length of the geodesic defining the pair of pants decomposition. Such a relation was conjectured by Verlinde \cite{Verlinde:1989ua} and later explained by Teschner \cite{Teschner:2003at}. We give a different derivation.

The main point to note is that Wilson lines measure monodromies around cycles of the three-dimensional manifold in a given representation. Classically, it is convenient to use the two-dimensional representation. Then the monodromy around a geodesic of boundary length $\ell$ is given by
\be 
\tr \begin{pmatrix}
    \mathrm{e}^{\frac{\ell}{2}} & 0 \\ 0 & \mathrm{e}^{-\frac{\ell}{2}} 
\end{pmatrix}=2 \cosh\left(\frac{\ell}{2}\right)\ .
\ee
This is to be compared with the expectation value of the Verlinde line operator. The two-dimensional representation of $\mathfrak{sl}(2,\RR)$ maps to a Virasoro representation with a second order null vector, this can either be $p_{\langle 2,1 \rangle}$ or $p_{\langle 1,2 \rangle}$, see eq.~\eqref{eq:Liouville momenta degenerate representations}. Which one we pick depends on the quantization scheme we use. We could either have $b$ or $b^{-1}$ take the role of $\hbar$. We will go with the first choice, in which case we should consider $p_{\langle 2,1 \rangle}$.

As in any TQFT, we can remove loops around Wilson lines as follows \cite{Witten:1988hf}:
\be 
\begin{tikzpicture}[baseline={([yshift=-2.2ex]current bounding box.center)}, xscale=.85, yscale=.8]
        \draw[very thick, Maroon] (-.8,0) to (0,0);
        \fill[white] (-.5,0) circle (.1);
        \draw[Maroon, very thick] (0,0) circle (.5 and 1);
        \fill[white] (.5,0) circle (.1);
        \draw[Maroon, very thick] (0,0) to (.8,0);
        \node at (0,.35) {$P$};
        \node at (0,1.4) {$P_\text{V}$};
        \end{tikzpicture}\, = \frac{\mathbb{S}_{P_\text{V},P}[\id]}{\mathbb{S}_{\id,P}[\id]} 
        \ \begin{tikzpicture}[baseline={([yshift=-2ex]current bounding box.center)}, xscale=.85, yscale=.8]
        \draw[very thick, Maroon] (-.8,0) to (.8,0);
        \node at (0,.35) {$P$};
        \end{tikzpicture}\ .
\ee
This follows by interpreting the modular S-matrix as the partition function of Virasoro TQFT on the Hopf link.

Working out the ratio of modular S-matrices from eq.~\eqref{eq:SP1,P2[1]} (while being careful about subtracting out null vectors as in \eqref{eq:S1P[1] computation}), we obtain the relation
\be 
2 \cosh\left(\frac{\ell}{2}\right)\sim \frac{\mathbb{S}_{P_{\langle 2,1 \rangle},P}[\id]}{\mathbb{S}_{\id, P}[\id]}=2 \cosh(2\pi b P)\ ,
\ee
where the first equation holds semiclassically in the $b \to 0$ limit. Thus we have semiclassically,
\be 
P\sim \frac{\ell}{4\pi b}\ . \label{eq:Liouville momentum length relation}
\ee
We note as a consistency condition that $\ell=2\pi i$ formally describes trivial monodromy, which indeed maps to the vacuum in a $b \to 0$ limit.
For the external momenta, \eqref{eq:Liouville momentum length relation} also follows from the conformal anomaly. Indeed, we already mentioned the relation of the conformal weight and the geodesic boundary length in eq.~\eqref{eq:conformal weight geodesic length relation}, which is equivalent to \eqref{eq:Liouville momentum length relation} in the semi-classical limit.
\paragraph{Quantization in Fenchel-Nielsen coordinates.}
One could have quantized Teichm\"uller space in a much simpler way. Since the Weil-Petersson form \eqref{eq:Weil-Petersson form Fenchel-Nielsen coordinates} takes a simple form in Fenchel-Nielsen coordinates, one could have simply considered wave functions depending only on the lengths $\ell_i$ or only on the twists $\theta_i$. The symplectic form takes the canonical form and thus the inner product is simply given by the usual $L^2$-inner product on $\RR_{\ge 0}^{3g-3+n}$.

We have seen that conformal blocks intertwine with both pictures. The values of the internal Liouville momenta may be identified with the hyperbolic lengths via \eqref{eq:Liouville momentum length relation} and appropriately normalized conformal blocks realize indeed the $L^2$-inner product on $\mathbb{R}_{\ge 0}^{3g-3+n}$ according to \eqref{eq:explicit inner product four-punctured sphere}. Thus, we can view the isomorphism \eqref{eq:isomorphism Hilbert spaces} as an isomorphism between the Hilbert space as obtained by quantizing in Fenchel-Nielsen coordinates and the Hilbert space when quantizing in K\"ahler quantization.

In Fenchel-Nielsen coordinates, the action of a change of pair-of-pants decomposition is somewhat complicated, but this problem is solved at the quantum level precisely via the crossing kernels.

\section*{Acknowledgements}
I thank Scott Collier and Mengyang Zhang for discussions and collaboration on related topics. I thank Xin Sun for discussions and particularly Guillaume Remy and J\"org Teschner for discussions and comments on the draft. I also acknowledge useful comments on the draft by Sylvain Ribault.
LE is supported by the grant DE-SC0009988 from the U.S. Department of Energy.

\appendix 

\section{Explicit crossing moves for the four-punctured sphere} \label{app:four-punctured sphere}
In this appendix, we make the crossing transformations somewhat more explicit for the four-punctured sphere. This illustrates the general theory explained in Section~\ref{subsec:crossing transformations}.
We in particular exhibit all the symmetries of conformal blocks on the four-punctured sphere.
\subsection{The expansion of the conformal block}
Let us record the first few orders in the conformal block expansion \eqref{eq:conformal block four-punctured sphere abstract expansion}. As in Section~\ref{sec:Virasoro conformal blocks}, we choose $z_1=1$, $z_2=\infty$, $z_3=0$ and $z_4=z$. We then have
\begin{align}
      & \begin{tikzpicture}[baseline={([yshift=-.5ex]current bounding box.center)}]
        \begin{scope}[xscale=.7, yscale=.7]
        \draw[very thick, Maroon, bend left=30] (-1.8,1.2) to (-.6,0);
        			\draw[very thick, Maroon, bend right=30] (-1.8,-1.2) to (-.6,0);
        			\draw[very thick, Maroon, bend right=30] (2.2,1.2) to (.6,0);
        			\draw[very thick, Maroon, bend left=30] (2.2,-1.2) to (.6,0);
        			\draw[very thick, Maroon] (-.6,0) to (.6,0);
        			\fill[Maroon] (-.6,0) circle (.07);
        			\fill[Maroon] (.6,0) circle (.07);
        			\draw[very thick] (-2,1.2) circle (.2 and .5);
        			\draw[very thick] (-2,-1.2) circle (.2 and .5);
        			\begin{scope}[shift={(2,1.2)}, xscale=.4]
            			\draw[very thick] (0,-.5) arc (-90:90:.5);
            			\draw[very thick, dashed] (0,.5) arc (90:270:.5);
        			\end{scope}
        			\begin{scope}[shift={(2,-1.2)}, xscale=.4]
            			\draw[very thick] (0,-.5) arc (-90:90:.5);
            			\draw[very thick, dashed] (0,.5) arc (90:270:.5);
        			\end{scope}
        			\begin{scope}[xscale=.4]
            			\draw[very thick, RoyalBlue] (0,-.9) arc (-90:90:.9);
            			\draw[very thick, dashed, RoyalBlue] (0,.9) arc (90:270:.9);
        			\end{scope}
        			\draw[very thick, out=0, in=180] (-2,1.7) to (0,.9) to (2,1.7);
        			\draw[very thick, out=0, in=180] (-2,-1.7) to (0,-.9) to (2,-1.7);
        			\draw[very thick, out=0, in=0, looseness=2.5] (-2,-.7) to (-2,.7);
        			\draw[very thick, out=180, in=180, looseness=2.5] (2,-.7) to (2,.7);
           \node at (2.4,-1.2) {$1$};
           \node at (2.4,1.2) {$2$};
           \node at (-2,-1.2) {$4$};
           \node at (-2,1.2) {$3$};
           \node at (0,.3) {$\Delta$};
		\end{scope}
   \end{tikzpicture}\!\!\!\! (z)=z^{\Delta -\Delta_3-\Delta_4} \bigg(1+\frac{(\Delta +\Delta_1-\Delta_2) (\Delta -\Delta_3+\Delta_4)}{2 \Delta }\, z\nonumber\\
   &\qquad +\frac{1}{2 \Delta  (2 c \Delta +c+16 \Delta^2-10 \Delta )}\, \Big(\big(\tfrac{c}{2}+4 \Delta \big) (\Delta +\Delta_1-\Delta_2)_2 (\Delta -\Delta_3+\Delta_4)_2\nonumber\\
   &\qquad\qquad+4 \Delta (2 \Delta +1)(\Delta +2 \Delta_1-\Delta_2) (\Delta -\Delta_3+2 \Delta_4)\nonumber\\
   &\qquad\qquad-6 \Delta (\Delta +\Delta_1-\Delta_2)_2 (\Delta -\Delta_3+2 \Delta_4) \nonumber\\
   &\qquad\qquad-6 \Delta  (\Delta +2 \Delta_1-\Delta_2) (\Delta -\Delta_3+\Delta_4)_2\Big) \, z^2+\mathcal{O}(z^3)\bigg)\ ,
\end{align}
where $(a)_n=a(a+1) \cdots (a+n-1)$ is the rising Pochhammer symbol.

\subsection{Basic symmetries and braiding transformations} 
To discuss the braiding moves, it is useful to reinstate the dependence on all four coordinates on the sphere,
\begin{multline} 
\begin{tikzpicture}[baseline={([yshift=-.5ex]current bounding box.center)}]
        \begin{scope}[xscale=.7, yscale=.7]
        \draw[very thick, Maroon, bend left=30] (-1.8,1.2) to (-.6,0);
        			\draw[very thick, Maroon, bend right=30] (-1.8,-1.2) to (-.6,0);
        			\draw[very thick, Maroon, bend right=30] (2.2,1.2) to (.6,0);
        			\draw[very thick, Maroon, bend left=30] (2.2,-1.2) to (.6,0);
        			\draw[very thick, Maroon] (-.6,0) to (.6,0);
        			\fill[Maroon] (-.6,0) circle (.07);
        			\fill[Maroon] (.6,0) circle (.07);
        			\draw[very thick] (-2,1.2) circle (.2 and .5);
        			\draw[very thick] (-2,-1.2) circle (.2 and .5);
        			\begin{scope}[shift={(2,1.2)}, xscale=.4]
            			\draw[very thick] (0,-.5) arc (-90:90:.5);
            			\draw[very thick, dashed] (0,.5) arc (90:270:.5);
        			\end{scope}
        			\begin{scope}[shift={(2,-1.2)}, xscale=.4]
            			\draw[very thick] (0,-.5) arc (-90:90:.5);
            			\draw[very thick, dashed] (0,.5) arc (90:270:.5);
        			\end{scope}
        			\begin{scope}[xscale=.4]
            			\draw[very thick, RoyalBlue] (0,-.9) arc (-90:90:.9);
            			\draw[very thick, dashed, RoyalBlue] (0,.9) arc (90:270:.9);
        			\end{scope}
        			\draw[very thick, out=0, in=180] (-2,1.7) to (0,.9) to (2,1.7);
        			\draw[very thick, out=0, in=180] (-2,-1.7) to (0,-.9) to (2,-1.7);
        			\draw[very thick, out=0, in=0, looseness=2.5] (-2,-.7) to (-2,.7);
        			\draw[very thick, out=180, in=180, looseness=2.5] (2,-.7) to (2,.7);
           \node at (2.4,-1.2) {$1$};
           \node at (2.4,1.2) {$2$};
           \node at (-2,-1.2) {$4$};
           \node at (-2,1.2) {$3$};
           \node at (0,.3) {$\Delta$};
		\end{scope}
   \end{tikzpicture}\!\!\!\! (z_1,z_2,z_3,z_4)=z_{13}^{-\Delta_1+\Delta_2-\Delta_3-\Delta_4}z_{21}^{-\Delta_1-\Delta_2+\Delta_3+\Delta_4}\\
   \times z_{23}^{\Delta_1-\Delta_2-\Delta_3+\Delta_4}z_{24}^{-2 \Delta_4}
   \begin{tikzpicture}[baseline={([yshift=-.5ex]current bounding box.center)}]
        \begin{scope}[xscale=.7, yscale=.7]
        \draw[very thick, Maroon, bend left=30] (-1.8,1.2) to (-.6,0);
        			\draw[very thick, Maroon, bend right=30] (-1.8,-1.2) to (-.6,0);
        			\draw[very thick, Maroon, bend right=30] (2.2,1.2) to (.6,0);
        			\draw[very thick, Maroon, bend left=30] (2.2,-1.2) to (.6,0);
        			\draw[very thick, Maroon] (-.6,0) to (.6,0);
        			\fill[Maroon] (-.6,0) circle (.07);
        			\fill[Maroon] (.6,0) circle (.07);
        			\draw[very thick] (-2,1.2) circle (.2 and .5);
        			\draw[very thick] (-2,-1.2) circle (.2 and .5);
        			\begin{scope}[shift={(2,1.2)}, xscale=.4]
            			\draw[very thick] (0,-.5) arc (-90:90:.5);
            			\draw[very thick, dashed] (0,.5) arc (90:270:.5);
        			\end{scope}
        			\begin{scope}[shift={(2,-1.2)}, xscale=.4]
            			\draw[very thick] (0,-.5) arc (-90:90:.5);
            			\draw[very thick, dashed] (0,.5) arc (90:270:.5);
        			\end{scope}
        			\begin{scope}[xscale=.4]
            			\draw[very thick, RoyalBlue] (0,-.9) arc (-90:90:.9);
            			\draw[very thick, dashed, RoyalBlue] (0,.9) arc (90:270:.9);
        			\end{scope}
        			\draw[very thick, out=0, in=180] (-2,1.7) to (0,.9) to (2,1.7);
        			\draw[very thick, out=0, in=180] (-2,-1.7) to (0,-.9) to (2,-1.7);
        			\draw[very thick, out=0, in=0, looseness=2.5] (-2,-.7) to (-2,.7);
        			\draw[very thick, out=180, in=180, looseness=2.5] (2,-.7) to (2,.7);
           \node at (2.4,-1.2) {$1$};
           \node at (2.4,1.2) {$2$};
           \node at (-2,-1.2) {$4$};
           \node at (-2,1.2) {$3$};
           \node at (0,.3) {$\Delta$};
		\end{scope}
   \end{tikzpicture}\!\!\!\! \left(\frac{z_{21}z_{43}}{z_{24}z_{13}}\right)\ . \label{eq:four-punctured sphere block all coordinates}
\end{multline}
This follows from solving the three conformal Ward identities in the sphere.

Let us first discuss the braiding move of the punctures `3' and `4'. For this, we will interchange $z_3 \leftrightarrow z_4$ and hence send $z_{43} \to z_{43} \, \mathrm{e}^{\pi i}$. On the level of the cross-ratio, we have
\be 
z \to \frac{\mathrm{e}^{\pi i}z}{1-z}\ .
\ee
We have
\be 
\begin{tikzpicture}[baseline={([yshift=-.5ex]current bounding box.center)}]
        \begin{scope}[xscale=.7, yscale=.7]
        \draw[very thick, Maroon, bend left=30] (-1.8,1.2) to (-.6,0);
        			\draw[very thick, Maroon, bend right=30] (-1.8,-1.2) to (-.6,0);
        			\draw[very thick, Maroon, bend right=30] (2.2,1.2) to (.6,0);
        			\draw[very thick, Maroon, bend left=30] (2.2,-1.2) to (.6,0);
        			\draw[very thick, Maroon] (-.6,0) to (.6,0);
        			\fill[Maroon] (-.6,0) circle (.07);
        			\fill[Maroon] (.6,0) circle (.07);
        			\draw[very thick] (-2,1.2) circle (.2 and .5);
        			\draw[very thick] (-2,-1.2) circle (.2 and .5);
        			\begin{scope}[shift={(2,1.2)}, xscale=.4]
            			\draw[very thick] (0,-.5) arc (-90:90:.5);
            			\draw[very thick, dashed] (0,.5) arc (90:270:.5);
        			\end{scope}
        			\begin{scope}[shift={(2,-1.2)}, xscale=.4]
            			\draw[very thick] (0,-.5) arc (-90:90:.5);
            			\draw[very thick, dashed] (0,.5) arc (90:270:.5);
        			\end{scope}
        			\begin{scope}[xscale=.4]
            			\draw[very thick, RoyalBlue] (0,-.9) arc (-90:90:.9);
            			\draw[very thick, dashed, RoyalBlue] (0,.9) arc (90:270:.9);
        			\end{scope}
        			\draw[very thick, out=0, in=180] (-2,1.7) to (0,.9) to (2,1.7);
        			\draw[very thick, out=0, in=180] (-2,-1.7) to (0,-.9) to (2,-1.7);
        			\draw[very thick, out=0, in=0, looseness=2.5] (-2,-.7) to (-2,.7);
        			\draw[very thick, out=180, in=180, looseness=2.5] (2,-.7) to (2,.7);
           \node at (2.4,-1.2) {$1$};
           \node at (2.4,1.2) {$2$};
           \node at (-2,-1.2) {$3$};
           \node at (-2,1.2) {$4$};
           \node at (0,.3) {$\Delta$};
		\end{scope}
   \end{tikzpicture}\!\!\!\! \Big(\frac{\mathrm{e}^{\pi i} z}{1-z}\Big)=\mathbb{B}^{P_3,P_4}_P (1-z)^{\Delta_1-\Delta_2+\Delta_3+\Delta_4}\!\!\!\begin{tikzpicture}[baseline={([yshift=-.5ex]current bounding box.center)}]
        \begin{scope}[xscale=.7, yscale=.7]
        \draw[very thick, Maroon, bend left=30] (-1.8,1.2) to (-.6,0);
        			\draw[very thick, Maroon, bend right=30] (-1.8,-1.2) to (-.6,0);
        			\draw[very thick, Maroon, bend right=30] (2.2,1.2) to (.6,0);
        			\draw[very thick, Maroon, bend left=30] (2.2,-1.2) to (.6,0);
        			\draw[very thick, Maroon] (-.6,0) to (.6,0);
        			\fill[Maroon] (-.6,0) circle (.07);
        			\fill[Maroon] (.6,0) circle (.07);
        			\draw[very thick] (-2,1.2) circle (.2 and .5);
        			\draw[very thick] (-2,-1.2) circle (.2 and .5);
        			\begin{scope}[shift={(2,1.2)}, xscale=.4]
            			\draw[very thick] (0,-.5) arc (-90:90:.5);
            			\draw[very thick, dashed] (0,.5) arc (90:270:.5);
        			\end{scope}
        			\begin{scope}[shift={(2,-1.2)}, xscale=.4]
            			\draw[very thick] (0,-.5) arc (-90:90:.5);
            			\draw[very thick, dashed] (0,.5) arc (90:270:.5);
        			\end{scope}
        			\begin{scope}[xscale=.4]
            			\draw[very thick, RoyalBlue] (0,-.9) arc (-90:90:.9);
            			\draw[very thick, dashed, RoyalBlue] (0,.9) arc (90:270:.9);
        			\end{scope}
        			\draw[very thick, out=0, in=180] (-2,1.7) to (0,.9) to (2,1.7);
        			\draw[very thick, out=0, in=180] (-2,-1.7) to (0,-.9) to (2,-1.7);
        			\draw[very thick, out=0, in=0, looseness=2.5] (-2,-.7) to (-2,.7);
        			\draw[very thick, out=180, in=180, looseness=2.5] (2,-.7) to (2,.7);
           \node at (2.4,-1.2) {$1$};
           \node at (2.4,1.2) {$2$};
           \node at (-2,-1.2) {$4$};
           \node at (-2,1.2) {$3$};
           \node at (0,.3) {$\Delta$};
		\end{scope}
   \end{tikzpicture}\!\!\!\! (z)\  ,
\ee
where we put again $z_1=1$, $z_2=\infty$, $z_3=0$ and $z_4=z$.

Similarly, we can braid the punctures `1' and `2'. This also gives a phase from the prefactor in \eqref{eq:four-punctured sphere block all coordinates} and we have
\be 
\begin{tikzpicture}[baseline={([yshift=-.5ex]current bounding box.center)}]
        \begin{scope}[xscale=.7, yscale=.7]
        \draw[very thick, Maroon, bend left=30] (-1.8,1.2) to (-.6,0);
        			\draw[very thick, Maroon, bend right=30] (-1.8,-1.2) to (-.6,0);
        			\draw[very thick, Maroon, bend right=30] (2.2,1.2) to (.6,0);
        			\draw[very thick, Maroon, bend left=30] (2.2,-1.2) to (.6,0);
        			\draw[very thick, Maroon] (-.6,0) to (.6,0);
        			\fill[Maroon] (-.6,0) circle (.07);
        			\fill[Maroon] (.6,0) circle (.07);
        			\draw[very thick] (-2,1.2) circle (.2 and .5);
        			\draw[very thick] (-2,-1.2) circle (.2 and .5);
        			\begin{scope}[shift={(2,1.2)}, xscale=.4]
            			\draw[very thick] (0,-.5) arc (-90:90:.5);
            			\draw[very thick, dashed] (0,.5) arc (90:270:.5);
        			\end{scope}
        			\begin{scope}[shift={(2,-1.2)}, xscale=.4]
            			\draw[very thick] (0,-.5) arc (-90:90:.5);
            			\draw[very thick, dashed] (0,.5) arc (90:270:.5);
        			\end{scope}
        			\begin{scope}[xscale=.4]
            			\draw[very thick, RoyalBlue] (0,-.9) arc (-90:90:.9);
            			\draw[very thick, dashed, RoyalBlue] (0,.9) arc (90:270:.9);
        			\end{scope}
        			\draw[very thick, out=0, in=180] (-2,1.7) to (0,.9) to (2,1.7);
        			\draw[very thick, out=0, in=180] (-2,-1.7) to (0,-.9) to (2,-1.7);
        			\draw[very thick, out=0, in=0, looseness=2.5] (-2,-.7) to (-2,.7);
        			\draw[very thick, out=180, in=180, looseness=2.5] (2,-.7) to (2,.7);
           \node at (2.4,-1.2) {$2$};
           \node at (2.4,1.2) {$1$};
           \node at (-2,-1.2) {$4$};
           \node at (-2,1.2) {$3$};
           \node at (0,.3) {$\Delta$};
		\end{scope}
   \end{tikzpicture}\!\!\!\! \Big(\frac{\mathrm{e}^{\pi i} z}{1-z}\Big)=\mathbb{B}^{P_1,P_2}_P (1-z)^{2\Delta_4}\!\!\!\begin{tikzpicture}[baseline={([yshift=-.5ex]current bounding box.center)}]
        \begin{scope}[xscale=.7, yscale=.7]
        \draw[very thick, Maroon, bend left=30] (-1.8,1.2) to (-.6,0);
        			\draw[very thick, Maroon, bend right=30] (-1.8,-1.2) to (-.6,0);
        			\draw[very thick, Maroon, bend right=30] (2.2,1.2) to (.6,0);
        			\draw[very thick, Maroon, bend left=30] (2.2,-1.2) to (.6,0);
        			\draw[very thick, Maroon] (-.6,0) to (.6,0);
        			\fill[Maroon] (-.6,0) circle (.07);
        			\fill[Maroon] (.6,0) circle (.07);
        			\draw[very thick] (-2,1.2) circle (.2 and .5);
        			\draw[very thick] (-2,-1.2) circle (.2 and .5);
        			\begin{scope}[shift={(2,1.2)}, xscale=.4]
            			\draw[very thick] (0,-.5) arc (-90:90:.5);
            			\draw[very thick, dashed] (0,.5) arc (90:270:.5);
        			\end{scope}
        			\begin{scope}[shift={(2,-1.2)}, xscale=.4]
            			\draw[very thick] (0,-.5) arc (-90:90:.5);
            			\draw[very thick, dashed] (0,.5) arc (90:270:.5);
        			\end{scope}
        			\begin{scope}[xscale=.4]
            			\draw[very thick, RoyalBlue] (0,-.9) arc (-90:90:.9);
            			\draw[very thick, dashed, RoyalBlue] (0,.9) arc (90:270:.9);
        			\end{scope}
        			\draw[very thick, out=0, in=180] (-2,1.7) to (0,.9) to (2,1.7);
        			\draw[very thick, out=0, in=180] (-2,-1.7) to (0,-.9) to (2,-1.7);
        			\draw[very thick, out=0, in=0, looseness=2.5] (-2,-.7) to (-2,.7);
        			\draw[very thick, out=180, in=180, looseness=2.5] (2,-.7) to (2,.7);
           \node at (2.4,-1.2) {$1$};
           \node at (2.4,1.2) {$2$};
           \node at (-2,-1.2) {$4$};
           \node at (-2,1.2) {$3$};
           \node at (0,.3) {$\Delta$};
		\end{scope}
   \end{tikzpicture}\!\!\!\! (z)\  .
\ee
We can also rotate the picture by 180 degrees, which corresponds to interchanging $z_1 \leftrightarrow z_3$ and $z_2 \leftrightarrow z_4$. This keeps the cross ratio unchanged. Including the contributions from the prefactors in \eqref{eq:four-punctured sphere block all coordinates}, we have
\be 
\begin{tikzpicture}[baseline={([yshift=-.5ex]current bounding box.center)}]
        \begin{scope}[xscale=.7, yscale=.7]
        \draw[very thick, Maroon, bend left=30] (-1.8,1.2) to (-.6,0);
        			\draw[very thick, Maroon, bend right=30] (-1.8,-1.2) to (-.6,0);
        			\draw[very thick, Maroon, bend right=30] (2.2,1.2) to (.6,0);
        			\draw[very thick, Maroon, bend left=30] (2.2,-1.2) to (.6,0);
        			\draw[very thick, Maroon] (-.6,0) to (.6,0);
        			\fill[Maroon] (-.6,0) circle (.07);
        			\fill[Maroon] (.6,0) circle (.07);
        			\draw[very thick] (-2,1.2) circle (.2 and .5);
        			\draw[very thick] (-2,-1.2) circle (.2 and .5);
        			\begin{scope}[shift={(2,1.2)}, xscale=.4]
            			\draw[very thick] (0,-.5) arc (-90:90:.5);
            			\draw[very thick, dashed] (0,.5) arc (90:270:.5);
        			\end{scope}
        			\begin{scope}[shift={(2,-1.2)}, xscale=.4]
            			\draw[very thick] (0,-.5) arc (-90:90:.5);
            			\draw[very thick, dashed] (0,.5) arc (90:270:.5);
        			\end{scope}
        			\begin{scope}[xscale=.4]
            			\draw[very thick, RoyalBlue] (0,-.9) arc (-90:90:.9);
            			\draw[very thick, dashed, RoyalBlue] (0,.9) arc (90:270:.9);
        			\end{scope}
        			\draw[very thick, out=0, in=180] (-2,1.7) to (0,.9) to (2,1.7);
        			\draw[very thick, out=0, in=180] (-2,-1.7) to (0,-.9) to (2,-1.7);
        			\draw[very thick, out=0, in=0, looseness=2.5] (-2,-.7) to (-2,.7);
        			\draw[very thick, out=180, in=180, looseness=2.5] (2,-.7) to (2,.7);
           \node at (2.4,-1.2) {$3$};
           \node at (2.4,1.2) {$4$};
           \node at (-2,-1.2) {$2$};
           \node at (-2,1.2) {$1$};
           \node at (0,.3) {$\Delta$};
		\end{scope}
   \end{tikzpicture}\!\!\!\! (z)=z^{\Delta_3+\Delta_4-\Delta_1-\Delta_2}(1-z)^{\Delta_1-\Delta_2-\Delta_3+\Delta_4}\begin{tikzpicture}[baseline={([yshift=-.5ex]current bounding box.center)}]
        \begin{scope}[xscale=.7, yscale=.7]
        \draw[very thick, Maroon, bend left=30] (-1.8,1.2) to (-.6,0);
        			\draw[very thick, Maroon, bend right=30] (-1.8,-1.2) to (-.6,0);
        			\draw[very thick, Maroon, bend right=30] (2.2,1.2) to (.6,0);
        			\draw[very thick, Maroon, bend left=30] (2.2,-1.2) to (.6,0);
        			\draw[very thick, Maroon] (-.6,0) to (.6,0);
        			\fill[Maroon] (-.6,0) circle (.07);
        			\fill[Maroon] (.6,0) circle (.07);
        			\draw[very thick] (-2,1.2) circle (.2 and .5);
        			\draw[very thick] (-2,-1.2) circle (.2 and .5);
        			\begin{scope}[shift={(2,1.2)}, xscale=.4]
            			\draw[very thick] (0,-.5) arc (-90:90:.5);
            			\draw[very thick, dashed] (0,.5) arc (90:270:.5);
        			\end{scope}
        			\begin{scope}[shift={(2,-1.2)}, xscale=.4]
            			\draw[very thick] (0,-.5) arc (-90:90:.5);
            			\draw[very thick, dashed] (0,.5) arc (90:270:.5);
        			\end{scope}
        			\begin{scope}[xscale=.4]
            			\draw[very thick, RoyalBlue] (0,-.9) arc (-90:90:.9);
            			\draw[very thick, dashed, RoyalBlue] (0,.9) arc (90:270:.9);
        			\end{scope}
        			\draw[very thick, out=0, in=180] (-2,1.7) to (0,.9) to (2,1.7);
        			\draw[very thick, out=0, in=180] (-2,-1.7) to (0,-.9) to (2,-1.7);
        			\draw[very thick, out=0, in=0, looseness=2.5] (-2,-.7) to (-2,.7);
        			\draw[very thick, out=180, in=180, looseness=2.5] (2,-.7) to (2,.7);
           \node at (2.4,-1.2) {$1$};
           \node at (2.4,1.2) {$2$};
           \node at (-2,-1.2) {$4$};
           \node at (-2,1.2) {$3$};
           \node at (0,.3) {$\Delta$};
		\end{scope}
   \end{tikzpicture}\!\!\!\! (z)\ .
\ee
In particular, this makes the symmetries
\be 
\mathbb{F}_{P_{21},P_{32}}\!\begin{bmatrix}
    P_3 & P_2 \\ P_4 & P_1
\end{bmatrix}
=
\mathbb{F}_{P_{21},P_{32}}\!\begin{bmatrix}
    P_4 & P_1 \\ P_3 & P_2
\end{bmatrix}
=\mathbb{F}_{P_{21},P_{32}}\!\begin{bmatrix}
    P_1 & P_4 \\ P_2 & P_3
\end{bmatrix}
\ee
very concrete.
\subsection{Relation to AGT correspondence}
Conformal blocks are linked to instanton partition functions of 4d $\mathcal{N}=2$ gauge theories via the AGT correspondence \cite{Alday:2009aq}. In particular, the four-point block has a particularly simple description in terms of the Nekrasov instanton partition function of $\mathcal{N}=2$ $\mathrm{SU}(2)$ SQCD with $N_f=4$ \cite{Nekrasov:2002qd}. The basic claim of the AGT correspondence is
\be 
\begin{tikzpicture}[baseline={([yshift=-.5ex]current bounding box.center)}]
        \begin{scope}[xscale=.7, yscale=.7]
        \draw[very thick, Maroon, bend left=30] (-1.8,1.2) to (-.6,0);
        			\draw[very thick, Maroon, bend right=30] (-1.8,-1.2) to (-.6,0);
        			\draw[very thick, Maroon, bend right=30] (2.2,1.2) to (.6,0);
        			\draw[very thick, Maroon, bend left=30] (2.2,-1.2) to (.6,0);
        			\draw[very thick, Maroon] (-.6,0) to (.6,0);
        			\fill[Maroon] (-.6,0) circle (.07);
        			\fill[Maroon] (.6,0) circle (.07);
        			\draw[very thick] (-2,1.2) circle (.2 and .5);
        			\draw[very thick] (-2,-1.2) circle (.2 and .5);
        			\begin{scope}[shift={(2,1.2)}, xscale=.4]
            			\draw[very thick] (0,-.5) arc (-90:90:.5);
            			\draw[very thick, dashed] (0,.5) arc (90:270:.5);
        			\end{scope}
        			\begin{scope}[shift={(2,-1.2)}, xscale=.4]
            			\draw[very thick] (0,-.5) arc (-90:90:.5);
            			\draw[very thick, dashed] (0,.5) arc (90:270:.5);
        			\end{scope}
        			\begin{scope}[xscale=.4]
            			\draw[very thick, RoyalBlue] (0,-.9) arc (-90:90:.9);
            			\draw[very thick, dashed, RoyalBlue] (0,.9) arc (90:270:.9);
        			\end{scope}
        			\draw[very thick, out=0, in=180] (-2,1.7) to (0,.9) to (2,1.7);
        			\draw[very thick, out=0, in=180] (-2,-1.7) to (0,-.9) to (2,-1.7);
        			\draw[very thick, out=0, in=0, looseness=2.5] (-2,-.7) to (-2,.7);
        			\draw[very thick, out=180, in=180, looseness=2.5] (2,-.7) to (2,.7);
           \node at (2.4,-1.2) {$1$};
           \node at (2.4,1.2) {$2$};
           \node at (-2,-1.2) {$4$};
           \node at (-2,1.2) {$3$};
           \node at (0,.3) {$\Delta$};
		\end{scope}
   \end{tikzpicture}\!\!\!\! (z)=z^{-\Delta_1-\Delta_2+\Delta} Z_\text{inst}\big(\varepsilon_1=b, \varepsilon_2=b^{-1}; z=\lambda(\tau_\text{IR})\big)\ ,
\ee
where $\tau_\text{IR}$ is the complexified gauge coupling in the infrared, $\lambda(\tau)=\frac{\vartheta_2(\tau)^4}{\vartheta_3(\tau)^4}$ is the modular lambda-function and $\varepsilon_1$ and $\varepsilon_2$ are the two deformation parameters, related to the squashing of the four-sphere $\mathrm{S}_b^4$ on which we compute the Nekrasov partition function.
We have
\be 
p_j=r m_j\ ,
\ee
where $m_j \in i \RR/\mathbb{Z}_2$ are the four purely imaginary mass-parameters of the theory and $r$ is the radius of $\mathrm{S}_b^4$. $\Delta$ is related to the vev of the adjoint scalar in the vector multiplet. 

Localization of the partition function on $\mathrm{S}_b^4$ \cite{Pestun:2007rz} localizes the path integral to a one-dimensional integral over the scalar expectation value. Combining the instanton partition function with the tree-level and one-loop piece gives the full partition function on $\mathrm{S}_b^4$, which precisely coincides with the Liouville four-point function.

\subsection{Regge symmetry}
The conformal block on the four-punctured sphere enjoys an additional symmetry beyond the ones mentioned above. We have
\be 
z^{2P_3P_4}(1-z)^{2P_1P_4}\!\begin{tikzpicture}[baseline={([yshift=-.5ex]current bounding box.center)}]
        \begin{scope}[xscale=.7, yscale=.7]
        \draw[very thick, Maroon, bend left=30] (-1.8,1.2) to (-.6,0);
        			\draw[very thick, Maroon, bend right=30] (-1.8,-1.2) to (-.6,0);
        			\draw[very thick, Maroon, bend right=30] (2.2,1.2) to (.6,0);
        			\draw[very thick, Maroon, bend left=30] (2.2,-1.2) to (.6,0);
        			\draw[very thick, Maroon] (-.6,0) to (.6,0);
        			\fill[Maroon] (-.6,0) circle (.07);
        			\fill[Maroon] (.6,0) circle (.07);
        			\draw[very thick] (-2,1.2) circle (.2 and .5);
        			\draw[very thick] (-2,-1.2) circle (.2 and .5);
        			\begin{scope}[shift={(2,1.2)}, xscale=.4]
            			\draw[very thick] (0,-.5) arc (-90:90:.5);
            			\draw[very thick, dashed] (0,.5) arc (90:270:.5);
        			\end{scope}
        			\begin{scope}[shift={(2,-1.2)}, xscale=.4]
            			\draw[very thick] (0,-.5) arc (-90:90:.5);
            			\draw[very thick, dashed] (0,.5) arc (90:270:.5);
        			\end{scope}
        			\begin{scope}[xscale=.4]
            			\draw[very thick, RoyalBlue] (0,-.9) arc (-90:90:.9);
            			\draw[very thick, dashed, RoyalBlue] (0,.9) arc (90:270:.9);
        			\end{scope}
        			\draw[very thick, out=0, in=180] (-2,1.7) to (0,.9) to (2,1.7);
        			\draw[very thick, out=0, in=180] (-2,-1.7) to (0,-.9) to (2,-1.7);
        			\draw[very thick, out=0, in=0, looseness=2.5] (-2,-.7) to (-2,.7);
        			\draw[very thick, out=180, in=180, looseness=2.5] (2,-.7) to (2,.7);
           \node at (2.4,-1.2) {$1$};
           \node at (2.4,1.2) {$2$};
           \node at (-2,-1.2) {$4$};
           \node at (-2,1.2) {$3$};
           \node at (0,.3) {$\Delta$};
		\end{scope}
   \end{tikzpicture}\!\!\!\! (z)=
   z^{2\tilde{P}_3\tilde{P}_4}(1-z)^{2\tilde{P}_1\tilde{P}_4}\!\begin{tikzpicture}[baseline={([yshift=-.5ex]current bounding box.center)}]
        \begin{scope}[xscale=.7, yscale=.7]
        \draw[very thick, Maroon, bend left=30] (-1.8,1.2) to (-.6,0);
        			\draw[very thick, Maroon, bend right=30] (-1.8,-1.2) to (-.6,0);
        			\draw[very thick, Maroon, bend right=30] (2.2,1.2) to (.6,0);
        			\draw[very thick, Maroon, bend left=30] (2.2,-1.2) to (.6,0);
        			\draw[very thick, Maroon] (-.6,0) to (.6,0);
        			\fill[Maroon] (-.6,0) circle (.07);
        			\fill[Maroon] (.6,0) circle (.07);
        			\draw[very thick] (-2,1.2) circle (.2 and .5);
        			\draw[very thick] (-2,-1.2) circle (.2 and .5);
        			\begin{scope}[shift={(2,1.2)}, xscale=.4]
            			\draw[very thick] (0,-.5) arc (-90:90:.5);
            			\draw[very thick, dashed] (0,.5) arc (90:270:.5);
        			\end{scope}
        			\begin{scope}[shift={(2,-1.2)}, xscale=.4]
            			\draw[very thick] (0,-.5) arc (-90:90:.5);
            			\draw[very thick, dashed] (0,.5) arc (90:270:.5);
        			\end{scope}
        			\begin{scope}[xscale=.4]
            			\draw[very thick, RoyalBlue] (0,-.9) arc (-90:90:.9);
            			\draw[very thick, dashed, RoyalBlue] (0,.9) arc (90:270:.9);
        			\end{scope}
        			\draw[very thick, out=0, in=180] (-2,1.7) to (0,.9) to (2,1.7);
        			\draw[very thick, out=0, in=180] (-2,-1.7) to (0,-.9) to (2,-1.7);
        			\draw[very thick, out=0, in=0, looseness=2.5] (-2,-.7) to (-2,.7);
        			\draw[very thick, out=180, in=180, looseness=2.5] (2,-.7) to (2,.7);
           \node at (2.4,-1.2) {$\tilde{1}$};
           \node at (2.4,1.2) {$\tilde{2}$};
           \node at (-2,-1.2) {$\tilde{4}$};
           \node at (-2,1.2) {$\tilde{3}$};
           \node at (0,.3) {$\Delta$};
		\end{scope}
   \end{tikzpicture}\!\!\!\! (z)\ , \label{eq:conformal block four-punctured sphere Regge symmetry}
\ee
where $\tilde{P}_j=\frac{1}{2} P_{1234}-P_j$. This equation implies in particular that the conformal blocks appearing on the LHS and RHS have the same crossing properties and thus
\be 
\mathbb{F}_{P_{21},P_{32}}\!\begin{bmatrix}
    P_3 & P_2 \\
    P_4 & P_1
\end{bmatrix}=\mathbb{F}_{P_{21},P_{32}}\!\begin{bmatrix}
    \frac{1}{2}P_{1234}-P_3 & \frac{1}{2}P_{1234}-P_2 \\
    \frac{1}{2}P_{1234}-P_4 & \frac{1}{2}P_{1234}-P_1
\end{bmatrix}\ ,
\ee
which is eq.~\eqref{eq:Regge symmetry}. 
We now give two explanations of this surprising identity, one technical and one conceptual. We also show below that the symmetry extends to the full Liouville four-point function. This was explained in \cite{Giribet:2009hm}.
\paragraph{Explanation from the AGT correspondence.} The main point for us is $\mathcal{N}=2$ $\mathrm{SU}(2)$ SQCD has an extended $\mathrm{SO}(8)$ flavour symmetry. In particular the partition function on $\mathrm{S}_b^4$ exhibits the triality symmetry of $\mathrm{SO}(8)$, which acts as
\be 
m_j \mapsto \frac{1}{2} m_{1234}-m_j\ .
\ee
This in particular implies that the Liouville four-point function has the same symmetry, which in our conventions read
\begin{multline}
    \prod_{1\le j<k \le 4}|z_j-z_k|^{4P_jP_k} \prod_{j=1}^4 \Gamma_b(Q \pm 2 P_j) \Big \langle \prod_{j=1}^4 V_{P_j}(z_j) \Big \rangle  \\
    \prod_{1\le j<k \le 4}|z_j-z_k|^{4\tilde{P}_j\tilde{P}_k} \prod_{j=1}^4 \Gamma_b(Q \pm 2 \tilde{P}_j) \Big \langle \prod_{j=1}^4 V_{\tilde{P}_j}(z_j) \Big \rangle   
\end{multline}
with $\tilde{P}_j=\frac{1}{2} P_{1234}-P_j$. 
The factor $\prod_{1\le j<k \le 4}|z_j-z_k|^{4P_jP_k}$ has to do with the correct decoupling of the $\mathrm{U}(1)$ factor of $\mathrm{U}(2)$ gauge theory.
Since conformal blocks are all linearly independent, this equation translates into the equation \eqref{eq:conformal block four-punctured sphere Regge symmetry} for the conformal blocks.

This explanation gives a natural reason why the four-point function of Liouville theory has such a symmetry, but e.g.\ the five-point function lacks it. The associated class $\mathcal{S}$ gauge theory on the five-punctured sphere has gauge group $\mathrm{SU}(2) \times \mathrm{SU}(2)$ and flavour symmetry $\mathrm{SU}(2)^5$ which does not enhance further \cite{Gaiotto:2009we}.
\paragraph{Explanation from the Coulomb gas formalism.} One can also prove such a statement directly in CFT, but it is not particularly straightforward and does not give a conceptual explanation. Let us sketch the argument.
It is more convenient to work with $\alpha_j=\frac{Q}{2}+i P_j$. Let us first consider the special case $\alpha_4=0$ of eq.~\eqref{eq:conformal block four-punctured sphere Regge symmetry}. The fusion rules then  necessarily also impose $P=P_3$. Then the LHS of \eqref{eq:conformal block four-punctured sphere Regge symmetry} simply becomes
\be 
z^{Q(\alpha_3-\frac{Q}{2})} (1-z)^{Q(\alpha_1-\frac{Q}{2})}\ .
\ee
Notice now that
\be 
\tilde{\alpha}_1+\tilde{\alpha}_2+\tilde{\alpha}_3+(Q-\tilde{\alpha}_4)=Q+2\alpha_4=Q\ . \label{eq:Coulomb gas background charge saturation}
\ee
Notice also that we can choose the Liouville momentum of the internal state as
\be 
\alpha=Q-\alpha_3=\tilde{\alpha}_3+Q-\tilde{\alpha}_4\ . \label{eq:Coulomb gas momentum conservation}
\ee
We can construct a stress tensor out of a free boson with a background charge $Q$. Since the conformal block on the RHS obeys momentum conservation at both vertices, it is actually a $\mathrm{U}(1)$ conformal block with momenta $\tilde{\alpha}_1$, $\tilde{\alpha}_2$, $\tilde{\alpha}_3$ and $Q-\tilde{\alpha}_4$ and thus takes the form
\begin{multline} 
z^{-2(\tilde{\alpha}_3-\frac{Q}{2})(\tilde{\alpha}_4-\frac{Q}{2})}(1-z)^{-2(\tilde{\alpha}_1-\frac{Q}{2})(\tilde{\alpha}_4-\frac{Q}{2})}z^{-2\tilde{\alpha}_3 (Q-\tilde{\alpha}_4)}(1-z)^{-2\tilde{\alpha}_1 (Q-\tilde{\alpha}_4)} \\
=z^{Q(\alpha_3-\frac{Q}{2})} (1-z)^{Q(\alpha_1-\frac{Q}{2})}\ ,
\end{multline}
which verifies \eqref{eq:conformal block four-punctured sphere Regge symmetry} in the simplest case. One can similarly consider other degenerate conformal blocks, such as $\alpha_4=-\frac{b}{2}$, in which case, the conformal blocks on the RHS admit a Coulomb gas integral representation and is thus a generalized hypergeometric function. The conformal block on the LHS satisfies a BPZ differential equation and is thus also a generalized hypergeometric function. It was shown in \cite{Fateev:2007tt} that the Coulomb gas integrals exhibit the desired symmetry \eqref{eq:conformal block four-punctured sphere Regge symmetry}. This can be used to show \eqref{eq:conformal block four-punctured sphere Regge symmetry} for all degenerate conformal blocks.

This is then enough to conclude that \eqref{eq:conformal block four-punctured sphere Regge symmetry} holds for \emph{all} conformal blocks. Indeed, considering the expansion in small $z$ of the conformal block as in eq.~\eqref{eq:conformal block four-punctured sphere abstract expansion}, every coefficient is a rational function of $\Delta_j$, $\Delta$ and $c$. Agreement of these rational functions between the LHS and RHS of \eqref{eq:conformal block four-punctured sphere Regge symmetry} for infinitely degenerate conformal blocks implies that the two rational functions must in fact be equal. Thus all terms in the small crossratio expansion agree which implies that the conformal blocks agree and thus \eqref{eq:conformal block four-punctured sphere Regge symmetry} indeed holds.

\section{Special functions} \label{app:special functions}
This appendix contains many details about the necessary special functions $\Gamma_b(z)$ and $S_b(z)$. Since they are ubiquitous in 2d CFT, we have tried to make this appendix somewhat more complete than necessary for our goals, in the hope that it can serve as a useful reference. In particular, we prove all of the properties we state.
\subsection{Double Gamma function}
Recall that we are assuming $\Re(b)>0$ and that the double Gamma function is uniquely characterized by the functional equation \eqref{eq:Gammab functional equations} together with the normalization $\Gamma_b(\frac{Q}{2})=1$.
\begin{enumerate}
    \item $\Gamma_b(z)$ has the integral representation in the half plane $\Re(z)>0$
    \be  
        \log \Gamma_b(z)=\int_0^\infty \frac{\mathrm{d}t}{t}\left(\frac{\mathrm{e}^{\frac{t}{2}(Q-2z)}-1}{4 \sinh(\frac{bt}{2})\sinh(\frac{t}{2b})}-\frac{1}{8}\left(Q-2z\right)^2 \mathrm{e}^{-t}-\frac{Q-2z}{2t}\right)\ . \label{eq:Gammab integral representation}
    \ee
    \begin{proof}
    It is trivial to check that the integral converges for $\Re(z)>0$.
    The integral representation is also manifestly invariant under $b \to b^{-1}$ and depends continuously on $b$. Thus we only need to check the functional equation in $b$ since this determines $\Gamma_b(z)$ fully up to overall normalization as explained in Section~\ref{subsec:special functions}. The normalization is fixed by requiring $\Gamma_b(\frac{Q}{2})=1$, which is obvious from the integral \eqref{eq:Gammab integral representation}.
    Thus it remains to show the functional equation. From the integral \eqref{eq:Gammab integral representation}, we have
    \begin{align}
        \log \Gamma_b(z+b)-\log \Gamma_b(z)=\int_0^\infty \frac{\mathrm{d}t}{t} \left(-\frac{\mathrm{e}^{\frac{t}{2}(\frac{1}{b}-2z)}}{2 \sinh(\frac{t}{2b})}+\frac{1-2b z}{2} \mathrm{e}^{-t}+\frac{b}{t}\right) \label{eq:difference Gammabz+b Gammab}
    \end{align}
    We can remove the $b$-dependence of the integral as follows. Let us first consider the simpler integral
    \be  
        f(b)=\int_0^\infty \frac{\mathrm{d}t}{t} \ \left(\mathrm{e}^{-b t}-\mathrm{e}^{-t}\right)\ , \label{eq:f integral}
    \ee
    which is clearly absolutely convergent. We have $f(1)=0$ and
    \be  
        f'(b)=-\int_0^\infty\!\! \mathrm{d}t\ \mathrm{e}^{-b t}=-\frac{1}{b}\ .
    \ee
    Thus $f(b)=-\log(b)$.
    Let us now rescale $t \to bt$ in the right hand side of \eqref{eq:difference Gammabz+b Gammab}, leading to
    \begin{align}  
        \log \Gamma_b(z+b)-\log \Gamma_b(z)&=\int_0^\infty \frac{\mathrm{d}t}{t} \left(-\frac{\mathrm{e}^{\frac{t}{2}(1-2bz)}}{2 \sinh(\frac{t}{2})}+\frac{1-2b z}{2} \mathrm{e}^{-bt}+\frac{1}{t}\right) \\
        &=\int_0^\infty \frac{\mathrm{d}t}{t} \left(-\frac{\mathrm{e}^{\frac{t}{2}(1-2bz)}}{2 \sinh(\frac{t}{2})}+\frac{1-2b z}{2} \mathrm{e}^{-t}+\frac{1}{t}\right) \nonumber\\
        &\qquad+\log(b)\left(bz-\frac{1}{2}\right)\ .
    \end{align}
    The remaining integral only depends on the combination $bz$. Let us denote
    \be  
        g(z)=\int_0^\infty \frac{\mathrm{d}t}{t} \left(-\frac{\mathrm{e}^{\frac{t}{2}(1-2z)}}{2 \sinh(\frac{t}{2})}+\frac{1-2z}{2} \mathrm{e}^{-t}+\frac{1}{t}\right)\ . \label{eq:g integral}
    \ee
    We will be done if we show that $g(z)=\frac{1}{2}\log(2\pi)-\log \Gamma(z)$. A convenient way to do this is by using the so-called Bohr-Mollerup theorem, which asserts that $\Gamma(z)$ is the unique function that satisfies $\Gamma(z)=1$, $\Gamma(z+1)=z\Gamma(z)$ and $\log \Gamma(z)$ is convex for $z>0$. Let us check these three properties. We have
    \be  
        g(z+1)-g(z)=\int_0^\infty \frac{\mathrm{d}t}{t} \left(\mathrm{e}^{-z t}-\mathrm{e}^{-t}\right)=-\log z\ ,
    \ee
    where we used the evaluation of the integral \eqref{eq:f integral} above. Let us next check convexity of $-g(z)$. We have
    \be  
        -g''(z)=\int_0^\infty \frac{\mathrm{d}t\ t \,\mathrm{e}^{t(1-z)}}{\mathrm{e}^t-1}>0\ ,
    \ee
    since the integrand is positive. This shows already that $g(z)=a-\log\Gamma(z)$ for some constant $a$. To fix it, we compute $g(\frac{1}{2})$, which gives
    \begin{align}
        g(\tfrac{1}{2})&=\int_0^\infty \frac{\mathrm{d}t}{t} \left(-\frac{1}{2 \sinh(\frac{t}{2})}+\frac{1}{t}\right) \\
        &=\int_{-\infty+i 0^+}^{\infty+i 0^+} \frac{\mathrm{d}t}{t} \left(-\frac{1}{4 \sinh(\frac{t}{2})}+\frac{1}{2t}\right) \\
        &=-\frac{\pi i}{2} \sum_{n=1}^\infty \Res_{t=2\pi i n} \frac{1}{t\, \sinh(\frac{t}{2})} \\
        &=\frac{1}{2} \sum_{n=1}^\infty \frac{(-1)^{n+1}}{n}=\frac{1}{2}\log(2)\ ,
    \end{align}
    which indeed shows that $g(z)=\frac{1}{2}\log(2\pi)-\log\Gamma(z)$. This shows that $\Gamma_b(z)$ as defined by the right hand side of \eqref{eq:Gammab integral representation} satisfies the functional equation
    \be  
        \frac{\Gamma_b(z+b)}{\Gamma_b(z)}=\frac{\sqrt{2\pi} b^{bz-\frac{1}{2}}}{\Gamma(bz)}\ ,
    \ee
    which shows the equality to the double Gamma function.
    \end{proof}
    \item $\Gamma_b(z)$ satisfies
        \be  
            \frac{\Gamma_b(z+m b+n b^{-1})}{\Gamma_b(z)}=\frac{(2\pi)^{\frac{m+n}{2}}b^{(mb-n b^{-1})z+\frac{1}{2}m(m-1)b^2-\frac{1}{2}n(n-1) b^{-2}+m n+\frac{1}{2}(n-m)}}{\prod_{j=0}^{m-1} \Gamma(bz+b^2j+n)\prod_{k=0}^{n-1}\Gamma(b^{-1} z+b^{-2}k)}\ . \label{eq:Gammab iterated functional equation}
        \ee
        \begin{proof}
            This formula follows directly from induction. We first apply the functional equation $m$ times to remove the $+m b$ shift and then the functional equation in $b^{-1}$ $n$ times to remove the shift $+n b^{-1}$.
        \end{proof}
    \item $\Gamma_b(z)$ has poles for $z=-m b-n b^{-1}$ ($m,\,n \in \ZZ_{\ge 0}$) and no zeros. The residues are
    \begin{multline}  
        \Res_{z=-m b-n b^{-1}} \Gamma_b(z)
            =(2\pi)^{-\frac{m+n+2}{2}}b^{\frac{1}{2}m(m+1)b^2-\frac{1}{2}n(n+1) b^{-2}-m n-\frac{1}{2}(n-m)}\\
            \times \prod_{j=1}^{m} \Gamma(-b^2j)\prod_{k=1}^{n}\Gamma(-b^{-2}k-m)\Gamma_b(Q)\ . \label{eq:Gammab residues}
    \end{multline}
    \begin{proof}
        From the integral representation \eqref{eq:Gammab integral representation} we see immediately that $\Gamma_b(z)$ has neither poles nor zeros in the half plane $\Re(z)>0$, since the integral is finite and hence gives a non-zero finite number upon exponentiation. Solving \eqref{eq:Gammab iterated functional equation} for $\Gamma_b(z)$ shows also that $\Gamma_b(z)$ cannot be zero for any $z$, since the term on the right hand side of \eqref{eq:Gammab iterated functional equation} does not have any poles. The poles of the Gamma function on the right hand side of \eqref{eq:Gammab iterated functional equation} lead to the poles of $\Gamma_b(z)$, which are hence indeed at $z=-m b-n b^{-1}$. Let us set $z=-m b-n b^{-1}+\varepsilon$ in \eqref{eq:Gammab iterated functional equation} and let $\varepsilon \to 0$. Then we obtain
        \begin{align} 
            \frac{\Res_{z=0} \Gamma_b(z)}{\Res_{z=-m b-n b^{-1}} \Gamma_b(z)}&=\frac{(2\pi)^{\frac{m+n}{2}}b^{-\frac{1}{2}m(m+1)b^2+\frac{1}{2}n(n+1) b^{-2}+m n+\frac{1}{2}(n-m)}}{\prod_{j=0}^{m-1} \Gamma(b^2(j-m))\prod_{k=0}^{n-1}\Gamma(b^{-2}(k-n) -m)} \\
            &=\frac{(2\pi)^{\frac{m+n}{2}}b^{-\frac{1}{2}m(m+1)b^2+\frac{1}{2}n(n+1) b^{-2}+m n+\frac{1}{2}(n-m)}}{\prod_{j=1}^{m} \Gamma(-b^2j)\prod_{k=1}^{n}\Gamma(-b^{-2}k-m)} \ . \label{eq:Gammab residue ratio}
        \end{align}
        It remains to compute $\Res_{z=0} \Gamma_b(z)$. For this, we again set $z=\varepsilon$ in \eqref{eq:Gammab iterated functional equation}, but also put $m=n=1$. This immediately leads to 
        \be  
            \Res_{z=0} \Gamma_b(z)=\frac{\Gamma_b(Q)}{2\pi}\ . \label{eq:Gammab residue 0}
        \ee
        Taking \eqref{eq:Gammab residue ratio} and \eqref{eq:Gammab residue 0} together finishes the computation.
    \end{proof}
    \item Semiclassical limit:
    \be  
        \log \Gamma_b(b^{-1}\zeta)=\frac{1}{b^2}\left(\frac{2\zeta-1}{4} \log(2\pi)+\psi^{(-2)}(\tfrac{1}{2})-\psi^{(-2)}(z)\right)+\mathcal{O}\left(\frac{\log b}{b^2}\right)\ . \label{eq:Gammab semiclassical limit}
    \ee
    for $b \to 0$. Here we use the polygamma function $\psi^{(n)}(z)$.
    \begin{proof}
        We plug in $z=b^{-1}\zeta$ in the integral representation \eqref{eq:Gammab integral representation}. We then rescale $t \to tb$ and change the exponential in the second term using \eqref{eq:f integral}. Using dominated convergence, we can then exchange the limit with the integral:
        \begin{align}
            \log \Gamma_b(b^{-1}\zeta)&=\int_0^\infty \frac{\mathrm{d}t}{t}\left(\frac{\mathrm{e}^{\frac{t}{2}(Qb-2\zeta)}-1}{4 \sinh(\frac{b^2t}{2})\sinh(\frac{t}{2})}-\frac{(Qb-2\zeta)^2}{8b^2} \mathrm{e}^{-t b}-\frac{Qb-2\zeta}{2t b^2}\right) \\
            &=\int_0^\infty \frac{\mathrm{d}t}{t}\left(\frac{\mathrm{e}^{\frac{t}{2}(Qb-2\zeta)}-1}{4 \sinh(\frac{b^2t}{2})\sinh(\frac{t}{2})}-\frac{(Qb-2\zeta)^2}{8b^2} \mathrm{e}^{-t}-\frac{Qb-2\zeta}{2t b^2}\right) \nonumber\\
            &\qquad+\frac{\left(Qb-2\zeta\right)^2}{8b^2}\log(b) \\
            &=\frac{1}{b^2}\int_0^\infty \frac{\mathrm{d}t}{t}\left(\frac{\mathrm{e}^{\frac{t}{2}(1-2\zeta)}-1}{2t\sinh(\frac{t}{2})}-\frac{(1-2\zeta)^2}{8} \mathrm{e}^{-t}-\frac{1-2\zeta}{2t}\right)\nonumber\\
            &\qquad+\frac{\left(1-2\zeta\right)^2}{8b^2}\log(b)-\frac{2\zeta-1}{4}\log(b)+\mathcal{O}(1)\ .
        \end{align}
        It hence remains to compute the integral
        \be  
            h(z)=\int_0^\infty \frac{\mathrm{d}t}{t}\left(\frac{\mathrm{e}^{\frac{t}{2}(1-2z)}-1}{2\sinh(\frac{t}{2})}-\frac{(1-2z)^2}{8} \mathrm{e}^{-t}-\frac{1-2z}{2t}\right)\ .
        \ee
        We have
        \be  
            h'(z)=\int_0^\infty\frac{\mathrm{d}t}{t}\left(-\frac{\mathrm{e}^{\frac{t}{2}(1-2z)}}{2 \sinh(\frac{t}{2})}+\frac{1-2z}{2} \mathrm{e}^{-t}+\frac{1}{t}\right)=\frac{\log(2\pi)}{2}-\log \Gamma(z)\ ,
        \ee
        where we used the integral \eqref{eq:g integral} that appeared in the proof above as well as its evaluation as discussed there. The integral of the logarithm of the Gamma function is by definition the polygamma function $\psi^{(-2)}(z)$. $\psi^{(-2)}(z)$ is normalized such that
        It thus follows
        \begin{align}
            h(z)=\frac{z}{2}\log(2\pi)-\psi^{(-2)}(z)+a
        \end{align}
        for some constant $a$. To determine it, we can put $z=\frac{1}{2}$. We have $h(\tfrac{1}{2})=0$ by definition. Hence
        \be  
            h(z)=\frac{2z-1}{4} \log(2\pi)+\psi^{(-2)}(\tfrac{1}{2})-\psi^{(-2)}(z)\ ,
        \ee
        thus finishing the proof.
    \end{proof}
    \item Multiplication formula:
    \be 
        \Gamma_b(z)=\lambda_{m,n,b} \, (mn)^{\frac{1}{4}z(Q-z)} \prod_{r=0}^{m-1} \prod_{s=0}^{n-1} \Gamma_{\frac{b \sqrt{m}}{\sqrt{n}}}\left(\frac{z+r b+s b^{-1}}{\sqrt{mn}} \right)\ . \label{eq:Gamma_b multiplication formula}
    \ee
    Notice in particular the special case $m=n$.
    \begin{proof}
    This is simple to prove, since it merely amounts to showing that the RHS satisfies the functional equation \eqref{eq:Gammab functional equations}. Let us write temporarily $\hat{\Gamma}_b$ for the RHS. Then
        \begin{align} 
            \frac{\hat{\Gamma}_b(z+b)}{\hat{\Gamma}_b(z)}&=(mn)^{\frac{1}{4}-\frac{b z}{2}} \prod_{s=0}^{n-1} \frac{\Gamma_{\frac{b\sqrt{m}}{\sqrt{n}}}(\frac{z+m b+s b^{-1}}{\sqrt{mn}})}{\Gamma_{\frac{b\sqrt{m}}{\sqrt{n}}}(\frac{z+s b^{-1}}{\sqrt{mn}})} \\
            &=(mn)^{\frac{1}{4}-\frac{b z}{4}} (2\pi)^{\frac{n}{2}} \left(\frac{b^2m}{n}\right)^{\frac{bz}{2}-\frac{1}{4}} \prod_{s=0}^{n-1} \frac{1}{\Gamma(\frac{bz+s}{n})} \\
            &=\frac{\sqrt{2\pi} b^{bz-\frac{1}{2}}}{\Gamma(bz)}\ ,
        \end{align}
        where we used the multiplication theorem of the Euler Gamma function in the last line. The same argument holds for $b \to b^{-1}$ and since this characterizes the double Gamma function unique, the claim follows.
    \end{proof}
\end{enumerate}
\subsection{Double Sine function}
The most important special function for our purposes is the double sine function. We can define it directly by
\be  
S_b(z)=\frac{\Gamma_b(z)}{\Gamma_b(Q-z)}\ .
\ee 
The function $S_b(z)$ satisfies the following properties:
\begin{enumerate}
    \item Integral representation for $0<\Re(z)<Q$:
    \begin{align} 
    \log S_b(z)&=\int_0^\infty \frac{\mathrm{d}t}{t}\left(\frac{\sinh((\frac{Q}{2}-z)t)}{2 \sinh(\frac{b t}{2}) \sinh(\frac{t}{2b})}-\frac{Q-2z}{t}\right) \\
    &=\frac{1}{4}\int_{\RR+i 0^+} \frac{\mathrm{d}t}{t}\frac{\sinh((\frac{Q}{2}-z)t)}{ \sinh(\frac{b t}{2}) \sinh(\frac{t}{2b})}\ , \label{eq:S_b definition}
    \end{align}
    \begin{proof}
        The first line follows directly from the integral representation of the double Gamma function \eqref{eq:Gammab integral representation}.    
        In the second line, we extended the integral to the whole real by using the the original function was even and shifted the contour slightly into the complex plane. The second term then gives a vanishing contribution and can be omitted.
    \end{proof}
    \item Reflection property: $S_b(Q-z)=S_b(z)^{-1}$.
    \item Reality: $\overline{S_b(z)}=S_b(\overline{z})$.
    \item Functional equation:
    \be 
        S_b(z+b^{\pm 1})=2 \sin(\pi b^{\pm 1} z) S_b(z)\ . \label{eq:S_b functional equation}
    \ee
    This gives the double sine function its name.
    \begin{proof}
        From the functional equation of the double Gamma function, we immediately obtain
        \be  
            \frac{S_b(z+b)}{S_b(z)}=\frac{\Gamma_b(z+b)\Gamma_b(b+b^{-1}-z)}{\Gamma_b(z)\Gamma_b(b^{-1}-z)}=\frac{2\pi\, b^{bz+b(b^{-1}-z)-1}}{\Gamma(\pi b z)\Gamma(\pi b(b^{-1}-z))}=2\sin(\pi b z)\ ,
        \ee
        where we used the Euler reflection formula for the Gamma function.
    \end{proof}
    \item Zeros: $z \in Q+b \ZZ_{\ge 0}+b^{-1} \ZZ_{\ge 0}$.
    \item Poles: $z \in -b \ZZ_{\ge 0}-b^{-1} \ZZ_{\ge 0}$.
    \item Classical limit: $z=b^{-1} \zeta$, $0<\Re \zeta<1$, $b \to 0$:
    \be 
        \log S_b(b^{-1} \zeta) = \frac{i}{4\pi b^2}\left(\Li_2(\mathrm{e}^{2\pi i \zeta})-\Li_2(\mathrm{e}^{-2\pi i \zeta})\right)+\mathcal{O}(1)\ , \label{eq:S_b classical limit}
    \ee
    with the standard choice of branch cut (running from $(1,\infty)$) for the dilogarithm. To access other values, we use the classical limit of the functional equation:
    \be 
      S_b(b^{-1}(\zeta+1))= S_b(b^{-1}\zeta)\begin{cases}
      i \mathrm{e}^{-\pi i b^{-2} \zeta}\ , \qquad \Im \zeta>0\ , \\
      -i \mathrm{e}^{\pi i b^{-2} \zeta}\ , \qquad \Im \zeta<0
      \end{cases}
    \ee
    \begin{proof}
           This is a direct consequence of \eqref{eq:Gammab semiclassical limit}. We should however rederive it to make sure that the error term is of order $1$. First note that
\begin{align}
    \log S_b(b^{-1} \zeta)&=\frac{1}{2}\int_{\RR+i 0^+} \frac{\mathrm{d}t}{t} \frac{\sinh((1-2\zeta)\frac{t}{2b})}{b t \sinh(\frac{t}{2b})}\, \left(1+\mathcal{O}(b^2)\right) \\
    &=\frac{1}{4b^2} \int_{\RR+i 0^+} \frac{\mathrm{d}t}{t^2} \frac{\sinh((1-2\zeta) t)}{\sinh(t)}+\mathcal{O}(1)\ .
\end{align}
We could relate this to the integral \eqref{eq:g integral}, but let us compute it another way. We can pull off the contour to $t \to +i \infty$, which causes us to pick up all the poles at $t \in  \pi i \ZZ_{\ge 1}$. In this process, we should be careful with the choice of branch cut. Assume first that $\Im \zeta<0$ and write $\log S_b(b^{-1} \zeta) \sim \frac{1}{b^2}\left(f(\zeta)-f(1-\zeta)\right)$ with  
\be 
f(\zeta)=\frac{1}{8} \int_{\RR+i 0^+} \frac{\mathrm{d}t}{t^2} \frac{\mathrm{e}^{(1-2\zeta) t}}{\sinh(t)}\ .
\ee
Let us determine $f(\zeta)$. Let us first assume that $\Im \zeta<0$, then we can pull the contour off to $t \to +i \infty$ and pick up all the poles in the process (including at the origin.
We obtain
\begin{align}
    f(\zeta)&=-\frac{i}{4\pi} \sum_{n=1}^\infty \frac{\mathrm{e}^{-2\pi i n \zeta}}{n^2} +\mathcal{O}(1)\\
    &=-\frac{i}{4\pi} \Li_2\left(\mathrm{e}^{-2\pi i \zeta}\right)\ ,
\end{align}
For $\Im \zeta<0$, we should pull off the contour instead to $t \to -i \infty$ and pick up all the other poles. This gives
\be 
    f(\zeta)=-\frac{\pi i}{12}(1-6 \zeta+6 \zeta^2)+\frac{i}{4\pi} \Li_2\left(\mathrm{e}^{-2\pi i \zeta}\right)\ .
\ee
We also have the standard dilogarithm identity with the standard choice of branch cut running from $(1,\infty)$
\be 
\Li_2\left(\frac{1}{z}\right)=-\Li_2(z)-\frac{\pi^2}{6}-\frac{1}{2}\log^2(-z)\ ,
\ee
which gives $f(\zeta)=-\frac{i}{4\pi} \Li_2(\mathrm{e}^{-2\pi i \zeta})$ also for $\Im \zeta>0$. Thus the result \eqref{eq:S_b classical limit} follows.
    \end{proof}
    \item Normalization: $\Res_{z=0} S_b(z)=\frac{1}{2\pi}$ and $S_b(\frac{Q}{2})=1$.  
    \begin{proof}
        The residue follows immediately from the corresponding residue of $\Gamma_b(z)$ at $z=0$.
    \end{proof}
    \item Infinite product representation:
    \be 
        S_b(z)=\mathrm{e}^{-\frac{\pi i}{2}(z^2-Qz+\frac{1}{6}(Q^2+1))}\frac{\prod_{m=1}^\infty 1-\mathrm{e}^{2\pi i b^{-1} z-2\pi i m b^{-2}}}{\prod_{m=0}^\infty 1-\mathrm{e}^{2\pi i b z +2\pi i m b^2}} \ . \label{eq:S_b product representation}
    \ee
    This representation is valid for $\Im (b^2)>0$.
    \begin{proof}
Let us denote the function defined by \eqref{eq:S_b product representation} by $\hat{S}_b(z)$. Let us check the functional equation. We have
\begin{align}
    \frac{\hat{S}_b(z+b)}{\hat{S}_b(z)}=i\, \mathrm{e}^{-\pi i b z} (1-\mathrm{e}^{2\pi i bz})=2 \sin(\pi b z)\ , \\
    \frac{\hat{S}_b(z+b^{-1})}{\hat{S}_b(z)}=i\, \mathrm{e}^{-\pi i b^{-1} z}(1-\mathrm{e}^{2\pi i b^{-1}z})=2 \sin(\pi b^{-1} z)\ .
\end{align}
This shows that $\frac{\hat{S}_b(z)}{S_b(z)}$ is an elliptic function with periods $b$ and $b^{-1}$. From the product representation, it is also obvious that $\hat{S}_b(z)$ has the correct set of zeros and poles. Thus $\frac{\hat{S}_b(z)}{S_b(z)}$ is an elliptic function without zeros and poles and thus constant. To determine the constant, we compute
\begin{align}
    \hat{S}_b\left(\frac{Q}{2}\right)&=\mathrm{e}^{\frac{\pi i}{24}(b^2+b^{-2})} \prod_{m=1}^\infty \frac{1+ \mathrm{e}^{-2\pi i b^{-2} (m+\frac{1}{2})}}{1+\mathrm{e}^{2\pi i b^2(m+\frac{1}{2})}} \\
    &= \sqrt{\frac{\vartheta_3(-b^{-2})\eta(b^{2})}{\vartheta_3(b^{2})\eta(-b^{-2})}}=1\ ,
\end{align}
where we used
\be 
\frac{\vartheta_3(\tau)}{\eta(\tau)}=\mathrm{e}^{-\frac{\pi i \tau }{12}}\prod_{n=0}^\infty (1+\mathrm{e}^{2\pi i(n+\frac{1}{2})\tau})^2
\ee
and the S-modular transformation of $\vartheta_3(\tau)$ and $\eta(\tau)$. This shows that $S_b(z)=\hat{S}_b(z)$.
    \end{proof}
    \item Asymptotics (for $b>0$):
    \be 
        S_b(z) \to \mathrm{e}^{\pm \frac{\pi i}{2}(z^2-Qz+\frac{1}{6}(Q^2+1))} \quad\text{for} \Im(z) \to \mp \infty\ . \label{eq:S_b asymptotics}
    \ee
    \begin{proof}
        We next analyze the asymptotics \eqref{eq:S_b asymptotics}. For this we use the product representation \eqref{eq:S_b product representation}. Thus rotate $b$ slightly such that $\Im(b^2)>0$ and we can use the product representation. When taking the limit $\Im(z) \to +\infty$, all the terms in the infinite product become 1 and we conclude immediately
\be 
\Im(z) \to \infty:\qquad S_b(z) \to \mathrm{e}^{-\frac{\pi i}{2}(z^2-Qz+\frac{1}{6}(Q^2+1))}
\ee
For $\Im(z) \to -\infty$, we can use the reflection relation, which gives immediately the result. 
    \end{proof}
    \item Alternative integral representation \cite{Woronowicz}:
    \be 
        \log S_b(z)=\frac{i}{2\pi} \int_{\RR+\frac{\pi i}{b^2}} \frac{\mathrm{d}t}{1-\mathrm{e}^t}\, \log \big(1-\mathrm{e}^{-2\pi i b z+t b^2}\big)+\tfrac{\pi i}{2}\big(z^2-Qz+\tfrac{Q^2+1}{6}\big)\ .  \label{eq:S_b alternative integral representation}
    \ee
    The contour is valid for $z$ purely imaginary and small positive $b$, otherwise it has to be deformed smoothly as we vary $b$ and $z$.
    \begin{proof}
        It is again sufficient to check that this expression satisfies the correct functional equation and the correct normalization. Let us temporarily denote the function defined by \eqref{eq:S_b alternative integral representation} by $\hat{S}_b(z)$. We want to demonstrate that it equals $S_b(z)$. Notice that the integrand has branch cuts for
\be 
t \in \frac{2\pi i z}{b}+\frac{2\pi i n}{b^2}+\RR_{\ge 0}\ .
\ee
By definition, the contour $t \in \RR+\frac{\pi i}{b^2}$ is valid for $z$ purely imaginary and small $b$. This means that it passes in between the branch cuts labelled by $n=0$ and $n=1$.
It also has poles for $t \in 2\pi i \ZZ$. Let us check the functional equation in $z \to z+b^{-1}$. This formally leave the integrand invariant, but the contour gets shifted. When varying $z$, the branch cuts move downward in the complex plane and after the replacement $z \to z+b^{-1}$, the contour now runs in between the branch cuts that we numbered before with $n=-1$ and $n=0$. Thus $\log \hat{S}_b(z+b^{-1}) -\log \hat{S}_b(z)$ is computed by the Pochhammer contour surrounding the branch cut $n=0$ counterclockwise. It gives the discontinuity of the log, which is $-2\pi i$. Thus
\begin{align}
   \frac{\hat{S}_b(z+b^{-1})}{\hat{S}_b(z)}&= \exp\left(\int_{\frac{2\pi i z}{b}+\RR_{\ge 0}} \frac{\mathrm{d}t}{1-\mathrm{e}^t}-\frac{\pi i}{2}+\pi i b^{-1}z \right) \\
    &=2 \sin(\pi b^{-1} z)
\end{align}
as desired. We also have to check the functional equation in $z \to z+b$. We note that the replacement $z \to z+b $ can be compensated by shifting the contour in $t$ downward, which picks up the pole at $t=0$. Thus we have
\begin{align}
    \frac{\hat{S}_b(z+b)}{\hat{S}_b(z)}&=\exp\left(-\log(1-\mathrm{e}^{-2\pi i b z})-\frac{\pi i}{2}+\pi i b z\right) \\
    &=2 \sin(\pi b z)\ .
\end{align}
It remains to compute the normalization constant. For this, we compute the asymptotics as $\Im(z) \to -\infty$ and compare with the known asymptotics of $S_b(z)$ of eq.~\eqref{eq:S_b asymptotics}. We have trivially for $b>0$
\be 
\Im(z) \to -\infty:\qquad \hat{S}_b(z)\to \mathrm{e}^{\frac{\pi i}{2}(z^2-Qz+\frac{1}{6}(Q^2+1))}\ .
\ee
Comparing with \eqref{eq:S_b asymptotics}, we conclude that the normalizations agree and hence $S_b(z)=\hat{S}_b(z)$.
    \end{proof}
    \item Multiplication formula:
    \be 
		S_b(z)=  \prod_{r=0}^{m-1} \prod_{s=0}^{n-1} S_{\frac{b \sqrt{m}}{\sqrt{n}}}\left(\frac{z+r b+s b^{-1}}{\sqrt{mn}} \right)  \ .\label{eq:S_b multiplication formula}
    \ee
    \begin{proof}
    This is a direct consequence of \eqref{eq:Gamma_b multiplication formula}.
    \end{proof}
    \item Relation with Faddeev quantum dilogarithm \cite{Faddeev:1995nb}
    \be 
        \log \Phi_b(z) \equiv \int_C \frac{\mathrm{d}w\, \mathrm{e}^{-2izw}}{4 \sinh(wb) \sinh(w b^{-1}) w}\ ,
    \ee
    where the contour passes above the pole at the origin. We have
    \be 
        \Phi_b(z)=S_b\big(iz+\tfrac{Q}{2}\big)\mathrm{e}^{\frac{\pi i z^2}{2}+\frac{\pi i}{24}(b^2+b^{-2})}\ .
    \ee
    The relation to the dilogarithm in the classical limit \eqref{eq:S_b classical limit} gives the Faadeev quantum dilogarithm its name.
    \begin{proof}
        This follows immediately by comparing the integral representations.
    \end{proof}
    \item Log convexity:
    \be  
        0<z<Q\,:\quad \frac{\mathrm{d}^2}{\mathrm{d}z^2} \log S_b(z)>0
    \ee
    \begin{proof}
        This is trivial to show from the integral representation. For $0<z<Q$ the integral converges absolutely and we can compute
        \be  
            \frac{\mathrm{d}^2}{\mathrm{d}z^2} \log S_b(z)=\int_0^\infty \mathrm{d}t\ \frac{t \sinh((\frac{Q}{2}-z)t)}{2 \sinh(\frac{bt}{2})\sinh(\frac{t}{2b})}>0\ ,
        \ee
        since the integrand is everywhere non-negative.
    \end{proof}
    \item Rational limit:
    \be  
        \lim_{b \to 0} S_b(b \zeta)2\pi b(2\pi b^2)^{-\zeta}=\, \Gamma(\zeta)\ . \label{eq:S_b rational limit}
    \ee
    \begin{proof}
        To show this, we again use the Bohr-Mollerup theorem. Let us denote
        \be  
        f(\zeta)=S_b(b\zeta) 2\pi b(2\pi b^2)^{-\zeta}
        \ee
        We first notice that the functional equation becomes
        \be  
            \frac{f(\zeta+1)}{f(\zeta)}=\frac{S_b(b\zeta+b)}{2\pi b^2S_b(b \zeta)}=\frac{2\sin(\pi b^2 \zeta)}{2\pi b^2} \longrightarrow \zeta\ ,   
        \ee
        which is indeed the functional equation of the Gamma function. Log convexity of $f(\zeta)$ is a direct consequence of log convexity of $S_b(z)$. Thus it only remains to compute the normalization, which we do by comparing the residue at $\zeta=0$. 
    \end{proof}
  
\end{enumerate}

\section{More about \texorpdfstring{$\boldsymbol{b}$}{b}-deformed hypergeometric functions} \label{app:hypergeometric functions}
\subsection{The \texorpdfstring{$b$}{b}-deformed Euler integral}
Following Spiridonov \cite{SpiridonovElliptic}, we consider the following family of integrals,
\be 
\int \frac{\d z}{2i}\  \frac{\prod_{i=1}^6 S_b(a_i \pm z)}{S_b(\pm 2z)}\ , \label{eq:b-deformed Euler integral}
\ee
where the parameters $a_1,\dots,a_6$ are constrained to obey the balancing condition
\be 
\sum_{i=1}^6 a_i=Q\ ,
\ee
but are otherwise free. The contour runs from $-i\infty$ to $+i \infty$, while avoiding all wedges of poles along the way.\footnote{Most of the subject is developed for a further generalization where all double sine functions are replaced by elliptic Gamma functions. Almost everything of the following discussion carries through to that case without modifications. In that context, $S_b(z)$ is often called the hyperbolic gamma function $\gamma^{(2)}(z;b,b^{-1})$. The elliptic version has played a very important role in the study of supersymmetric indices of $4d$ gauge theories \cite{Kinney:2005ej, Dolan:2008qi}.}

The integral \eqref{eq:b-deformed Euler integral} admits an explicit evaluation,
\be 
\int \frac{\d z}{2i}\  \frac{\prod_{i=1}^6 S_b(a_i \pm z)}{S_b(\pm 2z)}= \prod_{1 \le j <k \le 6} S_b(a_j+a_k)\ . \label{eq:Euler integral formula}
\ee
We refer to this integral formula as the ($b$-deformed) Euler integral identity. Indeed, it can be degenerated to the classical Euler integral formula using the rational limit \eqref{eq:S_b rational limit} of $S_b(z)$.
Let us explain the proof of this miraculous identity.
\begin{proof}
The proof we give is the trigonometric analogue of the proof given in \cite{Spiridonov:proofs}. 
Let us denote the integrand divided by the RHS by $\rho(z;a_1,\dots,a_5)$. We set $a_6=Q-a_{12345}$. As in the main text, $a_I=\sum_{i \in I} a_i$ for some subset $I \subset \{1,2,\dots,6\}$. Let us analyze $\rho(z;a_1)$ while keeping all other variables fixed. We claim that $\rho(z;a_1)$ satisfies
\begin{align}
    \rho(z;a_1+b)-\rho(z;a_1)=g(z-b) \rho(z-b;a_1)-g(z) \rho(z;a_1)
\end{align}
with 
\be 
g(z)=\frac{\prod_{j=1}^5 \sin(\pi b(z+a_j)) \sin(2\pi b a_{12345})}{\prod_{j=2}^5 \sin(\pi ba_{1j}) \sin(2\pi b z) \sin(\pi b (z+a_{12345}))}\ .
\ee
This is straightforward to check e.g.\ in Mathematica, since after dividing both sides of the equation by $\rho(z;a_1)$, this equality boils down to a trigonometric identity. We thus have
\be 
\int_{i \RR} \frac{\mathrm{d}z}{2i}\ \rho(z;a_1+b)-\int_{i \RR} \frac{\mathrm{d}z}{2i}\ \rho(z;a_1)=\left(\int_{i\RR-b}-\int_{i \RR}\right) \frac{\d z}{2i}\ g(z) \rho(z;a_1)\ ,
\ee
The contour on the right hand side can then be shifted back, so that the right hand side vanishes.
 Essentially by definition we do not pick up any poles on the RHS, since our contour avoids by assumption all poles. 

Thus we conclude that the ratio of LHS to RHS is unchanged upon replacing $a_1 \to a_1+b$. By symmetry, the same is also true when we replace $a_1 \to a_1+b^{-1}$. If we assume that $b^2 \in \RR \setminus \QQ$, this shows by continuity that the integral cannot depend on the real part of $a_1$ and by analyticity can hence not depend on $a_1$ at all. Again invoking continuity in $b$, the same is true for all $b>0$. Similarly, we show that the ratio of LHS and RHS is independent of all $a_i$. 

It remains to compute the constant. For this, we set $a \equiv a_5=a_6$ and consider $a \to 0$. As an additional cross check, we keep $a_1$, $a_2$, $a_3$ and $a_4$ arbitrary, but the constraint tends to $\sum_{i=1}^4 a_i=Q$.
 In this limit, the contour gets pinched from the double poles at $\pm z =a$. All poles associated to $a_j$ for $1 \le j \le 4$ stay a finite distance away from $z=0$ and are thus not relevant. The integrand becomes close to $z=0$:
 \begin{align}
     \frac{\prod_{j=1}^6 S_b(a_j \pm z)}{S_b(\pm 2 z)}\to -\prod_{j=1}^4 S_b(a_j)^2 \times \frac{z^2}{\pi^2(a-z)^2(a+z)^2}\ .
 \end{align}
 It is then simple to perform the integral over $z$, which gives for the leading behaviour of the LHS of the integral identity
 \be 
\text{LHS}=\frac{1}{4\pi a} \prod_{j=1}^4 S_b(a_j)^2\ .
 \ee
 This should be contrasted with the RHS of eq.~\eqref{eq:Euler integral formula}. $S_b(a_5+a_6)$ gives the same factor $\frac{1}{4\pi a}$ and the terms with $k=5$ or $k=6$ give the desired $\prod_{j=1}^4 S_b(a_j)^2$. The remaining factors cancel out thanks to the constraint $\sum_{j=1}^4 a_j=Q$:
 \begin{align}
    \prod_{1\le j<k\le 4}^4 S_b(a_j+a_k)=\prod_{1 \le j<k \le 3} S_b(a_j+a_k) \prod_{1 \le j<k\le 3} S_b(Q-a_j-a_k)=1\ .
 \end{align}
Thus we proved the Euler integral identity.
\end{proof}

\subsection{The \texorpdfstring{$b$}{b}-deformed hypergeometric function}
We define the $b$-deformed hypergeometric function as follows \cite{Spridonov:theta},
\be 
V_b(a_1,\dots,a_8)= \int \frac{\d z}{2i}\ \frac{\prod_{j=1}^8 S_b(a_j \pm z)}{S_b(\pm 2z)}\ . \label{eq:b-deformed hypergeometric function}
\ee
The parameters $a_1,\dots,a_8$ satisfy now the balancing condition
\be 
\sum_{j=1}^8 a_j=2Q\ .
\ee
Of course, upon setting $a_8=Q-a_7$, this reduces back to the Euler integral \eqref{eq:b-deformed Euler integral}.

\paragraph{Symmetries.} A surprising property of the hypergeometric function is its large symmetry group. Consider the double integral
\be 
I(a_j,\tilde{a}_j)=\int \frac{\d z \, \d w}{(2i)^2}\ \frac{\prod_{j=1}^4 S_b(a_j \pm z) S_b(\tilde{a}_j \pm w) S_b(A \pm z \pm w)}{S_b(\pm 2 z) S_b(\pm 2w)S_b(2A)}\ ,
\ee
subject to the constraints
\be 
2A+\sum_{j=1}^4 a_j=2A+\sum_{j=1}^4 \tilde{a}_j=Q\ .
\ee
Let us integrate out $w$ using the Euler integral identity \eqref{eq:Euler integral formula}. This leads to
\begin{align} 
I(a_j,\tilde{a}_j)&=\prod_{1 \le j \le k \le 4} S_b(\tilde{a}_j+\tilde{a}_k) \int \frac{\d z}{2i}\ \frac{\prod_{j=1}^4 S_b(a_j \pm z) S_b(\tilde{a}_j+A \pm z)}{S_b(\pm 2z)} \\
&=\prod_{1 \le j \le k \le 4} S_b(\tilde{a}_j+\tilde{a}_k) V_b(a_1,\dots,a_4,\tilde{a}_1+A,\dots,\tilde{a}_4+A)\ .
\end{align}
We could have equally well integrated out $z$ and would have found the same formula with the roles of $a_j$ and $\tilde{a}_j$ reversed.
Let us rename $a_{j+4}=\tilde{a}_j+A$. Then we have have shown the identity
\be 
 V_b(a_1,\dots,a_8)=\prod_{1 \le j \le k \le 4} S_b(a_j+a_k)\prod_{5 \le j \le k \le 8} S_b(a_j+a_k) V_b(\tilde{a}_1,\dots,\tilde{a}_8) \label{eq:E7 symmetry hypergeometric function}
\ee
with
\be 
\tilde{a}_j=\begin{cases}
    \frac{1}{2}(Q-a_{1234})+a_j\ , \quad j=1,\dots,4\ , \\
    \frac{1}{2}(Q-a_{5678})+a_j\ , \quad j=5,\dots,8\ .    
\end{cases} \label{eq:a parameter E7 transformation}
\ee
\eqref{eq:E7 symmetry hypergeometric function} means that the hypergeometric function has in fact a higher symmetry than what is visible from its definition \eqref{eq:b-deformed hypergeometric function}. 

To understand the relevant symmetry group, consider the shifted parameters $x_j=a_j-\frac{Q}{4}$, which satisfy $\sum_{j=1}^8 x_j=0$. The transformation \eqref{eq:a parameter E7 transformation} can then be written as
\be 
\boldsymbol{x} \longmapsto\boldsymbol{x}-2 (\boldsymbol{x} \cdot \boldsymbol{n}) \boldsymbol{n}
\ee
with
\be 
 \boldsymbol{n}=\tfrac{1}{2 \sqrt{2}}\begin{pmatrix}
    1 & 1 & 1 & 1 & -1 & -1 & -1 & -1
\end{pmatrix} \label{eq:exceptional vector E7}
\ee
Thus the transformation corresponds to a reflection along the plane normal to $\boldsymbol{n}$.
Exchanging $x_1$ and $x_2$ corresponds to a reflection along the plane normal to the vector
\be 
\boldsymbol{n}=\frac{1}{\sqrt{2}} \begin{pmatrix}
    1 & -1 & 0 & 0 & 0 & 0 & 0 & 0
\end{pmatrix}\ .
\ee
Together with the other exchanges of $x_j$ and $x_{j+1}$, these are precisely the simple roots of the Lie algebra $\text{E}_7$ and correspondingly the symmetry group of the hypergeometric function is the Weyl group $W(\text{E}_7)$.

We should also note the following useful corollary, which follows by repeatedly applying the $W(\mathrm{E}_7)$ symmetry. We can consider the longest word in $W(\mathrm{E}_7)$, which flips all coordinates. Thus we get the identity
\be 
V_b(a_1,\dots,a_8)=\prod_{1 \le j<k \le 8} S_b(a_j+a_k) V_b(\tfrac{Q}{2}-a_1,\dots,\tfrac{Q}{2}-a_8)\ . \label{eq:E7 symmetry total reflection}
\ee
\paragraph{Pole structure.} The $b$-deformed hypergeometric function has a large set of poles. We get a pole whenever the contour gets pinched between two poles of the integrand. This happens when
\be 
a_j+a_k=-mb-n b^{-1}
\ee
for $j \ne k$ and $m,\, n \in \ZZ_{\ge 0}$. This set of poles is invariant under the $W(\mathrm{E}_7)$ symmetry.

\subsection{Degenerations} \label{subapp:degenerations}
The integrals encountered in this paper do not directly need the Euler integral identity \eqref{eq:Euler integral formula} and the $b$-deformed hypergeometric function \eqref{eq:b-deformed hypergeometric function}, but rather special limits of those that we call degenerations.
For the hypergeometric function, we only explain the two degenerations that are relevant to our discussion of the crossing kernels, but other degenerations can also be considered. See e.g.\ \cite{VanDeBult} for a more complete discussion.

The possible degenerations of the Euler integral and the hypergeometric function $V_b$ can be worked out as follows. The idea is to set
\be 
a_i \to a_i+v_i x \label{eq:degeneration scaling}
\ee
for some vector $v$ and take $x \to +i \infty$. One can then use the asymptotics \eqref{eq:S_b asymptotics} to simplify the integrand, possibly after shifting $z$ also by a multiple of $x$.
Provided that this leads to a sensible limit, the resulting function than has a remaining symmetry given by the stabilizer group of the vector $v$.\footnote{This group may not act faithfully on the actual function.} 
\paragraph{Euler integral degenerations.} We now list a number of degenerations of the Euler integral \eqref{eq:Euler integral formula}. 
\begin{enumerate}
    \item Four term Euler integral, corresponding to the vector $v=\begin{pmatrix}
        0 & 0 & 0 & 0 & 1 & -1
    \end{pmatrix}$:
    \begin{align}
        \int\frac{\mathrm{d} z}{2i}\  \frac{\prod_{j=1}^4 S_b(a_j \pm z)}{S_b(\pm 2z)}=\frac{\prod_{1 \le j<k \le 4} S_b(a_j+a_k)}{S_b(a_{1234})}\ , \label{eq:S_b beta integral identity 4 a}
    \end{align}
    \item Three term Euler integral, corresponding to the vector $v=\begin{pmatrix}
        0 & 0 & 0 & 2 & -1 & -1
    \end{pmatrix}$:
    \begin{align}
    \int \frac{\mathrm{d} z}{2i}\  \mathrm{e}^{\pi i z^2} \frac{\prod_{j=1}^3 S_b(a_j \pm z)}{S_b(\pm 2z)}=\prod_{1 \le j<k \le 3} \mathrm{e}^{-\pi i a_j a_k}\, S_b(a_j+a_k)\ . \label{eq:S_b beta integral identity 3 a}
    \end{align}
    A similar identity with all signs in the exponents reversed also exists. It follows from taking the negative of the vector.
    \item Two term Euler integral, corresponding to the vector $v=\begin{pmatrix}
        0 & 0 & 3 & -1 & -1 & -1
    \end{pmatrix}$:
    \begin{align}
     \int \frac{\mathrm{d} z}{2i}\  \mathrm{e}^{2\pi i z^2} \frac{\prod_{j=1}^2 S_b(a_j \pm z)}{S_b(\pm 2z)}=\mathrm{e}^{\frac{\pi i}{2}\left((a_1-a_2)^2-Q(a_1+a_2)\right)} S_b(a_1+a_2)\ . \label{eq:S_b beta integral identity 2 a}
    \end{align}
    \item One term Euler integral, corresponding to the vector $v=\begin{pmatrix}
        0 & 4 & -1 & -1 & -1 & -1
    \end{pmatrix}$:
    \begin{align}
     \int\frac{\mathrm{d} z}{2i}\  \mathrm{e}^{3\pi i z^2} \frac{S_b(a \pm z)}{S_b(\pm 2z)}=\mathrm{e}^{\pi i\left(a(a-Q)-\frac{1}{12}(Q^2+1)\right)} \ . \label{eq:S_b beta integral identity 1 a}
    \end{align}
    \item Asymmetric Euler integral, corresponding to the vector $\begin{pmatrix}
        1 & 1 & 1 & -1 & -1 & -1
    \end{pmatrix}$:
        \be 
            \int \frac{\mathrm{d}z}{i}\ \prod_{j=1}^3 S_b(a_j+z) \prod_{j=4}^6 S_b(a_j-z)=\prod_{j=1}^3 \prod_{k=4}^6 S_b(a_j+a_k)\ , \qquad \sum_{j=1}^6 a_j=Q\ . \label{eq:S_b beta integral identity asymmetric}
        \ee
    \item Asymmetric 4 term Euler integral, corresponding to further degenerating the asymmetric Euler integral by the the vector $\begin{pmatrix}
        0 & 0 & 1 & 0 & 0 & -1
    \end{pmatrix}$:
    \begin{multline} 
         \int \frac{\mathrm{d}z}{i}\  \mathrm{e}^{-\pi i za_{1234}}\prod_{j=1}^2 S_b(a_j+z) \prod_{j=3}^4 S_b(a_j-z) \\
         =\frac{\mathrm{e}^{\pi i(a_1a_2-a_3a_4)} \prod_{j=1}^2 \prod_{k=3}^4 S_b(a_j+a_k)}{S_b(a_{1234})}\ . \label{eq:S_b beta integral identity asymmetric 4 terms}
    \end{multline}
    \item Asymmetric beta function 3 term identity:
    \begin{multline} 
         \int \frac{\mathrm{d}z}{i}\  \mathrm{e}^{\frac{\pi i}{2}z(z-Q+2a_{123})}  \prod_{j=1}^2 S_b(a_j-z)S_b(a_3+z)\\ 
         =\mathrm{e}^{\frac{\pi i}{2}a_3(Q-2a_{12}-a_3)-2\pi i a_1a_2} \prod_{j=1}^2 S_b(a_j+a_3)\ . \label{eq:S_b beta integral identity asymmetric 3 terms}
    \end{multline}
    \item Asymmetric beta function 2 term identity:
    \be 
        \int \frac{\mathrm{d}z}{i}\  \mathrm{e}^{2\pi i z u} S_b(a_1-z)S_b(a_2+z) =\mathrm{e}^{\pi i (a_1-a_2)u} S_b(a_{12}) S_b(\tfrac{1}{2}(Q-a_{12}) \pm u)\ , \label{eq:S_b beta integral identity asymmetric 2 terms}
    \ee
    where $u$ is an arbitrary parameter.
\end{enumerate}
\paragraph{The $W(\mathrm{D}_6)$ hypergeometric function.} We now discuss a particular degeneration of the hypergeometric function that appears as the solution for the crossing kernel on the four-punctured sphere. Consider the orbit
\be 
\{(2,-2,0,0,0,0,0,0),\ (1,1,1,1,-1,-1,-1,-1)\}+\text{perm} \label{eq:D6 orbit}
\ee
under the Weyl group $W(\mathrm{E}_7)$. It has order \num{126} and the stabilizer group of a given vector is isomorphic to the Weyl group of $\mathrm{D}_6$.

We can hence degenerate the hypergeometric function $V_b$ to a hypergeometric function $V_b^{\mathrm{D}_6}$ with $W(\mathrm{D}_6)$ symmetry. It has two essentially different integral representations, since the orbit \eqref{eq:D6 orbit} consists of permutations of two vectors. We will now discuss this degeneration in some detail. This degenerated hypergeometric function is also called Ruijsenaars' hypergeometric function \cite{Ruijsenaars1, Ruijsenaars2}.

Let us first consider the case where we replace $a_7 \mapsto a_7+x$, $a_8 \mapsto a_8 -x$ and take $x \to +i \infty$, corresponding to the first vector in the orbit. It is easy to see that the integral only receives appreciable contribution from the region where $z$ does not become asymptotically large. We may thus use the asymptotic formula \eqref{eq:S_b asymptotics} to simplify the factors $S_b(a_7+x \pm z)$ and $S_b(a_8-x \pm z)$ in the integrand. This leads to an oscillating prefactor, which however can be cancelled as follows:
\be 
\int \frac{\d z}{2i}\  \frac{\prod_{j=1}^6 S_b(a_j \pm z)}{S_b(\pm 2z)}=\lim_{x \to +i \infty} \mathrm{e}^{\pi i (a_{78}-Q)(2x+a_{7|8})} V_b(a_1,\dots,a_6,a_7+x,a_8-x)\ .
\label{eq:hypergeometric function D6 degeneration symmetric}
\ee
As in the main text, we use the notation \eqref{eq:definition pI|J}.
The LHS is the same as the $b$-deformed Euler integral \eqref{eq:b-deformed Euler integral}, but without any constraint on $\sum_{j=1}^6 a_j$.

We get a qualitatively different degeneration when we shift $a_j \mapsto a_j+x$ for $j=1,\dots,4$ and $a_j \mapsto a_j-x$ for $j=5,\dots,8$ and then take the limit $x \to +i \infty$, corresponding to the second vector in the orbit \eqref{eq:D6 orbit}. The integral is now strongly dominated from the regions $z \sim x$ and $z \sim -x$. Since the integral is symmetric in $z \mapsto -z$, the contributions from both regions are equal. We thus have for large $\Im(x)$:
\begin{multline}
    \int \frac{\d z}{2i}\ \frac{\prod_{j=1}^4 S_b(a_j +x \pm z) \prod_{j=5}^8 S_b(a_j-x \pm z)}{S_b(\pm 2z)}\\
    =
    2\int_{z \sim 0} \frac{\d z}{2i}\ \frac{\prod_{j=1}^4 S_b(a_j +x \pm (z-x)) \prod_{j=5}^8 S_b(a_j-x \pm (z-x))}{S_b(\pm 2(z-x))}\ .
\end{multline}
We can then insert the asymptotic expansion of the double sine function \eqref{eq:S_b asymptotics}. This yields
\begin{multline}
    \int \frac{\d z}{i}\  \prod_{j=1}^4 S_b(a_j+z) \prod_{j=5}^8 S_b(a_j-z)=\lim_{x \to +i \infty} \mathrm{e}^{\frac{\pi i}{2}(\sum_{j=1}^4 a_j(a_j-Q)-\sum_{j=5}^8 a_j(a_j-Q))-2\pi i Q x} \\
    \times V_b(a_1+x,\dots,a_4+x,a_5-x,\dots,a_8-x)\ .\label{eq:hypergeometric function D6 degeneration asymmetric}
\end{multline}
Here, the parameters $a_j$ are still constrained to obey $\sum_{j=1}^8 a_j=2Q$, but of course we may shift the first four $a_j$'s in an opposite way to the last four $a_j$'s and keep the definition invariant.
As a consequence of the $W(\text{E}_7)$ symmetry of the hypergeometric function, the left-hand sides of \eqref{eq:hypergeometric function D6 degeneration symmetric} and \eqref{eq:hypergeometric function D6 degeneration asymmetric} are related to each other and both enjoy $W(\mathrm{D}_6)$ symmetry which is in both cases larger than the obvious $\mathrm{S}_6$ and $\mathrm{S}_4 \times \mathrm{S}_4$ symmetry, respectively. The precise relation is
    \begin{multline} 
        \int \frac{\d z}{i}\  \prod_{j=1}^4 S_b(a_j+z) \prod_{j=5}^8 S_b(a_j-z)\\
        =\prod_{j=1}^3 S_b(a_j+a_8) S_b(a_{j+4}+a_4) \int \frac{\d z}{2i} \frac{\prod_{j=1}^6 S_b(\tilde{a}_j \pm z)}{S_b(\pm 2z)}\ . \label{eq:alternative representation Gb}
    \end{multline}
    with
    \be 
        \tilde{a}_j=a_j+\frac{Q-a_{1238}}{2}\ , \qquad \tilde{a}_{j+3}=a_{j+4}+\frac{Q-a_{4567}}{2}\ , \quad j=1,\,2,\, 3\ ,
    \ee
    and $\sum_{j=1}^8 a_j=2Q$.

\paragraph{The $\mathrm{S}_4 \times \ZZ_2$ hypergeometric function.} There is another degeneration of the hypergeometric function that appears in the modular crossing kernel. It corresponds to the orbit
\begin{multline} 
\{\pm (6,-2,-2,-2,0,0,0,0),\ \pm (3,3,1,1,-1,-1,-1,-5),\\
\pm (3,3,1,1,1,-3,-3,-3), \ \pm (2,1,1,0,0,-1,-1,-2)\}+\text{perm}\ .
\end{multline}
The orbit has length \num{10080}. There is an obvious $\mathrm{S}_3 \times \mathrm{S}_4$ subgroup of the symmetry group. It acts by permuting the three $-2$'s of the first vector and the four $0$'s of the first vector. The first vector is also preserved by the reflection along the additional vector \eqref{eq:exceptional vector E7} that generates the full $W(\mathrm{E}_7)$ and hence a given vector in the orbit is preserved by the symmetry group $\mathrm{S}_3 \times \mathrm{S}_4 \times \ZZ_2$. The $\mathrm{S}_3$ acts trivially and thus the actual symmetry group will just be $\mathrm{S}_4 \times \ZZ_2$, which explains our naming conventions.

There are three essentially different formulas for the corresponding hypergeometric function, corresponding to the degenerations along the different looking vectors in the orbit. The second and third vector lead to identical formulas. The three formulas are
\begin{subequations}
\begin{align}
    &\int \frac{\d z}{2i} \ \mathrm{e}^{-2\pi i z^2}\,  \frac{\prod_{j=1}^4 S_b(a_j \pm z)}{S_b(\pm 2z)}\nonumber\\
    &\quad=\lim_{x \to i \infty} \mathrm{e}^{\frac{\pi i}{3}(Q^2+1)+3\pi i x(x-Q)-\pi i(3x+a_{1234}-Q)(3x+a_{1234}-2Q)} \nonumber\\
    &\quad\quad\times V_b(a_1,a_2,a_3,a_4,x,x,x,2Q-a_{1234}-3x)\ , \label{eq:S4Z2 hypergeometric function 1st form}\\
    &\int \frac{\d z}{i}\ \mathrm{e}^{\pi i uz} \prod_{j=1}^2 S_b(a_j+z) S_b(a_{j+2}-z) \nonumber\\
    &\quad=\mathrm{e}^{\frac{\pi i}{2} (\sum_{j=1}^2 a_j(a_j-Q)-\sum_{j=3}^4 a_j(a_j-Q))-\frac{\pi i}{4} (a_{1234}+u)(a_{1234}+u-2Q)-\pi i (a_{1234}+2Q+u)x} \nonumber\\
    &\quad\quad\times V_b(a_1+x,a_2+x,a_3-x,a_4-x,2x,\tfrac{1}{2}(2Q-u-a_{1234})-2x,\nonumber\\
    &\hspace{9.2cm} 0,\tfrac{1}{2}(2Q-a_{1234}+u))\ , \label{eq:S4Z2 hypergeometric function 2nd form}\\
    &\int \frac{\d z}{i} \ \mathrm{e}^{\frac{\pi i}{2}(2a_{12345}-Q-z)z} \prod_{j=1}^2 S_b(a_j+z)\prod_{j=3}^5 S_b(a_j-z)\nonumber \\
    &\quad=\mathrm{e}^{\frac{\pi}{12}(Q^2+1)+\frac{\pi i}{2}(\sum_{j=1}^2 a_j(a_j-Q)-\sum_{j=3}^5 a_j(a_j-Q))-2(2Q-a_{12345})(Q-a_{12345})-2\pi ix(4x-3Q+4a_{12345})}\nonumber \\
    &\quad\quad\ \times V_b(a_1+x,a_2+x,a_3-x,a_4-x,a_5-x,3x,3x,2Q-a_{12345}-5x)\ . \label{eq:S4Z2 hypergeometric function 3rd form}%
\end{align}
\end{subequations}
The choice of $\mathrm{e}^{-2\pi i z^2}$ in the first integrand was of course arbitrary, we can relate it to the opposite choice by using the total reflection formula \eqref{eq:E7 symmetry total reflection} of the hypergeometric function. This gives
\begin{multline} 
\int \frac{\d z}{2i}\  \mathrm{e}^{-2\pi i z^2} \frac{\prod_{j=1}^4 S_b(a_j \pm z)}{S_b(\pm 2z)} 
=\mathrm{e}^{\frac{\pi i}{2} (a_{1234}(a_{1234}-Q)+Q^2)-\pi i \sum_j a_j^2} \prod_{1 \le j<k \le 4} S_b(a_{jk})\\
\times \int \frac{\d z}{2i} \ \mathrm{e}^{2\pi i z^2} \frac{\prod_{j=1}^4 S_b(\frac{Q}{2}-a_j \pm z)}{S_b(\pm 2z)}\ .\label{eq:S4Z2 hypergeometric function opposite sign exponent}
\end{multline}

The first form \eqref{eq:S4Z2 hypergeometric function 1st form} makes the $\mathrm{S}_4$ symmetry obvious, but not the $\ZZ_2$ symmetry. It follows from the $W(\mathrm{E}_7)$ symmetry of the hypergeometric function and reads
\begin{multline} 
    \int \frac{\d z}{2i} \ \mathrm{e}^{-2\pi i z^2}\,  \frac{\prod_{j=1}^4 S_b(a_j \pm z)}{S_b(\pm 2z)} 
    =\mathrm{e}^{\frac{\pi iQ}{2}(a_{1234}-Q)} \prod_{1 \le j<k \le 4} S_b(a_{jk}) \\
    \times \int \frac{\d z}{2i} \ \mathrm{e}^{-2\pi i z^2}\,  \frac{\prod_{j=1}^4 S_b(a_j+\frac{1}{2}(Q-a_{1234}) \pm z)}{S_b(\pm 2z)} \label{eq:S4Z2 hypergeometric function Z2 symmetry}\ .
\end{multline}
The relation between these different forms also follows from the $W(\mathrm{E}_7)$ symmetry and we have
\begin{subequations}
\begin{align}
     &\int \frac{\d z}{2i} \ \mathrm{e}^{-2\pi i z^2}\,  \frac{\prod_{j=1}^4 S_b(a_j \pm z)}{S_b(\pm 2z)} \nonumber \\
     &\ =\mathrm{e}^{\pi i(a_1a_2+a_3a_4)} S_b(a_{12})S_b(a_{34}) \int  \frac{\d z}{i}\ \mathrm{e}^{\pi i a_{12|34}z} \prod_{j=1}^2 S_b(a_j+z) S_b(a_{j+2}-z)\ , \label{eq:S4Z2 hypergeometric function relation first form second form} \\
     &\ =\mathrm{e}^{\frac{\pi i}{2} a_1(Q-a_1)}\!\!\!\! \prod_{2 \le j<k \le 4}\!\!\!\! \!S_b(a_{jk}) \int \frac{\d z}{i} \ \mathrm{e}^{\frac{\pi i}{2}(2a_1+Q-z)z}\prod_{j=1}^2 S_b(\tilde{a}_j+z)\prod_{j=3}^5 S_b(\tilde{a}_j-z)\ . \label{eq:S4Z2 hypergeometric function relation first form third form}%
\end{align}
\end{subequations}
In the last equation we have
\be 
\tilde{a}_1=a_1\ , \qquad \tilde{a}_2=Q-a_{234}\ , \qquad \tilde{a}_j=a_{j-1}\,, \ j \ge 3\ .
\ee

\bibliographystyle{JHEP}
\bibliography{bib}
\end{document}